\documentclass[longauth]{aaEC}

\usepackage{graphicx}
\usepackage{natbib}
\usepackage{scalerel}

\usepackage[table]{xcolor}
\usepackage{tabularx}
\usepackage[titles]{tocloft}

\usepackage{txfonts}
\usepackage{textcomp}
\usepackage[pdfencoding=auto,psdextra]{hyperref}
\hypersetup{
    colorlinks=true,
    linkcolor=blue,
    filecolor=magenta,      
    urlcolor=blue,
    citecolor=blue
}
\usepackage[nameinlink,capitalise]{cleveref}
\crefname{section}{Sect.}{Sects.}
\Crefname{section}{Section}{Sections}
\crefname{figure}{Fig.}{Figs.}
\Crefname{figure}{Figure}{Figures}
\crefname{equation}{Eq.}{Eqs.}
\Crefname{equation}{Equation}{Equations}

\newcommand{\Sersic}{S{\'e}rsic}

\newcommand{\oiii}[1]{[\ion{O}{iii}]#1}
\newcommand{\oii}[1]{[\ion{O}{ii}]#1}
\newcommand{\ha}{\ensuremath{\mathrm{H}\alpha}}
\newcommand{\hb}{\ensuremath{\mathrm{H}\beta}}
\newcommand{\nev}[1]{[\textrm{Ne}\textsc{\,v}]#1}
\newcommand{\nii}[1]{[\ion{N}{ii}]#1}

\newcommand{\hi}{H\,\textsc{i}}

\newcommand{\Pb}{\ensuremath{\mathrm{Pa}\beta}}

\makeatletter
\renewcommand*\aa@pageof{, page \thepage{} of \pageref*{LastPage}}
\makeatother

\usepackage[utf8]{inputenc}

\usepackage[switch, modulo]{lineno}

\usepackage{orcidlink} 
\newcommand{\orcid}[1]{\orcidlink{#1}}

\usepackage{euclid}
\usepackage[nopostdot, nonumberlist, acronym]{glossaries}
\makeglossaries
\glsdisablehyper

\newacronym{2dFGRS}{2dFGRS}{2-degree Field Galaxy Redshift Survey}
\newacronym{2PCF}{2PCF}{two-point correlation function}
\newacronym{3PCF}{3PCF}{three-point correlation function}
\newacronym{AA}{AA}{alpha angle}
\newacronym{ACT}{ACT}{Atacama Cosmology Telescope}
\newacronym{ADS}{ADS}{Airbus Defence and Space}
\newacronym[longplural={active galactic nuclei}]{AGN}{AGN}{active galactic nucleus}
\newacronym{AM}{AM}{abundance matching}
\newacronym{AOCS}{AOCS}{attitude and orbit-control system}
\newacronym{AP}{AP}{Alcock-Paczynski}
\newacronym{APE}{APE}{absolute pointing error}
\newacronym{BAO}{BAO}{baryon acoustic oscillation}
\newacronym{BBN}{BBN}{big bang nucleosynthesis}
\newacronym{BFE}{BFE}{brighter-fatter effect}
\newacronym{BG}{BG}{blue grism}
\newacronym{BGS}{BGS}{Bright Galaxy Sample}
\newacronym{BNT}{BNT}{Bernardeau--Nishimichi--Taruya}
\newacronym{BOSS}{BOSS}{Baryon Oscillation Spectroscopic Survey}
\newacronym{BPT}{BPT}{Baldwin--Phillips--Terlevich}
\newacronym{CAMB}{CAMB}{Code for Anisotropies in the Microwave Background}
\newacronym{CAS}{CAS}{concentration, asymmetry and smoothness}
\newacronym{CCD}{CCD}{charge-coupled device}
\newacronym{CDPU}{CDPU}{Control and Data Processing Unit}
\newacronym{CDS}{CDS}{Cosmic Dawn Survey}
\newacronym{CFHT}{CFHT}{Canada-France-Hawaii Telescope}
\newacronym{CIB}{CIB}{cosmic infrared background}
\newacronym{CXB}{CXB}{cosmic X-ray background}
\newacronym{CLASS}{CLASS}{Cosmic Linear Anisotropy Solving System}
\newacronym{CLOE}{CLOE}{Cosmology Likelihood for Observables in Euclid}
\newacronym{CMB}{CMB}{cosmic microwave background}
\newacronym{CMD}{CMD}{colour-magnitude diagram}
\newacronym{COSEBIs}{COSEBIs}{Complete Orthogonal Sets of E/B-Integrals}
\newacronym{CPC}{CPC}{completeness and purity calibration}
\newacronym{CPL}{CPL}{Chevallier--Polarski--Linder}
\newacronym{CPU}{CPU}{central processing unit}
\newacronym{CTI}{CTI}{charge-transfer inefficiency}
\newacronym{CU}{CU}{calibration unit}
\newacronym{DECam}{DECam}{Dark Energy Camera}
\newacronym{DES}{DES}{Dark Energy Survey}
\newacronym{DESI}{DESI}{Dark Energy Spectroscopic Instrument experiment}
\newacronym{DI}{DI}{direct integration}
\newacronym{DPU}{DPU}{data-processing unit}
\newacronym{DPS}{DPS}{data-processing system}
\newacronym{DR}{DR}{data release}
\newacronym{DR1}{DR1}{first data release}
\newacronym{DTCP}{DTCP}{daily telemetry communication period}
\newacronym{DUNE}{DUNE}{Dark Universe Explorer}
\newacronym{EAF}{EAF}{Euclid Auxiliary Field}
\newacronym{EC}{EC}{Euclid Consortium}
\newacronym{EDF}{EDF}{Euclid Deep Field}
\newacronym{EDS}{EDS}{Euclid Deep Survey}
\newacronym{EDF-S}{EDF-S}{Euclid Deep Field South}
\newacronym{EDF-N}{EDF-N}{Euclid Deep Field North}
\newacronym{EDF-F}{EDF-F}{Euclid Deep Field Fornax}
\newacronym{EFS}{EFS}{Euclid Flagship Simulation}
\newacronym{EFT}{EFT}{effective field theory}
\newacronym{EGC}{EGC}{extragalactic globular cluster}
\newacronym{ELG}{ELG}{emission line galaxy}
\newacronym{EMC}{EMC}{electromagnetic compatibility}
\newacronym{EMDS}{EMDS}{Euclid Medium-Deep Survey}
\newacronym{EOR}{EoR}{epoch of reionisation}
\newacronym{ERO}{ERO}{Early Release Observations}
\newacronym{eROSITA}{eROSITA}{extended ROentgen Survey with and Imaging Telescope Array}
\newacronym{ESA}{ESA}{European Space Agency}
\newacronym{ESAC}{ESAC}{European Space Astronomy Centre}
\newacronym{ESOC}{ESOC}{European Science Operations Centre}
\newacronym{ESOP}{ESOP}{early science operations}
\newacronym{EUDF}{EUDF}{Euclid Ultra-Deep Field}
\newacronym{EW}{EW}{equivalent width}
\newacronym{EWS}{EWS}{Euclid Wide Survey}
\newacronym{FFP}{FFP}{free-floating planet}
\newacronym{FFT}{FFT}{fast Fourier transform}
\newacronym{FGS}{FGS}{fine-guidance sensor}
\newacronym{FITS}{FITS}{flexible image transport system}
\newacronym{FKP}{FKP}{Feldman--Kaiser--Peacock}
\newacronym{FOM}{FoM}{figure of merit}
\newacronym{FOV}{FoV}{field of view}
\newacronym{FPA}{FPA}{focal-plane array}
\newacronym{FWHM}{FWHM}{full width at half maximum}
\newacronym{G3L}{G3L}{galaxy-galaxy-galaxy lensing}
\newacronym{GBTDS}{GBTDS}{Galactic Bulge Time Domain Survey}
\newacronym{GC}{GC}{globular cluster}
\newacronym{GF}{GF}{Gaussian fit}
\newacronym{GGL}{GGL}{galaxy-galaxy lensing}
\newacronym{GP}{GP}{Gaussian process}
\newacronym{GR}{GR}{general relativity}
\newacronym{HGA}{HGA}{high-gain antenna}
\newacronym{HK}{HK}{housekeeping}
\newacronym{HOD}{HOD}{halo-occupation distribution}
\newacronym{HOS}{HOS}{higher-order statistics}
\newacronym{HSC}{HSC}{Hyper Suprime-Cam}
\newacronym{HSC-SSP}{HSC-SSSP}{Hyper Suprime-Cam Subaru Strategic Program}
\newacronym{HST}{HST}{\HST}
\newacronym{IA}{IA}{intrinsic alignment}
\newacronym{ICRS}{ICRS}{International Celestial Reference System}
\newacronym{IGM}{IGM}{intergalactic medium}
\newacronym{IOT}{IOT}{Instrument Operation Team}
\newacronym{ISCS}{ISCS}{IRAC Shallow Cluster Survey}
\newacronym{ISW}{ISW}{integrated Sachs--Wolfe}
\newacronym{JWST}{JWST}{James Webb Space Telescope}
\newacronym{KiDS}{KiDS}{Kilo-Degree Survey}
\newacronym{KS}{KS}{Kaiser--Squires}
\newacronym{KSB}{KSB}{Kaiser--Squires--Broadhurst}
\newacronym{kSZ}{kSZ}{kinetic Sunyaev--Zeldovich}
\newacronym{LBT}{LBT}{Large Binocular Telescope}
\newacronym{LED}{LED}{light-emitting diode}
\newacronym{LF}{LF}{luminosity function}
\newacronym{LOS}{LoS}{line of sight}
\newacronym{LRG}{LRG}{luminous red galaxy}
\newacronym{LSB}{LSB}{low-surface brightness}
\newacronym{LSS}{LSS}{large-scale structure}
\newacronym{LSST}{LSST}{Legacy Survey of Space and Time}
\newacronym{M2M}{M2M}{M2 mechanism}
\newacronym{MACC}{MACC}{multi-accumulate}
\newacronym{MAMBO}{MAMBO}{Mocks with Abundance Matching in BOlogna}
\newacronym{MCMC}{MCMC}{Markov chain Monte Carlo}
\newacronym{MLI}{MLI}{multi-layer thermal insulation}
\newacronym{MOC}{MOC}{mission operation centre}
\newacronym{MS}{MS}{main sequence}
\newacronym{NEP}{NEP}{north ecliptic pole}
\newacronym{NFW}{NFW}{Navarro--Frenk--White}
\newacronym{NIR}{NIR}{near-infrared}
\newacronym{NISP}{NISP}{Near Infrared Spectrometer and Photometer}
\newacronym{NLA}{NLA}{nonlinear linear alignment}
\newacronym{OU}{OU}{Organisation Unit}
\newacronym{Pan-STARRS}{Pan-STARRS}{Panchromatic Survey Telescope and Rapid Response System}
\newacronym{PDC}{PDC}{phase-diversity calibration}
\newacronym{PDF}{PDF}{probability density function}
\newacronym{PF-SPE}{PF-SPE}{Spectroscopy Processing Function}
\newacronym{PLM}{PLM}{payload module}
\newacronym{pRF}{pRF}{probabilistic random forest}
\newacronym{PRNU}{PRNU}{pixel-response non-uniformity}
\newacronym{PSF}{PSF}{point spread function}
\newacronym{PTC}{PTC}{photon-transfer curve}
\newacronym{PV}{PV}{performance-verification}
\newacronym{QE}{QE}{quantum efficiency}
\newacronym{QSO}{QSO}{quasi-stellar object}
\newacronym{RG}{RG}{red grism}
\newacronym{ROI}{RoI}{region of interest}
\newacronym{RMS}{RMS}{root mean square}
\newacronym{ROS}{ROS}{reference observing sequence}
\newacronym{RPE}{RPE}{relative pointing error}
\newacronym{RSD}{RSD}{redshift-space distortion}
\newacronym{RSU}{RSU}{readout shutter unit}
\newacronym{SAA}{SAA}{Solar aspect angle}
\newacronym{SBF}{SBF}{surface brightness 
fluctuation}
\newacronym{SDC}{SDC}{Science Data Centre}
\newacronym{SDSS}{SDSS}{Sloan Digital Sky Survey}
\newacronym{SED}{SED}{spectral energy distribution}
\newacronym{SEF}{SEF}{single-epoch frame}
\newacronym{SEP}{SEP}{South Ecliptic Pole}
\newacronym{SFE}{SFE}{surface figure error}
\newacronym{SFR}{SFR}{star-formation rate}
\newacronym{SGS}{SGS}{Science Ground Segment}
\newacronym{SHMR}{SHMR}{stellar mass-halo mass relation}
\newacronym{SiC}{SiC}{silicon carbide}
\newacronym{SLICS}{SLICS}{Scinet LIghtCone Simulations}
\newacronym{SMBH}{SMBH}{supermassive black hole}
\newacronym{SMF}{SMF}{stellar mass function}
\newacronym{S/N}{S/N}{signal-to-noise ratio}
\newacronym{SNe}{SNe}{supernovae}
\newacronym{SO}{SO}{Simons Observatory}
\newacronym{SOC}{SOC}{Science Operations Centre}
\newacronym{SOM}{SOM}{self-organising map}
\newacronym{SPACE}{SPACE}{Spectroscopic All-Sky Cosmic Explorer}
\newacronym{SPT}{SPT}{South Pole Telescope}
\newacronym{SPV}{SPV}{Science Performance Verification}
\newacronym{SSO}{SSO}{Solar System object}
\newacronym{STOP}{STOP}{structural thermal optical performance}
\newacronym{STR}{STR}{star tracker}
\newacronym{SUDARE}{SUDARE}{SUpernovae Diversity And Rate Evolution}
\newacronym{SVM}{SVM}{service module}
\newacronym{SZ}{SZ}{Sunyaev--Zeldovich}
\newacronym{TCM}{TCM}{transfer correction manoeuvre}
\newacronym{TDE}{TDE}{tidal disruption event}
\newacronym{tSZ}{tSZ}{thermal Sunyaev--Zeldovich}
\newacronym{UCD}{UCD}{ultra-cool dwarf}
\newacronym{UNIONS}{UNIONS}{Ultraviolet Near Infrared Optical Northern Survey}
\newacronym{VIPERS}{VIPERS}{VIMOS Public Extragalactic Redshift Survey}
\newacronym{VIS}{VIS}{visible imaging instrument}
\newacronym{VISTA}{VISTA}{Visible and Infrared Survey Telescope for Astronomy}
\newacronym{VLT}{VLT}{Very Large Telescope}
\newacronym{WHIGS}{WHIGS}{Waterloo-Hawaii-IfA $g$-band Survey}
\newacronym{WISHES}{WISHES}{Wide Imaging with Subaru-Hyper Suprime-Cam Euclid Sky}
\newacronym{WL}{WL}{weak lensing}
\newacronym{zPDF}{$z$PDF}{redshift probability density function}
\newacronym{ZP}{ZP}{zero point}
\newacronym{NIRCAM}{NIRCAM}{Near-InfraRed Camera}
\newacronym{WFC3}{WFC3}{Wide Field Camera 3}
\newacronym{ALMA}{ALMA}{Atacama Large Millimeter/Sub-millimetre Array}

\definecolor{Orange}{rgb}{.8,0.4,0.1}

\begin{document}
\include{acronyms}

%
%
   \title{\Euclid. I. Overview of the \Euclid\ mission\thanks{Dedicated to our friends and colleagues Olivier Le F{\`e}vre (1960--2020) and Nick Kaiser (1954--2023), who contributed so much to the \Euclid mission and its underlying science.}}


\maxdeadcycles=500 

		   
\author{Euclid Collaboration:\thanks{\email{mellier@iap.fr}}   Y.~Mellier\inst{\ref{aff1},\ref{aff2}}
\and Abdurro'uf\orcid{0000-0002-5258-8761}\inst{\ref{aff3}}
\and J.~A.~Acevedo~Barroso\orcid{0000-0002-9654-1711}\inst{\ref{aff4}}
\and A.~Ach\'ucarro\inst{\ref{aff5},\ref{aff6}}
\and J.~Adamek\orcid{0000-0002-0723-6740}\inst{\ref{aff7}}
\and R.~Adam\orcid{0009-0000-5380-1109}\inst{\ref{aff8}}
\and G.~E.~Addison\orcid{0000-0002-2147-2248}\inst{\ref{aff3}}
\and N.~Aghanim\orcid{0000-0002-6688-8992}\inst{\ref{aff9}}
\and M.~Aguena\inst{\ref{aff10}}
\and V.~Ajani\orcid{0000-0001-9442-2527}\inst{\ref{aff11},\ref{aff12},\ref{aff13}}
\and Y.~Akrami\orcid{0000-0002-2407-7956}\inst{\ref{aff14},\ref{aff15}}
\and A.~Al-Bahlawan\inst{\ref{aff16}}
\and A.~Alavi\orcid{0000-0002-8630-6435}\inst{\ref{aff17}}
\and I.~S.~Albuquerque\orcid{0000-0002-5369-1226}\inst{\ref{aff18}}
\and G.~Alestas\orcid{0000-0003-1790-4914}\inst{\ref{aff14}}
\and G.~Alguero\inst{\ref{aff19}}
\and A.~Allaoui\inst{\ref{aff20}}
\and S.~W.~Allen\orcid{0000-0003-0667-5941}\inst{\ref{aff21},\ref{aff22},\ref{aff23}}
\and V.~Allevato\orcid{0000-0001-7232-5152}\inst{\ref{aff24}}
\and A.~V.~Alonso-Tetilla\orcid{0000-0002-6916-9133}\inst{\ref{aff25}}
\and B.~Altieri\orcid{0000-0003-3936-0284}\inst{\ref{aff26}}
\and A.~Alvarez-Candal\orcid{0000-0002-5045-9675}\inst{\ref{aff27},\ref{aff28}}
\and S.~Alvi\orcid{0000-0001-5779-8568}\inst{\ref{aff29}}
\and A.~Amara\inst{\ref{aff30}}
\and L.~Amendola\orcid{0000-0002-0835-233X}\inst{\ref{aff31}}
\and J.~Amiaux\inst{\ref{aff11}}
\and I.~T.~Andika\orcid{0000-0001-6102-9526}\inst{\ref{aff32},\ref{aff33}}
\and S.~Andreon\orcid{0000-0002-2041-8784}\inst{\ref{aff34}}
\and A.~Andrews\inst{\ref{aff35}}
\and G.~Angora\orcid{0000-0002-0316-6562}\inst{\ref{aff24},\ref{aff29}}
\and R.~E.~Angulo\orcid{0000-0003-2953-3970}\inst{\ref{aff36},\ref{aff37}}
\and F.~Annibali\inst{\ref{aff35}}
\and A.~Anselmi\orcid{0000-0001-9897-9779}\inst{\ref{aff38}}
\and S.~Anselmi\orcid{0000-0002-3579-9583}\inst{\ref{aff39},\ref{aff40},\ref{aff41}}
\and S.~Arcari\orcid{0000-0002-0551-1315}\inst{\ref{aff29},\ref{aff42}}
\and M.~Archidiacono\orcid{0000-0003-4952-9012}\inst{\ref{aff43},\ref{aff44}}
\and G.~Aric\`o\orcid{0000-0002-2802-2928}\inst{\ref{aff7}}
\and M.~Arnaud\inst{\ref{aff45},\ref{aff46}}
\and S.~Arnouts\inst{\ref{aff20}}
\and M.~Asgari\orcid{0000-0002-3064-083X}\inst{\ref{aff47}}
\and J.~Asorey\orcid{0000-0002-6211-499X}\inst{\ref{aff48}}
\and L.~Atayde\orcid{0000-0001-6373-9193}\inst{\ref{aff18}}
\and H.~Atek\orcid{0000-0002-7570-0824}\inst{\ref{aff2}}
\and F.~Atrio-Barandela\orcid{0000-0002-2130-2513}\inst{\ref{aff49}}
\and M.~Aubert\inst{\ref{aff50},\ref{aff51}}
\and E.~Aubourg\orcid{0000-0002-5592-023X}\inst{\ref{aff10},\ref{aff45}}
\and T.~Auphan\orcid{0009-0008-9988-3646}\inst{\ref{aff52}}
\and N.~Auricchio\orcid{0000-0003-4444-8651}\inst{\ref{aff35}}
\and B.~Aussel\orcid{0000-0003-2592-6806}\inst{\ref{aff53}}
\and H.~Aussel\orcid{0000-0002-1371-5705}\inst{\ref{aff11}}
\and P.~P.~Avelino\orcid{0000-0002-1440-6963}\inst{\ref{aff54},\ref{aff55}}
\and A.~Avgoustidis\inst{\ref{aff56}}
\and S.~Avila\orcid{0000-0001-5043-3662}\inst{\ref{aff57}}
\and S.~Awan\inst{\ref{aff16}}
\and R.~Azzollini\orcid{0000-0002-0438-0886}\inst{\ref{aff16}}
\and C.~Baccigalupi\orcid{0000-0002-8211-1630}\inst{\ref{aff58},\ref{aff59},\ref{aff60},\ref{aff61}}
\and E.~Bachelet\orcid{0000-0002-6578-5078}\inst{\ref{aff62}}
\and D.~Bacon\orcid{0000-0002-2562-8537}\inst{\ref{aff63}}
\and M.~Baes\orcid{0000-0002-3930-2757}\inst{\ref{aff64}}
\and M.~B.~Bagley\orcid{0000-0002-9921-9218}\inst{\ref{aff65}}
\and B.~Bahr-Kalus\orcid{0000-0002-4578-4019}\inst{\ref{aff51},\ref{aff66}}
\and A.~Balaguera-Antolinez\orcid{0000-0001-5028-3035}\inst{\ref{aff67},\ref{aff68}}
\and E.~Balbinot\orcid{0000-0002-1322-3153}\inst{\ref{aff69},\ref{aff70}}
\and M.~Balcells\orcid{0000-0002-3935-9235}\inst{\ref{aff67},\ref{aff68},\ref{aff71}}
\and M.~Baldi\orcid{0000-0003-4145-1943}\inst{\ref{aff72},\ref{aff35},\ref{aff73}}
\and I.~Baldry\inst{\ref{aff74}}
\and A.~Balestra\orcid{0000-0002-6967-261X}\inst{\ref{aff75}}
\and M.~Ballardini\orcid{0000-0003-4481-3559}\inst{\ref{aff29},\ref{aff35},\ref{aff42}}
\and O.~Ballester\orcid{0000-0002-7126-5300}\inst{\ref{aff57}}
\and M.~Balogh\orcid{0000-0003-4849-9536}\inst{\ref{aff76},\ref{aff77}}
\and E.~Ba\~nados\orcid{0000-0002-2931-7824}\inst{\ref{aff78}}
\and R.~Barbier\inst{\ref{aff51}}
\and S.~Bardelli\orcid{0000-0002-8900-0298}\inst{\ref{aff35}}
\and M.~Baron\inst{\ref{aff51}}
\and T.~Barreiro\orcid{0000-0001-8542-2066}\inst{\ref{aff18},\ref{aff79}}
\and R.~Barrena\orcid{0000-0003-4969-232X}\inst{\ref{aff80}}
\and J.-C.~Barriere\inst{\ref{aff81}}
\and B.~J.~Barros\orcid{0000-0002-2335-0703}\inst{\ref{aff18}}
\and A.~Barthelemy\orcid{0000-0003-1060-3959}\inst{\ref{aff82}}
\and N.~Bartolo\inst{\ref{aff40},\ref{aff39},\ref{aff75}}
\and A.~Basset\inst{\ref{aff83}}
\and P.~Battaglia\orcid{0000-0002-7337-5909}\inst{\ref{aff35}}
\and A.~J.~Battisti\orcid{0000-0003-4569-2285}\inst{\ref{aff84},\ref{aff85},\ref{aff86}}
\and C.~M.~Baugh\orcid{0000-0002-9935-9755}\inst{\ref{aff87}}
\and L.~Baumont\orcid{0000-0002-1518-0150}\inst{\ref{aff11}}
\and L.~Bazzanini\orcid{0000-0003-0727-0137}\inst{\ref{aff29},\ref{aff35}}
\and J.-P.~Beaulieu\inst{\ref{aff1},\ref{aff88}}
\and V.~Beckmann\orcid{0000-0002-2778-8569}\inst{\ref{aff89}}
\and A.~N.~Belikov\inst{\ref{aff69},\ref{aff90}}
\and J.~Bel\inst{\ref{aff91}}
\and F.~Bellagamba\inst{\ref{aff72},\ref{aff35}}
\and M.~Bella\orcid{0000-0002-6406-4789}\inst{\ref{aff92}}
\and E.~Bellini\inst{\ref{aff61},\ref{aff58},\ref{aff59},\ref{aff60}}
\and K.~Benabed\inst{\ref{aff2}}
\and R.~Bender\orcid{0000-0001-7179-0626}\inst{\ref{aff93},\ref{aff82}}
\and G.~Benevento\orcid{0000-0002-6999-2429}\inst{\ref{aff94}}
\and C.~L.~Bennett\orcid{0000-0001-8839-7206}\inst{\ref{aff3}}
\and K.~Benson\inst{\ref{aff16}}
\and P.~Bergamini\orcid{0000-0003-1383-9414}\inst{\ref{aff43},\ref{aff35}}
\and J.~R.~Bermejo-Climent\inst{\ref{aff67},\ref{aff68}}
\and F.~Bernardeau\inst{\ref{aff95},\ref{aff2}}
\and D.~Bertacca\orcid{0000-0002-2490-7139}\inst{\ref{aff40},\ref{aff75},\ref{aff39}}
\and M.~Berthe\inst{\ref{aff11}}
\and J.~Berthier\orcid{0000-0003-1846-6485}\inst{\ref{aff96}}
\and M.~Bethermin\orcid{0000-0002-3915-2015}\inst{\ref{aff97},\ref{aff20}}
\and F.~Beutler\orcid{0000-0003-0467-5438}\inst{\ref{aff98}}
\and C.~Bevillon\inst{\ref{aff99}}
\and S.~Bhargava\orcid{0000-0003-3851-7219}\inst{\ref{aff100}}
\and R.~Bhatawdekar\orcid{0000-0003-0883-2226}\inst{\ref{aff26}}
\and D.~Bianchi\inst{\ref{aff43}}
\and L.~Bisigello\orcid{0000-0003-0492-4924}\inst{\ref{aff101},\ref{aff40}}
\and A.~Biviano\orcid{0000-0002-0857-0732}\inst{\ref{aff59},\ref{aff58}}
\and R.~P.~Blake\inst{\ref{aff98}}
\and A.~Blanchard\orcid{0000-0001-8555-9003}\inst{\ref{aff92}}
\and J.~Blazek\orcid{0000-0002-4687-4657}\inst{\ref{aff102}}
\and L.~Blot\orcid{0000-0002-9622-7167}\inst{\ref{aff103},\ref{aff41}}
\and A.~Bosco\inst{\ref{aff38}}
\and C.~Bodendorf\inst{\ref{aff93}}
\and T.~Boenke\inst{\ref{aff104}}
\and H.~B\"ohringer\orcid{0000-0001-8241-4204}\inst{\ref{aff93},\ref{aff105},\ref{aff106}}
\and P.~Boldrini\inst{\ref{aff2}}
\and M.~Bolzonella\orcid{0000-0003-3278-4607}\inst{\ref{aff35}}
\and A.~Bonchi\orcid{0000-0002-2667-5482}\inst{\ref{aff107}}
\and M.~Bonici\orcid{0000-0002-8430-126X}\inst{\ref{aff108},\ref{aff77},\ref{aff109}}
\and D.~Bonino\orcid{0000-0002-3336-9977}\inst{\ref{aff110}}
\and L.~Bonino\inst{\ref{aff38}}
\and C.~Bonvin\orcid{0000-0002-5318-4064}\inst{\ref{aff111}}
\and W.~Bon\inst{\ref{aff20}}
\and J.~T.~Booth\inst{\ref{aff112}}
\and S.~Borgani\orcid{0000-0001-6151-6439}\inst{\ref{aff113},\ref{aff58},\ref{aff59},\ref{aff60}}
\and A.~S.~Borlaff\orcid{0000-0003-3249-4431}\inst{\ref{aff114},\ref{aff115}}
\and E.~Borsato\inst{\ref{aff40},\ref{aff39}}
\and A.~Bosco\inst{\ref{aff38}}
\and B.~Bose\orcid{0000-0003-1965-8614}\inst{\ref{aff98}}
\and M.~T.~Botticella\orcid{0000-0002-3938-692X}\inst{\ref{aff24}}
\and A.~Boucaud\orcid{0000-0001-7387-2633}\inst{\ref{aff10}}
\and F.~Bouche\orcid{0000-0002-4663-1786}\inst{\ref{aff116},\ref{aff117}}
\and J.~S.~Boucher\inst{\ref{aff16}}
\and D.~Boutigny\orcid{0000-0003-4887-2150}\inst{\ref{aff118}}
\and T.~Bouvard\orcid{0009-0002-7959-312X}\inst{\ref{aff119}}
\and R.~Bouwens\orcid{0000-0002-4989-2471}\inst{\ref{aff70}}
\and H.~Bouy\orcid{0000-0002-7084-487X}\inst{\ref{aff120},\ref{aff121}}
\and R.~A.~A.~Bowler\orcid{0000-0003-3917-1678}\inst{\ref{aff122}}
\and V.~Bozza\orcid{0000-0003-4590-0136}\inst{\ref{aff123},\ref{aff124}}
\and E.~Bozzo\orcid{0000-0002-8201-1525}\inst{\ref{aff125}}
\and E.~Branchini\orcid{0000-0002-0808-6908}\inst{\ref{aff126},\ref{aff127},\ref{aff34}}
\and G.~Brando\orcid{0000-0003-0805-1905}\inst{\ref{aff128}}
\and S.~Brau-Nogue\inst{\ref{aff92}}
\and P.~Brekke\inst{\ref{aff129}}
\and M.~N.~Bremer\inst{\ref{aff130}}
\and M.~Brescia\orcid{0000-0001-9506-5680}\inst{\ref{aff131},\ref{aff24},\ref{aff117}}
\and M.-A.~Breton\inst{\ref{aff132},\ref{aff133},\ref{aff41}}
\and J.~Brinchmann\orcid{0000-0003-4359-8797}\inst{\ref{aff55}}
\and T.~Brinckmann\orcid{0000-0002-1492-5181}\inst{\ref{aff29},\ref{aff42}}
\and C.~Brockley-Blatt\inst{\ref{aff16}}
\and M.~Brodwin\inst{\ref{aff134}}
\and L.~Brouard\inst{\ref{aff99}}
\and M.~L.~Brown\orcid{0000-0002-0370-8077}\inst{\ref{aff122}}
\and S.~Bruton\orcid{0000-0002-6503-5218}\inst{\ref{aff135}}
\and J.~Bucko\orcid{0000-0002-1662-1042}\inst{\ref{aff7},\ref{aff12}}
\and H.~Buddelmeijer\orcid{0000-0001-8001-0089}\inst{\ref{aff69},\ref{aff70}}
\and G.~Buenadicha\inst{\ref{aff26}}
\and F.~Buitrago\orcid{0000-0002-2861-9812}\inst{\ref{aff136},\ref{aff137}}
\and P.~Burger\orcid{0000-0001-8637-6305}\inst{\ref{aff138},\ref{aff77}}
\and C.~Burigana\orcid{0000-0002-3005-5796}\inst{\ref{aff101},\ref{aff139}}
\and V.~Busillo\orcid{0009-0000-6049-1073}\inst{\ref{aff24},\ref{aff131},\ref{aff117}}
\and D.~Busonero\orcid{0000-0002-3903-7076}\inst{\ref{aff110}}
\and R.~Cabanac\orcid{0000-0001-6679-2600}\inst{\ref{aff92}}
\and L.~Cabayol-Garcia\orcid{0000-0002-9498-2572}\inst{\ref{aff57},\ref{aff140}}
\and M.~S.~Cagliari\orcid{0000-0002-2912-9233}\inst{\ref{aff43}}
\and A.~Caillat\inst{\ref{aff20}}
\and L.~Caillat\inst{\ref{aff52}}
\and M.~Calabrese\orcid{0000-0002-2637-2422}\inst{\ref{aff141},\ref{aff108}}
\and A.~Calabro\orcid{0000-0003-2536-1614}\inst{\ref{aff142}}
\and G.~Calderone\orcid{0000-0002-7738-5389}\inst{\ref{aff59}}
\and F.~Calura\orcid{0000-0002-6175-0871}\inst{\ref{aff35}}
\and B.~Camacho~Quevedo\orcid{0000-0002-8789-4232}\inst{\ref{aff143},\ref{aff132}}
\and S.~Camera\orcid{0000-0003-3399-3574}\inst{\ref{aff144},\ref{aff13},\ref{aff110}}
\and L.~Campos\inst{\ref{aff104}}
\and G.~Ca$\tilde{\rm n}$as-Herrera\orcid{0000-0003-2796-2149}\inst{\ref{aff104},\ref{aff5}}
\and G.~P.~Candini\orcid{0000-0001-9481-8206}\inst{\ref{aff16}}
\and M.~Cantiello\orcid{0000-0003-2072-384X}\inst{\ref{aff145}}
\and V.~Capobianco\orcid{0000-0002-3309-7692}\inst{\ref{aff110}}
\and E.~Cappellaro\orcid{0000-0001-5008-8619}\inst{\ref{aff75}}
\and N.~Cappelluti\orcid{0000-0002-1697-186X}\inst{\ref{aff146}}
\and A.~Cappi\inst{\ref{aff35},\ref{aff8}}
\and K.~I.~Caputi\orcid{0000-0001-8183-1460}\inst{\ref{aff69},\ref{aff147}}
\and C.~Cara\inst{\ref{aff11}}
\and C.~Carbone\orcid{0000-0003-0125-3563}\inst{\ref{aff108}}
\and V.~F.~Cardone\inst{\ref{aff142},\ref{aff148}}
\and E.~Carella\inst{\ref{aff108},\ref{aff43}}
\and R.~G.~Carlberg\inst{\ref{aff149}}
\and M.~Carle\inst{\ref{aff20}}
\and L.~Carminati\inst{\ref{aff99}}
\and F.~Caro\inst{\ref{aff142}}
\and J.~M.~Carrasco\orcid{0000-0002-3029-5853}\inst{\ref{aff150},\ref{aff151},\ref{aff143}}
\and J.~Carretero\orcid{0000-0002-3130-0204}\inst{\ref{aff152},\ref{aff140}}
\and P.~Carrilho\orcid{0000-0003-1339-0194}\inst{\ref{aff98}}
\and J.~Carron~Duque\orcid{0000-0003-4554-395X}\inst{\ref{aff14}}
\and B.~Carry\orcid{0000-0001-5242-3089}\inst{\ref{aff8}}
\and A.~Carvalho\orcid{0000-0002-9301-262X}\inst{\ref{aff153},\ref{aff18}}
\and C.~S.~Carvalho\inst{\ref{aff137}}
\and R.~Casas\orcid{0000-0002-8165-5601}\inst{\ref{aff143},\ref{aff132}}
\and S.~Casas\orcid{0000-0002-4751-5138}\inst{\ref{aff154}}
\and P.~Casenove\inst{\ref{aff83}}
\and C.~M.~Casey\orcid{0000-0002-0930-6466}\inst{\ref{aff65},\ref{aff155}}
\and P.~Cassata\orcid{0000-0002-6716-4400}\inst{\ref{aff40},\ref{aff75}}
\and F.~J.~Castander\orcid{0000-0001-7316-4573}\inst{\ref{aff132},\ref{aff143}}
\and D.~Castelao\inst{\ref{aff18}}
\and M.~Castellano\orcid{0000-0001-9875-8263}\inst{\ref{aff142}}
\and L.~Castiblanco\orcid{0000-0002-2324-7335}\inst{\ref{aff47},\ref{aff156}}
\and G.~Castignani\orcid{0000-0001-6831-0687}\inst{\ref{aff35}}
\and T.~Castro\orcid{0000-0002-6292-3228}\inst{\ref{aff59},\ref{aff60},\ref{aff58},\ref{aff157}}
\and C.~Cavet\orcid{0000-0001-8988-9800}\inst{\ref{aff10}}
\and S.~Cavuoti\orcid{0000-0002-3787-4196}\inst{\ref{aff24},\ref{aff117}}
\and P.-Y.~Chabaud\inst{\ref{aff20}}
\and K.~C.~Chambers\orcid{0000-0001-6965-7789}\inst{\ref{aff158}}
\and Y.~Charles\inst{\ref{aff20}}
\and S.~Charlot\orcid{0000-0003-3458-2275}\inst{\ref{aff2}}
\and N.~Chartab\orcid{0000-0003-3691-937X}\inst{\ref{aff159}}
\and R.~Chary\orcid{0000-0001-7583-0621}\inst{\ref{aff17}}
\and F.~Chaumeil\inst{\ref{aff99}}
\and H.~Cho\inst{\ref{aff112}}
\and G.~Chon\inst{\ref{aff105},\ref{aff106}}
\and E.~Ciancetta\orcid{0000-0002-5470-8405}\inst{\ref{aff38}}
\and P.~Ciliegi\inst{\ref{aff35}}
\and A.~Cimatti\inst{\ref{aff160}}
\and M.~Cimino\inst{\ref{aff104}}
\and M.-R.~L.~Cioni\orcid{0000-0002-6797-696X}\inst{\ref{aff161}}
\and R.~Claydon\orcid{0000-0003-2117-7895}\inst{\ref{aff98}}
\and C.~Cleland\inst{\ref{aff10}}
\and B.~Cl\'ement\orcid{0000-0002-7966-3661}\inst{\ref{aff4},\ref{aff162}}
\and D.~L.~Clements\orcid{0000-0002-9548-5033}\inst{\ref{aff163}}
\and N.~Clerc\orcid{0000-0002-0296-1011}\inst{\ref{aff92}}
\and S.~Clesse\orcid{0000-0001-5079-1785}\inst{\ref{aff164}}
\and S.~Codis\inst{\ref{aff11}}
\and F.~Cogato\orcid{0000-0003-4632-6113}\inst{\ref{aff165},\ref{aff35}}
\and J.~Colbert\orcid{0000-0001-6482-3020}\inst{\ref{aff17}}
\and R.~E.~Cole\orcid{0000-0002-7093-7320}\inst{\ref{aff16}}
\and P.~Coles\orcid{0000-0002-5535-2850}\inst{\ref{aff166}}
\and T.~E.~Collett\orcid{0000-0001-5564-3140}\inst{\ref{aff63}}
\and R.~S.~Collins\orcid{0000-0001-8437-1703}\inst{\ref{aff98}}
\and C.~Colodro-Conde\inst{\ref{aff67}}
\and C.~Colombo\inst{\ref{aff104}}
\and F.~Combes\orcid{0000-0003-2658-7893}\inst{\ref{aff167},\ref{aff168}}
\and V.~Conforti\orcid{0000-0002-0007-3520}\inst{\ref{aff35}}
\and G.~Congedo\orcid{0000-0003-2508-0046}\inst{\ref{aff98}}
\and S.~Conseil\orcid{0000-0002-3657-4191}\inst{\ref{aff51}}
\and C.~J.~Conselice\orcid{0000-0003-1949-7638}\inst{\ref{aff122}}
\and S.~Contarini\orcid{0000-0002-9843-723X}\inst{\ref{aff93}}
\and T.~Contini\orcid{0000-0003-0275-938X}\inst{\ref{aff92}}
\and L.~Conversi\orcid{0000-0002-6710-8476}\inst{\ref{aff169},\ref{aff26}}
\and A.~R.~Cooray\orcid{0000-0002-3892-0190}\inst{\ref{aff170}}
\and Y.~Copin\orcid{0000-0002-5317-7518}\inst{\ref{aff51}}
\and P.-S.~Corasaniti\orcid{0000-0002-6386-7846}\inst{\ref{aff41}}
\and P.~Corcho-Caballero\orcid{0000-0001-6327-7080}\inst{\ref{aff69}}
\and L.~Corcione\orcid{0000-0002-6497-5881}\inst{\ref{aff110}}
\and O.~Cordes\inst{\ref{aff138}}
\and O.~Corpace\inst{\ref{aff81}}
\and M.~Correnti\orcid{0000-0001-6464-3257}\inst{\ref{aff142},\ref{aff107}}
\and M.~Costanzi\orcid{0000-0001-8158-1449}\inst{\ref{aff113},\ref{aff59},\ref{aff58}}
\and A.~Costille\inst{\ref{aff20}}
\and F.~Courbin\orcid{0000-0003-0758-6510}\inst{\ref{aff4}}
\and L.~Courcoult~Mifsud\inst{\ref{aff99}}
\and H.~M.~Courtois\orcid{0000-0003-0509-1776}\inst{\ref{aff66}}
\and M.-C.~Cousinou\inst{\ref{aff52}}
\and G.~Covone\orcid{0000-0002-2553-096X}\inst{\ref{aff131},\ref{aff24},\ref{aff117}}
\and T.~Cowell\inst{\ref{aff171}}
\and C.~Cragg\inst{\ref{aff172}}
\and G.~Cresci\orcid{0000-0002-5281-1417}\inst{\ref{aff173}}
\and S.~Cristiani\orcid{0000-0002-2115-5234}\inst{\ref{aff59},\ref{aff60},\ref{aff58}}
\and M.~Crocce\orcid{0000-0002-9745-6228}\inst{\ref{aff132},\ref{aff133}}
\and M.~Cropper\orcid{0000-0003-4571-9468}\inst{\ref{aff16}}
\and P.~E~Crouzet\inst{\ref{aff104}}
\and B.~Csizi\orcid{0000-0003-3227-6581}\inst{\ref{aff174}}
\and J.-G.~Cuby\orcid{0000-0002-8767-1442}\inst{\ref{aff175},\ref{aff20}}
\and E.~Cucchetti\orcid{0000-0002-5548-4351}\inst{\ref{aff83}}
\and O.~Cucciati\orcid{0000-0002-9336-7551}\inst{\ref{aff35}}
\and J.-C.~Cuillandre\orcid{0000-0002-3263-8645}\inst{\ref{aff11}}
\and P.~A.~C.~Cunha\orcid{0000-0002-9454-859X}\inst{\ref{aff176},\ref{aff55}}
\and V.~Cuozzo~\inst{\ref{aff177},\ref{aff94}}
\and E.~Daddi\orcid{0000-0002-3331-9590}\inst{\ref{aff11}}
\and M.~D'Addona\orcid{0000-0003-3445-0483}\inst{\ref{aff24},\ref{aff123}}
\and C.~Dafonte\orcid{0000-0003-4693-7555}\inst{\ref{aff178}}
\and N.~Dagoneau\orcid{0000-0002-1361-2562}\inst{\ref{aff179}}
\and E.~Dalessandro\orcid{0000-0003-4237-4601}\inst{\ref{aff35}}
\and G.~B.~Dalton\orcid{0000-0002-3031-2588}\inst{\ref{aff172}}
\and G.~D'Amico\orcid{0000-0002-8183-1214}\inst{\ref{aff180},\ref{aff181}}
\and H.~Dannerbauer\orcid{0000-0001-7147-3575}\inst{\ref{aff182}}
\and P.~Danto\inst{\ref{aff83}}
\and I.~Das\orcid{0009-0007-7088-2044}\inst{\ref{aff62}}
\and A.~Da~Silva\orcid{0000-0002-6385-1609}\inst{\ref{aff153},\ref{aff18}}
\and R.~da~Silva\orcid{0000-0003-4788-677X}\inst{\ref{aff142},\ref{aff107}}
\and W.~d'Assignies~Doumerg\orcid{0000-0002-9719-1717}\inst{\ref{aff57}}
\and G.~Daste\inst{\ref{aff20}}
\and J.~E.~Davies\orcid{0000-0002-5079-9098}\inst{\ref{aff78}}
\and S.~Davini\orcid{0000-0003-3269-1718}\inst{\ref{aff127}}
\and P.~Dayal\orcid{0000-0001-8460-1564}\inst{\ref{aff69}}
\and T.~de~Boer\orcid{0000-0001-5486-2747}\inst{\ref{aff158}}
\and R.~Decarli\orcid{0000-0002-2662-8803}\inst{\ref{aff35}}
\and B.~De~Caro\inst{\ref{aff108}}
\and H.~Degaudenzi\orcid{0000-0002-5887-6799}\inst{\ref{aff125}}
\and G.~Degni\orcid{0009-0001-4912-1087}\inst{\ref{aff183},\ref{aff184}}
\and J.~T.~A.~de~Jong\orcid{0000-0002-9511-1357}\inst{\ref{aff70},\ref{aff69}}
\and L.~F.~de~la~Bella\orcid{0000-0002-1064-3400}\inst{\ref{aff63}}
\and S.~de~la~Torre\inst{\ref{aff20}}
\and F.~Delhaise\inst{\ref{aff171}}
\and D.~Delley\orcid{0000-0002-4958-7469}\inst{\ref{aff93}}
\and G.~Delucchi\inst{\ref{aff126},\ref{aff127}}
\and G.~De~Lucia\orcid{0000-0002-6220-9104}\inst{\ref{aff59}}
\and J.~Denniston\inst{\ref{aff16}}
\and F.~De~Paolis\orcid{0000-0001-6460-7563}\inst{\ref{aff185},\ref{aff186},\ref{aff187}}
\and M.~De~Petris\orcid{0000-0001-7859-2139}\inst{\ref{aff188},\ref{aff142}}
\and A.~Derosa\inst{\ref{aff35}}
\and S.~Desai\orcid{0000-0002-0466-3288}\inst{\ref{aff189}}
\and V.~Desjacques\orcid{0000-0003-2062-8172}\inst{\ref{aff190}}
\and G.~Despali\orcid{0000-0001-6150-4112}\inst{\ref{aff165},\ref{aff35},\ref{aff73}}
\and G.~Desprez\inst{\ref{aff191}}
\and J.~De~Vicente-Albendea\inst{\ref{aff152}}
\and Y.~Deville\orcid{0000-0002-8769-2446}\inst{\ref{aff92}}
\and J.~D.~F.~Dias\orcid{0000-0002-1505-7734}\inst{\ref{aff192},\ref{aff55},\ref{aff176}}
\and A.~D\'iaz-S\'anchez\orcid{0000-0003-0748-4768}\inst{\ref{aff193}}
\and J.~J.~Diaz\inst{\ref{aff182}}
\and S.~Di~Domizio\orcid{0000-0003-2863-5895}\inst{\ref{aff126},\ref{aff127}}
\and J.~M.~Diego\orcid{0000-0001-9065-3926}\inst{\ref{aff194}}
\and D.~Di~Ferdinando\inst{\ref{aff73}}
\and A.~M.~Di~Giorgio\orcid{0000-0002-4767-2360}\inst{\ref{aff195}}
\and P.~Dimauro\orcid{0000-0001-7399-2854}\inst{\ref{aff142},\ref{aff196}}
\and J.~Dinis\orcid{0000-0001-5075-1601}\inst{\ref{aff153},\ref{aff18}}
\and K.~Dolag\inst{\ref{aff82}}
\and C.~Dolding\inst{\ref{aff16}}
\and H.~Dole\orcid{0000-0002-9767-3839}\inst{\ref{aff9}}
\and H.~Dom\'inguez~S\'anchez\orcid{0000-0002-9013-1316}\inst{\ref{aff197}}
\and O.~Dor\'e\orcid{0000-0001-7432-2932}\inst{\ref{aff198},\ref{aff112}}
\and F.~Dournac\inst{\ref{aff92}}
\and M.~Douspis\orcid{0000-0003-4203-3954}\inst{\ref{aff9}}
\and H.~Dreihahn\inst{\ref{aff171}}
\and B.~Droge\orcid{0000-0002-8279-868X}\inst{\ref{aff199}}
\and B.~Dryer\orcid{0000-0001-7925-9768}\inst{\ref{aff200}}
\and F.~Dubath\orcid{0000-0002-6533-2810}\inst{\ref{aff125}}
\and P.-A.~Duc\orcid{0000-0003-3343-6284}\inst{\ref{aff97}}
\and F.~Ducret\inst{\ref{aff20}}
\and C.~Duffy\orcid{0000-0001-6662-0200}\inst{\ref{aff201}}
\and F.~Dufresne\inst{\ref{aff20}}
\and C.~A.~J.~Duncan\inst{\ref{aff122}}
\and X.~Dupac\inst{\ref{aff26}}
\and V.~Duret\orcid{0009-0009-0383-4960}\inst{\ref{aff52}}
\and R.~Durrer\orcid{0000-0001-9833-2086}\inst{\ref{aff111}}
\and F.~Durret\orcid{0000-0002-6991-4578}\inst{\ref{aff1}}
\and S.~Dusini\orcid{0000-0002-1128-0664}\inst{\ref{aff39}}
\and A.~Ealet\orcid{0000-0003-3070-014X}\inst{\ref{aff51}}
\and A.~Eggemeier\orcid{0000-0002-1841-8910}\inst{\ref{aff138}}
\and P.~R.~M.~Eisenhardt\inst{\ref{aff112}}
\and D.~Elbaz\orcid{0000-0002-7631-647X}\inst{\ref{aff11}}
\and M.~Y.~Elkhashab\orcid{0000-0001-9306-2603}\inst{\ref{aff39},\ref{aff40}}
\and A.~Ellien\orcid{0000-0002-1038-3370}\inst{\ref{aff100}}
\and J.~Endicott\inst{\ref{aff202}}
\and A.~Enia\orcid{0000-0002-0200-2857}\inst{\ref{aff72},\ref{aff35}}
\and T.~Erben\inst{\ref{aff138}}
\and J.~A.~Escartin~Vigo\inst{\ref{aff93}}
\and S.~Escoffier\orcid{0000-0002-2847-7498}\inst{\ref{aff52}}
\and I.~Escudero~Sanz\inst{\ref{aff104}}
\and J.~Essert\inst{\ref{aff171}}
\and S.~Ettori\orcid{0000-0003-4117-8617}\inst{\ref{aff35},\ref{aff139}}
\and M.~Ezziati\orcid{0009-0003-6065-1585}\inst{\ref{aff20}}
\and G.~Fabbian\orcid{0000-0002-3255-4695}\inst{\ref{aff203},\ref{aff204}}
\and M.~Fabricius\orcid{0000-0002-7025-6058}\inst{\ref{aff93},\ref{aff82}}
\and Y.~Fang\inst{\ref{aff82}}
\and A.~Farina\orcid{0009-0000-3420-929X}\inst{\ref{aff126},\ref{aff34},\ref{aff127}}
\and M.~Farina\orcid{0000-0002-3089-7846}\inst{\ref{aff195}}
\and R.~Farinelli\inst{\ref{aff35}}
\and S.~Farrens\orcid{0000-0002-9594-9387}\inst{\ref{aff11}}
\and F.~Faustini\orcid{0000-0001-6274-5145}\inst{\ref{aff107},\ref{aff142}}
\and A.~Feltre\orcid{0000-0001-6865-2871}\inst{\ref{aff173}}
\and A.~M.~N.~Ferguson\inst{\ref{aff98}}
\and P.~Ferrando\orcid{0000-0002-0558-9155}\inst{\ref{aff11}}
\and A.~G.~Ferrari\orcid{0009-0005-5266-4110}\inst{\ref{aff160},\ref{aff73}}
\and A.~Ferr\'e-Mateu\orcid{0000-0002-6411-220X}\inst{\ref{aff68},\ref{aff67}}
\and P.~G.~Ferreira\orcid{0000-0002-3021-2851}\inst{\ref{aff172}}
\and I.~Ferreras\orcid{0000-0003-4584-3127}\inst{\ref{aff205},\ref{aff67},\ref{aff68}}
\and I.~Ferrero\orcid{0000-0002-1295-1132}\inst{\ref{aff206}}
\and S.~Ferriol\inst{\ref{aff51}}
\and P.~Ferruit\inst{\ref{aff26}}
\and D.~Filleul\inst{\ref{aff99}}
\and F.~Finelli\orcid{0000-0002-6694-3269}\inst{\ref{aff35},\ref{aff139}}
\and S.~L.~Finkelstein\orcid{0000-0001-8519-1130}\inst{\ref{aff65}}
\and A.~Finoguenov\orcid{0000-0002-4606-5403}\inst{\ref{aff207}}
\and B.~Fiorini\orcid{0000-0002-0092-4321}\inst{\ref{aff63},\ref{aff208}}
\and F.~Flentge\inst{\ref{aff171}}
\and P.~Focardi\inst{\ref{aff72}}
\and J.~Fonseca\orcid{0000-0003-0549-1614}\inst{\ref{aff55},\ref{aff176}}
\and A.~Fontana\orcid{0000-0003-3820-2823}\inst{\ref{aff142}}
\and F.~Fontanot\orcid{0000-0003-4744-0188}\inst{\ref{aff59},\ref{aff58}}
\and F.~Fornari\orcid{0000-0003-2979-6738}\inst{\ref{aff139}}
\and P.~Fosalba\orcid{0000-0002-1510-5214}\inst{\ref{aff143},\ref{aff133}}
\and M.~Fossati\orcid{0000-0002-9043-8764}\inst{\ref{aff34},\ref{aff209}}
\and S.~Fotopoulou\orcid{0000-0002-9686-254X}\inst{\ref{aff130}}
\and D.~Fouchez\orcid{0000-0002-7496-3796}\inst{\ref{aff52}}
\and N.~Fourmanoit\inst{\ref{aff52}}
\and M.~Frailis\orcid{0000-0002-7400-2135}\inst{\ref{aff59}}
\and D.~Fraix-Burnet\orcid{0000-0002-7418-0513}\inst{\ref{aff210}}
\and E.~Franceschi\orcid{0000-0002-0585-6591}\inst{\ref{aff35}}
\and A.~Franco\orcid{0000-0002-4761-366X}\inst{\ref{aff186},\ref{aff185},\ref{aff187}}
\and P.~Franzetti\inst{\ref{aff108}}
\and J.~Freihoefer\inst{\ref{aff171}}
\and C.~.S.~Frenk\inst{\ref{aff87}}
\and G.~Frittoli\orcid{0009-0001-3882-2355}\inst{\ref{aff177}}
\and P.-A.~Frugier\inst{\ref{aff11}}
\and N.~Frusciante\orcid{0000-0002-7375-1230}\inst{\ref{aff131}}
\and A.~Fumagalli\orcid{0009-0004-0300-2535}\inst{\ref{aff105},\ref{aff59},\ref{aff58}}
\and M.~Fumagalli\orcid{0000-0001-6676-3842}\inst{\ref{aff59},\ref{aff209}}
\and M.~Fumana\orcid{0000-0001-6787-5950}\inst{\ref{aff108}}
\and Y.~Fu\orcid{0000-0002-0759-0504}\inst{\ref{aff70},\ref{aff69}}
\and L.~Gabarra\orcid{0000-0002-8486-8856}\inst{\ref{aff172}}
\and S.~Galeotta\orcid{0000-0002-3748-5115}\inst{\ref{aff59}}
\and L.~Galluccio\orcid{0000-0002-8541-0476}\inst{\ref{aff8}}
\and K.~Ganga\orcid{0000-0001-8159-8208}\inst{\ref{aff10}}
\and H.~Gao\orcid{0000-0003-1015-5367}\inst{\ref{aff158}}
\and J.~Garc\'ia-Bellido\orcid{0000-0002-9370-8360}\inst{\ref{aff14}}
\and K.~Garcia\orcid{0000-0001-7145-549X}\inst{\ref{aff211}}
\and J.~P.~Gardner\orcid{0000-0003-2098-9568}\inst{\ref{aff212}}
\and B.~Garilli\orcid{0000-0001-7455-8750}\inst{\ref{aff108}}
\and L.-M.~Gaspar-Venancio\inst{\ref{aff104}}
\and T.~Gasparetto\orcid{0000-0002-7913-4866}\inst{\ref{aff59}}
\and V.~Gautard\inst{\ref{aff213}}
\and R.~Gavazzi\orcid{0000-0002-5540-6935}\inst{\ref{aff20},\ref{aff2}}
\and E.~Gaztanaga\orcid{0000-0001-9632-0815}\inst{\ref{aff132},\ref{aff143},\ref{aff63}}
\and L.~Genolet\inst{\ref{aff125}}
\and R.~Genova~Santos\orcid{0000-0001-5479-0034}\inst{\ref{aff67},\ref{aff182}}
\and F.~Gentile\orcid{0000-0002-8008-9871}\inst{\ref{aff72},\ref{aff35}}
\and K.~George\orcid{0000-0002-1734-8455}\inst{\ref{aff82}}
\and M.~Gerbino\orcid{0000-0002-3538-1283}\inst{\ref{aff42}}
\and Z.~Ghaffari\orcid{0000-0002-6467-8078}\inst{\ref{aff59},\ref{aff58}}
\and F.~Giacomini\orcid{0000-0002-3129-2814}\inst{\ref{aff73}}
\and F.~Gianotti\orcid{0000-0003-4666-119X}\inst{\ref{aff35}}
\and G.~P.~S.~Gibb\inst{\ref{aff98}}
\and W.~Gillard\orcid{0000-0003-4744-9748}\inst{\ref{aff52}}
\and B.~Gillis\orcid{0000-0002-4478-1270}\inst{\ref{aff98}}
\and M.~Ginolfi\orcid{0000-0002-9122-1700}\inst{\ref{aff214},\ref{aff173}}
\and C.~Giocoli\orcid{0000-0002-9590-7961}\inst{\ref{aff35},\ref{aff215}}
\and M.~Girardi\orcid{0000-0003-1861-1865}\inst{\ref{aff113},\ref{aff59}}
\and S.~K.~Giri\orcid{0000-0002-2560-536X}\inst{\ref{aff7},\ref{aff216}}
\and L.~W.~K.~Goh\orcid{0000-0002-0104-8132}\inst{\ref{aff11}}
\and P.~G\'omez-Alvarez\orcid{0000-0002-8594-5358}\inst{\ref{aff217},\ref{aff26}}
\and V.~Gonzalez-Perez\orcid{0000-0001-9938-2755}\inst{\ref{aff218}}
\and A.~H.~Gonzalez\orcid{0000-0002-0933-8601}\inst{\ref{aff211}}
\and E.~J.~Gonzalez\orcid{0000-0002-0226-9893}\inst{\ref{aff57},\ref{aff140},\ref{aff219}}
\and J.~C.~Gonzalez\orcid{0000-0003-4447-8658}\inst{\ref{aff26},\ref{aff220}}
\and S.~Gouyou~Beauchamps\inst{\ref{aff143},\ref{aff133}}
\and G.~Gozaliasl\orcid{0000-0002-0236-919X}\inst{\ref{aff221},\ref{aff207}}
\and J.~Gracia-Carpio\inst{\ref{aff93}}
\and S.~Grandis\orcid{0000-0002-4577-8217}\inst{\ref{aff174}}
\and B.~R.~Granett\orcid{0000-0003-2694-9284}\inst{\ref{aff34}}
\and M.~Granvik\orcid{0000-0002-5624-1888}\inst{\ref{aff207},\ref{aff222}}
\and A.~Grazian\orcid{0000-0002-5688-0663}\inst{\ref{aff75}}
\and A.~Gregorio\orcid{0000-0003-4028-8785}\inst{\ref{aff113},\ref{aff59},\ref{aff60}}
\and C.~Grenet\inst{\ref{aff2}}
\and C.~Grillo\orcid{0000-0002-5926-7143}\inst{\ref{aff43},\ref{aff108}}
\and F.~Grupp\inst{\ref{aff93},\ref{aff82}}
\and C.~Gruppioni\orcid{0000-0002-5836-4056}\inst{\ref{aff35}}
\and A.~Gruppuso\inst{\ref{aff35},\ref{aff73}}
\and C.~Guerbuez\inst{\ref{aff171}}
\and S.~Guerrini\orcid{0009-0004-3655-4870}\inst{\ref{aff223}}
\and M.~Guidi\orcid{0000-0001-9408-1101}\inst{\ref{aff72},\ref{aff35}}
\and P.~Guillard\orcid{0000-0002-2421-1350}\inst{\ref{aff121},\ref{aff2}}
\and C.~M.~Gutierrez\inst{\ref{aff182}}
\and P.~Guttridge\inst{\ref{aff16}}
\and L.~Guzzo\orcid{0000-0001-8264-5192}\inst{\ref{aff43},\ref{aff34}}
\and S.~Gwyn\orcid{0000-0001-8221-8406}\inst{\ref{aff224}}
\and J.~Haapala\inst{\ref{aff225}}
\and J.~Haase\orcid{0000-0001-7809-3671}\inst{\ref{aff93}}
\and C.~R.~Haddow\inst{\ref{aff171}}
\and M.~Hailey\inst{\ref{aff16}}
\and A.~Hall\orcid{0000-0002-3139-8651}\inst{\ref{aff98}}
\and D.~Hall\inst{\ref{aff226}}
\and N.~Hamaus\orcid{0000-0002-0876-2101}\inst{\ref{aff82},\ref{aff227}}
\and B.~S.~Haridasu\orcid{0000-0002-9153-1258}\inst{\ref{aff61},\ref{aff60},\ref{aff58}}
\and J.~Harnois-D\'eraps\orcid{0000-0002-4864-1240}\inst{\ref{aff47}}
\and C.~Harper\inst{\ref{aff228}}
\and W.~G.~Hartley\inst{\ref{aff125}}
\and G.~Hasinger\orcid{0000-0002-0797-0646}\inst{\ref{aff229},\ref{aff230}}
\and F.~Hassani\orcid{0000-0003-2640-4460}\inst{\ref{aff206}}
\and N.~A.~Hatch\orcid{0000-0001-5600-0534}\inst{\ref{aff56}}
\and S.~V.~H.~Haugan\orcid{0000-0001-9648-7260}\inst{\ref{aff206}}
\and B.~H\"au\ss~ler\orcid{0000-0002-1857-2088}\inst{\ref{aff231}}
\and A.~Heavens\orcid{0000-0003-1586-2773}\inst{\ref{aff163}}
\and L.~Heisenberg\inst{\ref{aff31}}
\and A.~Helmi\orcid{0000-0003-3937-7641}\inst{\ref{aff69}}
\and G.~Helou\orcid{0000-0003-3367-3415}\inst{\ref{aff62}}
\and S.~Hemmati\orcid{0000-0003-2226-5395}\inst{\ref{aff62}}
\and K.~Henares\inst{\ref{aff26}}
\and O.~Herent\inst{\ref{aff2}}
\and C.~Hern\'andez-Monteagudo\orcid{0000-0001-5471-9166}\inst{\ref{aff197},\ref{aff68},\ref{aff67}}
\and T.~Heuberger\inst{\ref{aff232}}
\and P.~C.~Hewett\orcid{0000-0002-6528-1937}\inst{\ref{aff233}}
\and S.~Heydenreich\orcid{0000-0002-7273-4076}\inst{\ref{aff138},\ref{aff234}}
\and H.~Hildebrandt\orcid{0000-0002-9814-3338}\inst{\ref{aff235}}
\and M.~Hirschmann\orcid{0000-0002-3301-3321}\inst{\ref{aff236},\ref{aff59}}
\and J.~Hjorth\orcid{0000-0002-4571-2306}\inst{\ref{aff237}}
\and J.~Hoar\inst{\ref{aff26}}
\and H.~Hoekstra\orcid{0000-0002-0641-3231}\inst{\ref{aff70}}
\and A.~D.~Holland\inst{\ref{aff226},\ref{aff202}}
\and M.~S.~Holliman\inst{\ref{aff98}}
\and W.~Holmes\inst{\ref{aff112}}
\and I.~Hook\orcid{0000-0002-2960-978X}\inst{\ref{aff201}}
\and B.~Horeau\inst{\ref{aff11}}
\and F.~Hormuth\inst{\ref{aff238}}
\and A.~Hornstrup\orcid{0000-0002-3363-0936}\inst{\ref{aff239},\ref{aff155}}
\and S.~Hosseini\inst{\ref{aff92}}
\and D.~Hu\inst{\ref{aff16}}
\and P.~Hudelot\inst{\ref{aff2}}
\and M.~J.~Hudson\orcid{0000-0002-1437-3786}\inst{\ref{aff76},\ref{aff77},\ref{aff109}}
\and M.~Huertas-Company\orcid{0000-0002-1416-8483}\inst{\ref{aff67},\ref{aff182},\ref{aff240},\ref{aff241}}
\and E.~M.~Huff\orcid{0000-0002-9378-3424}\inst{\ref{aff112}}
\and A.~C.~N.~Hughes\orcid{0000-0001-9294-3089}\inst{\ref{aff163}}
\and A.~Humphrey\inst{\ref{aff55},\ref{aff242}}
\and L.~K.~Hunt\orcid{0000-0001-9162-2371}\inst{\ref{aff173}}
\and D.~D.~Huynh\inst{\ref{aff213}}
\and R.~Ibata\orcid{0000-0002-3292-9709}\inst{\ref{aff97}}
\and K.~Ichikawa\orcid{0000-0002-4377-903X}\inst{\ref{aff243}}
\and S.~Iglesias-Groth\inst{\ref{aff67}}
\and O.~Ilbert\orcid{0000-0002-7303-4397}\inst{\ref{aff20}}
\and S.~Ili\'c\orcid{0000-0003-4285-9086}\inst{\ref{aff244},\ref{aff92}}
\and L.~Ingoglia\orcid{0000-0002-7587-0997}\inst{\ref{aff165}}
\and E.~Iodice\orcid{0000-0003-4291-0005}\inst{\ref{aff24}}
\and H.~Israel\orcid{0000-0002-3045-4412}\inst{\ref{aff245}}
\and U.~E.~Israelsson\inst{\ref{aff112}}
\and L.~Izzo\orcid{0000-0001-9695-8472}\inst{\ref{aff24}}
\and P.~Jablonka\orcid{0000-0002-9655-1063}\inst{\ref{aff4}}
\and N.~Jackson\inst{\ref{aff122}}
\and J.~Jacobson\inst{\ref{aff62}}
\and M.~Jafariyazani\orcid{0000-0001-8019-6661}\inst{\ref{aff62}}
\and K.~Jahnke\orcid{0000-0003-3804-2137}\inst{\ref{aff78}}
\and B.~Jain\inst{\ref{aff246}}
\and H.~Jansen\orcid{0009-0002-1332-7742}\inst{\ref{aff174}}
\and M.~J.~Jarvis\orcid{0000-0001-7039-9078}\inst{\ref{aff172}}
\and J.~Jasche\orcid{0000-0002-4677-5843}\inst{\ref{aff247},\ref{aff1}}
\and M.~Jauzac\orcid{0000-0003-1974-8732}\inst{\ref{aff248},\ref{aff87},\ref{aff249},\ref{aff250}}
\and N.~Jeffrey\orcid{0000-0003-2927-1800}\inst{\ref{aff205}}
\and M.~Jhabvala\inst{\ref{aff251}}
\and Y.~Jimenez-Teja\orcid{0000-0002-6090-2853}\inst{\ref{aff27},\ref{aff196}}
\and A.~Jimenez~Mu\~noz\orcid{0009-0004-5252-185X}\inst{\ref{aff19}}
\and B.~Joachimi\orcid{0000-0001-7494-1303}\inst{\ref{aff205}}
\and P.~H.~Johansson\orcid{0000-0001-8741-8263}\inst{\ref{aff207}}
\and S.~Joudaki\orcid{0000-0001-8820-673X}\inst{\ref{aff63}}
\and E.~Jullo\orcid{0000-0002-9253-053X}\inst{\ref{aff20}}
\and J.~J.~E.~Kajava\orcid{0000-0002-3010-8333}\inst{\ref{aff252},\ref{aff253}}
\and Y.~Kang\inst{\ref{aff125}}
\and A.~Kannawadi\orcid{0000-0001-8783-6529}\inst{\ref{aff254}}
\and V.~Kansal\orcid{0000-0002-4008-6078}\inst{\ref{aff255},\ref{aff256}}
\and D.~Karagiannis\orcid{0000-0002-4927-0816}\inst{\ref{aff208},\ref{aff257}}
\and M.~K{\"a}rcher\orcid{0000-0001-5868-647X}\inst{\ref{aff20},\ref{aff91}}
\and A.~Kashlinsky\orcid{0000-0003-2156-078X}\inst{\ref{aff212},\ref{aff258},\ref{aff259}}
\and M.~V.~Kazandjian\inst{\ref{aff70},\ref{aff260}}
\and F.~Keck\inst{\ref{aff171}}
\and E.~Keih\"anen\orcid{0000-0003-1804-7715}\inst{\ref{aff225}}
\and E.~Kerins\orcid{0000-0002-1743-4468}\inst{\ref{aff122}}
\and S.~Kermiche\orcid{0000-0002-0302-5735}\inst{\ref{aff52}}
\and A.~Khalil\inst{\ref{aff16}}
\and A.~Kiessling\orcid{0000-0002-2590-1273}\inst{\ref{aff112}}
\and K.~Kiiveri\inst{\ref{aff225}}
\and M.~Kilbinger\orcid{0000-0001-9513-7138}\inst{\ref{aff11}}
\and J.~Kim\orcid{0000-0003-2776-2761}\inst{\ref{aff172}}
\and R.~King\inst{\ref{aff228}}
\and C.~C.~Kirkpatrick\inst{\ref{aff225}}
\and T.~Kitching\orcid{0000-0002-4061-4598}\inst{\ref{aff16}}
\and M.~Kluge\orcid{0000-0002-9618-2552}\inst{\ref{aff93}}
\and M.~Knabenhans\orcid{0000-0002-2886-9838}\inst{\ref{aff7}}
\and J.~H.~Knapen\orcid{0000-0003-1643-0024}\inst{\ref{aff67},\ref{aff68}}
\and A.~Knebe\orcid{0000-0003-4066-8307}\inst{\ref{aff218},\ref{aff261},\ref{aff262}}
\and J.-P.~Kneib\orcid{0000-0002-4616-4989}\inst{\ref{aff4}}
\and R.~Kohley\inst{\ref{aff26}}
\and L.~V.~E.~Koopmans\orcid{0000-0003-1840-0312}\inst{\ref{aff69}}
\and H.~Koskinen\orcid{0000-0003-3839-6461}\inst{\ref{aff207}}
\and E.~Koulouridis\inst{\ref{aff11},\ref{aff263}}
\and R.~Kou\orcid{0000-0003-3408-3062}\inst{\ref{aff10},\ref{aff264}}
\and A.~Kov\'acs\orcid{0000-0002-5825-579X}\inst{\ref{aff265},\ref{aff266}}
\and I.~Kova{\v{c}}i{\'{c}}\orcid{0000-0001-6751-3263}\inst{\ref{aff64}}
\and A.~Kowalczyk\inst{\ref{aff171}}
\and K.~Koyama\orcid{0000-0001-6727-6915}\inst{\ref{aff63}}
\and K.~Kraljic\orcid{0000-0001-6180-0245}\inst{\ref{aff97}}
\and O.~Krause\inst{\ref{aff78}}
\and S.~Kruk\orcid{0000-0001-8010-8879}\inst{\ref{aff26}}
\and B.~Kubik\orcid{0009-0006-5823-4880}\inst{\ref{aff51}}
\and U.~Kuchner\orcid{0000-0002-0035-5202}\inst{\ref{aff267}}
\and K.~Kuijken\orcid{0000-0002-3827-0175}\inst{\ref{aff70}}
\and M.~K\"ummel\orcid{0000-0003-2791-2117}\inst{\ref{aff82}}
\and M.~Kunz\orcid{0000-0002-3052-7394}\inst{\ref{aff111}}
\and H.~Kurki-Suonio\orcid{0000-0002-4618-3063}\inst{\ref{aff207},\ref{aff268}}
\and F.~Lacasa\orcid{0000-0002-7268-3440}\inst{\ref{aff164},\ref{aff9},\ref{aff111}}
\and C.~G.~Lacey\orcid{0000-0001-9016-5332}\inst{\ref{aff87}}
\and F.~La~Franca\orcid{0000-0002-1239-2721}\inst{\ref{aff183}}
\and N.~Lagarde\inst{\ref{aff120}}
\and O.~Lahav\orcid{0000-0002-1134-9035}\inst{\ref{aff205}}
\and C.~Laigle\orcid{0009-0008-5926-818X}\inst{\ref{aff2}}
\and A.~La~Marca\orcid{0000-0002-7217-5120}\inst{\ref{aff269},\ref{aff69}}
\and O.~La~Marle\inst{\ref{aff83}}
\and B.~Lamine\orcid{0000-0002-9416-2320}\inst{\ref{aff92}}
\and M.~C.~Lam\orcid{0000-0002-9347-2298}\inst{\ref{aff98}}
\and A.~Lan\c{c}on\orcid{0000-0002-7214-8296}\inst{\ref{aff97}}
\and H.~Landt\orcid{0000-0001-8391-6900}\inst{\ref{aff248}}
\and M.~Langer\orcid{0000-0002-9088-2718}\inst{\ref{aff9}}
\and A.~Lapi\orcid{0000-0002-4882-1735}\inst{\ref{aff61},\ref{aff58},\ref{aff101},\ref{aff60}}
\and C.~Larcheveque\inst{\ref{aff270}}
\and S.~S.~Larsen\orcid{0000-0003-0069-1203}\inst{\ref{aff271}}
\and M.~Lattanzi\inst{\ref{aff42}}
\and F.~Laudisio\inst{\ref{aff39}}
\and D.~Laugier\orcid{0000-0002-2517-0204}\inst{\ref{aff52}}
\and R.~Laureijs\inst{\ref{aff104}}
\and V.~Laurent\inst{\ref{aff9}}
\and G.~Lavaux\orcid{0000-0003-0143-8891}\inst{\ref{aff2}}
\and A.~Lawrenson\inst{\ref{aff272},\ref{aff16}}
\and A.~Lazanu\orcid{0000-0002-8061-9828}\inst{\ref{aff122}}
\and T.~Lazeyras\orcid{0000-0002-7080-9839}\inst{\ref{aff209}}
\and Q.~Le~Boulc'h\inst{\ref{aff273}}
\and A.~M.~C.~Le~Brun\orcid{0000-0002-0936-4594}\inst{\ref{aff41},\ref{aff2}}
\and V.~Le~Brun\orcid{0000-0002-5027-1939}\inst{\ref{aff20}}
\and F.~Leclercq\orcid{0000-0002-9339-1404}\inst{\ref{aff2}}
\and S.~Lee\orcid{0000-0002-8289-740X}\inst{\ref{aff112}}
\and J.~Le~Graet\orcid{0000-0001-6523-7971}\inst{\ref{aff52}}
\and L.~Legrand\orcid{0000-0003-0610-5252}\inst{\ref{aff274}}
\and K.~N.~Leirvik\inst{\ref{aff206}}
\and M.~Le~Jeune\orcid{0000-0002-1008-3394}\inst{\ref{aff10}}
\and M.~Lembo\orcid{0000-0002-5271-5070}\inst{\ref{aff29},\ref{aff42}}
\and D.~Le~Mignant\orcid{0000-0002-5339-5515}\inst{\ref{aff20}}
\and M.~D.~Lepinzan\orcid{0000-0003-1287-9801}\inst{\ref{aff113},\ref{aff59}}
\and F.~Lepori\inst{\ref{aff7}}
\and A.~Le~Reun\inst{\ref{aff1}}
\and G.~Leroy\orcid{0009-0004-2523-4425}\inst{\ref{aff248},\ref{aff87}}
\and G.~F.~Lesci\orcid{0000-0002-4607-2830}\inst{\ref{aff165},\ref{aff35}}
\and J.~Lesgourgues\orcid{0000-0001-7627-353X}\inst{\ref{aff154}}
\and L.~Leuzzi\inst{\ref{aff165},\ref{aff35}}
\and M.~E.~Levi\orcid{0000-0003-1887-1018}\inst{\ref{aff275}}
\and T.~I.~Liaudat\orcid{0000-0002-9104-314X}\inst{\ref{aff45}}
\and G.~Libet\inst{\ref{aff83}}
\and P.~Liebing\inst{\ref{aff16}}
\and S.~Ligori\orcid{0000-0003-4172-4606}\inst{\ref{aff110}}
\and P.~B.~Lilje\orcid{0000-0003-4324-7794}\inst{\ref{aff206}}
\and C.-C.~Lin\orcid{0000-0002-7272-5129}\inst{\ref{aff158}}
\and D.~Linde\inst{\ref{aff181}}
\and E.~Linder\orcid{0000-0001-5536-9241}\inst{\ref{aff276}}
\and V.~Lindholm\orcid{0000-0003-2317-5471}\inst{\ref{aff207},\ref{aff268}}
\and L.~Linke\orcid{0000-0002-2622-8113}\inst{\ref{aff174}}
\and S.-S.~Li\orcid{0000-0001-9952-7408}\inst{\ref{aff70}}
\and S.~J.~Liu\orcid{0000-0001-7680-2139}\inst{\ref{aff195}}
\and I.~Lloro\inst{\ref{aff277}}
\and F.~S.~N.~Lobo\orcid{0000-0002-9388-8373}\inst{\ref{aff153},\ref{aff18}}
\and N.~Lodieu\orcid{0000-0002-3612-8968}\inst{\ref{aff67},\ref{aff182}}
\and M.~Lombardi\orcid{0000-0002-3336-4965}\inst{\ref{aff43}}
\and L.~Lombriser\inst{\ref{aff111}}
\and P.~Lonare\orcid{0009-0000-0028-0493}\inst{\ref{aff177},\ref{aff145}}
\and G.~Longo\orcid{0000-0002-9182-8414}\inst{\ref{aff131},\ref{aff117}}
\and M.~L\'opez-Caniego\orcid{0000-0003-1016-9283}\inst{\ref{aff278},\ref{aff279}}
\and X.~Lopez~Lopez\inst{\ref{aff72},\ref{aff280}}
\and J.~Lorenzo~Alvarez\orcid{0000-0002-6845-993X}\inst{\ref{aff104}}
\and A.~Loureiro\orcid{0000-0002-4371-0876}\inst{\ref{aff281},\ref{aff163}}
\and J.~Loveday\orcid{0000-0001-5290-8940}\inst{\ref{aff264}}
\and E.~Lusso\orcid{0000-0003-0083-1157}\inst{\ref{aff173},\ref{aff214}}
\and J.~Macias-Perez\orcid{0000-0002-5385-2763}\inst{\ref{aff19}}
\and T.~Maciaszek\inst{\ref{aff83}}
\and G.~Maggio\orcid{0000-0003-4020-4836}\inst{\ref{aff59}}
\and M.~Magliocchetti\orcid{0000-0001-9158-4838}\inst{\ref{aff195}}
\and F.~Magnard\orcid{0009-0005-5087-8270}\inst{\ref{aff2}}
\and E.~A.~Magnier\orcid{0000-0002-7965-2815}\inst{\ref{aff158}}
\and A.~Magro\orcid{0000-0002-8935-0618}\inst{\ref{aff282}}
\and G.~Mahler\orcid{0000-0003-3266-2001}\inst{\ref{aff283},\ref{aff248},\ref{aff87}}
\and G.~Mainetti\inst{\ref{aff273}}
\and D.~Maino\inst{\ref{aff43},\ref{aff108},\ref{aff44}}
\and E.~Maiorano\orcid{0000-0003-2593-4355}\inst{\ref{aff35}}
\and E.~Maiorano\inst{\ref{aff104}}
\and N.~Malavasi\orcid{0000-0001-9033-7958}\inst{\ref{aff93}}
\and G.~A.~Mamon\orcid{0000-0001-8956-5953}\inst{\ref{aff2},\ref{aff1}}
\and C.~Mancini\orcid{0000-0002-4297-0561}\inst{\ref{aff108}}
\and R.~Mandelbaum\orcid{0000-0003-2271-1527}\inst{\ref{aff284}}
\and M.~Manera\orcid{0000-0003-4962-8934}\inst{\ref{aff57},\ref{aff285}}
\and A.~Manj\'on-Garc\'ia\orcid{0000-0002-7413-8825}\inst{\ref{aff193}}
\and F.~Mannucci\orcid{0000-0002-4803-2381}\inst{\ref{aff173}}
\and O.~Mansutti\orcid{0000-0001-5758-4658}\inst{\ref{aff59}}
\and M.~Manteiga~Outeiro\orcid{0000-0002-7711-5581}\inst{\ref{aff178}}
\and R.~Maoli\orcid{0000-0002-6065-3025}\inst{\ref{aff188},\ref{aff142}}
\and C.~Maraston\orcid{0000-0001-7711-3677}\inst{\ref{aff63}}
\and S.~Marcin\inst{\ref{aff232}}
\and P.~Marcos-Arenal\orcid{0000-0003-1549-9396}\inst{\ref{aff286}}
\and B.~Margalef-Bentabol\orcid{0000-0001-8702-7019}\inst{\ref{aff269}}
\and O.~Marggraf\orcid{0000-0001-7242-3852}\inst{\ref{aff138}}
\and D.~Marinucci\inst{\ref{aff177}}
\and M.~Marinucci\orcid{0000-0003-1159-3756}\inst{\ref{aff190},\ref{aff40},\ref{aff39}}
\and K.~Markovic\orcid{0000-0001-6764-073X}\inst{\ref{aff112}}
\and F.~R.~Marleau\inst{\ref{aff174}}
\and J.~Marpaud\inst{\ref{aff19}}
\and J.~Martignac\inst{\ref{aff11}}
\and J.~Mart\'{i}n-Fleitas\orcid{0000-0002-8594-569X}\inst{\ref{aff278}}
\and P.~Martin-Moruno\orcid{0000-0001-8073-4896}\inst{\ref{aff48}}
\and E.~L.~Martin\orcid{0000-0002-1208-4833}\inst{\ref{aff182}}
\and M.~Martinelli\orcid{0000-0002-6943-7732}\inst{\ref{aff142},\ref{aff148}}
\and N.~Martinet\orcid{0000-0003-2786-7790}\inst{\ref{aff20}}
\and H.~Martin\orcid{0009-0009-1137-5880}\inst{\ref{aff77},\ref{aff76}}
\and C.~J.~A.~P.~Martins\orcid{0000-0002-4886-9261}\inst{\ref{aff192},\ref{aff55}}
\and F.~Marulli\orcid{0000-0002-8850-0303}\inst{\ref{aff165},\ref{aff35},\ref{aff73}}
\and D.~Massari\orcid{0000-0001-8892-4301}\inst{\ref{aff35}}
\and R.~Massey\orcid{0000-0002-6085-3780}\inst{\ref{aff248},\ref{aff87}}
\and D.~C.~Masters\orcid{0000-0001-5382-6138}\inst{\ref{aff17}}
\and S.~Matarrese\orcid{0000-0002-2573-1243}\inst{\ref{aff40},\ref{aff75},\ref{aff39},\ref{aff287}}
\and Y.~Matsuoka\orcid{0000-0001-5063-0340}\inst{\ref{aff288}}
\and S.~Matthew\orcid{0000-0001-8448-1697}\inst{\ref{aff98}}
\and B.~J.~Maughan\orcid{0000-0003-0791-9098}\inst{\ref{aff130}}
\and N.~Mauri\orcid{0000-0001-8196-1548}\inst{\ref{aff160},\ref{aff73}}
\and L.~Maurin\orcid{0000-0002-8406-0857}\inst{\ref{aff9}}
\and S.~Maurogordato\inst{\ref{aff8}}
\and K.~McCarthy\orcid{0000-0001-6857-018X}\inst{\ref{aff112}}
\and A.~W.~McConnachie\orcid{0000-0003-4666-6564}\inst{\ref{aff224}}
\and H.~J.~McCracken\orcid{0000-0002-9489-7765}\inst{\ref{aff2}}
\and I.~McDonald\orcid{0000-0003-0356-0655}\inst{\ref{aff122}}
\and J.~D.~McEwen\inst{\ref{aff16}}
\and C.~J.~R.~McPartland\orcid{0000-0003-0639-025X}\inst{\ref{aff155},\ref{aff289}}
\and E.~Medinaceli\orcid{0000-0002-4040-7783}\inst{\ref{aff35}}
\and V.~Mehta\orcid{0000-0001-7166-6035}\inst{\ref{aff62}}
\and S.~Mei\orcid{0000-0002-2849-559X}\inst{\ref{aff10}}
\and M.~Melchior\inst{\ref{aff232}}
\and J.-B.~Melin\inst{\ref{aff290}}
\and B.~M\'enard\orcid{0000-0003-3164-6974}\inst{\ref{aff3}}
\and J.~Mendes\inst{\ref{aff171}}
\and J.~Mendez-Abreu\orcid{0000-0002-8766-2597}\inst{\ref{aff68},\ref{aff67}}
\and M.~Meneghetti\orcid{0000-0003-1225-7084}\inst{\ref{aff35},\ref{aff73}}
\and A.~Mercurio\orcid{0000-0001-9261-7849}\inst{\ref{aff24},\ref{aff123},\ref{aff124}}
\and E.~Merlin\orcid{0000-0001-6870-8900}\inst{\ref{aff142}}
\and R.~B.~Metcalf\orcid{0000-0003-3167-2574}\inst{\ref{aff165},\ref{aff35}}
\and G.~Meylan\inst{\ref{aff4}}
\and M.~Migliaccio\inst{\ref{aff177},\ref{aff94}}
\and M.~Mignoli\orcid{0000-0002-9087-2835}\inst{\ref{aff35}}
\and L.~Miller\orcid{0000-0002-3376-6200}\inst{\ref{aff172}}
\and M.~Miluzio\inst{\ref{aff26},\ref{aff286}}
\and B.~Milvang-Jensen\orcid{0000-0002-2281-2785}\inst{\ref{aff147},\ref{aff289},\ref{aff155}}
\and J.~P.~Mimoso\orcid{0000-0002-9758-3366}\inst{\ref{aff153},\ref{aff18}}
\and R.~Miquel\orcid{0000-0002-6610-4836}\inst{\ref{aff57},\ref{aff291}}
\and H.~Miyatake\orcid{0000-0001-7964-9766}\inst{\ref{aff292},\ref{aff293},\ref{aff103}}
\and B.~Mobasher\orcid{0000-0001-5846-4404}\inst{\ref{aff294}}
\and J.~J.~Mohr\orcid{0000-0002-6875-2087}\inst{\ref{aff82},\ref{aff93}}
\and P.~Monaco\orcid{0000-0003-2083-7564}\inst{\ref{aff113},\ref{aff59},\ref{aff60},\ref{aff58}}
\and M.~Mongui\'{o}\orcid{0000-0002-4519-6700}\inst{\ref{aff150},\ref{aff295}}
\and A.~Montoro\orcid{0000-0003-4730-8590}\inst{\ref{aff132},\ref{aff143}}
\and A.~Mora\orcid{0000-0002-1922-8529}\inst{\ref{aff278}}
\and A.~Moradinezhad~Dizgah\orcid{0000-0001-8841-9989}\inst{\ref{aff296},\ref{aff111}}
\and M.~Moresco\orcid{0000-0002-7616-7136}\inst{\ref{aff165},\ref{aff35}}
\and C.~Moretti\orcid{0000-0003-3314-8936}\inst{\ref{aff61},\ref{aff157},\ref{aff59},\ref{aff58},\ref{aff60}}
\and G.~Morgante\inst{\ref{aff35}}
\and N.~Morisset\inst{\ref{aff125}}
\and T.~J.~Moriya\orcid{0000-0003-1169-1954}\inst{\ref{aff297},\ref{aff298}}
\and P.~W.~Morris\orcid{0000-0002-5186-4381}\inst{\ref{aff198}}
\and D.~J.~Mortlock\inst{\ref{aff163},\ref{aff299},\ref{aff247}}
\and L.~Moscardini\orcid{0000-0002-3473-6716}\inst{\ref{aff165},\ref{aff35},\ref{aff73}}
\and D.~F.~Mota\orcid{0000-0003-3141-142X}\inst{\ref{aff206}}
\and S.~Mottet\inst{\ref{aff2}}
\and L.~A.~Moustakas\orcid{0000-0003-3030-2360}\inst{\ref{aff112}}
\and T.~Moutard\orcid{0000-0002-3305-9901}\inst{\ref{aff26}}
\and T.~M\"{u}ller\orcid{0000-0002-0717-0462}\inst{\ref{aff93}}
\and E.~Munari\orcid{0000-0002-1751-5946}\inst{\ref{aff59},\ref{aff58}}
\and G.~Murphree\orcid{0009-0007-7266-8914}\inst{\ref{aff158}}
\and C.~Murray\inst{\ref{aff10}}
\and N.~Murray\inst{\ref{aff226}}
\and P.~Musi\inst{\ref{aff38}}
\and S.~Nadathur\orcid{0000-0001-9070-3102}\inst{\ref{aff63}}
\and B.~C.~Nagam\orcid{0000-0002-3724-7694}\inst{\ref{aff69}}
\and T.~Nagao\orcid{0000-0002-7402-5441}\inst{\ref{aff288}}
\and K.~Naidoo\orcid{0000-0002-9182-1802}\inst{\ref{aff205}}
\and R.~Nakajima\inst{\ref{aff138}}
\and C.~Nally\orcid{0000-0002-7512-1662}\inst{\ref{aff98}}
\and P.~Natoli\inst{\ref{aff29},\ref{aff42}}
\and A.~Navarro-Alsina\orcid{0000-0002-3173-2592}\inst{\ref{aff138}}
\and D.~Navarro~Girones\inst{\ref{aff132},\ref{aff143}}
\and C.~Neissner\orcid{0000-0001-8524-4968}\inst{\ref{aff57},\ref{aff140}}
\and A.~Nersesian\orcid{0000-0001-6843-409X}\inst{\ref{aff283},\ref{aff64}}
\and S.~Nesseris\orcid{0000-0002-0567-0324}\inst{\ref{aff14}}
\and H.~N.~Nguyen-Kim\inst{\ref{aff2}}
\and L.~Nicastro\orcid{0000-0001-8534-6788}\inst{\ref{aff35}}
\and R.~C.~Nichol\inst{\ref{aff30}}
\and M.~Nielbock\orcid{0000-0002-1271-1825}\inst{\ref{aff78}}
\and S.-M.~Niemi\inst{\ref{aff104}}
\and S.~Nieto\orcid{0009-0008-9315-5832}\inst{\ref{aff26}}
\and K.~Nilsson\inst{\ref{aff300}}
\and J.~Noller\orcid{0000-0003-2210-775X}\inst{\ref{aff205},\ref{aff63}}
\and P.~Norberg\orcid{0000-0002-5875-0440}\inst{\ref{aff248},\ref{aff87}}
\and A.~Nouri-Zonoz\orcid{0009-0006-6164-8670}\inst{\ref{aff111}}
\and P.~Ntelis\orcid{0000-0002-7849-2418}\inst{\ref{aff52}}
\and A.~A.~Nucita\inst{\ref{aff185},\ref{aff186},\ref{aff187}}
\and P.~Nugent\orcid{0000-0002-3389-0586}\inst{\ref{aff275}}
\and N.~J.~Nunes\orcid{0000-0002-3837-6914}\inst{\ref{aff153},\ref{aff18}}
\and T.~Nutma\inst{\ref{aff69},\ref{aff70}}
\and I.~Ocampo\orcid{0000-0001-7709-1930}\inst{\ref{aff14}}
\and J.~Odier\orcid{0000-0002-1650-2246}\inst{\ref{aff19}}
\and P.~A.~Oesch\orcid{0000-0001-5851-6649}\inst{\ref{aff125},\ref{aff289},\ref{aff147}}
\and M.~Oguri\orcid{0000-0003-3484-399X}\inst{\ref{aff301},\ref{aff302}}
\and D.~Magalhaes~Oliveira\orcid{0000-0003-1540-238X}\inst{\ref{aff18},\ref{aff153}}
\and M.~Onoue\orcid{0000-0003-2984-6803}\inst{\ref{aff103}}
\and T.~Oosterbroek\inst{\ref{aff104}}
\and F.~Oppizzi\orcid{0000-0003-3904-8370}\inst{\ref{aff39},\ref{aff40}}
\and C.~Ordenovic\orcid{0000-0003-0256-6596}\inst{\ref{aff8}}
\and K.~Osato\orcid{0000-0002-7934-2569}\inst{\ref{aff301},\ref{aff103}}
\and F.~Pacaud\orcid{0000-0002-6622-4555}\inst{\ref{aff138}}
\and F.~Pace\orcid{0000-0001-8039-0480}\inst{\ref{aff144},\ref{aff13},\ref{aff110}}
\and C.~Padilla\orcid{0000-0001-7951-0166}\inst{\ref{aff57}}
\and K.~Paech\orcid{0000-0003-0625-2367}\inst{\ref{aff93}}
\and L.~Pagano\orcid{0000-0003-1820-5998}\inst{\ref{aff29},\ref{aff42}}
\and M.~J.~Page\orcid{0000-0002-6689-6271}\inst{\ref{aff16}}
\and E.~Palazzi\orcid{0000-0002-8691-7666}\inst{\ref{aff35}}
\and S.~Paltani\orcid{0000-0002-8108-9179}\inst{\ref{aff125}}
\and S.~Pamuk\orcid{0009-0004-0852-8624}\inst{\ref{aff154}}
\and S.~Pandolfi\inst{\ref{aff104}}
\and D.~Paoletti\orcid{0000-0003-4761-6147}\inst{\ref{aff35},\ref{aff139}}
\and M.~Paolillo\orcid{0000-0003-4210-7693}\inst{\ref{aff303},\ref{aff24},\ref{aff117}}
\and P.~Papaderos\orcid{0000-0002-3733-8174}\inst{\ref{aff55}}
\and K.~Pardede\orcid{0000-0002-7728-8220}\inst{\ref{aff181}}
\and G.~Parimbelli\orcid{0000-0002-2539-2472}\inst{\ref{aff132},\ref{aff304},\ref{aff61}}
\and A.~Parmar\orcid{0009-0009-9390-9232}\inst{\ref{aff163}}
\and C.~Partmann\orcid{0009-0003-9985-0012}\inst{\ref{aff32}}
\and F.~Pasian\orcid{0000-0002-4869-3227}\inst{\ref{aff59}}
\and F.~Passalacqua\inst{\ref{aff40},\ref{aff39}}
\and K.~Paterson\orcid{0000-0001-8340-3486}\inst{\ref{aff78}}
\and L.~Patrizii\inst{\ref{aff73}}
\and C.~Pattison\inst{\ref{aff63}}
\and A.~Paulino-Afonso\orcid{0000-0002-0943-0694}\inst{\ref{aff192},\ref{aff55}}
\and R.~Paviot\orcid{0009-0002-8108-3460}\inst{\ref{aff11}}
\and J.~A.~Peacock\orcid{0000-0002-1168-8299}\inst{\ref{aff98}}
\and F.~R.~Pearce\orcid{0000-0002-2383-9250}\inst{\ref{aff56}}
\and K.~Pedersen\inst{\ref{aff305}}
\and A.~Peel\orcid{0000-0003-0488-8978}\inst{\ref{aff4}}
\and R.~F.~Peletier\orcid{0000-0001-7621-947X}\inst{\ref{aff69}}
\and M.~Pellejero~Ibanez\orcid{0000-0003-4680-7275}\inst{\ref{aff98},\ref{aff36}}
\and R.~Pello\orcid{0000-0003-0858-6109}\inst{\ref{aff20}}
\and M.~T.~Penny\orcid{0000-0001-7506-5640}\inst{\ref{aff306}}
\and W.~J.~Percival\orcid{0000-0002-0644-5727}\inst{\ref{aff77},\ref{aff76},\ref{aff109}}
\and A.~Perez-Garrido\orcid{0000-0002-5139-1975}\inst{\ref{aff193}}
\and L.~Perotto\orcid{0000-0001-6937-5052}\inst{\ref{aff19}}
\and V.~Pettorino\inst{\ref{aff104}}
\and A.~Pezzotta\orcid{0000-0003-0726-2268}\inst{\ref{aff93}}
\and S.~Pezzuto\orcid{0000-0001-7852-1971}\inst{\ref{aff195}}
\and A.~Philippon\inst{\ref{aff9}}
\and M.~Pierre\orcid{0000-0003-2648-2469}\inst{\ref{aff11}}
\and O.~Piersanti\inst{\ref{aff104}}
\and M.~Pietroni\orcid{0000-0001-5480-5996}\inst{\ref{aff180},\ref{aff181}}
\and L.~Piga\orcid{0000-0003-2221-7406}\inst{\ref{aff180},\ref{aff181},\ref{aff108}}
\and L.~Pilo\orcid{0000-0003-3554-2427}\inst{\ref{aff307}}
\and S.~Pires\orcid{0000-0002-0249-2104}\inst{\ref{aff11}}
\and A.~Pisani\orcid{0000-0002-6146-4437}\inst{\ref{aff52},\ref{aff308},\ref{aff204},\ref{aff309}}
\and A.~Pizzella\orcid{0000-0001-9585-417X}\inst{\ref{aff40},\ref{aff75}}
\and L.~Pizzuti\orcid{0000-0001-5654-7580}\inst{\ref{aff209}}
\and C.~Plana\inst{\ref{aff16}}
\and G.~Polenta\orcid{0000-0003-4067-9196}\inst{\ref{aff107}}
\and J.~E.~Pollack\inst{\ref{aff213},\ref{aff10}}
\and M.~Poncet\inst{\ref{aff83}}
\and M.~P\"ontinen\orcid{0000-0001-5442-2530}\inst{\ref{aff207}}
\and P.~Pool\inst{\ref{aff310}}
\and L.~A.~Popa\inst{\ref{aff311}}
\and V.~Popa\inst{\ref{aff311}}
\and J.~Popp\orcid{0000-0002-3724-1727}\inst{\ref{aff200}}
\and C.~Porciani\orcid{0000-0002-7797-2508}\inst{\ref{aff138}}
\and L.~Porth\orcid{0000-0003-1176-6346}\inst{\ref{aff138}}
\and D.~Potter\orcid{0000-0002-0757-5195}\inst{\ref{aff7}}
\and M.~Poulain\orcid{0000-0002-7664-4510}\inst{\ref{aff312}}
\and A.~Pourtsidou\orcid{0000-0001-9110-5550}\inst{\ref{aff98},\ref{aff313}}
\and L.~Pozzetti\orcid{0000-0001-7085-0412}\inst{\ref{aff35}}
\and I.~Prandoni\orcid{0000-0001-9680-7092}\inst{\ref{aff101}}
\and G.~W.~Pratt\inst{\ref{aff11}}
\and S.~Prezelus\inst{\ref{aff104}}
\and E.~Prieto\inst{\ref{aff20}}
\and A.~Pugno\inst{\ref{aff138}}
\and S.~Quai\orcid{0000-0002-0449-8163}\inst{\ref{aff165},\ref{aff35}}
\and L.~Quilley\orcid{0009-0008-8375-8605}\inst{\ref{aff314}}
\and G.~D.~Racca\inst{\ref{aff104}}
\and A.~Raccanelli\orcid{0000-0001-6726-0438}\inst{\ref{aff40},\ref{aff39},\ref{aff75},\ref{aff315}}
\and G.~R\'acz\orcid{0000-0003-3906-5699}\inst{\ref{aff112}}
\and S.~Radinovi\'c\inst{\ref{aff132},\ref{aff206}}
\and M.~Radovich\orcid{0000-0002-3585-866X}\inst{\ref{aff75}}
\and A.~Ragagnin\orcid{0000-0002-8106-2742}\inst{\ref{aff35},\ref{aff58},\ref{aff165},\ref{aff157}}
\and U.~Ragnit\inst{\ref{aff104}}
\and F.~Raison\orcid{0000-0002-7819-6918}\inst{\ref{aff93}}
\and N.~Ramos-Chernenko\inst{\ref{aff67},\ref{aff182}}
\and C.~Ranc\orcid{0000-0003-2388-4534}\inst{\ref{aff2}}
\and Y.~Rasera\orcid{0000-0003-3424-6941}\inst{\ref{aff41},\ref{aff121}}
\and N.~Raylet\inst{\ref{aff99}}
\and R.~Rebolo\inst{\ref{aff67},\ref{aff68}}
\and A.~Refregier\inst{\ref{aff12}}
\and P.~Reimberg\orcid{0000-0003-3410-0280}\inst{\ref{aff1}}
\and T.~H.~Reiprich\orcid{0000-0003-2047-2884}\inst{\ref{aff138}}
\and F.~Renk\inst{\ref{aff171}}
\and A.~Renzi\orcid{0000-0001-9856-1970}\inst{\ref{aff40},\ref{aff39}}
\and J.~Retre\orcid{0000-0002-8394-129X}\inst{\ref{aff137}}
\and Y.~Revaz\orcid{0000-0002-6227-0108}\inst{\ref{aff4}}
\and C.~Reyl\'e\orcid{0000-0003-2258-2403}\inst{\ref{aff316}}
\and L.~Reynolds\orcid{0000-0002-4732-3718}\inst{\ref{aff57}}
\and J.~Rhodes\orcid{0000-0002-4485-8549}\inst{\ref{aff112}}
\and F.~Ricci\orcid{0000-0001-5742-5980}\inst{\ref{aff183},\ref{aff142}}
\and M.~Ricci\orcid{0000-0002-3645-9652}\inst{\ref{aff8},\ref{aff10}}
\and G.~Riccio\inst{\ref{aff24}}
\and S.~O.~Ricken\inst{\ref{aff171}}
\and S.~Rissanen\inst{\ref{aff207}}
\and I.~Risso\orcid{0000-0003-2525-7761}\inst{\ref{aff304}}
\and H.-W.~Rix\orcid{0000-0003-4996-9069}\inst{\ref{aff78}}
\and A.~C.~Robin\orcid{0000-0001-8654-9499}\inst{\ref{aff316}}
\and B.~Rocca-Volmerange\inst{\ref{aff1},\ref{aff2}}
\and P.-F.~Rocci\inst{\ref{aff9}}
\and M.~Rodenhuis\inst{\ref{aff317}}
\and G.~Rodighiero\orcid{0000-0002-9415-2296}\inst{\ref{aff40},\ref{aff75}}
\and M.~Rodriguez~Monroy\orcid{0000-0001-6163-1058}\inst{\ref{aff14}}
\and R.~P.~Rollins\orcid{0000-0003-1291-1023}\inst{\ref{aff98}}
\and M.~Romanello\orcid{0000-0003-4563-4923}\inst{\ref{aff72},\ref{aff35}}
\and J.~Roman\orcid{0000-0002-3849-3467}\inst{\ref{aff68},\ref{aff67}}
\and E.~Romelli\orcid{0000-0003-3069-9222}\inst{\ref{aff59}}
\and M.~Romero-Gomez\orcid{0000-0003-3936-1025}\inst{\ref{aff150},\ref{aff143},\ref{aff151}}
\and M.~Roncarelli\orcid{0000-0001-9587-7822}\inst{\ref{aff35}}
\and P.~Rosati\orcid{0000-0002-6813-0632}\inst{\ref{aff29},\ref{aff35}}
\and C.~Rosset\orcid{0000-0003-0286-2192}\inst{\ref{aff10}}
\and E.~Rossetti\orcid{0000-0003-0238-4047}\inst{\ref{aff72}}
\and W.~Roster\orcid{0000-0002-9149-6528}\inst{\ref{aff93},\ref{aff235}}
\and H.~J.~A.~Rottgering\orcid{0000-0001-8887-2257}\inst{\ref{aff70}}
\and A.~Rozas-Fern\'andez\orcid{0000-0002-6131-2804}\inst{\ref{aff18}}
\and K.~Ruane\inst{\ref{aff16}}
\and J.~A.~Rubino-Martin\orcid{0000-0001-5289-3021}\inst{\ref{aff67},\ref{aff68}}
\and A.~Rudolph\inst{\ref{aff171}}
\and F.~Ruppin\orcid{0000-0002-0955-8954}\inst{\ref{aff51}}
\and B.~Rusholme\orcid{0000-0001-7648-4142}\inst{\ref{aff62}}
\and S.~Sacquegna\orcid{0000-0002-8433-6630}\inst{\ref{aff185},\ref{aff186},\ref{aff187}}
\and I.~S\'aez-Casares\orcid{0000-0003-0013-5266}\inst{\ref{aff41}}
\and S.~Saga\orcid{0000-0002-7387-7570}\inst{\ref{aff292}}
\and R.~Saglia\orcid{0000-0003-0378-7032}\inst{\ref{aff82},\ref{aff93}}
\and M.~Sahl\'en\orcid{0000-0003-0973-4804}\inst{\ref{aff318}}
\and T.~Saifollahi\orcid{0000-0002-9554-7660}\inst{\ref{aff97},\ref{aff69}}
\and Z.~Sakr\orcid{0000-0002-4823-3757}\inst{\ref{aff31},\ref{aff92},\ref{aff319}}
\and J.~Salvalaggio\orcid{0000-0002-1431-5607}\inst{\ref{aff113},\ref{aff59},\ref{aff58},\ref{aff60}}
\and R.~Salvaterra\orcid{0000-0002-9393-8078}\inst{\ref{aff108}}
\and L.~Salvati\inst{\ref{aff9}}
\and M.~Salvato\orcid{0000-0001-7116-9303}\inst{\ref{aff93}}
\and J.-C.~Salvignol\inst{\ref{aff104}}
\and A.~G.~S\'anchez\orcid{0000-0003-1198-831X}\inst{\ref{aff93}}
\and E.~Sanchez\orcid{0000-0002-9646-8198}\inst{\ref{aff152}}
\and D.~B.~Sanders\orcid{0000-0002-1233-9998}\inst{\ref{aff158}}
\and D.~Sapone\orcid{0000-0001-7089-4503}\inst{\ref{aff320}}
\and M.~Saponara\inst{\ref{aff38}}
\and E.~Sarpa\orcid{0000-0002-1256-655X}\inst{\ref{aff61},\ref{aff157},\ref{aff60}}
\and F.~Sarron\orcid{0000-0001-8376-0360}\inst{\ref{aff321},\ref{aff92}}
\and S.~Sartori\orcid{0009-0000-5585-8336}\inst{\ref{aff52}}
\and B.~Sartoris\orcid{0000-0003-1337-5269}\inst{\ref{aff82},\ref{aff59}}
\and B.~Sassolas\orcid{0000-0002-3077-8951}\inst{\ref{aff51}}
\and L.~Sauniere\inst{\ref{aff52}}
\and M.~Sauvage\orcid{0000-0002-0809-2574}\inst{\ref{aff11}}
\and M.~Sawicki\orcid{0000-0002-7712-7857}\inst{\ref{aff191}}
\and R.~Scaramella\orcid{0000-0003-2229-193X}\inst{\ref{aff142},\ref{aff148}}
\and C.~Scarlata\orcid{0000-0002-9136-8876}\inst{\ref{aff135}}
\and L.~Scharr\'e\orcid{0000-0003-2551-4430}\inst{\ref{aff236}}
\and J.~Schaye\orcid{0000-0002-0668-5560}\inst{\ref{aff70}}
\and J.~A.~Schewtschenko\inst{\ref{aff98}}
\and J.-T.~Schindler\orcid{0000-0002-4544-8242}\inst{\ref{aff322}}
\and E.~Schinnerer\orcid{0000-0002-3933-7677}\inst{\ref{aff78}}
\and M.~Schirmer\orcid{0000-0003-2568-9994}\inst{\ref{aff78}}
\and F.~Schmidt\orcid{0000-0002-6807-7464}\inst{\ref{aff32},\ref{aff227}}
\and F.~Schmidt\inst{\ref{aff171}}
\and M.~Schmidt\inst{\ref{aff171}}
\and A.~Schneider\orcid{0000-0001-7055-8104}\inst{\ref{aff7}}
\and M.~Schneider\inst{\ref{aff171}}
\and P.~Schneider\orcid{0000-0001-8561-2679}\inst{\ref{aff138}}
\and N.~Sch\"oneberg\orcid{0000-0002-7873-0404}\inst{\ref{aff150}}
\and T.~Schrabback\orcid{0000-0002-6987-7834}\inst{\ref{aff174}}
\and M.~Schultheis\inst{\ref{aff8}}
\and S.~Schulz\orcid{0000-0002-8235-9986}\inst{\ref{aff7}}
\and N.~Schuster\inst{\ref{aff82}}
\and J.~Schwartz\inst{\ref{aff171}}
\and D.~Sciotti\orcid{0009-0008-4519-2620}\inst{\ref{aff142},\ref{aff148}}
\and M.~Scodeggio\inst{\ref{aff108}}
\and D.~Scognamiglio\orcid{0000-0001-8450-7885}\inst{\ref{aff112}}
\and D.~Scott\orcid{0000-0002-6878-9840}\inst{\ref{aff323}}
\and V.~Scottez\inst{\ref{aff1},\ref{aff324}}
\and A.~Secroun\orcid{0000-0003-0505-3710}\inst{\ref{aff52}}
\and E.~Sefusatti\orcid{0000-0003-0473-1567}\inst{\ref{aff59},\ref{aff58},\ref{aff60}}
\and G.~Seidel\orcid{0000-0003-2907-353X}\inst{\ref{aff78}}
\and M.~Seiffert\orcid{0000-0002-7536-9393}\inst{\ref{aff112}}
\and E.~Sellentin\inst{\ref{aff325},\ref{aff70}}
\and M.~Selwood\inst{\ref{aff130}}
\and E.~Semboloni\inst{\ref{aff5}}
\and M.~Sereno\orcid{0000-0003-0302-0325}\inst{\ref{aff35},\ref{aff73}}
\and S.~Serjeant\orcid{0000-0002-0517-7943}\inst{\ref{aff200}}
\and S.~Serrano\orcid{0000-0002-0211-2861}\inst{\ref{aff143},\ref{aff326},\ref{aff132}}
\and G.~Setnikar\inst{\ref{aff51}}
\and F.~Shankar\orcid{0000-0001-8973-5051}\inst{\ref{aff327}}
\and R.~M.~Sharples\orcid{0000-0003-3449-8583}\inst{\ref{aff248}}
\and A.~Short\inst{\ref{aff104}}
\and A.~Shulevski\orcid{0000-0002-1827-0469}\inst{\ref{aff328},\ref{aff69},\ref{aff329}}
\and M.~Shuntov\orcid{0000-0002-7087-0701}\inst{\ref{aff1},\ref{aff155},\ref{aff289}}
\and M.~Sias\inst{\ref{aff38}}
\and G.~Sikkema\inst{\ref{aff69}}
\and A.~Silvestri\orcid{0000-0001-6904-5061}\inst{\ref{aff5}}
\and P.~Simon\inst{\ref{aff138}}
\and C.~Sirignano\orcid{0000-0002-0995-7146}\inst{\ref{aff40},\ref{aff39}}
\and G.~Sirri\orcid{0000-0003-2626-2853}\inst{\ref{aff73}}
\and J.~Skottfelt\orcid{0000-0003-1310-8283}\inst{\ref{aff226}}
\and E.~Slezak\orcid{0000-0003-4771-7263}\inst{\ref{aff8}}
\and D.~Sluse\orcid{0000-0001-6116-2095}\inst{\ref{aff283}}
\and G.~P.~Smith\orcid{0000-0003-4494-8277}\inst{\ref{aff330}}
\and L.~C.~Smith\orcid{0000-0002-3259-2771}\inst{\ref{aff233}}
\and R.~E.~Smith\orcid{0000-0001-9989-2149}\inst{\ref{aff264}}
\and S.~J.~A.~Smit\inst{\ref{aff16}}
\and F.~Soldano\inst{\ref{aff2}}
\and B.~G.~B.~Solheim\orcid{0009-0008-2307-2978}\inst{\ref{aff331}}
\and J.~G.~Sorce\orcid{0000-0002-2307-2432}\inst{\ref{aff332},\ref{aff9},\ref{aff161}}
\and F.~Sorrenti\orcid{0000-0001-7141-9659}\inst{\ref{aff111}}
\and E.~Soubrie\orcid{0000-0001-9295-1863}\inst{\ref{aff9}}
\and L.~Spinoglio\orcid{0000-0001-8840-1551}\inst{\ref{aff195}}
\and A.~Spurio~Mancini\orcid{0000-0001-5698-0990}\inst{\ref{aff333},\ref{aff16}}
\and J.~Stadel\orcid{0000-0001-7565-8622}\inst{\ref{aff7}}
\and L.~Stagnaro\inst{\ref{aff104}}
\and L.~Stanco\orcid{0000-0002-9706-5104}\inst{\ref{aff39}}
\and S.~A.~Stanford\orcid{0000-0003-0122-0841}\inst{\ref{aff334}}
\and J.-L.~Starck\orcid{0000-0003-2177-7794}\inst{\ref{aff11}}
\and P.~Stassi\orcid{0000-0001-5584-8410}\inst{\ref{aff19}}
\and J.~Steinwagner\inst{\ref{aff93}}
\and D.~Stern\orcid{0000-0003-2686-9241}\inst{\ref{aff112}}
\and C.~Stone\orcid{0000-0002-9086-6398}\inst{\ref{aff335}}
\and P.~Strada\inst{\ref{aff104}}
\and F.~Strafella\orcid{0000-0002-8757-9371}\inst{\ref{aff185},\ref{aff187},\ref{aff186}}
\and D.~Stramaccioni\inst{\ref{aff104}}
\and C.~Surace\orcid{0000-0003-2592-0113}\inst{\ref{aff20}}
\and F.~Sureau\inst{\ref{aff11}}
\and S.~H.~Suyu\orcid{0000-0001-5568-6052}\inst{\ref{aff33},\ref{aff32},\ref{aff336}}
\and I.~Swindells\inst{\ref{aff310}}
\and M.~Szafraniec\inst{\ref{aff104}}
\and I.~Szapudi\orcid{0000-0003-2274-0301}\inst{\ref{aff158}}
\and S.~Taamoli\orcid{0000-0003-0749-4667}\inst{\ref{aff294}}
\and M.~Talia\orcid{0000-0003-4352-2063}\inst{\ref{aff165},\ref{aff35}}
\and P.~Tallada-Cresp\'{i}\orcid{0000-0002-1336-8328}\inst{\ref{aff152},\ref{aff140}}
\and K.~Tanidis\inst{\ref{aff172}}
\and C.~Tao\orcid{0000-0001-7961-8177}\inst{\ref{aff52}}
\and P.~Tarr\'io\orcid{0000-0002-0915-0131}\inst{\ref{aff337},\ref{aff11}}
\and D.~Tavagnacco\orcid{0000-0001-7475-9894}\inst{\ref{aff59}}
\and A.~N.~Taylor\inst{\ref{aff98}}
\and J.~E.~Taylor\orcid{0000-0002-6639-4183}\inst{\ref{aff76},\ref{aff77}}
\and P.~L.~Taylor\inst{\ref{aff338},\ref{aff339}}
\and E.~M.~Teixeira\orcid{0000-0001-7417-0780}\inst{\ref{aff340}}
\and M.~Tenti\orcid{0000-0002-4254-5901}\inst{\ref{aff73}}
\and P.~Teodoro~Idiago\inst{\ref{aff278}}
\and H.~I.~Teplitz\orcid{0000-0002-7064-5424}\inst{\ref{aff17}}
\and I.~Tereno\inst{\ref{aff153},\ref{aff137}}
\and N.~Tessore\orcid{0000-0002-9696-7931}\inst{\ref{aff205}}
\and V.~Testa\orcid{0000-0003-1033-1340}\inst{\ref{aff142}}
\and G.~Testera\inst{\ref{aff127}}
\and M.~Tewes\orcid{0000-0002-1155-8689}\inst{\ref{aff138}}
\and R.~Teyssier\orcid{0000-0001-7689-0933}\inst{\ref{aff308}}
\and N.~Theret\inst{\ref{aff83}}
\and C.~Thizy\inst{\ref{aff341}}
\and P.~D.~Thomas\inst{\ref{aff16}}
\and Y.~Toba\orcid{0000-0002-3531-7863}\inst{\ref{aff342}}
\and S.~Toft\orcid{0000-0003-3631-7176}\inst{\ref{aff155},\ref{aff147},\ref{aff289}}
\and R.~Toledo-Moreo\orcid{0000-0002-2997-4859}\inst{\ref{aff343}}
\and E.~Tolstoy\inst{\ref{aff69}}
\and E.~Tommasi\inst{\ref{aff344}}
\and O.~Torbaniuk\orcid{0000-0003-4465-2564}\inst{\ref{aff165}}
\and F.~Torradeflot\orcid{0000-0003-1160-1517}\inst{\ref{aff140},\ref{aff152}}
\and C.~Tortora\orcid{0000-0001-7958-6531}\inst{\ref{aff24}}
\and S.~Tosi\orcid{0000-0002-7275-9193}\inst{\ref{aff126},\ref{aff127}}
\and S.~Tosti\inst{\ref{aff9}}
\and M.~Trifoglio\orcid{0000-0002-2505-3630}\inst{\ref{aff35}}
\and A.~Troja\orcid{0000-0003-0239-4595}\inst{\ref{aff40},\ref{aff39}}
\and T.~Trombetti\orcid{0000-0001-5166-2467}\inst{\ref{aff101}}
\and A.~Tronconi\orcid{0000-0003-1913-9654}\inst{\ref{aff73},\ref{aff160}}
\and M.~Tsedrik\orcid{0000-0002-0020-5343}\inst{\ref{aff98}}
\and A.~Tsyganov\inst{\ref{aff199}}
\and M.~Tucci\inst{\ref{aff125}}
\and I.~Tutusaus\orcid{0000-0002-3199-0399}\inst{\ref{aff92}}
\and C.~Uhlemann\orcid{0000-0001-7831-1579}\inst{\ref{aff156},\ref{aff47}}
\and L.~Ulivi\orcid{0009-0001-3291-5382}\inst{\ref{aff345},\ref{aff214},\ref{aff173}}
\and M.~Urbano\orcid{0000-0001-5640-0650}\inst{\ref{aff97}}
\and L.~Vacher\orcid{0000-0001-9551-1417}\inst{\ref{aff61}}
\and L.~Vaillon\inst{\ref{aff99}}
\and P.~Valageas\inst{\ref{aff95}}
\and I.~Valdes\orcid{0009-0002-8551-9372}\inst{\ref{aff158}}
\and E.~A.~Valentijn\inst{\ref{aff69}}
\and L.~Valenziano\orcid{0000-0002-1170-0104}\inst{\ref{aff35},\ref{aff139}}
\and C.~Valieri\inst{\ref{aff73}}
\and J.~Valiviita\orcid{0000-0001-6225-3693}\inst{\ref{aff207},\ref{aff268}}
\and M.~Van~den~Broeck\inst{\ref{aff171},\ref{aff346}}
\and T.~Vassallo\orcid{0000-0001-6512-6358}\inst{\ref{aff82},\ref{aff59}}
\and R.~Vavrek\inst{\ref{aff26}}
\and J.~Vega-Ferrero\orcid{0000-0003-2338-5567}\inst{\ref{aff197},\ref{aff136}}
\and B.~Venemans\orcid{0000-0001-9024-8322}\inst{\ref{aff70}}
\and A.~Venhola\orcid{0000-0001-6071-4564}\inst{\ref{aff312}}
\and S.~Ventura\inst{\ref{aff39}}
\and G.~Verdoes~Kleijn\orcid{0000-0001-5803-2580}\inst{\ref{aff69}}
\and D.~Vergani\orcid{0000-0003-0898-2216}\inst{\ref{aff35}}
\and A.~Verma\orcid{0000-0002-0730-0781}\inst{\ref{aff172}}
\and F.~Vernizzi\orcid{0000-0003-3426-2802}\inst{\ref{aff95}}
\and A.~Veropalumbo\orcid{0000-0003-2387-1194}\inst{\ref{aff34},\ref{aff127},\ref{aff304}}
\and G.~Verza\orcid{0000-0002-1886-8348}\inst{\ref{aff347},\ref{aff204}}
\and C.~Vescovi\inst{\ref{aff19}}
\and D.~Vibert\orcid{0009-0008-0607-631X}\inst{\ref{aff20}}
\and M.~Viel\orcid{0000-0002-2642-5707}\inst{\ref{aff58},\ref{aff59},\ref{aff61},\ref{aff60},\ref{aff157}}
\and P.~Vielzeuf\orcid{0000-0003-2035-9339}\inst{\ref{aff52}}
\and C.~Viglione\inst{\ref{aff143},\ref{aff132}}
\and A.~Viitanen\orcid{0000-0001-9383-786X}\inst{\ref{aff225},\ref{aff142}}
\and F.~Villaescusa-Navarro\orcid{0000-0002-4816-0455}\inst{\ref{aff204},\ref{aff308}}
\and S.~Vinciguerra\inst{\ref{aff20}}
\and F.~Visticot\inst{\ref{aff11}}
\and K.~Voggel\orcid{0000-0001-6215-0950}\inst{\ref{aff97}}
\and M.~von~Wietersheim-Kramsta\orcid{0000-0003-4986-5091}\inst{\ref{aff87},\ref{aff248},\ref{aff205}}
\and W.~J.~Vriend\inst{\ref{aff69}}
\and S.~Wachter\inst{\ref{aff159}}
\and M.~Walmsley\orcid{0000-0002-6408-4181}\inst{\ref{aff149},\ref{aff122}}
\and G.~Walth\orcid{0000-0002-6313-6808}\inst{\ref{aff62}}
\and D.~M.~Walton\inst{\ref{aff16}}
\and N.~A.~Walton\orcid{0000-0003-3983-8778}\inst{\ref{aff233}}
\and M.~Wander\inst{\ref{aff226}}
\and L.~Wang\orcid{0000-0002-6736-9158}\inst{\ref{aff269},\ref{aff69}}
\and Y.~Wang\orcid{0000-0002-4749-2984}\inst{\ref{aff17}}
\and J.~R.~Weaver\orcid{0000-0003-1614-196X}\inst{\ref{aff86}}
\and J.~Weller\orcid{0000-0002-8282-2010}\inst{\ref{aff82},\ref{aff93}}
\and M.~Wetzstein\inst{\ref{aff93}}
\and D.~J.~Whalen\orcid{0000-0001-6646-2337}\inst{\ref{aff63}}
\and I.~H.~Whittam\orcid{0000-0003-2265-5983}\inst{\ref{aff172},\ref{aff257}}
\and A.~Widmer\orcid{0009-0005-4111-2716}\inst{\ref{aff10}}
\and M.~Wiesmann\orcid{0009-0000-8199-5860}\inst{\ref{aff206}}
\and J.~Wilde\orcid{0000-0002-4460-7379}\inst{\ref{aff200}}
\and O.~R.~Williams\orcid{0000-0003-0274-1526}\inst{\ref{aff199}}
\and H.-A.~Winther\orcid{0000-0002-6325-2710}\inst{\ref{aff206}}
\and A.~Wittje\orcid{0000-0002-8173-3438}\inst{\ref{aff235}}
\and J.~H.~W.~Wong\orcid{0000-0001-7133-7741}\inst{\ref{aff122}}
\and A.~H.~Wright\orcid{0000-0001-7363-7932}\inst{\ref{aff235}}
\and V.~Yankelevich\orcid{0000-0001-8288-7335}\inst{\ref{aff348}}
\and H.~W.~Yeung\orcid{0000-0002-4993-9014}\inst{\ref{aff98}}
\and M.~Yoon\orcid{0000-0002-3683-9559}\inst{\ref{aff70}}
\and S.~Youles\orcid{0000-0002-7520-5911}\inst{\ref{aff63}}
\and L.~Y.~A.~Yung\orcid{0000-0003-3466-035X}\inst{\ref{aff251},\ref{aff349}}
\and A.~Zacchei\orcid{0000-0003-0396-1192}\inst{\ref{aff59},\ref{aff58}}
\and L.~Zalesky\orcid{0000-0001-5680-2326}\inst{\ref{aff158}}
\and G.~Zamorani\orcid{0000-0002-2318-301X}\inst{\ref{aff35}}
\and A.~Zamorano~Vitorelli\orcid{0000-0002-9740-4591}\inst{\ref{aff112}}
\and M.~Zanoni~Marc\inst{\ref{aff99}}
\and M.~Zennaro\orcid{0000-0002-4458-1754}\inst{\ref{aff172}}
\and F.~M.~Zerbi\inst{\ref{aff34}}
\and I.~A.~Zinchenko\inst{\ref{aff82}}
\and J.~Zoubian\inst{\ref{aff52}}
\and E.~Zucca\orcid{0000-0002-5845-8132}\inst{\ref{aff35}}
\and M.~Zumalacarregui\orcid{0000-0002-9943-6490}\inst{\ref{aff128}}}
										   
\institute{Institut d'Astrophysique de Paris, 98bis Boulevard Arago, 75014, Paris, France\label{aff1}
\and
Institut d'Astrophysique de Paris, UMR 7095, CNRS, and Sorbonne Universit\'e, 98 bis boulevard Arago, 75014 Paris, France\label{aff2}
\and
Johns Hopkins University 3400 North Charles Street Baltimore, MD 21218, USA\label{aff3}
\and
Institute of Physics, Laboratory of Astrophysics, Ecole Polytechnique F\'ed\'erale de Lausanne (EPFL), Observatoire de Sauverny, 1290 Versoix, Switzerland\label{aff4}
\and
Institute Lorentz, Leiden University, Niels Bohrweg 2, 2333 CA Leiden, The Netherlands\label{aff5}
\and
Departamento de F\'isica, Universidad del Pa\'is Vasco UPV-EHU, 48940 Leioa, Spain\label{aff6}
\and
Department of Astrophysics, University of Zurich, Winterthurerstrasse 190, 8057 Zurich, Switzerland\label{aff7}
\and
Universit\'e C\^{o}te d'Azur, Observatoire de la C\^{o}te d'Azur, CNRS, Laboratoire Lagrange, Bd de l'Observatoire, CS 34229, 06304 Nice cedex 4, France\label{aff8}
\and
Universit\'e Paris-Saclay, CNRS, Institut d'astrophysique spatiale, 91405, Orsay, France\label{aff9}
\and
Universit\'e Paris Cit\'e, CNRS, Astroparticule et Cosmologie, 75013 Paris, France\label{aff10}
\and
Universit\'e Paris-Saclay, Universit\'e Paris Cit\'e, CEA, CNRS, AIM, 91191, Gif-sur-Yvette, France\label{aff11}
\and
Institute for Particle Physics and Astrophysics, Dept. of Physics, ETH Zurich, Wolfgang-Pauli-Strasse 27, 8093 Zurich, Switzerland\label{aff12}
\and
INFN-Sezione di Torino, Via P. Giuria 1, 10125 Torino, Italy\label{aff13}
\and
Instituto de F\'isica Te\'orica UAM-CSIC, Campus de Cantoblanco, 28049 Madrid, Spain\label{aff14}
\and
CERCA/ISO, Department of Physics, Case Western Reserve University, 10900 Euclid Avenue, Cleveland, OH 44106, USA\label{aff15}
\and
Mullard Space Science Laboratory, University College London, Holmbury St Mary, Dorking, Surrey RH5 6NT, UK\label{aff16}
\and
Infrared Processing and Analysis Center, California Institute of Technology, Pasadena, CA 91125, USA\label{aff17}
\and
Instituto de Astrof\'isica e Ci\^encias do Espa\c{c}o, Faculdade de Ci\^encias, Universidade de Lisboa, Campo Grande, 1749-016 Lisboa, Portugal\label{aff18}
\and
Univ. Grenoble Alpes, CNRS, Grenoble INP, LPSC-IN2P3, 53, Avenue des Martyrs, 38000, Grenoble, France\label{aff19}
\and
Aix-Marseille Universit\'e, CNRS, CNES, LAM, Marseille, France\label{aff20}
\and
Kavli Institute for Particle Astrophysics \& Cosmology (KIPAC), Stanford University, Stanford, CA 94305, USA\label{aff21}
\and
Department of Physics, Stanford University, 382 Via Pueblo Mall, Stanford, CA 94305, USA\label{aff22}
\and
SLAC National Accelerator Laboratory, 2575 Sand Hill Road, Menlo Park, CA 94025, USA\label{aff23}
\and
INAF-Osservatorio Astronomico di Capodimonte, Via Moiariello 16, 80131 Napoli, Italy\label{aff24}
\and
Department of Physics and Astronomy, University of Southampton, Southampton, SO17 1BJ, UK\label{aff25}
\and
ESAC/ESA, Camino Bajo del Castillo, s/n., Urb. Villafranca del Castillo, 28692 Villanueva de la Ca\~nada, Madrid, Spain\label{aff26}
\and
Instituto de Astrof\'isica de Andaluc\'ia, CSIC, Glorieta de la Astronom\'\i a, 18080, Granada, Spain\label{aff27}
\and
Instituto de F\'isica Aplicada a las Ciencias y las Tecnolog\'ias, Universidad de Alicante, San Vicent del Raspeig, E03080, Alicante, Spain\label{aff28}
\and
Dipartimento di Fisica e Scienze della Terra, Universit\`a degli Studi di Ferrara, Via Giuseppe Saragat 1, 44122 Ferrara, Italy\label{aff29}
\and
School of Mathematics and Physics, University of Surrey, Guildford, Surrey, GU2 7XH, UK\label{aff30}
\and
Institut f\"ur Theoretische Physik, University of Heidelberg, Philosophenweg 16, 69120 Heidelberg, Germany\label{aff31}
\and
Max-Planck-Institut f\"ur Astrophysik, Karl-Schwarzschild-Str.~1, 85748 Garching, Germany\label{aff32}
\and
Technical University of Munich, TUM School of Natural Sciences, Physics Department, James-Franck-Str.~1, 85748 Garching, Germany\label{aff33}
\and
INAF-Osservatorio Astronomico di Brera, Via Brera 28, 20122 Milano, Italy\label{aff34}
\and
INAF-Osservatorio di Astrofisica e Scienza dello Spazio di Bologna, Via Piero Gobetti 93/3, 40129 Bologna, Italy\label{aff35}
\and
Donostia International Physics Center (DIPC), Paseo Manuel de Lardizabal, 4, 20018, Donostia-San Sebasti\'an, Guipuzkoa, Spain\label{aff36}
\and
IKERBASQUE, Basque Foundation for Science, 48013, Bilbao, Spain\label{aff37}
\and
Thales Alenia Space -- \Euclid satellite Prime contractor, Strada Antica di Collegno 253, 10146 Torino, Italy\label{aff38}
\and
INFN-Padova, Via Marzolo 8, 35131 Padova, Italy\label{aff39}
\and
Dipartimento di Fisica e Astronomia "G. Galilei", Universit\`a di Padova, Via Marzolo 8, 35131 Padova, Italy\label{aff40}
\and
Laboratoire Univers et Th\'eorie, Observatoire de Paris, Universit\'e PSL, Universit\'e Paris Cit\'e, CNRS, 92190 Meudon, France\label{aff41}
\and
Istituto Nazionale di Fisica Nucleare, Sezione di Ferrara, Via Giuseppe Saragat 1, 44122 Ferrara, Italy\label{aff42}
\and
Dipartimento di Fisica "Aldo Pontremoli", Universit\`a degli Studi di Milano, Via Celoria 16, 20133 Milano, Italy\label{aff43}
\and
INFN-Sezione di Milano, Via Celoria 16, 20133 Milano, Italy\label{aff44}
\and
IRFU, CEA, Universit\'e Paris-Saclay 91191 Gif-sur-Yvette Cedex, France\label{aff45}
\and
Universit\'e Paris Diderot, AIM, Sorbonne Paris Cit\'e, CEA, CNRS 91191 Gif-sur-Yvette Cedex, France\label{aff46}
\and
School of Mathematics, Statistics and Physics, Newcastle University, Herschel Building, Newcastle-upon-Tyne, NE1 7RU, UK\label{aff47}
\and
Departamento de F{\'\i}sica Te\'{o}rica and Instituto de F{\'\i}sica de Part{\'\i}-culas y del Cosmos (IPARCOS-UCM), Universidad Complutense de Madrid, 28040 Madrid, Spain\label{aff48}
\and
Departamento de F{\'\i}sica Fundamental. Universidad de Salamanca. Plaza de la Merced s/n. 37008 Salamanca, Spain\label{aff49}
\and
Universit\'e Clermont Auvergne, CNRS/IN2P3, LPC, F-63000 Clermont-Ferrand, France\label{aff50}
\and
Universit\'e Claude Bernard Lyon 1, CNRS/IN2P3, IP2I Lyon, UMR 5822, Villeurbanne, F-69100, France\label{aff51}
\and
Aix-Marseille Universit\'e, CNRS/IN2P3, CPPM, Marseille, France\label{aff52}
\and
Institut f\"{u}r Planetologie, Universit\"at M\"unster, Wilhelm-Klemm-Str. 10, 48149 M\"unster, Germany\label{aff53}
\and
Departamento de F\'{\i}sica e Astronomia, Faculdade de Ci\^encias, Universidade do Porto, Rua do Campo Alegre 687, PT4169-007 Porto, Portugal\label{aff54}
\and
Instituto de Astrof\'isica e Ci\^encias do Espa\c{c}o, Universidade do Porto, CAUP, Rua das Estrelas, PT4150-762 Porto, Portugal\label{aff55}
\and
School of Physics and Astronomy, University of Nottingham, University Park, Nottingham NG7 2RD, UK\label{aff56}
\and
Institut de F\'{i}sica d'Altes Energies (IFAE), The Barcelona Institute of Science and Technology, Campus UAB, 08193 Bellaterra (Barcelona), Spain\label{aff57}
\and
IFPU, Institute for Fundamental Physics of the Universe, via Beirut 2, 34151 Trieste, Italy\label{aff58}
\and
INAF-Osservatorio Astronomico di Trieste, Via G. B. Tiepolo 11, 34143 Trieste, Italy\label{aff59}
\and
INFN, Sezione di Trieste, Via Valerio 2, 34127 Trieste TS, Italy\label{aff60}
\and
SISSA, International School for Advanced Studies, Via Bonomea 265, 34136 Trieste TS, Italy\label{aff61}
\and
Caltech/IPAC, 1200 E. California Blvd., Pasadena, CA 91125, USA\label{aff62}
\and
Institute of Cosmology and Gravitation, University of Portsmouth, Portsmouth PO1 3FX, UK\label{aff63}
\and
Sterrenkundig Observatorium, Universiteit Gent, Krijgslaan 281 S9, 9000 Gent, Belgium\label{aff64}
\and
The University of Texas at Austin, Austin, TX, 78712, USA\label{aff65}
\and
UCB Lyon 1, CNRS/IN2P3, IUF, IP2I Lyon, 4 rue Enrico Fermi, 69622 Villeurbanne, France\label{aff66}
\and
Instituto de Astrof\'isica de Canarias, Calle V\'ia L\'actea s/n, 38204, San Crist\'obal de La Laguna, Tenerife, Spain\label{aff67}
\and
Departamento de Astrof\'isica, Universidad de La Laguna, 38206, La Laguna, Tenerife, Spain\label{aff68}
\and
Kapteyn Astronomical Institute, University of Groningen, PO Box 800, 9700 AV Groningen, The Netherlands\label{aff69}
\and
Leiden Observatory, Leiden University, Einsteinweg 55, 2333 CC Leiden, The Netherlands\label{aff70}
\and
Isaac Newton Group of Telescopes, Apartado 321, 38700 Santa Cruz de La Palma, Spain\label{aff71}
\and
Dipartimento di Fisica e Astronomia, Universit\`a di Bologna, Via Gobetti 93/2, 40129 Bologna, Italy\label{aff72}
\and
INFN-Sezione di Bologna, Viale Berti Pichat 6/2, 40127 Bologna, Italy\label{aff73}
\and
Astrophysics Research Institute, Liverpool John Moores University, 146 Brownlow Hill, Liverpool L3 5RF, UK\label{aff74}
\and
INAF-Osservatorio Astronomico di Padova, Via dell'Osservatorio 5, 35122 Padova, Italy\label{aff75}
\and
Department of Physics and Astronomy, University of Waterloo, Waterloo, Ontario N2L 3G1, Canada\label{aff76}
\and
Waterloo Centre for Astrophysics, University of Waterloo, Waterloo, Ontario N2L 3G1, Canada\label{aff77}
\and
Max-Planck-Institut f\"ur Astronomie, K\"onigstuhl 17, 69117 Heidelberg, Germany\label{aff78}
\and
ECEO, Universidade Lus\'ofona, Campo Grande 376, 1749-024 Lisboa, Portugal\label{aff79}
\and
Instituto de Astrof\'\i sica de Canarias, c/ Via Lactea s/n, La Laguna E-38200, Spain. Departamento de Astrof\'\i sica de la Universidad de La Laguna, Avda. Francisco Sanchez, La Laguna, E-38200, Spain\label{aff80}
\and
CEA-Saclay, DRF/IRFU, departement d'ingenierie des systemes, bat472, 91191 Gif sur Yvette cedex, France\label{aff81}
\and
Universit\"ats-Sternwarte M\"unchen, Fakult\"at f\"ur Physik, Ludwig-Maximilians-Universit\"at M\"unchen, Scheinerstrasse 1, 81679 M\"unchen, Germany\label{aff82}
\and
Centre National d'Etudes Spatiales -- Centre spatial de Toulouse, 18 avenue Edouard Belin, 31401 Toulouse Cedex 9, France\label{aff83}
\and
Research School of Astronomy and Astrophysics, Australian National University, Cotter Road, Weston Creek, ACT 2611, Australia\label{aff84}
\and
ARC Centre of Excellence for All Sky Astrophysics in 3 Dimensions (ASTRO 3D), Australia\label{aff85}
\and
Department of Astronomy, University of Massachusetts, Amherst, MA 01003, USA\label{aff86}
\and
Department of Physics, Institute for Computational Cosmology, Durham University, South Road, DH1 3LE, UK\label{aff87}
\and
School of Natural Sciences, University of Tasmania, Private Bag 37 Hobart, Tasmania 7001, Australia\label{aff88}
\and
Institut national de physique nucl\'eaire et de physique des particules, 3 rue Michel-Ange, 75794 Paris C\'edex 16, France\label{aff89}
\and
ATG Europe BV, Huygensstraat 34, 2201 DK Noordwijk, The Netherlands\label{aff90}
\and
Aix-Marseille Universit\'e, Universit\'e de Toulon, CNRS, CPT, Marseille, France\label{aff91}
\and
Institut de Recherche en Astrophysique et Plan\'etologie (IRAP), Universit\'e de Toulouse, CNRS, UPS, CNES, 14 Av. Edouard Belin, 31400 Toulouse, France\label{aff92}
\and
Max Planck Institute for Extraterrestrial Physics, Giessenbachstr. 1, 85748 Garching, Germany\label{aff93}
\and
INFN, Sezione di Roma 2, Via della Ricerca Scientifica 1, Roma, Italy\label{aff94}
\and
Institut de Physique Th\'eorique, CEA, CNRS, Universit\'e Paris-Saclay 91191 Gif-sur-Yvette Cedex, France\label{aff95}
\and
IMCCE, Observatoire de Paris, Universit\'{e} PSL, CNRS, Sorbonne Universit{\'e}, Univ. Lille, 77 av. Denfert-Rochereau, 75014 Paris, France\label{aff96}
\and
Universit\'e de Strasbourg, CNRS, Observatoire astronomique de Strasbourg, UMR 7550, 67000 Strasbourg, France\label{aff97}
\and
Institute for Astronomy, University of Edinburgh, Royal Observatory, Blackford Hill, Edinburgh EH9 3HJ, UK\label{aff98}
\and
Airbus Defence \& Space SAS, Toulouse, France\label{aff99}
\and
OCA, P.H.C Boulevard de l'Observatoire CS 34229, 06304 Nice Cedex 4, France\label{aff100}
\and
INAF, Istituto di Radioastronomia, Via Piero Gobetti 101, 40129 Bologna, Italy\label{aff101}
\and
Department of Physics, Northeastern University, Boston, MA, 02115, USA\label{aff102}
\and
Kavli Institute for the Physics and Mathematics of the Universe (WPI), University of Tokyo, Kashiwa, Chiba 277-8583, Japan\label{aff103}
\and
European Space Agency/ESTEC, Keplerlaan 1, 2201 AZ Noordwijk, The Netherlands\label{aff104}
\and
Ludwig-Maximilians-University, Schellingstrasse 4, 80799 Munich, Germany\label{aff105}
\and
Max-Planck-Institut f\"ur Physik, Boltzmannstr. 8, 85748 Garching, Germany\label{aff106}
\and
Space Science Data Center, Italian Space Agency, via del Politecnico snc, 00133 Roma, Italy\label{aff107}
\and
INAF-IASF Milano, Via Alfonso Corti 12, 20133 Milano, Italy\label{aff108}
\and
Perimeter Institute for Theoretical Physics, Waterloo, Ontario N2L 2Y5, Canada\label{aff109}
\and
INAF-Osservatorio Astrofisico di Torino, Via Osservatorio 20, 10025 Pino Torinese (TO), Italy\label{aff110}
\and
Universit\'e de Gen\`eve, D\'epartement de Physique Th\'eorique and Centre for Astroparticle Physics, 24 quai Ernest-Ansermet, CH-1211 Gen\`eve 4, Switzerland\label{aff111}
\and
Jet Propulsion Laboratory, California Institute of Technology, 4800 Oak Grove Drive, Pasadena, CA, 91109, USA\label{aff112}
\and
Dipartimento di Fisica - Sezione di Astronomia, Universit\`a di Trieste, Via Tiepolo 11, 34131 Trieste, Italy\label{aff113}
\and
NASA Ames Research Center, Moffett Field, CA 94035, USA\label{aff114}
\and
Bay Area Environmental Research Institute, Moffett Field, California 94035, USA\label{aff115}
\and
Scuola Superiore Meridionale, Via Mezzocannone 4, 80138, Napoli, Italy\label{aff116}
\and
INFN section of Naples, Via Cinthia 6, 80126, Napoli, Italy\label{aff117}
\and
Univ. Grenoble Alpes, Univ. Savoie Mont Blanc, CNRS, LAPP, 74000 Annecy, France\label{aff118}
\and
Thales~Services~S.A.S., 290 All\'ee du Lac, 31670 Lab\`ege, France\label{aff119}
\and
Laboratoire d'Astrophysique de Bordeaux, CNRS and Universit\'e de Bordeaux, All\'ee Geoffroy St. Hilaire, 33165 Pessac, France\label{aff120}
\and
Institut universitaire de France (IUF), 1 rue Descartes, 75231 PARIS CEDEX 05, France\label{aff121}
\and
Jodrell Bank Centre for Astrophysics, Department of Physics and Astronomy, University of Manchester, Oxford Road, Manchester M13 9PL, UK\label{aff122}
\and
Universita di Salerno, Dipartimento di Fisica "E.R. Caianiello", Via Giovanni Paolo II 132, I-84084 Fisciano (SA), Italy\label{aff123}
\and
INFN - Gruppo Collegato di Salerno - Sezione di Napoli, Dipartimento di Fisica "E.R. Caianiello", Universita di Salerno, via Giovanni Paolo II, 132 - I-84084 Fisciano (SA), Italy\label{aff124}
\and
Department of Astronomy, University of Geneva, ch. d'Ecogia 16, 1290 Versoix, Switzerland\label{aff125}
\and
Dipartimento di Fisica, Universit\`a di Genova, Via Dodecaneso 33, 16146, Genova, Italy\label{aff126}
\and
INFN-Sezione di Genova, Via Dodecaneso 33, 16146, Genova, Italy\label{aff127}
\and
Max Planck Institute for Gravitational Physics (Albert Einstein Institute), Am Muhlenberg 1, D-14476 Potsdam-Golm, Germany\label{aff128}
\and
Norwegian Space Agency,~Drammensveien 165,~0277 Oslo, Norway\label{aff129}
\and
School of Physics, HH Wills Physics Laboratory, University of Bristol, Tyndall Avenue, Bristol, BS8 1TL, UK\label{aff130}
\and
Department of Physics "E. Pancini", University Federico II, Via Cinthia 6, 80126, Napoli, Italy\label{aff131}
\and
Institute of Space Sciences (ICE, CSIC), Campus UAB, Carrer de Can Magrans, s/n, 08193 Barcelona, Spain\label{aff132}
\and
Institut de Ciencies de l'Espai (IEEC-CSIC), Campus UAB, Carrer de Can Magrans, s/n Cerdanyola del Vall\'es, 08193 Barcelona, Spain\label{aff133}
\and
Department of Physics and Astronomy, University of Missouri, 5110 Rockhill Road, Kansas City, MO 64110, USA\label{aff134}
\and
Minnesota Institute for Astrophysics, University of Minnesota, 116 Church St SE, Minneapolis, MN 55455, USA\label{aff135}
\and
Departamento de F\'{i}sica Te\'{o}rica, At\'{o}mica y \'{O}ptica, Universidad de Valladolid, 47011 Valladolid, Spain\label{aff136}
\and
Instituto de Astrof\'isica e Ci\^encias do Espa\c{c}o, Faculdade de Ci\^encias, Universidade de Lisboa, Tapada da Ajuda, 1349-018 Lisboa, Portugal\label{aff137}
\and
Universit\"at Bonn, Argelander-Institut f\"ur Astronomie, Auf dem H\"ugel 71, 53121 Bonn, Germany\label{aff138}
\and
INFN-Bologna, Via Irnerio 46, 40126 Bologna, Italy\label{aff139}
\and
Port d'Informaci\'{o} Cient\'{i}fica, Campus UAB, C. Albareda s/n, 08193 Bellaterra (Barcelona), Spain\label{aff140}
\and
Astronomical Observatory of the Autonomous Region of the Aosta Valley (OAVdA), Loc. Lignan 39, I-11020, Nus (Aosta Valley), Italy\label{aff141}
\and
INAF-Osservatorio Astronomico di Roma, Via Frascati 33, 00078 Monteporzio Catone, Italy\label{aff142}
\and
Institut d'Estudis Espacials de Catalunya (IEEC),  Edifici RDIT, Campus UPC, 08860 Castelldefels, Barcelona, Spain\label{aff143}
\and
Dipartimento di Fisica, Universit\`a degli Studi di Torino, Via P. Giuria 1, 10125 Torino, Italy\label{aff144}
\and
INAF - Osservatorio Astronomico d'Abruzzo, Via Maggini, 64100, Teramo, Italy\label{aff145}
\and
Department of Physics, University of Miami, Coral Gables, FL 33124, USA\label{aff146}
\and
Cosmic Dawn Center (DAWN)\label{aff147}
\and
INFN-Sezione di Roma, Piazzale Aldo Moro, 2 - c/o Dipartimento di Fisica, Edificio G. Marconi, 00185 Roma, Italy\label{aff148}
\and
David A. Dunlap Department of Astronomy \& Astrophysics, University of Toronto, 50 St George Street, Toronto, Ontario M5S 3H4, Canada\label{aff149}
\and
Institut de Ci\`{e}ncies del Cosmos (ICCUB), Universitat de Barcelona (IEEC-UB), Mart\'{i} i Franqu\`{e}s 1, 08028 Barcelona, Spain\label{aff150}
\and
Departament de F\'isica Qu\`antica i Astrof\'isica (FQA), Universitat de Barcelona (UB), Mart\'{\i} i Franqu\`es 1, E-08028 Barcelona, Spain\label{aff151}
\and
Centro de Investigaciones Energ\'eticas, Medioambientales y Tecnol\'ogicas (CIEMAT), Avenida Complutense 40, 28040 Madrid, Spain\label{aff152}
\and
Departamento de F\'isica, Faculdade de Ci\^encias, Universidade de Lisboa, Edif\'icio C8, Campo Grande, PT1749-016 Lisboa, Portugal\label{aff153}
\and
Institute for Theoretical Particle Physics and Cosmology (TTK), RWTH Aachen University, 52056 Aachen, Germany\label{aff154}
\and
Cosmic Dawn Center (DAWN), Denmark\label{aff155}
\and
Fakult\"at f\"ur Physik, Universit\"at Bielefeld, Postfach 100131, 33501 Bielefeld, Germany\label{aff156}
\and
ICSC - Centro Nazionale di Ricerca in High Performance Computing, Big Data e Quantum Computing, Via Magnanelli 2, Bologna, Italy\label{aff157}
\and
Institute for Astronomy, University of Hawaii, 2680 Woodlawn Drive, Honolulu, HI 96822, USA\label{aff158}
\and
Carnegie Observatories, Pasadena, CA 91101, USA\label{aff159}
\and
Dipartimento di Fisica e Astronomia "Augusto Righi" - Alma Mater Studiorum Universit\`a di Bologna, Viale Berti Pichat 6/2, 40127 Bologna, Italy\label{aff160}
\and
Leibniz-Institut f\"{u}r Astrophysik (AIP), An der Sternwarte 16, 14482 Potsdam, Germany\label{aff161}
\and
SCITAS, Ecole Polytechnique F\'ed\'erale de Lausanne (EPFL), 1015 Lausanne, Switzerland\label{aff162}
\and
Astrophysics Group, Blackett Laboratory, Imperial College London, London SW7 2AZ, UK\label{aff163}
\and
Universit\'e Libre de Bruxelles (ULB), Service de Physique Th\'eorique CP225, Boulevard du Triophe, 1050 Bruxelles, Belgium\label{aff164}
\and
Dipartimento di Fisica e Astronomia "Augusto Righi" - Alma Mater Studiorum Universit\`a di Bologna, via Piero Gobetti 93/2, 40129 Bologna, Italy\label{aff165}
\and
Department of Theoretical Physics, Maynooth University, Maynooth, Co. Kildare, Ireland\label{aff166}
\and
Coll\'ege de France, 11 Place Marcelin Berthelot, 75231 Paris, France\label{aff167}
\and
Sorbonne Universit\'e, Observatoire de Paris, Universit\'e PSL, CNRS, LERMA, 75014, Paris, France\label{aff168}
\and
European Space Agency/ESRIN, Largo Galileo Galilei 1, 00044 Frascati, Roma, Italy\label{aff169}
\and
Department of Physics \& Astronomy, University of California Irvine, Irvine CA 92697, USA\label{aff170}
\and
European Space Agency/ESOC, Robert-Bosch-Str. 5, 64293 Darmstadt, Germany\label{aff171}
\and
Department of Physics, Oxford University, Keble Road, Oxford OX1 3RH, UK\label{aff172}
\and
INAF-Osservatorio Astrofisico di Arcetri, Largo E. Fermi 5, 50125, Firenze, Italy\label{aff173}
\and
Universit\"at Innsbruck, Institut f\"ur Astro- und Teilchenphysik, Technikerstr. 25/8, 6020 Innsbruck, Austria\label{aff174}
\and
Canada-France-Hawaii Telescope, 65-1238 Mamalahoa Hwy, Kamuela, HI 96743, USA\label{aff175}
\and
Faculdade de Ci\^encias da Universidade do Porto, Rua do Campo de Alegre, 4150-007 Porto, Portugal\label{aff176}
\and
Dipartimento di Fisica, Universit\`a di Roma Tor Vergata, Via della Ricerca Scientifica 1, Roma, Italy\label{aff177}
\and
CIGUS CITIC, Centre for Information and Communications Technologies Research, Universidade da Coru\~na, Campus de Elvi\~na s/n, 15071 A Coru\~na, Spain\label{aff178}
\and
Universit\'e Paris-Saclay, CEA, D\'epartement d'\'Electronique des D\'etecteurs et d'Informatique pour la Physique, 91191, Gif-sur-Yvette, France\label{aff179}
\and
Dipartimento di Scienze Matematiche, Fisiche e Informatiche, Universit\`a di Parma, Viale delle Scienze 7/A 43124 Parma, Italy\label{aff180}
\and
INFN Gruppo Collegato di Parma, Viale delle Scienze 7/A 43124 Parma, Italy\label{aff181}
\and
Instituto de Astrof\'isica de Canarias (IAC); Departamento de Astrof\'isica, Universidad de La Laguna (ULL), 38200, La Laguna, Tenerife, Spain\label{aff182}
\and
Department of Mathematics and Physics, Roma Tre University, Via della Vasca Navale 84, 00146 Rome, Italy\label{aff183}
\and
INFN-Sezione di Roma Tre, Via della Vasca Navale 84, 00146, Roma, Italy\label{aff184}
\and
Department of Mathematics and Physics E. De Giorgi, University of Salento, Via per Arnesano, CP-I93, 73100, Lecce, Italy\label{aff185}
\and
INFN, Sezione di Lecce, Via per Arnesano, CP-193, 73100, Lecce, Italy\label{aff186}
\and
INAF-Sezione di Lecce, c/o Dipartimento Matematica e Fisica, Via per Arnesano, 73100, Lecce, Italy\label{aff187}
\and
Dipartimento di Fisica, Sapienza Universit\`a di Roma, Piazzale Aldo Moro 2, 00185 Roma, Italy\label{aff188}
\and
Dept. of Physics, IIT Hyderabad, Kandi, Telangana 502285, India\label{aff189}
\and
Technion Israel Institute of Technology, Israel\label{aff190}
\and
Department of Astronomy \& Physics and Institute for Computational Astrophysics, Saint Mary's University, 923 Robie Street, Halifax, Nova Scotia, B3H 3C3, Canada\label{aff191}
\and
Centro de Astrof\'{\i}sica da Universidade do Porto, Rua das Estrelas, 4150-762 Porto, Portugal\label{aff192}
\and
Departamento F\'isica Aplicada, Universidad Polit\'ecnica de Cartagena, Campus Muralla del Mar, 30202 Cartagena, Murcia, Spain\label{aff193}
\and
Instituto de F\'isica de Cantabria, Edificio Juan Jord\'a, Avenida de los Castros, 39005 Santander, Spain\label{aff194}
\and
INAF-Istituto di Astrofisica e Planetologia Spaziali, via del Fosso del Cavaliere, 100, 00100 Roma, Italy\label{aff195}
\and
Observatorio Nacional, Rua General Jose Cristino, 77-Bairro Imperial de Sao Cristovao, Rio de Janeiro, 20921-400, Brazil\label{aff196}
\and
Centro de Estudios de F\'isica del Cosmos de Arag\'on (CEFCA), Plaza San Juan, 1, planta 2, 44001, Teruel, Spain\label{aff197}
\and
California institute of Technology, 1200 E California Blvd, Pasadena, CA 91125, USA\label{aff198}
\and
Centre for Information Technology, University of Groningen, P.O. Box 11044, 9700 CA Groningen, The Netherlands\label{aff199}
\and
School of Physical Sciences, The Open University, Milton Keynes, MK7 6AA, UK\label{aff200}
\and
Department of Physics, Lancaster University, Lancaster, LA1 4YB, UK\label{aff201}
\and
XCAM Limited, 2 Stone Circle Road, Northampton, NN3 8RF, UK\label{aff202}
\and
School of Physics and Astronomy, Cardiff University, The Parade, Cardiff, CF24 3AA, UK\label{aff203}
\and
Center for Computational Astrophysics, Flatiron Institute, 162 5th Avenue, 10010, New York, NY, USA\label{aff204}
\and
Department of Physics and Astronomy, University College London, Gower Street, London WC1E 6BT, UK\label{aff205}
\and
Institute of Theoretical Astrophysics, University of Oslo, P.O. Box 1029 Blindern, 0315 Oslo, Norway\label{aff206}
\and
Department of Physics, P.O. Box 64, 00014 University of Helsinki, Finland\label{aff207}
\and
School of Physics and Astronomy, Queen Mary University of London, Mile End Road, London E1 4NS, UK\label{aff208}
\and
Dipartimento di Fisica ``G. Occhialini", Universit\`a degli Studi di Milano Bicocca, Piazza della Scienza 3, 20126 Milano, Italy\label{aff209}
\and
Univ. Grenoble Alpes, CNRS, IPAG, Grenoble, France\label{aff210}
\and
Department of Astronomy, University of Florida, Bryant Space Science Center, Gainesville, FL 32611, USA\label{aff211}
\and
Code 665, NASA/GSFC, 8800 Greenbelt Road, Greenbelt, MD 20771, USA\label{aff212}
\and
CEA Saclay, DFR/IRFU, Service d'Astrophysique, Bat. 709, 91191 Gif-sur-Yvette, France\label{aff213}
\and
Dipartimento di Fisica e Astronomia, Universit\`{a} di Firenze, via G. Sansone 1, 50019 Sesto Fiorentino, Firenze, Italy\label{aff214}
\and
Istituto Nazionale di Fisica Nucleare, Sezione di Bologna, Via Irnerio 46, 40126 Bologna, Italy\label{aff215}
\and
Nordita, KTH Royal Institute of Technology and Stockholm 1859 University, Hannes Alfv\'ens v\"{a}g 12, Stockholm, SE-106 91, Sweden\label{aff216}
\and
FRACTAL S.L.N.E., calle Tulip\'an 2, Portal 13 1A, 28231, Las Rozas de Madrid, Spain\label{aff217}
\and
Departamento de F\'isica Te\'orica, Facultad de Ciencias, Universidad Aut\'onoma de Madrid, 28049 Cantoblanco, Madrid, Spain\label{aff218}
\and
Instituto de Astronomia Teorica y Experimental (IATE-CONICET), Laprida 854, X5000BGR, C\'ordoba, Argentina\label{aff219}
\and
ASDEX Upgrade Team, Max-Planck-Institut fur Plasmaphysik, Garching, Germany\label{aff220}
\and
Department of Computer Science, Aalto University, PO Box 15400, Espoo, FI-00 076, Finland\label{aff221}
\and
Asteroid Engineering Laboratory, Lule\aa{} University of Technology, Box 848, 98128 Kiruna, Sweden\label{aff222}
\and
Universite Paris Cit\'e, Universite Paris-Saclay, CEA, CNRS, AIM, 91191, Gif-sur-Yvette, France\label{aff223}
\and
NRC Herzberg, 5071 West Saanich Rd, Victoria, BC V9E 2E7, Canada\label{aff224}
\and
Department of Physics and Helsinki Institute of Physics, Gustaf H\"allstr\"omin katu 2, 00014 University of Helsinki, Finland\label{aff225}
\and
Centre for Electronic Imaging, Open University, Walton Hall, Milton Keynes, MK7~6AA, UK\label{aff226}
\and
Excellence Cluster ORIGINS, Boltzmannstrasse 2, 85748 Garching, Germany\label{aff227}
\and
UK Space Agency, Swindon, SN2 1SZ, UK\label{aff228}
\and
Technische Universitat Dresden, Institut f\"ur Kern- und Teilchenphysik, Zellescher Weg 19, 01069 Dresden, Germany\label{aff229}
\and
Deutsches Elektronen-Synchrotron DESY, Platanenallee 6, 15738 Zeuthen, Germany\label{aff230}
\and
European Southern Observatory, Alonso de Cordova 3107, Casilla 19001, Santiago, Chile\label{aff231}
\and
University of Applied Sciences and Arts of Northwestern Switzerland, School of Engineering, 5210 Windisch, Switzerland\label{aff232}
\and
Institute of Astronomy, University of Cambridge, Madingley Road, Cambridge CB3 0HA, UK\label{aff233}
\and
Department of Astronomy and Astrophysics, University of California, Santa Cruz, 1156 High Street, Santa Cruz, CA 95064, USA\label{aff234}
\and
Ruhr University Bochum, Faculty of Physics and Astronomy, Astronomical Institute (AIRUB), German Centre for Cosmological Lensing (GCCL), 44780 Bochum, Germany\label{aff235}
\and
Institute of Physics, Laboratory for Galaxy Evolution, Ecole Polytechnique F\'ed\'erale de Lausanne, Observatoire de Sauverny, CH-1290 Versoix, Switzerland\label{aff236}
\and
DARK, Niels Bohr Institute, University of Copenhagen, Jagtvej 155, 2200 Copenhagen, Denmark\label{aff237}
\and
Felix Hormuth Engineering, Goethestr. 17, 69181 Leimen, Germany\label{aff238}
\and
Technical University of Denmark, Elektrovej 327, 2800 Kgs. Lyngby, Denmark\label{aff239}
\and
Universit\'e PSL, Observatoire de Paris, Sorbonne Universit\'e, CNRS, LERMA, 75014, Paris, France\label{aff240}
\and
Universit\'e Paris-Cit\'e, 5 Rue Thomas Mann, 75013, Paris, France\label{aff241}
\and
DTx -- Digital Transformation CoLAB, Building 1, Azur\'em Campus, University of Minho, 4800-058 Guimar\~aes, Portugal\label{aff242}
\and
Department of Physics, School of Advanced Science and Engineering, Faculty of Science and Engineering, Waseda University, 3-4-1 Okubo, Shinjuku, 169-8555 Tokyo, Japan\label{aff243}
\and
Universit\'e Paris-Saclay, CNRS/IN2P3, IJCLab, 91405 Orsay, France\label{aff244}
\and
Ernst-Reuter-Str. 4e, 31224 Peine, Germany\label{aff245}
\and
Department of Physics and Astronomy, University of Pennsylvania, Philadelphia, PA 19146, USA\label{aff246}
\and
Department of Astronomy, Stockholm University, Albanova, 10691 Stockholm, Sweden\label{aff247}
\and
Department of Physics, Centre for Extragalactic Astronomy, Durham University, South Road, DH1 3LE, UK\label{aff248}
\and
Astrophysics Research Centre, University of KwaZulu-Natal, Westville Campus, Durban 4041, South Africa\label{aff249}
\and
School of Mathematics, Statistics \& Computer Science, University of KwaZulu-Natal, Westville Campus, Durban 4041, South Africa\label{aff250}
\and
NASA Goddard Space Flight Center, Greenbelt, MD 20771, USA\label{aff251}
\and
Department of Physics and Astronomy, Vesilinnantie 5, 20014 University of Turku, Finland\label{aff252}
\and
Serco for European Space Agency (ESA), Camino bajo del Castillo, s/n, Urbanizacion Villafranca del Castillo, Villanueva de la Ca\~nada, 28692 Madrid, Spain\label{aff253}
\and
Department of Physics, Duke University, Box 90305, Durham, NC 27708, USA\label{aff254}
\and
ARC Centre of Excellence for Dark Matter Particle Physics, Melbourne, Australia\label{aff255}
\and
Centre for Astrophysics \& Supercomputing, Swinburne University of Technology,  Hawthorn, Victoria 3122, Australia\label{aff256}
\and
Department of Physics and Astronomy, University of the Western Cape, Bellville, Cape Town, 7535, South Africa\label{aff257}
\and
SSAI, Lanham, MD 20706, USA\label{aff258}
\and
Department of Astronomy, University of Maryland, College Park, MD 20742, USA\label{aff259}
\and
Center For Advanced Mathematical Sciences, American University of Beirut PO Box 11-0236, Riad El-Solh, Beirut 11097 2020, Lebanon\label{aff260}
\and
Centro de Investigaci\'{o}n Avanzada en F\'isica Fundamental (CIAFF), Facultad de Ciencias, Universidad Aut\'{o}noma de Madrid, 28049 Madrid, Spain\label{aff261}
\and
International Centre for Radio Astronomy Research, University of Western Australia, 35 Stirling Highway, Crawley, Western Australia 6009, Australia\label{aff262}
\and
Institute for Astronomy, Astrophysics, Space Applications and Remote Sensing, National Observatory of Athens, 15236, Penteli, Greece\label{aff263}
\and
Department of Physics \& Astronomy, University of Sussex, Brighton BN1 9QH, UK\label{aff264}
\and
MTA-CSFK Lend\"ulet Large-Scale Structure Research Group, Konkoly-Thege Mikl\'os \'ut 15-17, H-1121 Budapest, Hungary\label{aff265}
\and
Konkoly Observatory, HUN-REN CSFK, MTA Centre of Excellence, Budapest, Konkoly Thege Mikl\'os {\'u}t 15-17. H-1121, Hungary\label{aff266}
\and
University of Nottingham, University Park, Nottingham NG7 2RD, UK\label{aff267}
\and
Helsinki Institute of Physics, Gustaf H{\"a}llstr{\"o}min katu 2, University of Helsinki, Helsinki, Finland\label{aff268}
\and
SRON Netherlands Institute for Space Research, Landleven 12, 9747 AD, Groningen, The Netherlands\label{aff269}
\and
APCO Technologies, Chemin de Champex 10, 1860 Aigle, Switzerland\label{aff270}
\and
Department of Astrophysics/IMAPP, Radboud University, PO Box 9010, 6500 GL Nijmegen, The Netherlands\label{aff271}
\and
Surrey Satellite Technology Limited, Tycho House, 20 Stephenson Road, Surrey Research Park, Guildford, GU2 7YE, UK\label{aff272}
\and
Centre de Calcul de l'IN2P3/CNRS, 21 avenue Pierre de Coubertin 69627 Villeurbanne Cedex, France\label{aff273}
\and
ICTP South American Institute for Fundamental Research, Instituto de F\'{\i}sica Te\'orica, Universidade Estadual Paulista, S\~ao Paulo, Brazil\label{aff274}
\and
Lawrence Berkeley National Laboratory, One Cyclotron Road, Berkeley, CA 94720, USA\label{aff275}
\and
University of California, Berkeley, Berkeley, CA 94720, USA\label{aff276}
\and
NOVA optical infrared instrumentation group at ASTRON, Oude Hoogeveensedijk 4, 7991PD, Dwingeloo, The Netherlands\label{aff277}
\and
Aurora Technology for European Space Agency (ESA), Camino bajo del Castillo, s/n, Urbanizacion Villafranca del Castillo, Villanueva de la Ca\~nada, 28692 Madrid, Spain\label{aff278}
\and
Universidad Europea de Madrid, 28670, Madrid, Spain\label{aff279}
\and
INAF-IASF Bologna, Via Piero Gobetti 101, 40129 Bologna, Italy\label{aff280}
\and
Oskar Klein Centre for Cosmoparticle Physics, Department of Physics, Stockholm University, Stockholm, SE-106 91, Sweden\label{aff281}
\and
Institute of Space Sciences and Astronomy (ISSA), University of Malta, Msida, MSD 2080, Malta\label{aff282}
\and
STAR Institute, Quartier Agora - All\'ee du six Ao\^ut, 19c B-4000 Li\`ege, Belgium\label{aff283}
\and
McWilliams Center for Cosmology, Department of Physics, Carnegie Mellon University, Pittsburgh, PA 15213, USA\label{aff284}
\and
Serra H\'unter Fellow, Departament de F\'{\i}sica, Universitat Aut\`onoma de Barcelona, E-08193 Bellaterra, Spain\label{aff285}
\and
HE Space for European Space Agency (ESA), Camino bajo del Castillo, s/n, Urbanizacion Villafranca del Castillo, Villanueva de la Ca\~nada, 28692 Madrid, Spain\label{aff286}
\and
Gran Sasso Science Institute (GSSI), Viale F. Crispi 7, L'Aquila (AQ), 67100, Italy\label{aff287}
\and
Research Center for Space and Cosmic Evolution, Ehime University, 2-5 Bunkyo-cho, Matsuyama, Ehime 790-8577, Japan\label{aff288}
\and
Niels Bohr Institute, University of Copenhagen, Jagtvej 128, 2200 Copenhagen, Denmark\label{aff289}
\and
Universit\'e Paris-Saclay, CEA, D\'epartement de Physique des Particules, 91191, Gif-sur-Yvette, France\label{aff290}
\and
Instituci\'o Catalana de Recerca i Estudis Avan\c{c}ats (ICREA), Passeig de Llu\'{\i}s Companys 23, 08010 Barcelona, Spain\label{aff291}
\and
Kobayashi-Maskawa Institute for the Origin of Particles and the Universe, Nagoya University, Chikusa-ku, Nagoya, 464-8602, Japan\label{aff292}
\and
Institute for Advanced Research, Nagoya University, Chikusa-ku, Nagoya, 464-8601, Japan\label{aff293}
\and
Physics and Astronomy Department, University of California, 900 University Ave., Riverside, CA 92521, USA\label{aff294}
\and
Dribia Data Research S.L., Pg. de Gr\'acia, 55, 3r 4a, 08007 Barcelona, Spain\label{aff295}
\and
Laboratoire d'Annecy-le-Vieux de Physique Theorique, CNRS \& Universite Savoie Mont Blanc, 9 Chemin de Bellevue, BP 110, Annecy-le-Vieux, 74941 ANNECY Cedex, France\label{aff296}
\and
National Astronomical Observatory of Japan, National Institutes of Natural Sciences, 2-21-1 Osawa, Mitaka, Tokyo 181-8588, Japan\label{aff297}
\and
School of Physics and Astronomy, Faculty of Science, Monash University, Clayton, Victoria 3800, Australia\label{aff298}
\and
Department of Mathematics, Imperial College London, London SW7 2AZ, UK\label{aff299}
\and
Finnish Centre for Astronomy with ESO (FINCA)\label{aff300}
\and
Center for Frontier Science, Chiba University, 1-33 Yayoi-cho, Inage-ku, Chiba 263-8522, Japan\label{aff301}
\and
Department of Physics, Graduate School of Science, Chiba University, 1-33 Yayoi-Cho, Inage-Ku, Chiba 263-8522, Japan\label{aff302}
\and
Dipartimento di Fisica "E. Pancini", Universita degli Studi di Napoli Federico II, Via Cinthia 6, 80126, Napoli, Italy\label{aff303}
\and
Dipartimento di Fisica, Universit\`a degli studi di Genova, and INFN-Sezione di Genova, via Dodecaneso 33, 16146, Genova, Italy\label{aff304}
\and
Department of Physics and Astronomy, University of Aarhus, Ny Munkegade 120, DK-8000 Aarhus C, Denmark\label{aff305}
\and
Department of Physics \& Astronomy, Louisiana State University, 202 Nicholson Hall, Baton Rouge, LA 70803, USA\label{aff306}
\and
Dipartimento di Scienze Fisiche e Chimiche, University of L'Aquila, Italy\label{aff307}
\and
Department of Astrophysical Sciences, Peyton Hall, Princeton University, Princeton, NJ 08544, USA\label{aff308}
\and
The Cooper Union for the Advancement of Science and Art, 41 Cooper Square, New York, NY 10003, USA\label{aff309}
\and
Teledyne E2V, 106, Waterhouse Lane, Chelmsford, CM1 2QU, UK\label{aff310}
\and
Institute of Space Science, Str. Atomistilor, nr. 409 M\u{a}gurele, Ilfov, 077125, Romania\label{aff311}
\and
Space physics and astronomy research unit, University of Oulu, Pentti Kaiteran katu 1, FI-90014 Oulu, Finland\label{aff312}
\and
Higgs Centre for Theoretical Physics, School of Physics and Astronomy, The University of Edinburgh, Edinburgh EH9 3FD, UK\label{aff313}
\and
Centre de Recherche Astrophysique de Lyon, UMR5574, CNRS, Universit\'e Claude Bernard Lyon 1, ENS de Lyon, 69230, Saint-Genis-Laval, France\label{aff314}
\and
CERN, Theoretical Physics Department, Geneva, Switzerland\label{aff315}
\and
Universit\'e de Franche-Comt\'e, Institut UTINAM, CNRS UMR6213, OSU THETA Franche-Comt\'e-Bourgogne, Observatoire de Besan\c con, BP 1615, 25010 Besan\c con Cedex, France\label{aff316}
\and
NOVA, Netherlands Research School For Astronomy, Einsteinweg 55, NL-2333CC Leiden, The Netherlands\label{aff317}
\and
Theoretical astrophysics, Department of Physics and Astronomy, Uppsala University, Box 515, 751 20 Uppsala, Sweden\label{aff318}
\and
Universit\'e St Joseph; Faculty of Sciences, Beirut, Lebanon\label{aff319}
\and
Departamento de F\'isica, FCFM, Universidad de Chile, Blanco Encalada 2008, Santiago, Chile\label{aff320}
\and
Institut de Recherche en Informatique de Toulouse (IRIT) et Centre de Biologie Int\'egrative (CBI), Laboratoire MCD, CNRS, Universit\'e de Toulouse, 31062 Toulouse, France\label{aff321}
\and
Hamburger Sternwarte, University of Hamburg, Gojenbergsweg 112, 21029 Hamburg, Germany\label{aff322}
\and
Department of Physics and Astronomy, University of British Columbia, Vancouver, BC V6T 1Z1, Canada\label{aff323}
\and
Junia, EPA department, 41 Bd Vauban, 59800 Lille, France\label{aff324}
\and
Mathematical Institute, University of Leiden, Niels Bohrweg 1, 2333 CA Leiden, The Netherlands\label{aff325}
\and
Satlantis, University Science Park, Sede Bld 48940, Leioa-Bilbao, Spain\label{aff326}
\and
School of Physics \& Astronomy, University of Southampton, Highfield Campus, Southampton SO17 1BJ, UK\label{aff327}
\and
ASTRON, the Netherlands Institute for Radio Astronomy, Postbus 2, 7990 AA, Dwingeloo, The Netherlands\label{aff328}
\and
Anton Pannekoek Institute for Astronomy, University of Amsterdam, Postbus 94249, 1090 GE Amsterdam, The Netherlands\label{aff329}
\and
School of Physics and Astronomy, University of Birmingham, Birmingham, B15 2TT, UK\label{aff330}
\and
Clara Venture Labs AS, Fantoftvegen 38, 5072 Bergen, Norway\label{aff331}
\and
Univ. Lille, CNRS, Centrale Lille, UMR 9189 CRIStAL, 59000 Lille, France\label{aff332}
\and
Department of Physics, Royal Holloway, University of London, TW20 0EX, UK\label{aff333}
\and
Department of Physics and Astronomy, University of California, Davis, CA 95616, USA\label{aff334}
\and
Department of Physics, Universit\'{e} de Montr\'{e}al, 2900 Edouard Montpetit Blvd, Montr\'{e}al, Qu\'{e}bec H3T 1J4, Canada\label{aff335}
\and
Academia Sinica Institute of Astronomy and Astrophysics (ASIAA), 11F of ASMAB, No.~1, Section 4, Roosevelt Road, Taipei 10617, Taiwan\label{aff336}
\and
Observatorio Astron\'omico Nacional, IGN, Calle Alfonso XII 3, E-28014 Madrid, Spain\label{aff337}
\and
Center for Cosmology and AstroParticle Physics, The Ohio State University, 191 West Woodruff Avenue, Columbus, OH 43210, USA\label{aff338}
\and
Department of Physics, The Ohio State University, Columbus, OH 43210, USA\label{aff339}
\and
Laboratoire univers et particules de Montpellier, Universit\'e de Montpellier, CNRS, 34090 Montpellier, France\label{aff340}
\and
Centre Spatial de Liege, Universite de Liege, Avenue du Pre Aily, 4031 Angleur, Belgium\label{aff341}
\and
National Astronomical Observatory of Japan, 2-21-1 Osawa, Mitaka, Tokyo 181-8588, Japan\label{aff342}
\and
Universidad Polit\'ecnica de Cartagena, Departamento de Electr\'onica y Tecnolog\'ia de Computadoras,  Plaza del Hospital 1, 30202 Cartagena, Spain\label{aff343}
\and
Italian Space Agency, via del Politecnico snc, 00133 Roma, Italy\label{aff344}
\and
University of Trento, Via Sommarive 14, I-38123 Trento, Italy\label{aff345}
\and
Telespazio Germany GmbH, Europapl. 5, 64293 Darmstadt, Germany\label{aff346}
\and
Center for Cosmology and Particle Physics, Department of Physics, New York University, New York, NY 10003, USA\label{aff347}
\and
RAL Space, Rutherford Appleton Laboratory, STFC, UKRI, Harwell Campus, Oxfordshire, OX11 0QX, UK\label{aff348}
\and
Space Telescope Science Institute, 3700 San Martin Dr, Baltimore, MD 21218, USA\label{aff349}}

 \date\today

%
%
\abstract{The current standard model of cosmology successfully describes a variety of measurements, but the nature of its main ingredients, dark matter and dark energy, remains unknown.
\Euclid\ is a medium-class mission in the Cosmic Vision 2015--2025 programme of the European Space Agency (ESA)
that will provide high-resolution optical imaging, as well as near-infrared imaging and spectroscopy, over about 14\,000\,deg$^2$ of extragalactic sky. In addition to accurate weak lensing and clustering measurements that probe structure formation over half of the age of the Universe, its primary probes for cosmology, these exquisite data will enable a wide range of science. This paper provides a high-level overview of the mission, summarising the survey characteristics, the various data-processing steps, and data products. We also highlight the main science objectives and expected performance.}
%
%
%
%
   \titlerunning{Overview of the \Euclid mission}
   \authorrunning{Euclid Collaboration: Y.\ Mellier et al.}
   
   \maketitle
\addtolength{\cftsubsecnumwidth}{4pt}


\section{\label{sc:Intro}Introduction}

A century of ever improving observations has resulted in a concordance cosmological model that is surprisingly simple: only six numbers are currently needed to describe a wide variety of precise measurements \citep{PlanckParams2018}. The result, however, is also unsatisfactory because it highlights a serious problem for our understanding of fundamental physics and astronomy: it relies on assumptions about the initial conditions and the theory of gravity, while the nature of the main ingredients, dark matter and dark energy, remain great mysteries. The observational evidence for a largely `dark' universe is overwhelming, demonstrating that our theories of particle physics and/or gravity are either incomplete or incorrect. Moreover, we lack compelling theoretical guidance to solve this crisis in fundamental physics \citep{Albrecht2006,Amendola18}.

Arguably, the biggest challenge is the observation that the expansion of the Universe is accelerating \citep[e.g.][]{Riess98, Perlmutter99, Eisenstein2005, Betoule2014}.
Current explanations range from Einstein’s cosmological constant, dynamic mechanisms such as quintessence, or a modification of the laws of gravity on cosmological scales \citep[see][for an extensive overview of ideas]{Amendola18}. To robustly distinguish between these different theoretical ideas, the precision of the measurements needs to improve by at least an order of magnitude, whilst our ability to interpret the data correctly has to advance accordingly if we are to take advantage of the smaller statistical uncertainties.

The exquisite observations of the temperature fluctuations in the \gls{CMB} performed by WMAP \citep{Hinshaw2013} and {\it Planck} \citep{Planck:2018nkj} have been crucial to establish the baseline cosmological constant-dominated cold dark matter ($\Lambda$CDM) model.  This is because the physical interpretation of the measurements is relatively straightforward, and all-sky experiments from space benefit from a stable environment and superior control of instrumental effects. 
Unfortunately, the \gls{CMB} provides limited information about the nature of dark matter and dark energy, because it primarily probes the physical conditions at the time of recombination, when dark energy was negligible. 
To quantify how the balance between dark matter and dark energy evolved, the \gls{CMB} results need to be complemented by high-quality measurements of the cosmic expansion history and the growth of \gls{LSS} over the past eight billion years or more. In principle, such studies can provide complementary constraints on the initial conditions, explore modifications of the theory of general relativity on cosmological scales, and determine the neutrino mass scale \citep{Albrecht2006}. Research in observational cosmology is therefore shifting towards studies of the \gls{LSS}, and a number of large spectroscopic and imaging surveys will collect vast amounts of data in the coming decade. For instance, the \gls{LSST} by the Vera C. Rubin Observatory will repeatedly image the entire southern sky in multiple bands \citep{Ivezic2019}, while the \gls{DESI} will measure redshifts for about 40 million galaxies \citep{DESI2023}. These projects are major improvements over previous surveys, but a robust interpretation is essential.  
Accounting for the  complexities of ground-based data, in particular the variable sky conditions, presents a major challenge. Hence, further advances will come from space-based facilities, such as \Euclid \citep{Laureijs11} and the planned \textit{Nancy Grace Roman} Space Telescope \citep{Akeson2019}. 

This paper describes the background, instruments, performance, and planned science of \Euclid, a medium-class mission in the Cosmic Vision 2015--2025 programme of the \gls{ESA}. \Euclid resulted from a 2007 call by \gls{ESA} for the selection of a medium-sized space mission. Besides five other mission proposals, a dark energy mission concept was included for competitive down-selection. This new mission concept, ultimately named \Euclid, was based on a combination of two initial dark-energy mission proposals, the \acrlong{DUNE}\glsunset{DUNE} \citep[DUNE;][]{Refregier09} and the \acrlong{SPACE}\glsunset{SPACE} \cite[SPACE;][]{Cimatti09}. It envisioned
an extragalactic sky survey with visual imaging, and near-infrared photometry and spectroscopy, optimised for the measurement of the two primary cosmology probes, namely galaxy clustering and weak gravitational lensing, which are both powerful ways of measuring the evolution of the \gls{LSS}, whilst complementing each other in terms of constraining power.

The main science case of \Euclid was presented in \cite{Laureijs11}, which formed the basis for the official selection of the concept by \gls{ESA} in 2011, and adoption as a mission in 2012. Since then, the mission hardware and software have been built, culminating in the successful launch of \Euclid on 1st July 2023 on a Falcon-9 rocket from Cape Canaveral. Here we provide an up-to-date high-level overview of the mission during its initial time in orbit, describing the survey, the data products, and the science that the Euclid Consortium aims to perform.

The structure of this paper is as follows. In \cref{sec:primary} we summarise the main science objectives of the \Euclid mission and introduce the primary cosmological probes that drove the design. In \cref{sec:instruments} we provide a brief overview of the spacecraft and its instruments. The survey characteristics and supporting observations are described in \cref{sec:survey}, while early results from the commissioning and performance verification are presented in \cref{sec:preliminaryCommissioning}. The main data products that will be released publicly comprise simulated data (\cref{sec:sims}) and the actual data processed through the \Euclid \gls{SGS} pipeline (\cref{sec:data}). The cosmological inferences enabled by \Euclid are discussed in \cref{sec:cosmology}. Additional cosmological probes that are enabled or enhanced by the \Euclid data are reviewed in \cref{sec:additional}. Finally, the impact of \Euclid is not limited to cosmology, and a taste of the wide range of astrophysics that will be done is presented in more detail in \cref{sec:legacy}. To aid in the readability of this and accompanying papers, we include a glossary of acronyms in \cref{app:glossary}. Unless specified otherwise, magnitudes and surface brightness values are reported using the AB magnitude system.


\section{Primary probes for cosmology with \Euclid}
\label{sec:primary}

The biggest mysteries in cosmology are the nature of dark matter and dark energy. Indirect evidence for the existence of dark matter has come from a wide range of astronomical and cosmological observations, but if it is composed of new elementary particles outside the standard model of particle physics, direct detections in terrestrial experiments \citep[e.g.][]{Battaglieri17} are likely to play a prominent role in establishing its nature.  
The situation is markedly different for the observed accelerating expansion of the Universe, where progress will most likely come from advances in observational cosmology. To explain the observations, a component with a negative equation-of-state parameter\footnote{The equation-of-state parameter $w$ relates the pressure $P$ and the energy density $\rho c^2$ of a substance as $P\equiv w \rho c^2$. In the case of a cosmological constant $\Lambda\neq 0$, or a non-zero vacuum energy density, $w=-1$, and the time derivative is zero.} is required, which is commonly referred to as dark energy.
It makes up about 69\%  of the present-day energy density \citep{PlanckParams2018}, but we do not know if it is a cosmological constant, a field that evolves dynamically, or reflects a more profound modification of the gravitation laws at cosmological scales \citep[see e.g.][]{Amendola18}. One of the key observables to distinguish between such models is the way the equation-of-state parameter $w(z)$ varies with redshift. To allow for a convenient comparison between cosmological probes, we adopt
\begin{equation}
w(z) = w_{\rm 0} + w_{a}\, \frac{z}{1+z} \label{eq:w0wa}\;,
\end{equation}
which captures the dynamical nature of dark energy to first order in $(1-a)$, where $a=1/(1+z)$ is the scale factor. Here, $w_{\rm 0}$ is the present day equation of state, and $w_a$ quantifies the dependence with redshift. This extension corresponds to a basic non-clustering dynamical dark energy model, and constraints on the parameters $w_0$ and $w_{a}$ show how well \Euclid can test this scenario. 
For completeness, we note that the cosmological constant corresponds to the choice $w_0=-1$ and $w_{a} = 0$. 

We can quantify the performance of a particular survey by comparing the dark energy \gls{FOM}, which we defined as the inverse square root of the covariance matrix determinant for the dark energy parameters
\citep{Wang2008},
\begin{equation}
{\rm FoM}=\frac{1}{\sqrt{\det {\rm Cov}(w_{\rm 0},w_{a})}}\;,
\label{eq:FOM}
\end{equation}
so that a larger value implies a more precise measurement of the dark energy properties.\footnote{To evaluate the \gls{FOM} we marginalise over the cosmological and nuisance parameters, while imposing a Gaussian prior on the baryon density from \gls{BBN}. See \cref{sec:constraints} for more details.}
The challenge for the experimental design, however, is to establish a minimum value to achieve. \cite{Laureijs11} presented a statistical argument, concluding that the value of $w$ needs to be determined with a precision of about 1\% to robustly test the $\Lambda$CDM model. Specifically, one of the main objectives of \Euclid is to constrain the dark energy equation of state so that ${\rm FoM}>400$ for this baseline scenario. 

This is a challenging target for two reasons. First, to achieve such statistical constraining power requires surveying a considerable fraction of the cosmological volume out to $z=2$ \citep[see e.g.][]{Amara07}, thus covering the epoch during which dark energy became the dominant component in the Universe. In the case of \Euclid, the aim is to observe 14\,000\,deg$^2$ of extragalactic sky with low zodiacal background and low Galactic extinction (see \cref{sec:survey} for details). Second, systematic biases need to be sufficiently small as to not overwhelm the orders of magnitude improvement in precision. The observational signatures of dark energy are subtle, and unrecognised instrumental effects could be mistaken for new physics. To minimise this, exquisite experimental control is paramount. Although still challenging (and dependent on the specific probe used), this is best achieved from space.

A measurement of the expansion history via the distance-redshift relation provides the most direct constraints on the equation of state of dark energy. More generally, it provides constraints on the relative balance of ingredients and spatial curvature via the Friedmann equations, which link the time evolution of the scale factor $a(t)$ to these ingredients. The rate at which density fluctuations grow also depends on the background cosmology. Hence, studying the growth of \gls{LSS} provides another way to infer information about the composition of the Universe. It has the added benefit that it can constrain additional cosmological parameters and provide some key tests of the underlying theory of gravity \citep{Amendola18}.

The expansion history and growth of \gls{LSS} can be probed using a variety of techniques, each with their own advantages and limitations. Clearly, to achieve ${\rm FoM}>400$, the probes of interest should depend on the properties of the dark energy. Although some are more sensitive than others, no individual probe can reach the target \gls{FOM}, given the practical constraints on the mission design (see \cref{sec:instruments}). Instead, probes need to be combined. Ideally, the individual probes should have comparable sensitivity, but also complement each other in terms of observational needs and precision. That is, the sum should be more than its parts. 

Analogous to the use of the dark energy equation of state, the dimensionless linear growth rate $f_\mathrm{g}(z)$ depends on $\Omega_{\rm m}(z)$, the redshift-dependent ratio of  matter density divided by the critical density,
\begin{equation}
f_{\rm g}(z)\equiv \frac{{\rm d}\ln g_+(z)}{{\rm d}\ln a}\simeq\brackets{\Omega_{\rm m}(z)}^{\gamma_\mathrm{g}}\;,
\label{eq:gamma}
\end{equation}
where $g_+(z)$ is the linear growth factor that relates the amplitude of a linear density fluctuation, $\delta(\vec{x},z)$, to its present value via $\delta(\vec{x},z)=g_+(z)\,\delta(\vec{x},0)$. As \gls{GR} predicts a value of $\gamma_\mathrm{g}\simeq 0.55$ for a flat $\Lambda$CDM
cosmology \citep[e.g.][]{Peebles1980,Lahav1991}, a measurement of $f_\mathrm{g}(z)$ can be used to constrain the composition of the Universe, similar to what is done for the expansion history. However, a detailed measurement of the growth rate as a function of redshift and possibly of scale, can also shed light on the nature of dark energy and the underlying theory of gravity. In fact, a dynamical, clustering dark energy component, as well as modifications of \gls{GR}, would not only lead to a different $\Omega_{\rm m}(z)$ but also to $\gamma_\mathrm{g}\neq 0.55$ \citep{Linder:2005in}. 
Therefore, another key objective of \Euclid is to determine the value of $\gamma_\mathrm{g}$ with a precision better than 0.02 (68\% confidence), sufficient to distinguish between \gls{GR} and a wide range of modified gravity theories
\citep{Laureijs11}.
Moreover, modified gravity theories tend to affect dynamical and relativistic observables differently \citep[e.g.][]{Amendola18}, suggesting that one would like to combine such probes.
 
\begin{table}[t]
\caption{Fiducial values for the cosmological parameters of the baseline flat $\Lambda$CDM cosmological model.}
\label{tab:fiducials}
\begin{center}
\begin{tabular}{cc}
\hline\hline
\noalign{\vskip 1pt}
Cosmological parameter & Fiducial value\\
\hline
\noalign{\vskip 1pt}
$\Omega_{\rm m}$ & 0.32\\
$\Omega_{\rm b}$ & 0.05\\
$h$ & 0.67\\
$n_{\rm s}$ & 0.96\\
$\sigma_8$ & 0.816\\
$\sum m_{\nu}$\,[eV] & 0.06\\ 
$\tau$ & 0.058\\
\hline
\end{tabular}
\end{center}
\end{table}

These considerations, combined with specific mission constraints, led to the decision to optimise \Euclid for two powerful and highly complementary probes, namely weak gravitational lensing and galaxy clustering. These are the most sensitive probes of dark energy and gravity on cosmological scales \citep[see reviews in][]{Peacock2006,Albrecht2006,Albrecht2009,Weinberg2013}. Combined, they probe the cosmological expansion history, the growth of structure, and the relation between dark and luminous matter.

More detail is provided below, but in summary, to map the three-dimensional matter distribution, \Euclid aims to determine emission-line redshifts for more than 25 million galaxies over the redshift range $0.9<z<1.8$ using slitless spectroscopy at near-infrared wavelengths\footnote{The actual wavelength range covered by the grism is larger, allowing the detection of H$\alpha$ emitters over the redshift range $0.84<z<1.88$ (see \cref{sec:nisp-s}). Throughout this paper, however, we consider the more conservative as-required numbers.}. These data provide precise measurements of the growth of structure through the clustering of galaxies and redshift-space distortions, while on large scales the \glspl{BAO} probe the expansion history. Simultaneously with the slitless spectroscopy, \Euclid will collect diffraction-limited\footnote{As discussed in more detail in \cref{sec:psfmodel}, the telescope is not a perfect optical system. Hence, in this context, diffraction-limited is to be understood as referring to a telescope of extremely good image quality.} images at optical wavelengths over the same area. These enable accurate measurements of the shapes of about 1.5 billion galaxies that will be used to map the distribution of matter using weak gravitational lensing. Photometric redshifts for these sources will be determined by combining supporting ground-based observations with near-infrared images in three passbands from \Euclid over the same area. 

As discussed in \cite{Laureijs11} and more recently by \cite{Blanchard-EP7}, this particular probe combination can achieve a dark energy ${\rm FoM}>400$ and measure $\gamma_\mathrm{g}$ with an uncertainty of 0.02 ($1\,\sigma$), where we caution that, to achieve these objectives, the predictions for the observables on small scales need to be improved further (see \cref{sec:nonlinear}). This precision allows us to explore physics beyond the concordance
$\Lambda$CDM model. These scenarios, $\Lambda$CDM with and without spatial curvature, a dynamic dark energy model with an equation of state given by \cref{eq:w0wa}, and a modified gravity scenario based on \cref{eq:gamma}, provide the basic benchmark cases used to evaluate the performance of \Euclid. We present updated estimates on the precision that \Euclid aims to achieve in \cref{sec:likelihood}.

Apart from distinguishing dark energy and modified gravity models, improving constraints on the cosmological parameters that make up the $\Lambda$CDM model provides a crucial consistency test. Currently, local measurements \citep{Riess21} of the Hubble constant $H_0$ show disagreement with the values inferred from the analysis of \Planck data \citep{PlanckParams2018}. The high-quality measurements from \Euclid will help to settle the $H_0$ debate, because the primary probes of \Euclid can also provide state-of-the-art constraints on the cosmological parameters that form the baseline flat $\Lambda$CDM model; these are listed in \cref{tab:fiducials} alongside their fiducial values that are used to assess the performance of \Euclid. 

Historically, these values were chosen based on table~3 of the 2015  \textit{Planck} results \citep{PlanckParams2015}. We consider a baseline fixed sum of neutrino masses $\Sigma m_\nu=0.06\,$eV, a fixed optical depth to Thomson scattering from reionisation, $\tau=0.058$, and a spectral index of the primordial density power spectrum, 
$n_{\rm s}=0.96$. For the spatially flat $\Lambda$CDM model, the dark energy density parameter today, $\Omega_{\rm DE}$, is a derived parameter, but we allow it to vary when we consider a model with spatial curvature. For the dimensionless Hubble parameter $h$ (defined through $H_0=100\,h\,{\rm km}\,{\rm s}^{-1}\,{\rm Mpc}^{-1}$), we adopt the \gls{CMB} value of $h=0.67$.
The remaining parameters are $\Omega_{\rm b}$ and $\Omega_{\rm m}$, respectively, the baryon and total matter energy densities at the present time, divided by the critical density. Finally, $\sigma_8$ measures the amplitude of the relative linear density fluctuations within a sphere of radius $8\,h^{-1}$\,Mpc at the present day. We refer the reader to \citet{Blanchard-EP7} for more details on the fiducial choices. \Euclid will reduce the uncertainties on all these parameters significantly. 

\Euclid will also greatly advance our ability to constrain extensions of the standard cosmological model. In  
\cref{sec:beyondlcdm} we discuss some cases in more detail, but here we highlight two specific examples. 
First, \Euclid will improve constraints on the sum of neutrino masses. In combination with \Planck, we expect to reach a precision of $\sigma(\Sigma m_\nu)=0.02$\,eV. Second, \Euclid will
improve our understanding of the initial conditions. The concordance model assumes an initial Gaussian random field of perturbations. The parameter 
$f_{\rm NL}$ quantifies the quadratic term in the potential \citep[e.g.][]{Matarrese2000,Dalal2008}, and thus provides a measure for any initial non-Gaussianity, which is  encoded in the \gls{LSS} that \Euclid will map with unprecedented precision. The aim is 
to improve over the current measurements from \Planck \citep{Planck:2019kim}. Taken together, \Euclid will test many aspects of the $\Lambda$CDM model \citep{Amendola18, Blanchard-EP7}. 

In the remainder of this section we discuss the primary probes in more detail, but we note that the same data enable additional cosmological studies, which are discussed in \cref{sec:additional}. Although including this information consistently is not trivial, as highlighted in \cite{Laureijs11}, it is worthwhile to pursue; significant improvement is expected for a wide range of cosmological parameters, but the largest impact is foreseen on the dark energy constraints.

\subsection{Galaxy clustering}
\label{sec:galaxyclustering}

\begin{figure}
    \centering
    \includegraphics[width=\linewidth]{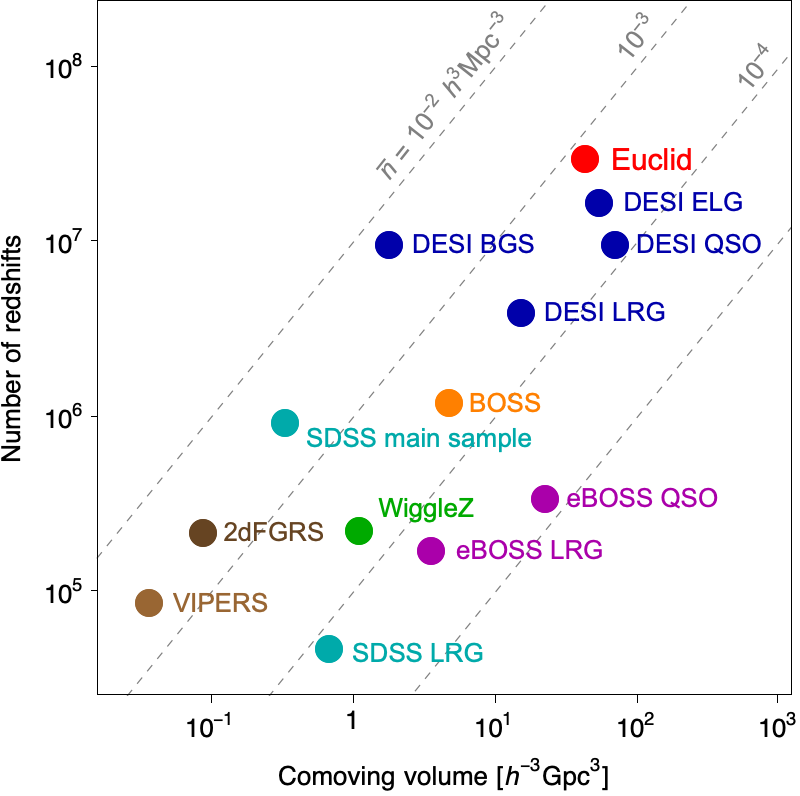}
    \caption{Comparison of the number of redshifts and comoving volume covered by various previous and ongoing spectroscopic surveys against the predictions for \Euclid (see text for details). The grey lines show lines of constant number density as labelled.}
    \label{fig:spectro-surveys}
\end{figure}

The large-scale clustering of galaxies is one of the most powerful probes of the Universe, carrying crucial information on its mass/energy budget and fundamental parameters.  
The cosmological information is best extracted by observing the 3D distribution of galaxies in space, combining angular positions with estimates of galaxy distances using the cosmological redshifts from their spectra as a proxy.  

This led to the start of systematic redshift surveys that, since the 1970s, have increasingly collected galaxy redshifts over larger and larger volumes. 
Following the pioneering years of surveys collecting individual spectra from ground-based telescopes \citep[see e.g.][for historical reviews]{Sandage1975,Rood1988,Giovanelli1991}, the 1990s saw the emergence of 
multi-object spectrographs (MOS), which allowed for a quantum leap in survey efficiency. Representative examples of surveys that, in different fashions, exploited MOS spectroscopy for large-scale structure work are the Las Campanas Redshift Survey (LCRS; \citealt{Shectman1996}), the \acrlong{2dFGRS}\glsunset{2dFGRS} (2dFGRS; \citealt{2dFGRS}), the \acrlong{SDSS}\glsunset{SDSS} (SDSS; \citealt{SDSS}), and, more recently and at higher redshifts, the \acrlong{VIPERS}\glsunset{VIPERS} (VIPERS; \citealt{Guzzo2014, Scodeggio18}) and the WiggleZ survey \citep{WiggleZ}. The SDSS encompasses a number of experiments, including the early \textit{main galaxy} and \gls{LRG} samples, and the subsequent \acrlong{BOSS}\glsunset{BOSS} (BOSS; \citealt{Dawson2013}) and extended-BOSS (eBOSS; \citealt{Dawson2016}). Similarly, VIPERS represented the final act, for studies of large-scale structure at $z\sim 1$, of the deep surveys enabled by the VIsible Multi-Object Spectrograph (VIMOS) at the
\acrlong{VLT}\glsunset{VLT} \citep[\gls{VLT};][]{LeFevre13}.

The \Euclid NISP slitless spectroscopic survey is part of a new generation of such surveys, which will bring the total number of measured redshifts close to a hundred million. While \Euclid will sample the $0.9<z<1.8$ redshift range from space, the \gls{DESI} project \citep{DESI}, already under way at the Kitt Peak 4-m Mayall telescope, is primarily targeting galaxies at lower redshifts (with significant overlap), using an innovative 5000-fibre automatic positioner.  In fact, the wavelength range of its red grisms (\cref{sec:nisp-s}) was specifically chosen as to make \Euclid complementary to existing and planned ground-based surveys.
The number of expected redshifts and the sampled volume of \Euclid and DESI are compared with some previous surveys in \cref{fig:spectro-surveys}.

The observed clustering of galaxies within a past lightcone encodes a wide range of physics through a number of processes.
The currently favoured picture envisages that the initial comoving pattern of the overdensities that grew to form today's galaxies and large-scale structures was set up in the early Universe, driven initially by inflation (e.g. \citealt{Inflation-review}). These were then modified before recombination, driven by physics that imprints scales related to the epoch of matter-radiation equality and the propagation of acoustic waves. These are commonly included in models of the power spectrum by a transfer function that multiplies the primordial power-law inflationary spectrum (e.g. \citealt{CAMB}). Shortly after recombination, the density contrast is still small over the scales of interest, perturbations are still linear and their growth is essentially scale-independent.
Departures from scale-independence might thus signal the effects of non-zero neutrino mass (an effect of about one per cent is expected with current mass limits -- see \citealt[][for a review]{Lesgourgues2006}), or more exotic physics. Starting from the smallest scales, at late times gravitational evolution 
becomes nonlinear, bringing in additional information, but also new complications \citep[see, e.g.][]{2002PhR...367....1B}. As we discuss below, the way that pattern is imprinted into the angles and redshifts measured by a galaxy survey allows us to extract important cosmological information.

Much of this
is encoded in the two-point statistics of the overdensity field: in configuration space,\footnote{With \textit{configuration space} we mean the space where we measure galaxy positions and distances, dual to \textit{Fourier space}. In turn, configuration space is distinguished into \textit{real} space, where one uses true galaxy distances and separations are indicated with $\vec r$, and \textit{redshift} space, where distances are derived from measured redshifts, and galaxy separations are typically indicated with $\vec s$. Correspondingly, the same distinction applies to the Fourier side, where wavenumbers $\vec k$ can be defined in real and redshift space.  Again, redshift-space quantities are usually indicated with the subscript $s$.} 
we measure the spatial \gls{2PCF} of galaxies, $\xi_{\rm gg}(\vec r),$ which quantifies the excess probability of finding two objects at a given separation $\vec r$ with respect to a random Poisson sample tracing the same volume,
\begin{equation}
  \langle N_{\rm g,1}(\vec x) \,N_{\rm g,2}(\vec x + \vec r) \rangle = 
  \bar{n}^2\,\left[1+\xi_{\rm gg}(\vec r)\right]\, \delta V_1\,\delta V_2\;.
\end{equation}
Here $\bar{n}$ is the mean number of galaxies per unit volume and we are considering two small regions, separated by a vector $\vec r$, with volumes $\delta V_1$ and $\delta V_2$,
containing $N_{\rm g,1}$ and $N_{\rm g,2}$ galaxies, respectively.  It is often convenient to measure clustering in Fourier space, and there we measure the power spectrum, which is the Fourier transform of the correlation function:
\begin{equation}
  P_{\rm gg}(\vec k) = \int_0^\infty\diff^3r\,\xi_{\rm gg}(\vec r)\,{\rm e}^{{\rm i}\,\vec k\cdot\vec r}\;.
\end{equation}

\begin{figure}
    \centering
    \includegraphics[width=\linewidth]{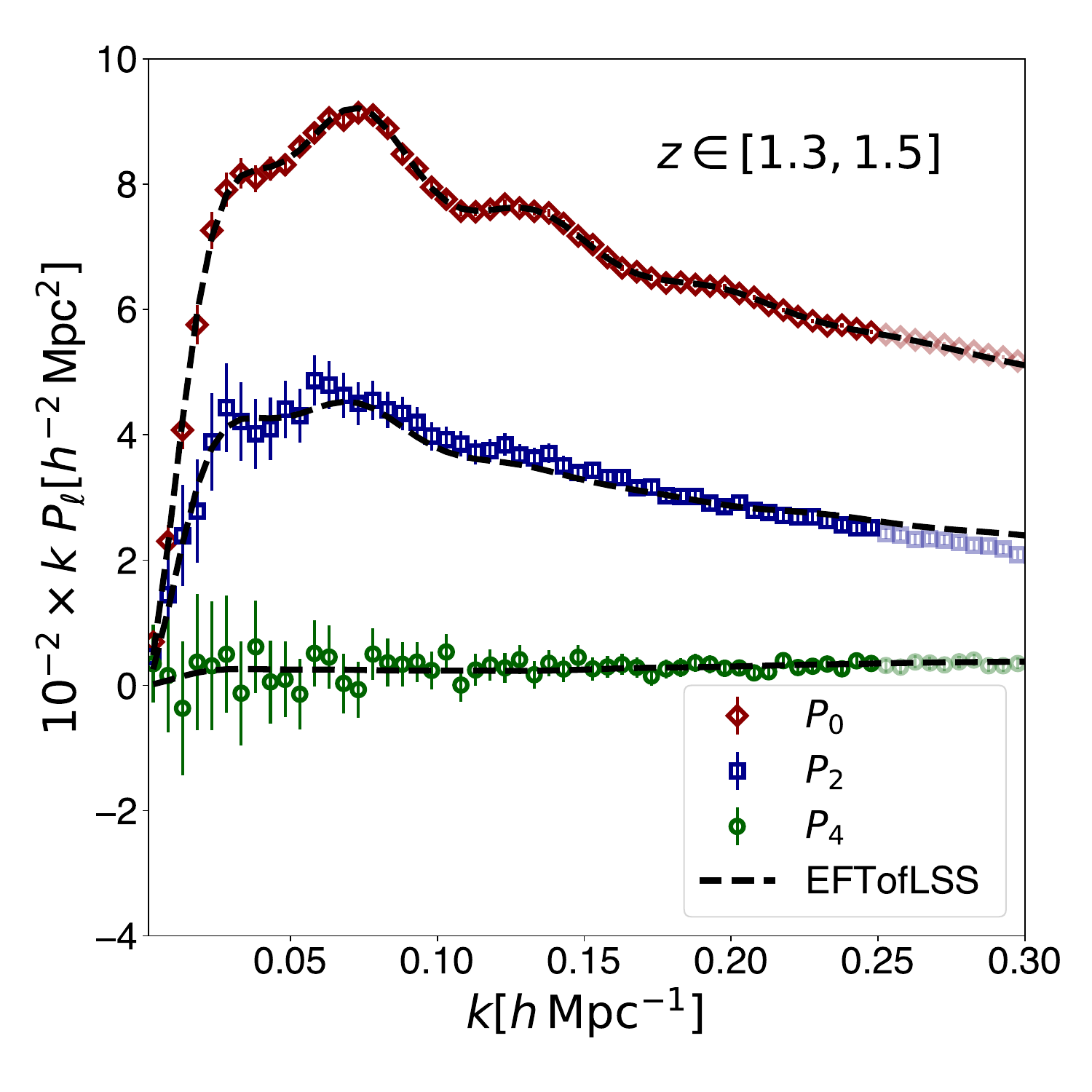}
    \caption{Measured galaxy power spectrum multipoles calculated from dedicated mocks (Pezzotta et al. in prep.) based on the Flagship simulation (see \cref{sec:flagship}) of the \Euclid emission-line sample for a redshift bin $1.3 < z < 1.5$, compared to a best-fit model based on \acrlong{EFT} (EFT, also called EFTofLSS, see \cref{sec:nonlinear} for further discussion) assuming $k_{\rm max} = 0.25\,h$\,Mpc$^{-1}$. Error bars here correspond to the \Euclid full mission volume for this redshift bin prior to observational effects.}
    \label{fig:Pk-prediction}
\end{figure}

Our clustering computations use galaxy redshifts to derive their distances. However, redshifts include, in addition to the pure cosmological Hubble flow, the line-of-sight contribution of the galaxy peculiar velocities, induced by density inhomogeneities. This leads to coherent \acrlong{RSD}s\glsunset{RSD} (RSDs; \citealt{Kaiser87}), introducing an anisotropy in the observed clustering between \gls{LOS} and transverse separations (see below for more details), such that we measure the clustering in redshift space with separations denoted by $\vec s$ rather than $\vec r$. Hence, to properly describe (and model) this effect we typically measure the clustering with respect to the \gls{LOS}. The cosmological information of interest is contained within the first three even power-law moments of the correlation function or power spectrum with respect to $\mu$, under the global plane-parallel approximation, where $\mu$ is constant across a survey and gives the cosine of the angle that a pair of galaxies, for $\xi(\vec s)$, or that the Fourier mode wave vector, for $P(\vec k)$, makes with respect to the \gls{LOS}. The first three even Legendre polynomials encode these power-law moments and form an orthonormal basis:
\begin{eqnarray}
  \mathcal{L}_0(\mu) &=& 1\,;\\
  \mathcal{L}_2(\mu) &=& \frac{1}{2}\,(3\,\mu^2-1)\,;\\
  \mathcal{L}_4(\mu) &=& \frac{1}{8}\,(35\,\mu^4-30\,\mu^2+3)\,.
\end{eqnarray}
We therefore typically decompose the clustering into the Legendre polynomial moments of the correlation
function and power spectrum  \citep{hamilton98}:
\begin{align}
  \xi_\ell(s)&= (2\,\ell+1) \,\int_{-1}^1 \diff\mu\,\xi_{\rm gg}(\vec s)\,\mathcal{L}_\ell(\mu)\,,\label{eq:xi_l}\\
  P_\ell(k)&= (2\,\ell+1) \,\int_{-1}^1 \diff\mu\,P_{\rm gg}(\vec k)\,\mathcal{L}_\ell(\mu)\,,
  \label{eq:Pelldef}
\end{align}
with $\ell=0$ corresponding to the monopole, $\ell=2$ the quadrupole, and $\ell=4$ the hexadecapole moment. In practice, the \gls{LOS} varies across a survey, and we typically make a local-plane parallel approximation, where we assume that the LOS is the same for each pair of galaxies analysed. In this case, the statistic we wish to measure is that given in \cref{eq:xi_l}, but the method by which we estimate it is not as simple as this equation suggests, as discussed in more detail in \cref{sec:clustering_stats}. A prediction for the power spectrum moments to be measured by \Euclid is given in \cref{fig:Pk-prediction}. This figure shows measurements on mock catalogues based on the Euclid Flagship simulation (see \cref{sec:flagship}), and a best-fit model that is able to predict the clustering into the nonlinear regime (see \cref{sec:nonlinear} for more details).

The most robust (and easy to isolate) signal in the pattern of galaxies are the so-called Baryonic Acoustic Oscillations (\glspl{BAO}), a series of peaks and troughs in the power spectrum caused by acoustic waves during the pre-recombination era \citep[e.g.][]{EisensteinHu1998,Meiksin1999}. These are evident in the monopole of \cref{fig:Pk-prediction}. The acoustic waves push baryonic material out from initial perturbations to the baryon-drag scale, which is linked to the comoving sound horizon at recombination. When analysed in Fourier space this results in a sinusoidal term in the transfer function, depending on whether the movement of material cancels or reinforces that from other perturbations separated by the wavenumber of interest. The \glspl{BAO} observed have the same physical origin as the oscillations seen in the CMB anisotropy power spectrum \citep{PlanckParams2018}, and were first observed in the 2dFGRS \citep{Percival2001,Cole2005} and SDSS \citep{Eisenstein2005} surveys. Hardware improvements made to the Sloan telescope enabled BOSS to provide the first 5\,$\sigma$ measurement of \glspl{BAO} from the largest volume of the Universe obtained at that point \citep{BOSS-DR9-BAO}. The extended-BOSS project pushed these \gls{BAO} observations to higher redshifts \citep{Alam2021}. The \gls{BAO} are largely insensitive to galaxy bias, simplifying their modelling. 

The power of \glspl{BAO} in a galaxy survey results from using them as a standard ruler undergoing expansion that is comoving with the average expansion of the Universe. The observed wavenumbers of \glspl{BAO} constrains the ratio $r_{\rm d}/D_{\rm H}(z)$ in the radial direction, and $r_{\rm d}/D_{\rm A}(z)$ in the transverse direction, where $r_{\rm d}$ is the sound horizon at the end of the drag epoch, $D_{\rm H}(z)=c/H(z)$ is the Hubble distance, and $D_{\rm A}(z)$ is the angular-diameter distance. The different dependencies along and across the LOS lead to a very clean geometrical measurement: the correct set of cosmological parameters will be the one leading to a statistically isotropic clustering requiring the product $H(z)\,D_{\rm A}(z)$ to match the truth. This is known as the \acrlong{AP}\glsunset{AP} \citep[AP;][]{AP} effect and the principle holds for features in the power spectrum other than \glspl{BAO}; for example it can also be applied to stacks of objects such as voids, which are under-dense regions defined by a specific threshold \citep[see][for a review]{Pisani2019}. 
With the unprecedented volume of the Wide survey, void statistics with \Euclid, such as the void size function and the void-galaxy cross-correlation function are expected to deliver competitive, complementary cosmological constraints \citep[see][for specific forecasts]{Verza2019, Hamaus2022, Contarini2022, Bonici2023, Radinovic2023}.

The \glspl{BAO} are just one feature within the full power spectrum of the galaxy distribution. The shape of the galaxy power spectrum underlying the \gls{BAO} depends on cosmology through the spectral index of the fluctuations coming from inflation $n_{\rm s}$ and the matter-radiation horizon scale, which depends on the parameter combination $\Omega_{\rm m}\,h$. The full power spectrum predicted to be observed by \Euclid is shown in Fig.~\ref{fig:Pk-prediction} in terms of its Legendre multipoles. 

While \glspl{BAO} carry information about the expansion history $H(z)$ and thus the equation-of-state parameter of dark energy $w(z)$, the anisotropy of the clustering pattern produced by \glspl{RSD}, mentioned earlier, provides us with complementary, potentially powerful information on the growth rate of structure $f_\mathrm{g}(z)$.  As discussed in the introduction to this section, combined precise measurements of $w(z)$ and $f_\mathrm{g}(z)$ are key to understand the origin of cosmic acceleration, potentially discriminating between dark energy and modifications of \gls{GR}, a major goal of \Euclid.
The confidence in the use of \glspl{RSD} as a test of dark energy consolidated at the time of the ESA Cosmic Vision 2020-2025 call \citep{guzzo08, Wang2008rsd}. In fact, the use of \glspl{RSD} as a primary cosmological probe was one key original ingredient in the \gls{SPACE} proposal \citep{Cimatti09}, which eventually became the spectroscopic experiment on board \Euclid \citep[also see][]{Wang2010}. Since then, virtually all redshift surveys have been including \glspl{RSD} as a standard probe of the growth rate of structure \citep[see][and references therein]{Alam2017,Alam2021}. In practice, the growth rate is derived from \glspl{RSD} by modelling the measured anisotropy of the correlation function or power spectrum, as quantified by their multipoles, which depends on the combination $f_\mathrm{g}(z)\,g_+(z)\,\sigma_8$ \citep{percival09}.

In galaxy clustering measurements, $g_+(z)\,\sigma_8$ is degenerate with the so-called `galaxy bias', that is the unknown amplification with respect to the clustering of the underlying matter density field.  This is expected to arise naturally if galaxies form at the peaks of the density-fluctuation field \citep{kaiser84}, with additional complications and scale-dependence on small scales \citep[see][for an extensive review]{Desjacques2018}. 
There, comoving galaxy separations no longer match those between the seed perturbations from which their hosting dark matter halos grew. Galaxy formation and evolution inside halos adds a further layer of complication, that is a galaxy-halo bias. On very large scales the galaxy bias signal can depend on the level of primordial non-Gaussianity, typically enabling a measurement of the $f_{\rm NL}$ parameter. Except for the deviation caused by the $f_{\rm NL}$ signal, we expect the large-scale bias of galaxies to tend towards the scale-independent deterministic linear value predicted by a pure statistical peak (halo) to background (matter) bias. In that case the galaxy power spectrum is simply proportional to the matter power spectrum, $P_{\rm gg}({\vec k})=b_1^2\,P_{\rm m}({\vec k})$. 

The statistical properties of a  Gaussian random field are completely described by its two-point correlation function or, equivalently, the power spectrum. This is not the case for the galaxy distribution which is highly non-Gaussian as it is shaped by several nonlinear processes, such as gravitational instability, redshift-space distortions, and galaxy biasing. Higher-order clustering statistics, starting from the galaxy three-point function in configuration space, or the bispectrum in Fourier space, are the direct result of these nonlinear effects. Recent studies have demonstrated that performing a joint analysis of two- and three-point statistics of the galaxy distribution is key to disentangling the impact of these nonlinearities from the signatures of new physics. This is expected to improve constraints on the cosmological parameters by 10-30\% \citep{YankelevichPorciani2019}. In addition, there are models of inflation that can only be constrained using higher-order statistics \citep[e.g.][]{DAmicoEtal2022A}. To take advantage of this additional information, we will measure the redshift-space multipoles for both the three-point function and the bispectrum (also see \cref{sec:additional}) and analyse them jointly with two-point statistics as a natural extension of all probes and methods mentioned above.

The signatures of the physical processes discussed above can lead to strong degeneracies between measurements, for example, between the \gls{AP} effect and \glspl{RSD} \citep{Ballinger1996}. Thus it is important to measure them together, and to mitigate the effects of galaxy bias (see \cref{sec:nonlinear}). The relative robustness of using the \gls{BAO} signature as a standard ruler means that it is often advantageous to extract this information separately. This is commonly achieved by fitting a model where polynomial or similar terms are added in order to isolate the \gls{BAO} feature in the power spectrum \citep{BOSS-BAO-fit-pk} or correlation function \citep{BOSS-BAO-fit-xi}. The scale of the \gls{BAO} signal is then extracted and used directly to constrain models. The precision with which the \gls{BAO} scale can be measured can be improved by a technique called reconstruction \citep{reconstruction}, where the nonlinear motions of galaxies are estimated from the galaxy field, and used to find the positions of the initial overdensities. This sharpens the \gls{BAO} signal, increasing the precision of the determination of the centroid.

The galaxy-clustering probe uses overdensities in the galaxy field as a direct probe of cosmology and thus is sensitive to non-cosmological effects that alter observed densities. In order to use galaxy surveys one has to understand their specific selection function, so as to define a mask or window that describes where galaxies can be found. Typically, because of the windows' complexity, this is usually quantified by making use of random catalogues, Poisson sampling the expected density. The overdensity is then extracted by comparing galaxy and random catalogues (see \cref{sec:clustering_stats} for a description of how these will be created). Typical problems in the analysis of galaxy surveys arise from the selection of the galaxy sample, which can be distorted by Galactic extinction or stellar density \citep{Ross2012}. For example, areas near bright stars are unusable and must be masked. If these changes in the observed density are not corrected by matching the weighted galaxy field to the weighted random field, then the spatial distribution of bright stars may be imprinted in the overdensities in our map of the Universe and misinterpreted as a cosmological signal. Space-based slitless spectroscopy helps to reduce the impact of many of these effects. In particular, no target sample is required to be selected from ground-based imaging data.  On the other hand, this requires careful understanding of the potential density-dependent systematic effects that could arise due to confusion among faint spectra in crowded areas. Considerable effort has been invested to understand and model these effects through end-to-end simulations.

The cosmological information available from the galaxy field is simplest to extract where the physics can be explained by linear processes. On small scales, gravity induces nonlinearities in the distribution of density. It is possible that by studying particular locations in the density field, such as voids or clusters, the linear information on small scales may be easier to extract than from the field as a whole. For example, the relation between the overdensity and velocity field near voids is thought to be close to linear \citep{Hamaus2014}. For the AP effect, we can extract additional information from the fact that a stack of voids should be spherical on average (e.g. \citealt{Woodfinden2023}). Thus, in addition to the galaxy field, \Euclid will study these special places in the Universe to obtain further, complementary cosmological information.

\subsection{Weak gravitational lensing}
\label{sec:weaklensing}

As demonstrated in the previous section, galaxies are powerful, but biased, tracers of the \gls{LSS}.
To fully exploit the statistical power of the density modes traced by the galaxies, we need to link their properties to the surrounding matter structures. Hence, a direct measurement of the (predominantly dark) matter distribution provides not only important information on how galaxies populate halos, but also probes the growth of large-scale structures. Conveniently, such a measurement is possible thanks 
to the observable effects of gravitational lensing: massive structures distort space-time, warping the paths of photons. If the deflections are sufficiently large, multiple images of distant galaxies are formed. Such cases of strong gravitational lensing can be used to study the matter distribution on small scales, or exploited as cosmic magnifying glasses (see \cref{sec:stronglensing} for some applications). 

Generally, the deflections of single objects cannot be measured. However, the matter distribution, via its tidal gravitational field, gives rise to small coherent distortions in the shapes of distant galaxies. 
This change in the observed galaxy shapes is called weak gravitational lensing \citep[for thorough introductions, see][]{1999ARA&A..37..127M,Bartelmann2001, Bartelmann2010}. In particular, the weak lensing shear, which describes the weak lensing-induced difference in the galaxies' observed ellipticity \footnote{As is common in the weak lensing literature, we refer to the quantity describing the shape of a galaxy as `ellipticity'. Mathematically, this ellipticity corresponds to the third flattening of the galaxy image.} contains information on the cosmic matter distribution. This shear
is small compared to the intrinsic ellipticities of galaxies, and ensembles of galaxies need to be averaged to reveal the coherent patterns that can be used to map the distribution of matter along the \gls{LOS}
\citep[also see \cref{sec:massmap}]{mm:KaiserSquires1993}.
The weak lensing signal was first detected around massive clusters of galaxies \citep{Tyson90} and its potential for cosmology was quickly recognised \citep[e.g.][]{Blandford1991, MiraldaEscude1991, Kaiser92, Bernardeau97}  as 
the shape correlations provide statistical information about the cosmic \gls{LSS}, which, in turn, depends on the cosmic expansion history and structure growth \citep[see e.g.][for a recent review]{Kilbinger2015}. This led to the first deep imaging campaigns to measure weak lensing by \gls{LSS}, or cosmic shear, with the first unambiguous detections reported nearly simultaneously by \citet{2000MNRAS.318..625B}, \citet{Kaiser00}, \citet{2000A&A...358...30V}, \citet{2000Natur.405..143W}. 
Since then, weak lensing has become an established tool to infer cosmological parameters and has been successfully applied to ever larger surveys, such as the \acrlong{HSC}\glsunset{HSC} (HSC; \citealp{Aihara2018}), the \acrlong{KiDS}\glsunset{KiDS} (KiDS; \citealp{Kuijken2015}), and the \acrlong{DES}\glsunset{DES} (DES; \citealp{Abbott2016}; \citealp{Becker2016}).

The weak lensing signal can be decomposed into a curl-free $E$-mode and a gradient-free $B$-mode component \citep{Crittenden2002, Schneider2002}.  To first order, weak lensing by the \gls{LSS} only causes $E$-modes, while $B$-modes can be caused by systematic and astrophysical effects \citep[e.g.][]{Heavens2000, Schneider2002b, Hoekstra2004}. Usually, the $E$-mode signal is used for cosmological analysis, while the $B$-mode signal is a probe of unmodelled systematic effects \citep[e.g.][]{Hildebrandt2017, Asgari2019}.

For cosmic shear studies, key observables are the two-point shear correlation functions $\xi_+(\theta)$ and $\xi_-(\theta)$, which correlate the estimates for the shear components of pairs of distant galaxies at angular positions $\Vec{\vartheta}$ and $\Vec{\vartheta'}$, respectively, so that 
\begin{equation}
    \xi_{\pm}(\theta) = \langle\gamma_\mathrm{t}(\Vec{\vartheta})\,\gamma_\mathrm{t}(\Vec{\vartheta'})\rangle \pm \langle\gamma_\times(\Vec{\vartheta})\,\gamma_\times(\Vec{\vartheta'})\rangle\;,\label{eq:ellcor}
    \end{equation}
where $\gamma_\mathrm{t}$ and $\gamma_\times$ are the tangential and cross-component of the shear, defined with respect to the separation $\Vec{\theta}$ between the galaxies. 
The shear correlation functions are related\footnote{To simplify the discussion, we ignore the fact that the observed ellipticities provide an estimate for the reduced shear $g\equiv \gamma/(1-\kappa)$, where $\kappa$ is the convergence \citep[see e.g.][for more details]{SchnSeitz95,Bartelmann2001}. We also implicitly ignore intrinsic alignments. In the actual analysis of \Euclid data, these, and several other subtle complications, will need to be accounted for \citep{Deshpande-EP28}.} to the $E$- and $B$-mode shear power spectra $C^{EE}(\ell)$ and $C^{BB}(\ell)$ as \citep{Chon04, Lemos17, Kilbinger2017, Kitching2017}
\begin{equation}
\label{eq: model xipm}
    \xi_{\pm}(\theta)
    = \sum_{\ell \ge 2} \frac{2\,\ell + 1}{4\pi} \, \left[C^{EE}(\ell) \pm C^{BB}(\ell)\right]
        \, d^\ell_{2\pm2}(\theta) \;,
\end{equation}
where $d^\ell_{2\pm2}$ is the Wigner-$d$ function. In turn, $C^{EE}$ is related to the matter power spectrum $P_{\rm m}(k,z)$. 
Under the flat-sky approximation this relation reduces to Hankel transforms,
\begin{equation}
        \xi_{\pm}(\theta)
    \simeq \int_0^\infty \frac{\mathrm{d}\ell\, \ell}{2\pi} J_{0/4}(\ell\,\theta) \, \left[C^{EE}(\ell) \pm C^{BB}(\ell)\right] \;,
\end{equation}
where the $J_\nu$ are Bessel-functions of the first kind and $J_0$ is used for $\xi_+$ and $J_4$ for $\xi_-$. 
Neglecting $B$-modes, a simple form of the relation between $C^{EE}$ and $P_\mathrm{m}$ can be derived by assuming the Limber approximation \citep{Limber1953, Kaiser1998}, and a spatially flat Universe \citep[see][for a generalisation beyond this assumption]{2018PhRvD..98b3522T} as 
\begin{equation}
\label{eq:limber}
    C^{EE}(\ell)   = \left[L^\gamma(\ell)\right]^2\,\int_0^\infty \mathrm{d}{z}\,\frac{c}{H(z)}\, \left[\frac{W^\gamma(z)}{\chi(z)}\right]^2\, P_{\rm m}\left[\frac{\ell+1/2}{\chi(z)},z\right]\;,
\end{equation}
where the prefactor,
\begin{equation}
    L^\gamma(\ell) = \sqrt{\frac{(\ell+2)!}{(\ell-2)!}}\,\left(\frac{2}{2\,\ell+1}\right)^2\;,\label{eq:shear_prefactor}
\end{equation}
comes partly from the conversion from the spectrum of the lensing potential to that of shear \(E\) modes, and partly from the Limber approximation \citep[see][for derivations of these expressions, and relaxation of the approximations]{Lemos17, Kilbinger2017, Kitching2017}. Above, $\chi(z)$ is the comoving distance at redshift $z$, and $W^\gamma(z)$ is the lensing efficiency kernel,
\begin{equation}
    W^\gamma(z) = \frac{3\,\Omega_\mathrm{m\,}H_0^2}{2\,c^2}\,\chi(z)\,(1+z)\,\int_z^\infty\diff z'\,n(z')\,\frac{\chi(z')-\chi(z)}{\chi(z')} \;,\label{eq:kernel}
\end{equation}
for a sample of sources with a redshift distribution $n(z)$.\footnote{Note that \(n(z)\) has to be normalised to unit area, that is, \(\int\diff z\,n(z)=1\).} Hence, the interpretation of the observed lensing signal depends on knowing the redshift distribution of the source galaxies. Although redshifts for individual galaxies are not required, the sensitivity to cosmological parameters is rather limited when a single set of sources is used. 

To exploit the information on the evolution of the \gls{LSS} over cosmic time, key to constraining the dark energy equation of state parameter, $w(z)$, the sources need to be divided into narrow redshift bins that, ideally, do not overlap. Combined, the bins provide tomographic information on the matter distribution along the \gls{LOS}. A finer tomographic binning increases the redshift resolution, but also leads to a decrease of the signal-to-noise ratio in each individual bin as it contains fewer galaxies. Moreover, as the sources in different bins probe the same structures at lower redshifts, the resulting lensing signals are highly correlated, and the statistical gain saturates quickly \citep{Ma06}. 

A tomographic analysis requires redshift estimates for the individual sources, which are too faint and too numerous for dedicated spectroscopic follow-up. Fortunately, photometric redshifts \citep{Koo1985, Loh1986, Newman22} can be used, provided their precision is substantially better than the width of the tomographic bins. In the case of \Euclid, we aim to divide the source sample into as many as 13 bins in the range $0.2 \leq z \leq 2.5$.  To achieve these objectives, the standard deviation $\sigma_z$ of the photometric redshift estimates needs to satisfy
$\sigma_z<0.05 (1+z)$, while the catastrophic failure rate needs to be less than 10\% \citep{Amara07, Laureijs11}. To meet these stringent requirements, \Euclid complements the VIS data with deep space-based \gls{NIR} photometry in three bands; we also take additional, uniform photometry from the ground (see \cref{sec:groundbased}). Moreover, to obtain accurate cosmological parameter estimates, the mean galaxy redshifts within the bins need to be known with an accuracy $\sigma_{\langle z\rangle}<0.002\,(1+z)$ \citep{Ma06, Amara07,2008MNRAS.389..173K}. As discussed in more detail in \cref{sec:photoz}, this drives the need for extensive spectroscopy that fully samples the colour-redshift space \citep[e.g.][]{Masters15}.

Such relatively narrow redshift bins offer a powerful handle on the time evolution of the cosmic shear spectra. It allows us also to take full advantage of the so-called  
\gls{BNT} transformation \citep[][also see \cref{sec:cosmology}]{2014MNRAS.445.1526B,2018PhRvD..98h3514T,2021OJAp....4E...6T} with which scale mixing from projection effects can be fully controlled. 
This has major benefits for the interpretation of the weak lensing signal from theoretical models and the modelling of astrophysical systematic effects \citep{2021OJAp....4E...6T}. Another motivation for the fine binning is the need to combine the lensing measurements with the angular positions of galaxies in a so-called `3\texttimes2pt' analysis, discussed in \cref{sec:3x2pt}.

The shear correlation functions (and convergence power spectra) primarily depend on the square of the parameter
$S_8\equiv \sigma_8\sqrt{\Omega_\mathrm{m}/0.3}$ in the linear regime, and to higher orders of $S_8$ at smaller scales.
\citep[e.g.][]{Hall2021}. The degeneracy between $\Omega_\mathrm{m}$ and $\sigma_8$ is broken by nonlinear corrections or with the use of higher-order statistics \citep{Bernardeau97}. Cosmic shear analyses of current surveys at the time of writing have already tightly constrained $S_8$ with a precision of $2\mbox{--}4\%$ \citep[e.g.][]{Asgari2021, Amon2022,Li2023}. Interestingly, the reported values are consistently lower than the one inferred from the cosmic microwave background with \Planck. The current level of disagreement for individual measurements ranges from 2 to $3\,\sigma$, but it remains to be seen if this points to a problem with the cosmological standard model \citep{diValentino2021}. 

Cosmic shear is also sensitive to the cosmic expansion history and the dark energy equation of state through the angular-diameter distances between the observer, the distorted source galaxies, and the lensing matter structures; changes in the projection of structures along the \gls{LOS}; and the decay in gravitational potentials as the expansion accelerates.
Current cosmic shear analyses lack the statistical constraining power to provide meaningful constraints on the dark energy equation of state. This will change with \Euclid, since it will achieve more than an order-of-magnitude increase in precision compared to previous surveys \citep{Blanchard-EP7}, owing to its depth, precise shear measurements, large area, and high galaxy density.

\begin{figure*}
    \centering
    \includegraphics[width=\linewidth]{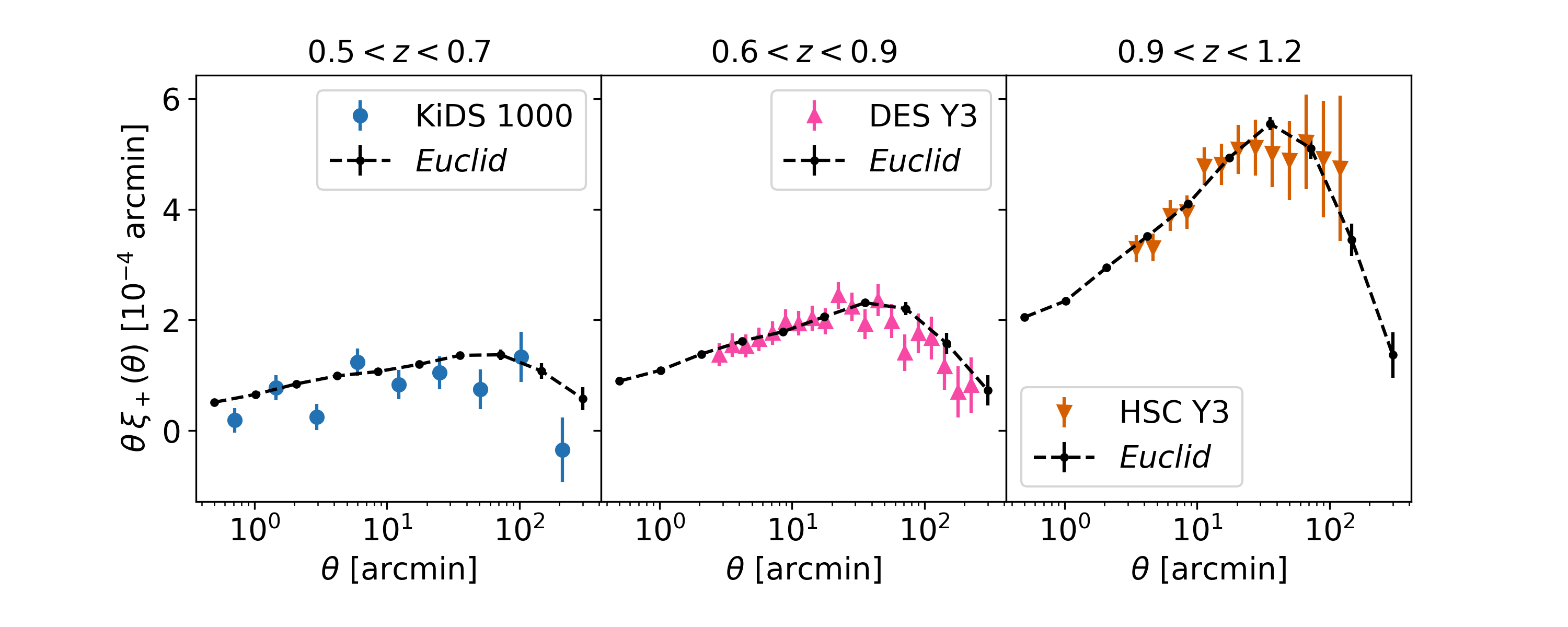}
    \caption{Shear correlation function $\xi_+(\theta)$ for \gls{KiDS}-1000 (left, from \citealp{Asgari2021}), \gls{DES} Y3 (middle, from \citealp{Amon2022}), and \gls{HSC} Y3 (right, from \citealp{Li2023}), and expected for \Euclid. Each panel uses sources distributed according to the tomographic bin with the highest \gls{S/N} of the respective survey. The \gls{S/N} for \Euclid is an order of magnitude larger than that of the most recent surveys. The other shear correlation function $\xi_-(\theta)$ shows a similar improvement in S/N (not shown). }
    \label{fig:3Tomo}
\end{figure*}

\begin{figure}
    \centering
    \includegraphics[width=\linewidth]{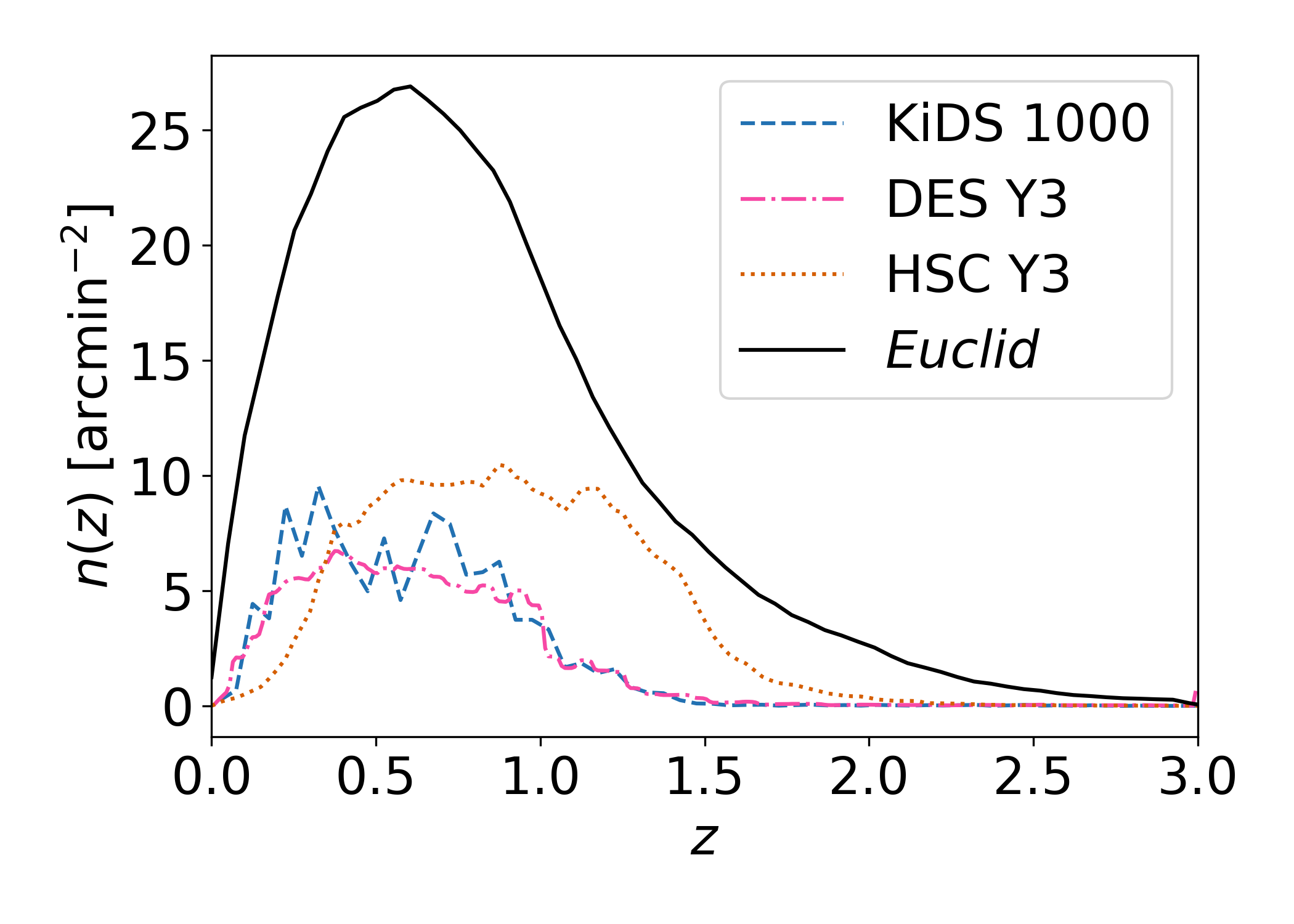}
    \caption{Source-redshift distributions $n(z)$ of the \gls{KiDS}, \gls{DES}, and \gls{HSC}, and as expected for \Euclid. Distributions are normalised to the mean number density of sources used in the lensing analyses.}
    \label{fig:nz}
\end{figure}

The statistical uncertainty of weak lensing measurements is primarily limited by sample variance and shape noise, both of which will be immensely reduced by \Euclid. The \acrlong{EWS}\glsunset{EWS} (EWS, see \cref{sec:survey}) will cover an area 3 times larger than the final \gls{DES} release, the largest deep imaging survey to date, suppressing sample variance. Additionally, since \Euclid has greater depth than previous surveys, it effectively probes a larger volume of the Universe. \Euclid is expected to detect about 2~billion source galaxies for which shapes can be measured, several orders-of-magnitude more than current surveys, thereby reducing shape noise. Given that many of these galaxies are at higher redshifts than those used in previous surveys, the observed galaxies typically exhibit larger lensing signals, further enhancing the signal-to-noise ratio. These strengths, particularly its depth, accurate shear measurements, and large galaxy numbers, will help \Euclid to achieve a significant increase in precision compared to previous surveys \citep[see][for the scaling of the \gls{FOM} with these survey parameters]{Amara07}, while the sharp \gls{PSF} reduces the detrimental impact of blending on the shape measurements \citep{MacCrann22,SSLi23}.

This is illustrated in \cref{fig:3Tomo}, where we show shear correlation functions $\xi_+(\theta)$ measured by the \gls{KiDS} \citep{Asgari2021}, \gls{DES} \citep{Amon2022}, and \gls{HSC} \citep{Li2023}. Each panel shows  $\theta\, \xi_+(\theta)$ for the tomographic bin with the highest \gls{S/N} for each survey. Error bars are the square root of the diagonal elements of the cosmic shear covariances, including shape noise and sample variance. For \Euclid, these were computed with the \texttt{OneCovariance}-code\footnote{\url{https://github.com/rreischke/OneCovariance}; it uses the same prescription as \citet{Joachimi2021}, which is similar to \citet{Sciotti23} and \citet{Upham22}}  (Reischke et al., in prep.). Comparing with the predictions for \Euclid in the same tomographic bin, we expect an order of magnitude increase in \gls{S/N} thanks to the combination of a higher galaxy number density and the increase in survey area. 

The increase in galaxy number density is also apparent from \cref{fig:nz}, which demonstrates another advantage of \Euclid compared to previous surveys, namely its increased redshift range. The amplitude of the lensing signal rapidly increases as a function of source redshift \citep{Bernardeau97}, so that even a modest increase 
in number density of higher redshift sources can lead to a significant improvement in constraining power. The figure compares the redshift distribution of \gls{KiDS} \citep{Hildebrandt2021}, \gls{DES} \citep{Myles2021}, and \gls{HSC} \citep{Rau2023}, to the expected redshift distribution for \Euclid, obtained by selecting galaxies with $\IE\leq 24.5$ in the Flagship~2 simulation (see \cref{sec:flagship}). Shape measurements in \gls{KiDS} and \gls{DES} are essentially limited to galaxies at $z<1.2$. The deep \gls{HSC} data allow for a significant number of galaxies at redshifts between 1 and 1.5, but only \Euclid will obtain a meaningful number of sources at higher redshifts. This larger redshift range is crucial for determining the evolution of dark energy.

To exploit the unprecedented statistical power of \Euclid, it is essential that instrumental sources of bias are much smaller than the measurement uncertainties. Moreover, the exquisite measurements need to be complemented with accurate modelling of cosmological and astrophysical effects. Systematic effects for weak lensing arise, for example, from imperfect shape measurements and biases in the estimation of the source redshift distribution. We refer the interested reader to the review by
\cite{Mandelbaum2018} for a more in-depth discussion.

The observed images are modified by the telescope optics: even for space-based observations the blurring by the \gls{PSF} is a dominant source of bias that needs to be accounted for. Moreover, imperfections in the detector introduce additional changes in the images, while cosmic rays pose another challenge. 
Considering the estimated shear $\hat{\gamma}$ and the true shear $\gamma$ as complex numbers, their difference can be expressed to first order \citep[see][]{2020OJAp....3E..14K} as
\begin{equation}
    \label{eq:mcbias}
    \hat{\gamma}-\gamma = m^{\rm bias}_0 \, \gamma + m^{\rm bias}_4 \, \gamma^* + c^\mathrm{bias} + n\;,
\end{equation}
where $m^{\rm bias}_0$ and $m^{\rm bias}_4$ are spin-0 and spin-4 complex operators, respectively, and the asterisk denotes complex conjugation. The value of $m^{\rm bias}_0$ quantifies the dilation and rotation of the true shear, whereas $m^{\rm bias}_4$ allows for a reflection around the axis determined by its phase \citep{Congedo2024}.
$c^{\rm bias}$ is the additive bias, while $n$ corresponds to the random (shape) noise in the shear estimate.

The biases depend on the instrument, the shape measurement method \citep[e.g.][]{Heymans+06, Hoekstra+17, Hoekstra+2021}, and galaxy properties, in particular the size and signal-to-noise ratio \citep[e.g.][]{Hoekstra+15}. They can also vary spatially, but 
$C^{EE}$ is affected principally only by the mean multiplicative bias\footnote{Recently, \citet{2022OJAp....5E...6K} highlighted that any nonlinearity in the relation between the true and estimated shear needs to be quantified and possibly accounted for.} and the correlation between the shear field and the additive bias \citep{Kitching2019, Kitching2021}. 

Given a survey design, it is possible to derive limits on the shear biases that can be tolerated \citep{Amara08}. These can be specified further by exploring how errors in the estimate of the \gls{PSF} propagate \citep{Paulin-Henriksson09, Massey+13}, as well as other sources of bias. The detailed breakdown presented in \cite{Cropper+13} formed the basis for the development of the shape measurement pipeline for \Euclid, which is discussed in more detail in \cref{sec:wlmeasurement}.

To realise the full statistical potential of \Euclid, we require the uncertainty on the multiplicative bias to be less than $2\times 10^{-3}$, and the uncertainty on the additive bias to be less than $1.5\times 10^{-4}$ \citep{Cropper+13}. Recent studies
\citep{Paykari-EP6, Kitching2019} have shown that we need to distinguish between sources of bias that are constant across the survey and spatially varying effects. Although these refinements can provide margin for specific instrumental effects, the baseline requirements provided an excellent basis for the hardware and software development needed for the mission.

Because we aim to push the limits of what can be done within the mission constraints, the measurements are challenging, despite the advantages that a space telescope brings. Compared to ground-based surveys the main benefit of \Euclid is that the \gls{PSF} residuals scale with the square of the \gls{PSF} size \citep{Paulin-Henriksson09,Massey+13}. This reduces the baseline multiplicative bias for \Euclid compared to ground-based telescopes, which cannot avoid the blurring of the images by atmospheric turbulence. Nonetheless, the \gls{PSF} needs to be known with unprecedented accuracy. A complication is that we need to account for the fact that the \gls{PSF} is a strong function of wavelength, and therefore depends on the \gls{SED} of each individual galaxy \citep{Cypriano10,Eriksen18}, which can also vary spatially \citep{Voigt+12,Semboloni+13,Er+18}. The challenges in modelling the \gls{PSF} and measuring the shapes of galaxies are discussed in \cref{sec:psfmodel,sec:shapemeasurement}, respectively.  

Since the weak lensing observables provide unbiased estimates of the projected matter distribution, it was believed that the interpretation of the lensing signal would be relatively immune to astrophysical processes. However, it has become clear that this is not the case, especially at the precision of \Euclid. First of all, the intrinsic shapes of galaxies are correlated with each other and their surrounding matter distribution due to tidal interactions during their formation. This \gls{IA} effect needs to be accounted for, because it 
causes spurious signals in the cosmic shear signal and the position-shear correlations \citep{Joachimi2015, Kirk2015, Troxel15}. 
On large scales, tidal alignment models provide a description of the scale dependence \citep[e.g.][]{Hirata04, Bridle2007, Blazek2019}, but the amplitude of the \gls{IA} signal cannot be predicted from first principles because it depends on the complex process of galaxy formation. Moreover, the signal depends on galaxy type, galaxy luminosity, and redshift. Although observations can be used to constrain the predicted amplitude \citep{Fortuna:2020vsz}, our current knowledge of the \gls{IA} signal is insufficient. \Euclid itself will be a great resource for direct measurements, but the modelling of the \gls{IA} signal is likely to remain an active area of research for the foreseeable future.

The second complication is that non-gravitational processes,
such as heating by active galactic nuclei, supernovae, or star formation, redistribute baryons. To explain current observations, models of galaxy formation require that a significant fraction of the baryons are expelled, leading to a suppression of the matter power spectrum
\citep{vanDaalen11, Debackere2020}. Neglecting these processes can significantly bias cosmological parameter constraints
\citep[e.g.][]{Semboloni2011, Semboloni13b, Chisari2019}. In principle, the changes in the matter distribution can be modelled using hydrodynamical simulations (e.g. \citealp{Schaye2010}, \citealp{LeBrun2014}, \citealp{McCarthy2017}), physically motivated modifications to analytical halo models (e.g. \citealp{Debackere2020}, \citealp{Mead2021}), or `baryonification' models that modify halos in gravity-only simulations according to prescribed gas content (\citealp{Schneider2015}, \citealp{Schneider2019}, \citealp{Arico2020}).
The challenge is to decide which models capture the feedback processes correctly, although the findings of 
\cite{vanDaalen20} suggest it may be possible to
describe the effect of feedback on the matter power spectrum with only a few nuisance parameters that need to be marginalised over when estimating the cosmological parameters. Finally, several simplifying assumptions in the modelling of the observed cosmic shear power spectrum need re-evaluation for \Euclid, such as an increased source galaxy density due to weak lensing magnification (magnification bias), the exclusion of blended galaxy pairs (source obscuration), and local over- or under-densities \citep{Deshpande-EP28}. 

\subsection{Photometric 3\texttimes2pt analysis}
\label{sec:3x2pt}

Although cosmic shear is a powerful probe of cosmology, it cannot reach a ${\rm FoM}>400$ by itself \citep{Laureijs11, Blanchard-EP7}. To unlock the full constraining power of \Euclid, we need to combine the cosmic shear correlation functions, $\xi_\pm(\theta)$, with the galaxy angular correlation function, $w(\theta)$, and the cross-correlations between galaxy angular positions and the tangential component of the ellipticities of background galaxies, $\langle \gamma_\mathrm{t}\rangle(\theta)$. These additional two-point functions are commonly referred to as `photometric galaxy clustering' and `galaxy-galaxy lensing' or `shear-clustering cross-correlation', respectively. 

Equivalently, angular power spectra can be measured and used for the analysis. Hence, \cref{eq:limber} can be generalised to encompass all three probes,
\begin{align}
    C^{AB}_{ij}(\ell) &= L^A(\ell)\, L^B(\ell) \\
    &\notag \quad \times \int_0^\infty \diff z \,\frac{c}{H(z)} \, \frac{W^A_i(z)\,W^B_j(z)}{\chi^2(z)}\, P_{\rm m}\left[\frac{\ell+1/2}{\chi(z)},z\right]\;.\label{eq:limber_cross}
\end{align}
In the expression above, \(A\) and \(B\) label the probes being correlated, while $i$ and $j$ indicate the tomographic bins considered. In the case of shear, \(L^\gamma\) is given by \cref{eq:shear_prefactor}, whereas \(L^\mathrm{g}(\ell)=1\). For instance, for \(A=B=\gamma\) and $i=j$, we obtain \cref{eq:limber}. Alternatively, it is possible to include intrinsic alignments by using the ellipticity power spectrum
\citep{Blanchard-EP7}. For photometric galaxy clustering, we take \(A=B={\rm g}\) and, assuming a linear galaxy bias,
\begin{equation}
    W^{\rm g}_i(z)=\frac{H(z)}{c}\,b_{\rm gal}(z)\,n_i(z)\;,
\end{equation}
while \(A=\gamma\) and \(B={\rm g}\) corresponds to the shear-clustering cross-correlation. Note that the galaxy bias, \(b_{\rm gal}\), will in general be different from that of the galaxies in the spectroscopic sample discussed in \cref{sec:galaxyclustering}. Moreover, as mentioned in this subsection, the (binned) redshift probability distribution of sources, \(n_i(z)\), used for clustering might differ from that employed for cosmic shear.

A combined analysis including all of such correlations is commonly referred to as a `$3\,{\times}\,2$pt analysis'. It can significantly enhance the constraining power: for \Euclid the \gls{FOM} is increased by roughly a factor 20, relative to a cosmic-shear-only scenario \citep{Tutusaus20,Blanchard-EP7}. Hence, it should not come as a surprise that such an approach has become the standard for current surveys \citep[e.g.][]{vanUitert2018, Joudaki18, desy1_3x2pt, Abbott2022, More2023}. Additional cross-correlations, for instance with CMB measurements (\cref{sec:cmbx}), can be included for $n\,{\times}\,2$pt analyses \citep[e.g.][]{desy1cmb, des6x2pt2023}.

The benefit of combining these three types of correlations lies in their ability to lift degeneracies between parameters (both cosmological and nuisance), such as the galaxy bias parameters that are needed to describe how galaxies trace the underlying matter distribution. In contrast, galaxy clustering is not affected by intrinsic-alignment and shape-measurement biases. Moreover, the clustering-shear cross-correlation is subject to the same systematic effects as cosmic shear and galaxy clustering, but has a different functional dependence. Combining these probes allows for the partial self-calibration of systematic effects \citep{Bernstein2004, Hu2004, Bernstein2009, Joachimi2010}. This self-calibration enables tight control over systematic effects. Thus, the main cosmic shear cosmological results from \Euclid will be derived from a $3\,{\times}\,2$pt analysis, made possible by the precise weak lensing measurements. In the cosmological parameter forecasts in \cref{sec:constraints}, we present the expected parameter constraints from this analysis.

The improved constraining power comes at a price: we need to accurately measure and consistently model a larger number of probes. In principle, photometric clustering includes all the 
physical effects described in \cref{sec:galaxyclustering}, including \glspl{BAO} and \glspl{RSD}. However, given the lack of resolution along the \gls{LOS}, the information is generally projected within each tomographic bin, similar to what is done for weak lensing. By limiting the analysis to the angular correlation function between galaxies, most of the \gls{RSD} signal is lost (although not completely, see Euclid Collaboration: Camera et al., in prep.). However, we need to include \glspl{RSD} in the modelling in order not to bias our cosmological results \citep[e.g.][]{2023arXiv230900052E}. 

Moreover, thanks to the low radial resolution of this probe, the \gls{BAO} signal is partially smoothed out and therefore we can use smaller scales than those considered in spectroscopic clustering analyses, even if our model is not as accurate. 
As a result, most of the information from photometric galaxy clustering comes from scales slightly smaller than those used in \cref{sec:galaxyclustering}, making them complementary probes. We note, however, that photometric galaxy clustering still uses biased tracers; therefore, we cannot use scales as small as for weak lensing, given the difficulty in modelling the galaxy bias at small scales.
Another physical effect that needs to be included in the modelling of the signal is magnification. This lensing effect does not add much constraining power \citep{Mahony22}, but ignoring its impact on the clustering signal leads to biased parameter estimates
\citep[see e.g.][]{Duncan22, 2022A&A...662A..93E,2022MNRAS.510.1964M}. For galaxy-galaxy lensing, all these effects also need to be considered, in addition to those relevant for weak lensing.

A robust measurement of photometric clustering poses new challenges, in part owing to its reliance on ground-based data that were obtained under varying observing conditions. If these variations are not accounted for, they can lead to spurious clustering \citep{2011MNRAS.417.1350R, 2018PhRvD..98d2006E, Johnston2021, RodriguesMonroy2022}. Further complications arise from contamination by stars and Galactic extinction, or zodiacal light, which can change the galaxy counts on relatively large scales.
In principle, these contributions are included in the visibility mask (see \cref{sec:weaklensingstatistics}), which is used to correct the measurements. Work to optimise the galaxy samples is ongoing. In particular, it may be advantageous to consider samples of bright galaxies, which are more immune against spurious clustering.


\section{Spacecraft and instruments}
\label{sec:instruments}

 The primary cosmological probes of \Euclid's core science case have defined the main survey characteristics (\cref{sec:survey}), as well as the requirements for the capabilities of the telescope and instrument. 
Besides a maximum cost, in accordance with its `medium' mission size, \gls{ESA} imposed a so-called technology readiness level of 5 or higher for the components of the proposed mission. This restriction allows only the use of technologies that have already been validated in the relevant environment. 

Before \Euclid's final design and scope were defined, initial assessment studies with industry and the science community resulted in a set of feasible science and mission requirements. The subsequent definition phase provided a detailed description of the scientific scope based on a mission design that could be developed within the programmatic constraints set by \gls{ESA} \citep{Laureijs11}. After its selection in 2011, the \Euclid mission was adopted by ESA's science programme committee in June 2012, to enter the implementation phase with industrial contracts for the development of the space segment and with a multilateral agreement between \gls{ESA} and the participating countries for the delivery of the two science instruments and the development of the science ground segment.

\Euclid was originally planned to be launched on a Soyuz ST-2.1B rocket \citep{Laureijs11}, but the geopolitical developments that unfolded in 2022 resulted in the cancellation of this possibility. Investigations revealed that a SpaceX Falcon-9 could provide a suitable alternative, which was ultimately confirmed with the successful launch on 1 July 2023 into an orbit around the second Lagrange point of the Sun-Earth system (hereafter L2). This orbit provides a thermally stable environment with unobstructed views of the sky, prerequisites for a successful use of the planned cosmological probes. The most salient details about the spacecraft are presented in \cref{sec:spacecraft}. Information about the transfer into the halo orbit around L2 and orbital maintenance is provided in \cref{sec:orbitmaintenance}, while pointing constrains are discussed in \cref{sec:techpointingconstraints}.

The spacecraft contains two main science instruments: the visible imaging instrument (VIS; \cref{sec:vis}); and the 
\acrlong{NISP}\glsunset{NISP} (\gls{NISP}; \cref{sec:nisp}). Both instruments were designed to provide high-quality data over a wide \gls{FOV} with a high degree of accuracy and precision. The resulting homogeneous, high-quality space-based observations benefit from the thermally very stable environment, and from the absence of atmospheric blurring and bright sky background, providing a data set of unrivalled fidelity.

\subsection{Spacecraft}
\label{sec:spacecraft}

The \Euclid spacecraft can be subdivided into three main parts: the \gls{SVM}; the \gls{PLM} including the telescope; and the scientific instruments (called collectively `extended PLM'). The design of the spacecraft is described in detail by \citet{Racca2016}; here, we present a summary overview and the most important changes until launch, and observations during the commissioning phase. 

\begin{table}[th!]
 \caption{System mass budget for launch.}
 \centering
\addtolength{\tabcolsep}{-0.25em}
\begin{tabular}{lc} 
 \hline\hline
  \noalign{\vskip 1pt}
  Component & \phantom{0}Mass [kg] \\
 \hline
  \noalign{\vskip 1pt}
 Service module (SVM) & \phantom{0}901.0 \\  
 (Warm instrument units) & \phantom{00}(64.5) \\ 
 \hline
  \noalign{\vskip 1pt}
 Payload module (PLM) &  \phantom{0}806.4 \\
 (Cold instrument units) & \phantom{0}(156.4) \\
 \hline
  \noalign{\vskip 1pt}
Propellant ($N_2H_4$ and $N_2$) & \phantom{0}210.7 \\ 
Launch vehicle adaptor and clamp band & \phantom{00}70.0 \\ 
 \hline
 \noalign{\vskip 1pt}
Total launch mass &1988.1 \\ 
 \hline
 \end{tabular}
 \footnotesize
 \tablefoot{The \gls{SVM} and the PLM masses include the instrument units located in each module, reported below each item.}
\label{tab:mass}
\end{table}

\subsubsection{Service module}
\label{sc_sec:SVM}

\begin{figure*}
\centering{\hbox{%
    \includegraphics[width=0.5\hsize]{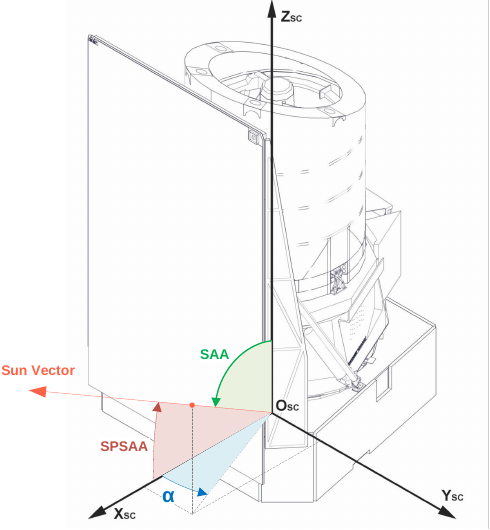}
    \includegraphics[angle=0,width=0.5\hsize]{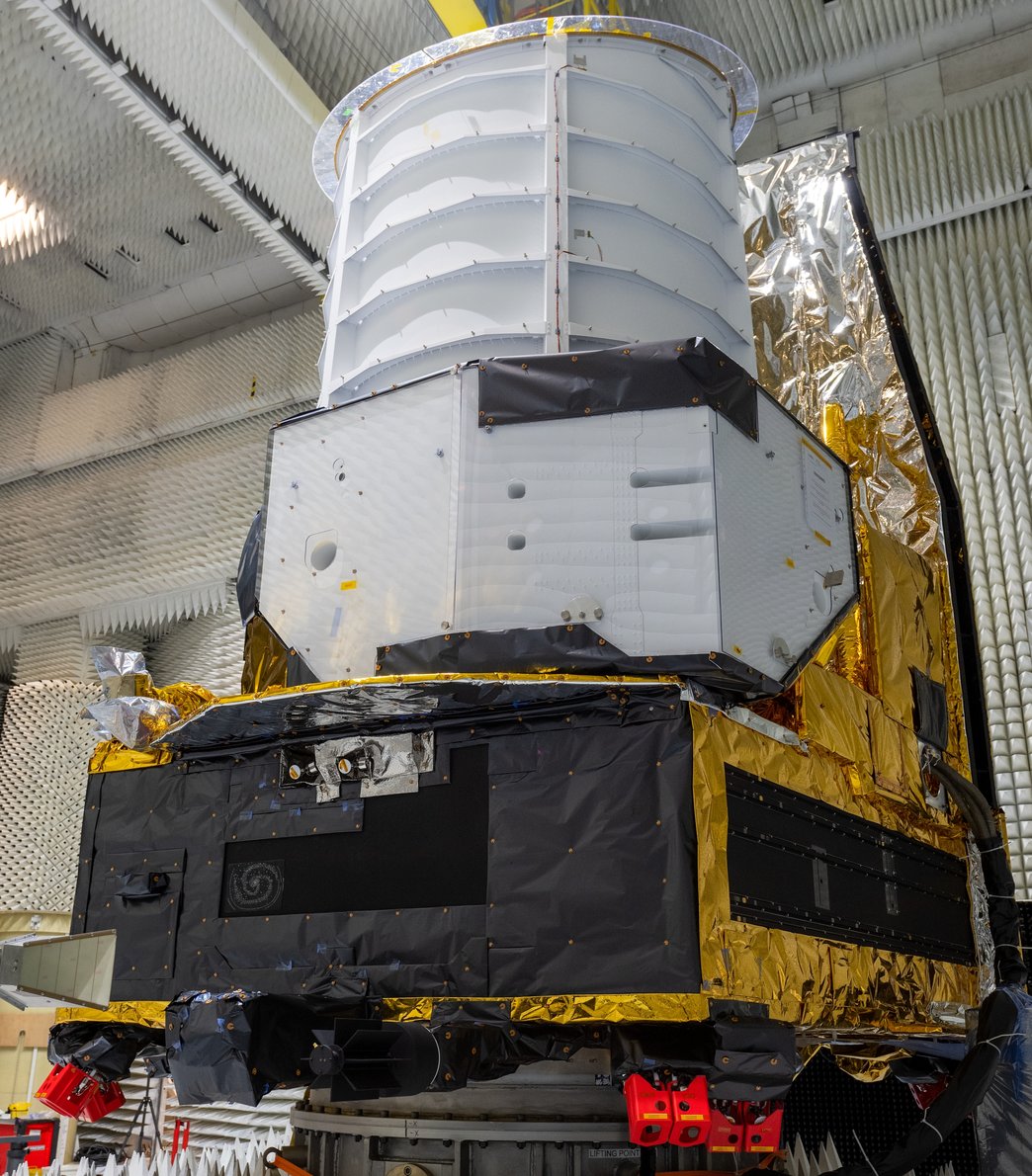}}}
    \caption[caption]{{\it Left:} Overview of the \Euclid spacecraft with the principal axes highlighted. {\it Right:} The fully assembled spacecraft on February 2023 in the anechoic chamber of Thales Alenia Space in France, after completing final 
    electromagnetic compatibility tests.
    The side shown here will always face away from the Sun. The large white structure below the cylindrical telescope baffle is the NISP radiator. The hydrazine thrusters still have their protective red covers on.
    The plaque with the miniaturised fingerprint galaxy created
    thanks to a \href{https://www.esa.int/ESA_Multimedia/Videos/2022/07/The_Fingertip_Galaxy_Reflecting_Euclid_in_art}{collaboration with visual artist Lisa Pettibone} and
    Euclid Consortium members can be seen at the lower left.
    Figure credit: \gls{ESA} -- M.~P\'edoussaut.
    \label{fig:SC}}
\end{figure*}

The \gls{SVM} comprises the spacecraft subsystems supporting the payload operations; it hosts the warm electronics of the payload, and provides structural interfaces to the PLM and the launch vehicle. The prominent sunshield is part of the SVM. It protects the PLM from illumination by the Sun and supports the photovoltaic assembly supplying electrical power to the spacecraft. On top of it, a triple blade sun-baffle is mounted  with the purpose to reduce the sunlight diffracted towards the PLM baffle aperture.
The overall spacecraft envelope fits within a diameter of 3.74\,m and a height of 4.8\,m. \Euclid's total launch mass budget, including propellant for operations, is 1988.1\,kg (see \cref{tab:mass}).
The left panel in \cref{fig:SC} shows an overview sketch of \Euclid in launch configuration, including the definition of its axes, while the right panel shows the fully assembled spacecraft during testing in February 2023.

\subsubsection{Payload module}
\label{sc_sec:PLM}

\begin{figure*}
\centering{\hbox{%
    \includegraphics[angle=0,width=0.48\hsize, trim=0 -1cm 0 0]{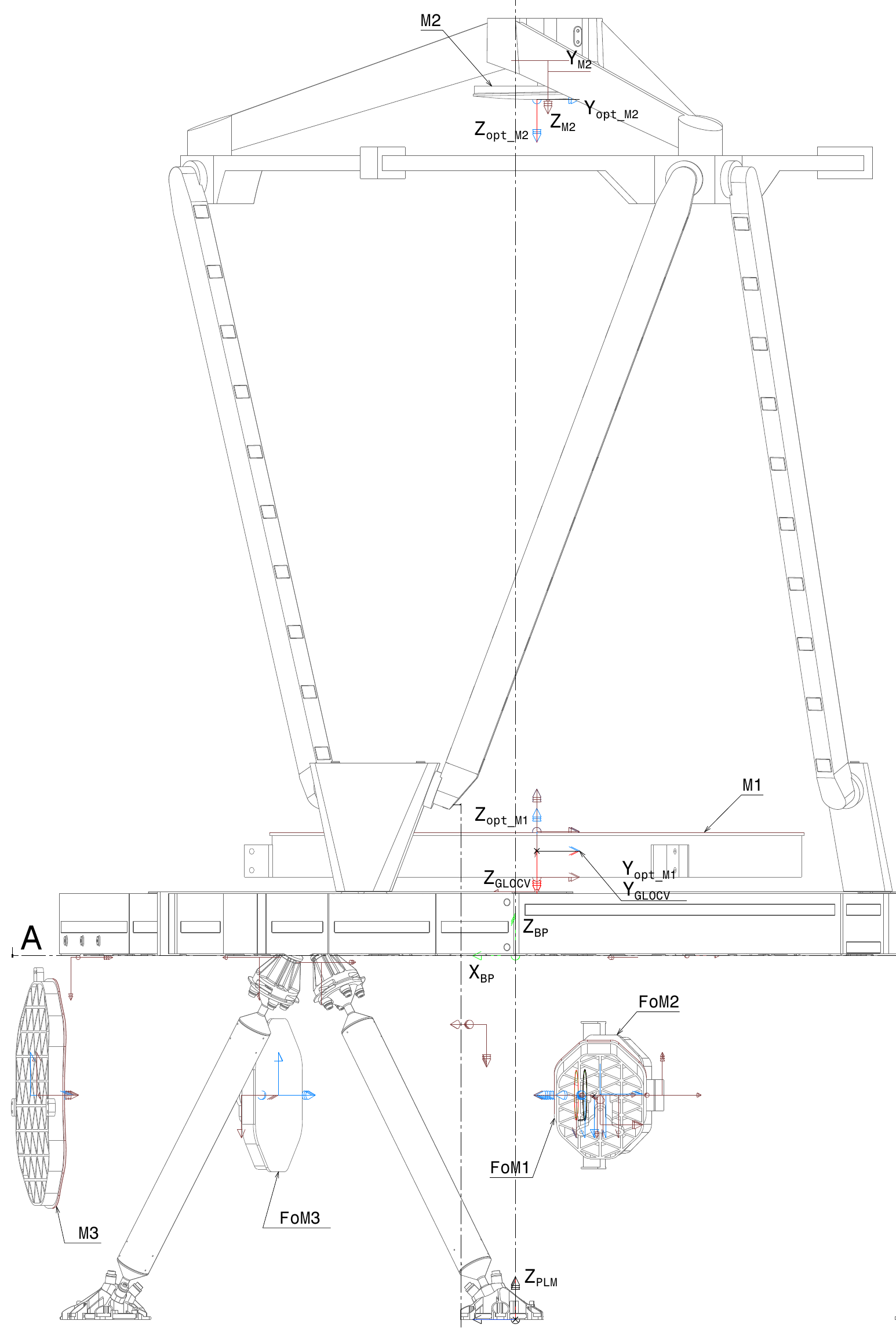} \includegraphics[width=0.5\hsize]{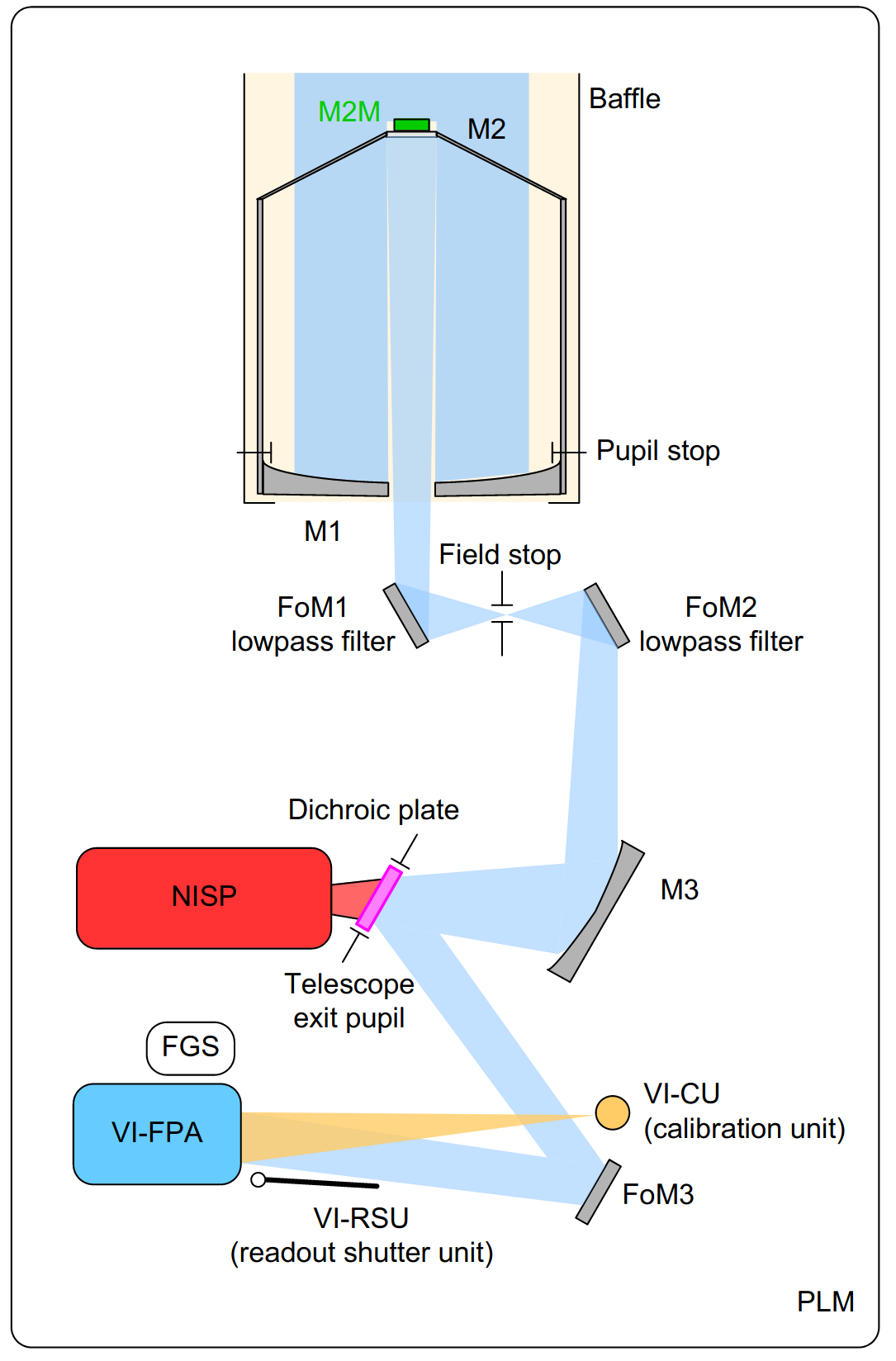}}}
    \caption{{\it Left:} Technical layout drawing of the \gls{PLM} optical surfaces to scale. Note that the dichroic plate and the field stop are not shown as covered by the FoM1 in this view. {\it Right:} Schematic functional view: light enters from the top onto the primary mirror M1. The secondary mirror M2 can be moved in 3 degrees of freedom by the \gls{M2M} to compensate launch and cool-down effects. Separated by a baffle, the light then enters the instrument cavity, where it gets relayed by two flat folding mirrors (FoM1, FoM2) whose coatings suppress photons below 0.5\,\micron. The tertiary mirror M3 directs the beam towards the dichroic plate. In transmission light enters NISP and in reflection VIS, by use of a third folding mirror (FoM3, silver coated). VIS consists of a separate \gls{FPA}, an \gls{RSU}, and a \gls{CU}. \Euclid's \gls{FGS} are co-mounted on the same structure as the VIS \gls{FPA}. Figure credit: \gls{ADS}.}
    \label{fig:korsch}
\end{figure*}

The \Euclid \gls{PLM} is designed around a three-mirror anastigmat Korsch telescope \citep{korsch1977} with a 1.2-m primary mirror and an effective collecting area of 0.9926\,m$^2$
\citep{GasparVenancio2014}. The telescope provides a common area between instruments of about 0.54\,deg$^2$ with minimal spherical aberration, astigmatism, coma, and field curvature. The mirrors and telescope structure are made from \acrlong{SiC}\glsunset{SiC} \citep[SiC;][]{bougoin2019}, with the optical path schematically shown in \cref{fig:korsch}. The light separation between the two instruments is performed by a dichroic plate located at the exit pupil of the telescope. The \gls{PLM} provides the mechanical and thermal interfaces to the instruments, consisting of radiating areas and heating lines. 

Whereas NISP is a stand-alone instrument with interface bipods, VIS is delivered in several separate parts. It consists of a focal plane assembly (FPA) containing all detectors, connected to proximity electronics, readout shutter unit, and calibration unit, each with their own dedicated mechanical and thermal interfaces with the PLM.

The secondary mirror (M2) is mounted on the \gls{M2M}, allowing adjustment in three degrees of freedom for focusing and some optical alignment. In addition, the \gls{PLM} hosts the \glspl{FGS}, used as pointing reference by the \gls{AOCS}. The \gls{FGS} detectors are mounted on the same structure carrying the VIS focal plane to ensure precise co-alignment. Except for the proximity electronics of the VIS and \gls{FGS} focal planes, all electronics are placed on the \gls{SVM} to minimise thermal disturbances of the \gls{PLM}.

\begin{figure}[ht]
\centering
\includegraphics[angle=0,width=1.0\hsize]{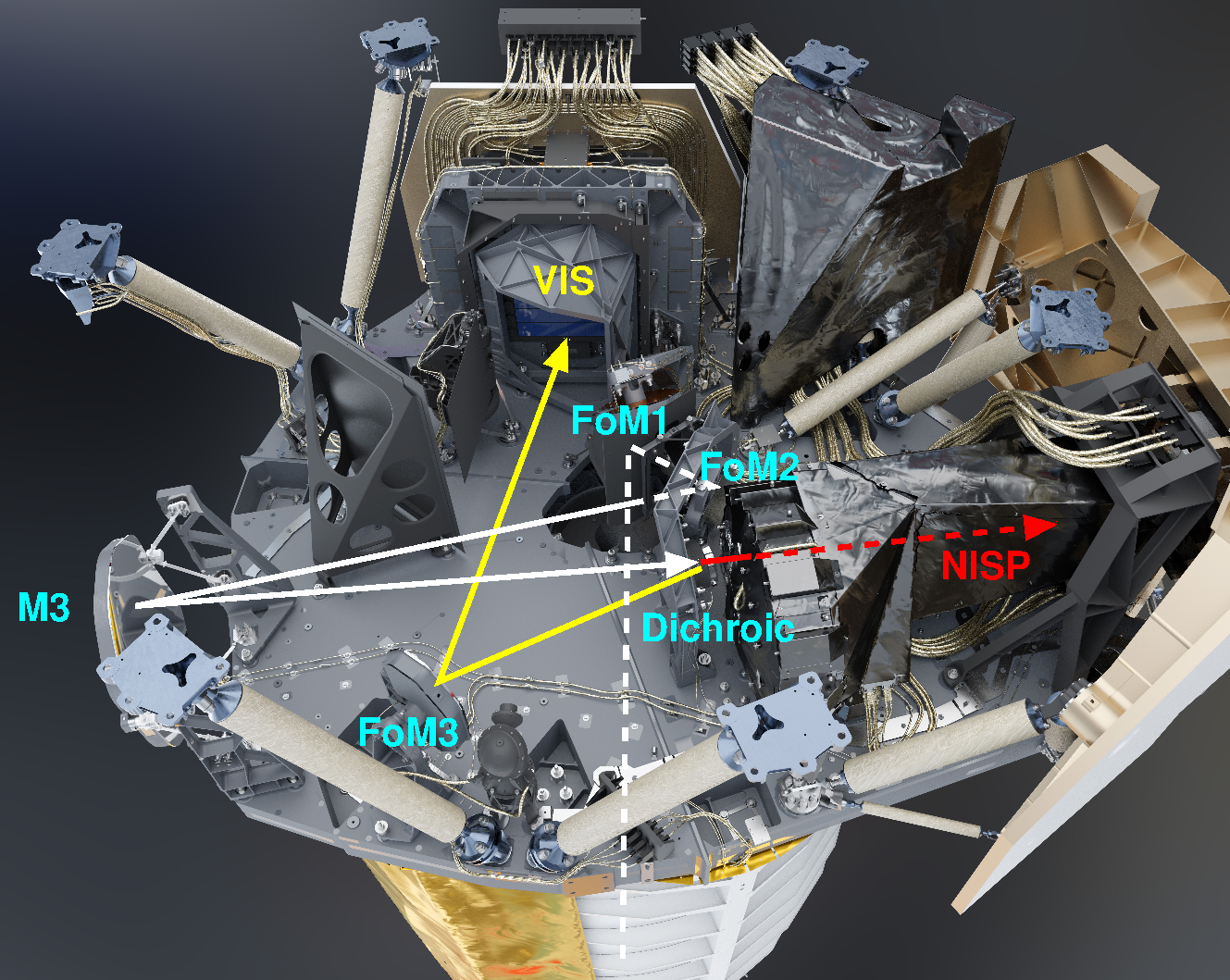}
\caption{3D digital rendering of the instrument cavity. In this orientation the telescope is below the assembly and observing towards the bottom. For clarity, we have added the principal light path and optical components to the rendering; dashed lines are obstructed from the chosen point of view. The large structure to the right of NISP is its outward-facing radiator. It can be clearly seen in the photograph shown in \cref{fig:SC}. Figure credit: \gls{ADS}, annotations by the authors.
}
\label{fig:euclid_CAD_annotated}
\end{figure}

The \gls{PLM} is divided into two cavities, separated by the baseplate. The front cavity includes the  primary and secondary mirrors of the telescope, as well as the \gls{M2M} and the associated support structure. This cavity is thermally insulated by a baffle that functions as a stray light shield and as a thermal radiator at the same time. \Cref{fig:euclid_CAD_annotated} shows an annotated overview of the instrument cavity, which includes the telescope folding mirrors, the tertiary mirror M3, the dichroic plate, the \gls{FGS}, and the two instruments NISP and VIS -- the latter with its separate \gls{FPA}, filter and grism wheels, and calibration source components.

The \gls{PLM}'s mechanical architecture is based on a common SiC baseplate that supports on one side M1 and M2, and on the other side the remaining optics and the two instruments. On the baseplate two planar low-pass coated folding mirrors, FoM1 and FoM2, fold the optical beam in the plane of the baseplate at the entrance of the instrument cavity between M2 and M3. A third, silver-coated folding mirror (FoM3) allows us to have the VIS instrument close to a radiator to efficiently remove the front-end electronics' heat. The telescope is cooled down to its equilibrium temperature (M1 temperature around 126\,K). This cold telescope offers high thermo-elastic stability, where the SiC's coefficient of thermal expansion is reduced to 0.4\,\micron\,m$^{-1}$\,K$^{-1}$, and provides a time-stable cold environment for the instruments. The baseplate temperature range of 130\,K to 135\,K is maintained constant during the mission to about 200\,mK and to a few tens of millikelvin during the observations. The main drivers of the baseplate temperature are the attitude of the spacecraft and the local heat dissipation generated by the instrument units operational status.  

\begin{figure}[t]
\centering
\includegraphics[angle=0,width=1.0\hsize]{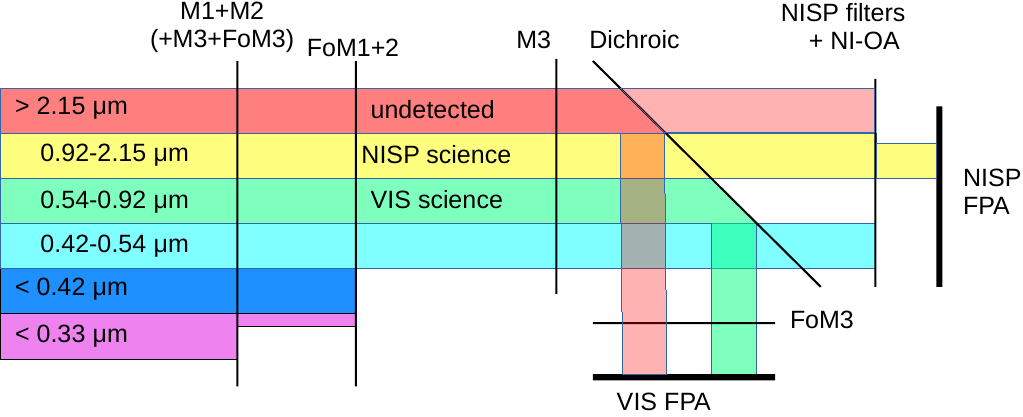}
\caption{Chromatic selection function of \Euclid's optical elements. Since the optical design minimises the number of refractive elements, mirror coatings and the dichroic element play a central role in preparing the passbands for the instruments. The VIS detectors have zero quantum efficiency for $\lambda\,{>}\,1.1$\,\micron. The behaviour of the dichroic element above $2.2$\,\micron~is not specified; longer wavelengths could enter NISP and would be blocked by the filters. Figure adapted from \cite{Schirmer-EP18}.
\label{fig:chromatic_selection}}
\end{figure}

\subsection{Transfer and orbital maintenance\label{sec:orbitmaintenance}}

\Euclid operates from a large-amplitude quasi-periodic halo orbit around L2, with a maximum Sun-Spacecraft-Earth angle of 35$^\circ$ and a period of about half a year. \Euclid travelled on a so-called stable manifold towards its operational orbit, which did not require an orbit-injection manoeuvre. The Falcon-9 launch vehicle injected \Euclid very accurately to this stable manifold. Of the three planned \glspl{TCM}, only two were required.  The first one was executed one day into the mission, providing a $\Delta v=2.14$\,m\,s$^{-1}$ to remove the launcher dispersion. The second \gls{TCM}, three weeks after launch, delivered a correction of $\Delta v=0.19$\,m\,s$^{-1}$. Overall, a total of $\Delta v=50$\,m\,s$^{-1}$ was budgeted for all three \glspl{TCM}, meaning that a considerable amount of propellant (about 43 kg) was saved for the scientific mission. However, it should be pointed out that the amount of hydrazine propellant is not a sizing parameter for the mission duration, which is limited by the amount of gaseous nitrogen used by the micro-propulsion system during the science observations.  

The launch day and lift-off times were selected such that the resulting operational orbit is eclipse-free, without excursions into the Earth and Moon shadows. Such an excursion would result in a considerable thermal disturbance and also power limitation. The quasi-periodic halo orbit around L2 is dynamically unstable, that is the perturbations grow exponentially over time. Perturbations can be non-gravitational, from sources such as outgassing \citep{schirmer2023}, imperfect thruster firings, leakage, variable solar radiation pressure, and offloading of wheel movements. Further perturbations occur in the time span from  the on-ground processing to the actual correction manoeuvre; that is, the latter does not perfectly fit the true orbit anymore. The time constant of the exponential decay of a typical wide halo orbit is 22--23 days, and escape from the operational orbit occurs after approximately 90 days. 

Station-keeping or orbital maintenance is achieved by thruster firings on a regular basis. A more frequent orbital maintenance keeps the exponentially growing perturbations better in check, meaning a smaller total velocity correction $\Delta v$ needs to be applied each year, saving propellant. \Euclid requires slots of about 6\,h per orbital maintenance, which would considerably reduce the time available for the survey if executed too frequently. The best compromise between propellant efficiency and survey efficiency for \Euclid is by scheduling orbital maintenance at fixed intervals of 28 days \citep[see also][]{Scaramella-EP1}. A total of 1000 orbital years were simulated by ESA, assuming various cases of residual acceleration from non-gravitational factors, spanning from $1\times10^{-9}$\,m\,s$^{-2}$ to $6\times10^{-8}$\,m\,s$^{-2}$. The yearly required $\Delta v$ with a 28-day maintenance schedule then ranges between 0.76\,m\,s$^{-1}$ and 7.00\,m\,s$^{-1}$, dependent on the assumed expected stochastic residual acceleration of the spacecraft.
Other \gls{ESA} missions at L2, such as \Gaia, \textit{Planck}, and \textit{Herschel}, were found to be in this range of acceleration in their respective orbit-assessment analyses.

At the time of writing, only a few orbital maintenance manoeuvres were executed, insufficient to make a reliable estimate of the actual long-term fuel consumption. The current budgeting for \Euclid is therefore necessarily conservative and for the worst case $\Delta v=7$\,m\,s$^{-1}$ per year assuming corrections every 28 days. These burns would last about 50\,s on average and consume approximately 1\,kg of hydrazine propellant. \Euclid carries a total of 137.5\,kg of hydrazine for transfer corrections into the L2 orbit, orbit maintenance for six years, and disposal into a heliocentric graveyard orbit \citep{Racca2016}. The latter is required by \gls{ESA}'s space-debris mitigation code of conduct, and requires up to $\Delta v=10$\,m\,s$^{-1}$.

\subsection{Pointing constraints and data downlink}
\label{sec:techpointingconstraints}
 
The \Euclid image quality requirements demand very precise pointing stability, while the survey requirements call for fast and accurate slews. The image quality requirements were translated into \gls{AOCS} requirements on the relative and absolute pointing errors (RPE and APE, respectively) at the 99.7\% confidence level. In science mode, the allowed RPE over a period of 700 seconds\footnote{The longest \Euclid science exposures are about 570\,s (see \cref{table:ROS}).} around the $X$- and $Y$-axes of the spacecraft is 75\,mas (milli-arcseconds), and \ang{;;1.5} around the $Z$-axis (roll angle). The allowed APE is \ang{;;7.5} around the $X$- and $Y$-axes, and \ang{;;22.5} around the $Z$-axis. 

An \gls{FGS} with four \gls{CCD} sensors co-located within the focal plane of the telescope at the side of the VIS imager provides the fine attitude measurement based on a pair of operational \glspl{CCD}. Cold-gas micro-propulsion thrusters with micro-Newton resolution provide the fine torque commands used to achieve the high-accuracy pointing. The gyro- and FGS-based 
attitude control corrects for low-frequency noise, ensuring that the RPE requirement is met. Three star tracker optical heads used in a 3:2 cold redundancy scheme provide the inertial attitude. The star trackers are mounted on the \gls{SVM} and are thus subject to thermo-elastic deformation when large slews are executed. The \gls{FGS} is also endowed with absolute pointing capabilities -- based on a reference star catalogue -- to comply with the APE requirement. This capability allows the autonomous cross-calibration of the star trackers and \gls{FGS} so that the commanded target attitude is achieved for the subsequent observation. A high-performance gyroscope is included to propagate the \gls{FGS} attitude between two measurements and during the temporary \gls{FGS} outages, for example, when operating the VIS shutter (see \cref{sec:vis} below). 

Three or four reaction wheels execute the science mode slews, specifically field, dither, and large slews between different sky zones. Before the end of the slew manoeuvre, the wheels' torques are commanded to zero and the wheels are left to brake on their own friction. Keeping the reaction wheels at rest during observations ensures noise-free science exposures by eliminating micro-vibrations and torque-noise effects.

Finally, to optimise the \gls{PSF} (\cref{sec:psfmodel}) and reduce its variability, thermal variations need to be minimised to avoid degrading the image quality as much as possible.
This places restrictions on the spacecraft attitude and internal power dissipation variations that can be tolerated. To quantify these, an analysis of the full \gls{STOP} of the satellite was performed, 
where the impact of the spacecraft attitude variations on the \gls{PSF} stability was studied (Anselmi et al., in prep.). This resulted in limiting the allowed range in \gls{SAA} and \gls{AA}. \gls{SAA} is defined as the angle between the spacecraft's $Z$-axis (telescope pointing direction; see \cref{fig:SC}) and the Sun vector (direction to the centre of the solar disk from the origin of the spacecraft reference frame). \gls{AA} is defined as the angle between the Sun vector projected onto the $X$--$Y$ plane and the $X$-axis, and it increases as the spacecraft rotates clockwise about its $+Z$-axis. In addition to the range limitation, further minimisation of the field-to-field variation in these angles is desirable. The range of \gls{AA} and \gls{SAA} considered and its implication for the survey design are discussed in \cref{sec:wide} and \cref{sec:straylight}.

\subsubsection{Data downlink}
\label{sec:downlink}

\Euclid can use a 4-hour long \gls{DTCP} to downlink its science data and recorded telemetry, about 826\,Gbit of compressed data per day, either to the Cebreros (Spain) or the Malarg\"ue (Argentina) ground station. The science data are transmitted using the consultative committee for space data systems file delivery protocol (CFDP), which simplifies science operations and processing. This is the first time this protocol has been used from L2. The downlink is performed with \Euclid's steerable 70-cm diameter $K$-band \gls{HGA}. From \Euclid's perspective, the Earth traces approximately a wide ellipse on the sky every 6 months, with an opening angle of about 35$^\circ$ and the Sun near the centre. The beam width of the \gls{HGA} is smaller than the apparent diameter of the Earth as seen from \Euclid, about \ang{0.5;;}, meaning that the \gls{HGA}'s position must be adjusted frequently. Adjustments are permitted during spacecraft slews only, so as not to disturb the pointing stability during an ongoing observation.

Within a \gls{DTCP} window the \gls{HGA} follows a ground station, accounting for Earth's rotation for maximum antenna gain.  The \gls{HGA} is repositioned with every dither slew, that is 3 times per survey field or approximately every 20 minutes. Outside a \gls{DTCP} window the \gls{HGA} is repointed to the position where a ground station is expected to appear at the next \gls{DTCP}. This happens every 72\,min when slewing to the next survey field, but an actual antenna movement might not be required every time.

To simplify spacecraft operations, the \gls{DTCP} windows are decoupled from the scientific observations, that is the \gls{HGA} must be allowed to repoint at most every 50 minutes. This means that \Euclid cannot stare at the same position on the sky for longer than 50\,min. Both the \Euclid survey and our calibration programme (\cref{sec:calibration}) are designed to fit within this constraint, which will likely also apply to future mission extensions and potential mini-surveys during windows of unallocated time (\cref{sec:unallocatedtime}).

\begin{figure*}[t]
\includegraphics[angle=0,width=1.0\hsize]{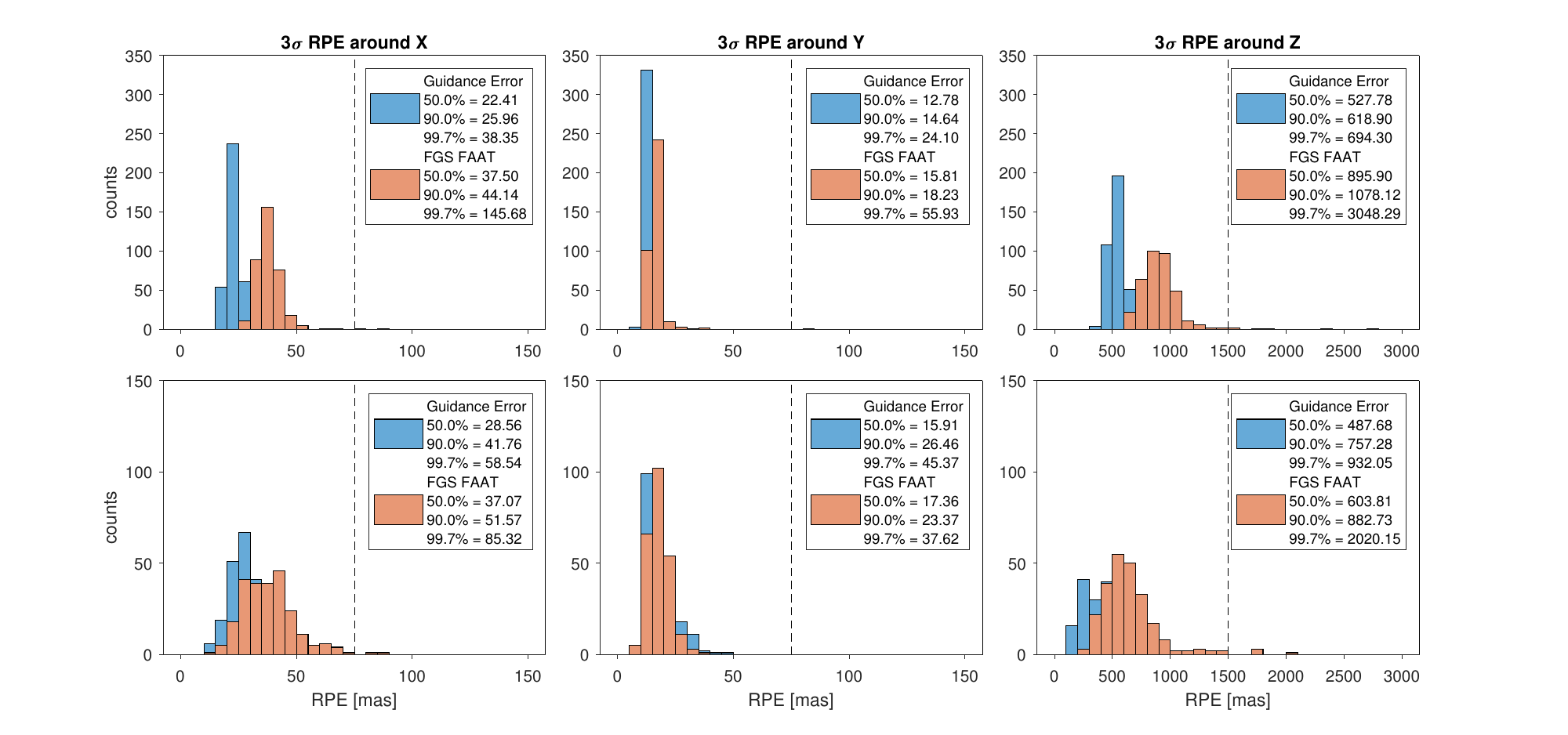}
\caption{\Gls{RPE} performance in 360 nominal (\textit{top row}) and 243 short (\textit{bottom row}) VIS science exposures. The blue histograms are based on the \gls{AOCS} controller-error vector, and the orange ones on the \gls{FGS}-provided absolute quaternion. The dashed vertical lines show the $3\,\sigma$ allocation by industry. Even though that requirement is not always met in practice, in particular about the $z$-axis, it does not mean that the \gls{PSF} requirements are violated, because there are also margins on the optical \gls{PSF}.}
\label{fig:RPE_VIS_nominal}
\end{figure*}

\subsubsection{Pointing accuracy and stability}

In-flight estimates of the \gls{APE} are larger than the requirements
by a factor 2 or more: \ang{;;3.5} around the spacecraft's $X$-axis, \ang{;;2.8} around the $Y$-axis, and \ang{;;7} around the $Z$-axis, with the axes shown in \cref{fig:SC}. This is sufficient for survey purposes, but it also means that \Euclid cannot place an object onto a specific pixel. 

Similarly, the \glspl{RPE} can be estimated from a range of orbital house-keeping parameters, providing slightly different results as they were intended for different purposes. We used the \gls{RPE} contained in the \gls{AOCS} `guidance error' vector and the absolute quaternion produced by the \gls{FGS}. In \cref{fig:RPE_VIS_nominal} we show our findings from an analysis of 360 VIS nominal science exposures and 243 VIS short science exposures taken from 5 to 9 December 2023, after an \gls{AOCS} software update to improve the pointing performance. These results should be representative for the remainder of the survey. 

Accordingly, for nominal exposures the  \gls{RPE} around the $X$-axis is poorer than around the $Y$-axis, but still fully compliant if the guidance error vector is considered (38\,mas at 99.7\% confidence level), while it would exceed the requirement (75\,mas) if the \gls{FGS} absolute quaternion is considered. Around the $Y$-axis the requirements are fulfilled using both indicators. Around the $Z$-axis the situation is similar to the $X$-axis and the requirement (1500\,mas) is exceeded if the \gls{FGS} absolute quaternion is considered. 
The \gls{RPE} for short science exposures is somewhat different due to the larger fraction of the stabilisation time in the exposure.  A full discussion of these results is beyond the scope of this paper. Concerning the \gls{PSF} reconstruction, the \gls{FGS} absolute quaternion should be used only for the $X$- and $Y$-axes, while for the $Z$-axis the pointing derived from the \gls{FGS} quaternion should be combined with the more accurate gyroscope vector.

The pointing stability during an exposure is about 35\,mas or 1/3 of a VIS pixel around the $X$- and $Y$-axes with 99.7\% confidence. This is achieved by continuous operation of cold-gas thrusters to counter non-gravitational accelerations, mainly from solar radiation pressure but also from outgassing. In the magnitude range 10 to 19, \Euclid's \gls{FGS} can find sufficient guide stars for the great majority of fields ($\ga$99.9\%) when the telescope is focused; exceptions are \Euclid's VIS wavefront-retrieval observations, where the telescope must be slightly defocused (\cref{sec:cal_PDC}).

\begin{figure*}[t]
\includegraphics[angle=0,width=1.0\hsize]{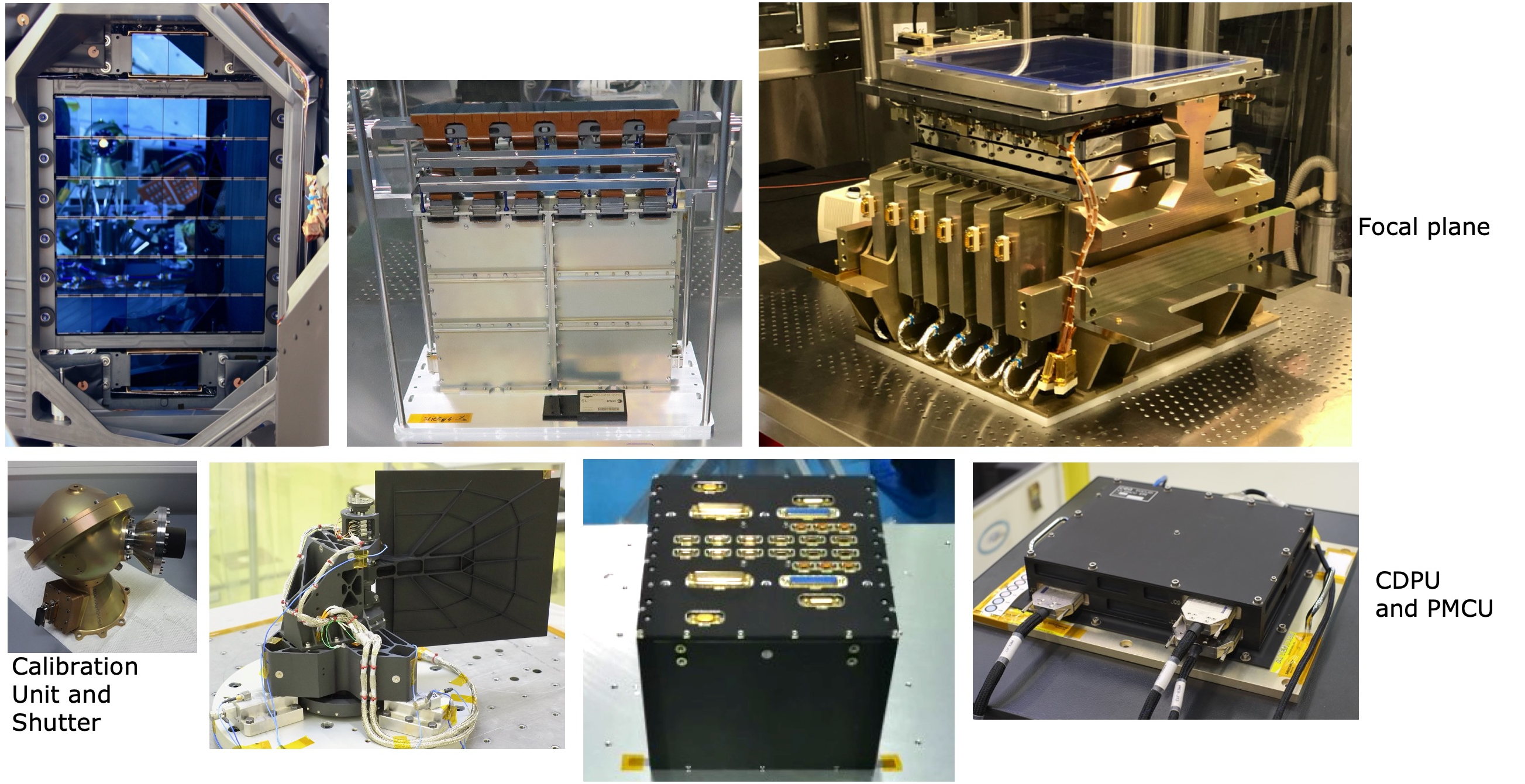}
\caption{Constituents of VIS. \textit{Top:} The VIS focal plane showing (from left to right) the array of 36 close-butted \glspl{CCD} within their SiC structure, as integrated on the \gls{PLM}; a `slice' of six \glspl{CCD} with a pair of ROEs to control them and digitise the signals; and the integrated focal plane with a protective cover for the \glspl{CCD} (six power supplies for each ROE can be seen on the left of the structure with the other six out of view behind it). \textit{Bottom:} From left to right: the \gls{CU} used for providing a flat illumination of the focal plane at six different wavelengths; the shutter; the Control and Data Processing Unit, which controls the instrument, sequences the 144 channels of data from the 36 \glspl{CCD}, compresses the image, and communicates with the spacecraft; and the Power and Mechanism Control Unit, which drives the shutter and the \gls{CU}. All of these have redundant halves except for the multiplexers on the \gls{CDPU} to the 12 ROEs.}
\label{fig:VIS_units}
\end{figure*}

\subsection{Visible instrument: VIS}
\label{sec:vis}

To enable the weak gravitational lensing science discussed in \cref{sec:weaklensing}, accurate shapes need to be measured for about 1.5 billion galaxies. This requires an instrument that can image large parts of the Universe with fine spatial resolution. Galaxies in the redshift range $0.2<z<2.5$, which will be used to map dark matter, have typical angular diameters of $0\farcs3$ and must be sampled with $0\farcs1$ resolution or better. The \Euclid instrument designed to meet these requirements, VIS, is a large-format imager with an \gls{FOV} of 0.54\,deg$^2$ sampled at $0\farcs 1$\,pixel$^{-1}$, operating in a single red passband. The considerations driving the VIS design, and its development and initial performance, are described in detail in \citet{EuclidSkyVIS}; a brief overview is provided here. A mosaic of the VIS subsystems is shown in \cref{fig:VIS_units}.

The cosmological lensing signal is extracted from a statistical analysis of a large number of coarsely sampled faint galaxies. With sample sizes to achieve the precision targeted in \cref{sec:weaklensing}, biases in the measurements become dominant. Obtaining meaningful results, therefore, depends critically on a deep understanding of the instrumental effects. VIS fits within a system of external optics, calibrations, and survey design, all of which have been highly and mutually optimised to meet the stringent performance required for weak gravitational lensing
\citep{Cropper+13}. VIS is therefore designed to be maximally stable and able to be calibrated. 

\begin{figure}[ht]
    \centering
    \includegraphics[width=1.0\hsize]{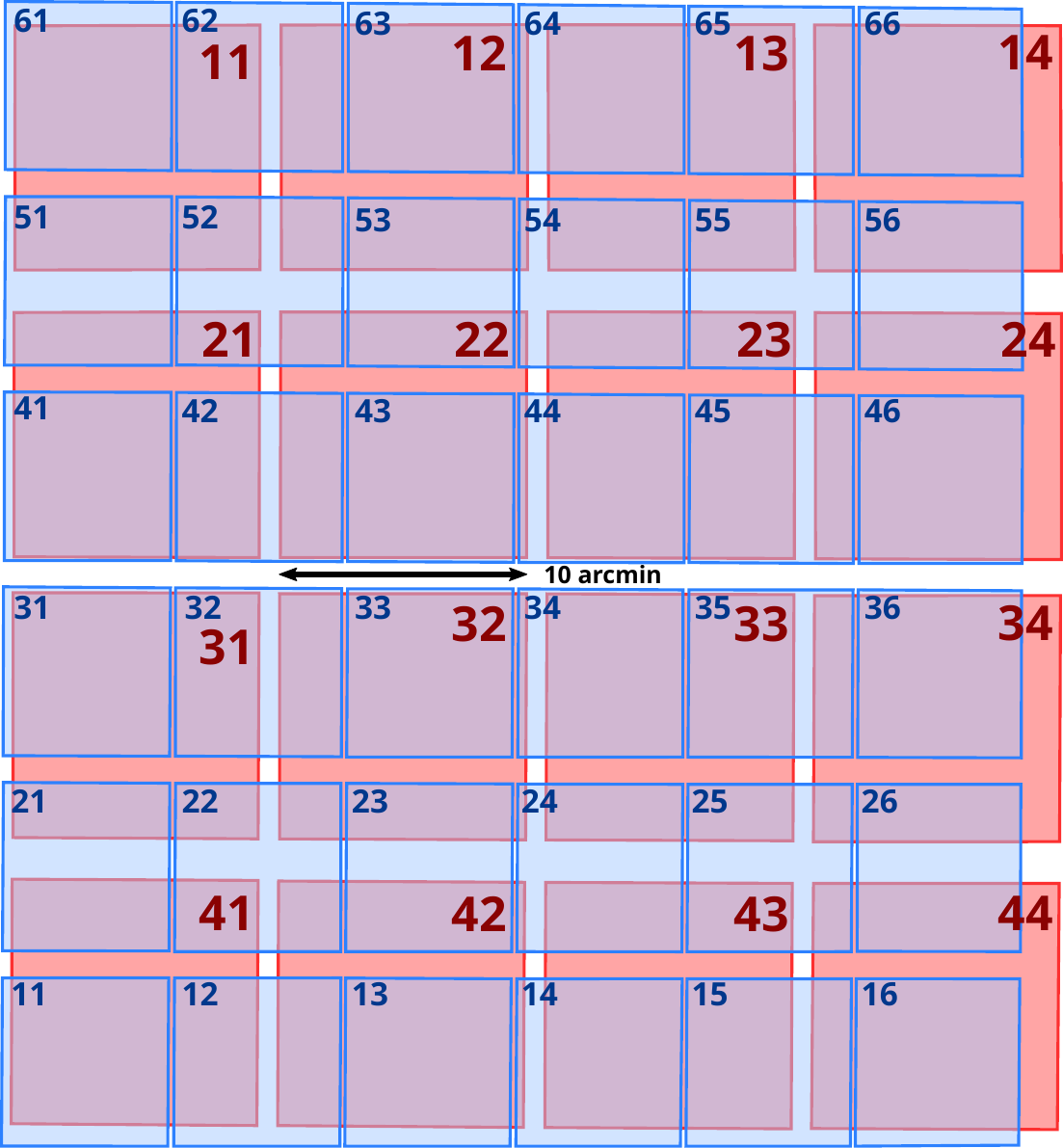}
    \caption{Common instrumental view to the sky of the VIS (blue) and NISP (red) instruments. The footprint was generated from two simultaneously taken VIS and NISP images, astrometrically calibrated and registered to a common pixel grid. 
    Small blue numbers refer to VIS, large red numbers to NISP detector IDs. Interchip gaps are evident. The VIS detectors have an additional thin horizontal gap (not shown here) from the charge-injection lines used to monitor radiation damage through charge-transfer inefficiency. The respective spatial and angular offsets between both instruments are \mbox{\ang{;;52.5}} and \ang{0.078}.}
    \label{fig:vis_nisp_fov}
\end{figure}

The VIS detectors are \glspl{CCD}, which were chosen because of the detailed understanding gained from past missions on their behaviour and performance in the space environment, and their stability, which results from the signals passing through a limited number of readout nodes. In order to cover the large \gls{FOV}, VIS has 36 CCD273-82 designed and manufactured by e2v \citep{Endicott2012} to a custom \Euclid specification \citep{Short2014} in a $6\times6$ array (see \cref{fig:vis_nisp_fov}). Each \gls{CCD} has $4132\times4096$ pixels in four quadrants, so the VIS images comprise $6.09\times10^8$ pixels. The pixels in the 144 separate quadrants are read out synchronously, to minimise noise, and digitised to 16-bit precision by 12 front-end electronics units. They are then passed to a control and data processing unit \citep[CDPU;][]{DiGiorgio2010,DiGiorgio2012,Galli2014}, where the image is constructed and losslessly compressed, then sent to the spacecraft for downlink to Earth.

The \glspl{CCD} are held in an SiC structure \citep{Martignac2014} within the PLM at 153\,K to minimise dark noise and optimise their performance in the presence of radiation damage (mainly by solar protons) to the Si-lattice within the devices. Although the pixel-to-pixel transfer of charge during
readout of the image is very good, electrons can be temporarily
trapped if they encounter a damaged site. When they escape, they can then be released into a charge packet of a later pixel. Hence galaxy shapes can be distorted because of missing charge in their pixels closer to the readout register or readout node, and spurious additional charge in pixels further from them. These distortions have a direct effect on the measured shape of a galaxy and are therefore of concern
\citep[e.g.][]{Massey+14,israel2015}. A number of enhancements to minimise these effects were incorporated into the \Euclid CCD273-82 design, including an enhanced capability to inject charge directly into the image area, even at low charge levels, to quantify the distortions from the traps. Its operational parameters were optimised for \Euclid in an extensive characterisation campaign \citep{Clarke2012, Prod'homme2014, Skottfelt2017a}, with novel tri-level clocking schemes and the capability to shuffle charge repeatedly backwards and forwards -- referred to as `trap pumping' -- to identify trap locations in both the image area and the readout register \citep{Skottfelt2017b}. Charge-injection and trap-pumping calibrations are run repeatedly throughout the survey (see \cref{table:ROS}).

Behind the SiC structure holding the 36 \glspl{CCD}, 12 sets of readout electronics units \citep[ROEs;][]{Cropper2016, Szafraniec2016} set the operating conditions for the \glspl{CCD}, clock them to read them out and digitise the pixel charge signal from the 144 \gls{CCD} quadrants. They in turn, with their 12 power supplies, are combined in an aluminium structure \citep{Martignac2014} that is interfaced to an external radiator to dissipate the heat from the ROEs. These operate at ~270\,K, so that thermal shields are used to minimise the parasitic heating of the \glspl{CCD} held nearby in their SiC structure, nearly 120\,K colder. These two halves comprise the \gls{FPA} shown in \cref{fig:VIS_units}. They are integrated on each side of a substantial SiC bracket on the \gls{PLM} baseplate (see \cref{fig:euclid_CAD_annotated}).

\begin{figure*}[t]
\includegraphics[angle=0,width=1.0\hsize]{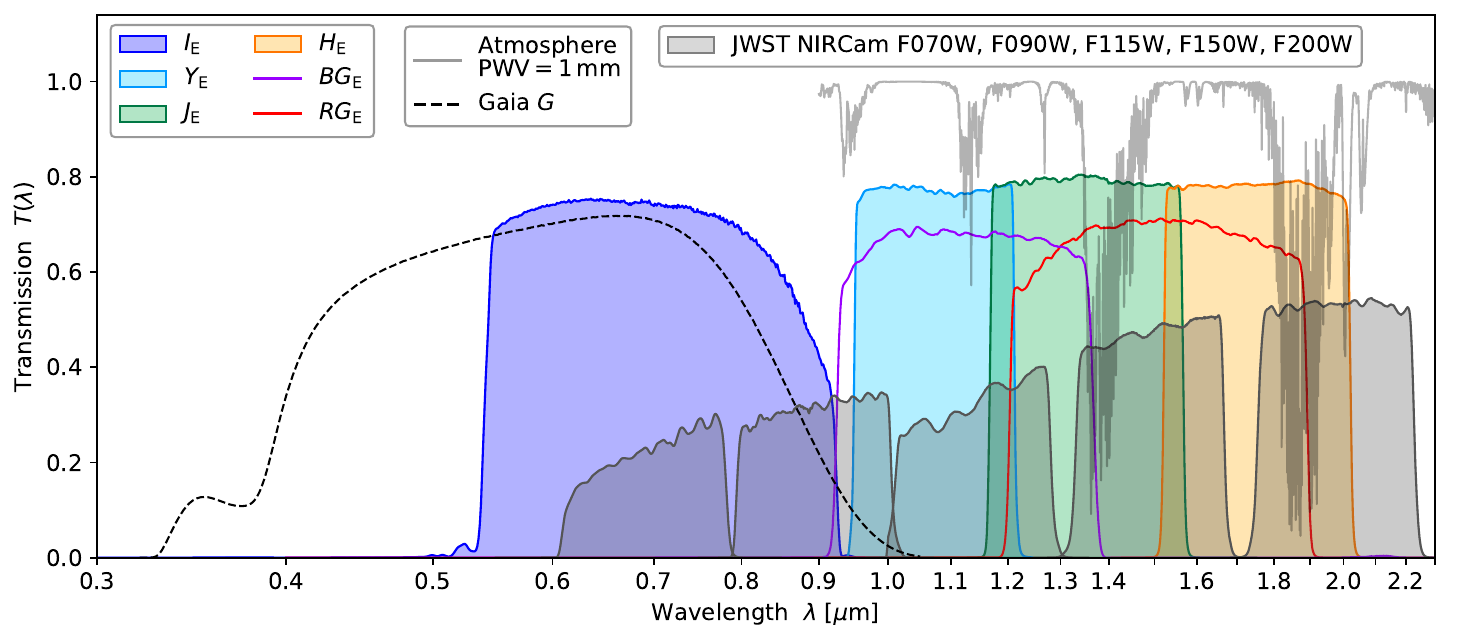}
\caption{Spectral response of \Euclid's imaging (VIS: \IE; NISP: \YE, \JE, \HE) and spectroscopic channels (NISP: \BGE, \RGE) at the beginning of the mission. The expected transmission loss at the end of the mission due to space weathering and non-volatile contamination is at most 0.05. For reference we show the \Gaia $G$ passband from their third data release \citep{Gaia-DR3}, the atmospheric transmission for a precipitable water vapour level of 1.0\,mm \citep{rothman2013},
and some of the \gls{JWST} passbands of their Near Infrared Camera \citep[NIRCam;][]{rieke2005}.}
\label{fig:euclid_passband_comparison}
\end{figure*}

In order to maximise the stability of the VIS imaging, the instrument does not have a filter wheel; at the level required by \Euclid this would not permit sufficiently repeatable image registration from exposure to exposure. The bandpass \IE\ for the instrument is therefore set by the \Euclid telescope, dichroic plate, the folding optics (\cref{fig:korsch}), and the \gls{QE} of the \glspl{CCD}, to be in the range 530--920\,nm, optimised for the spectral energy distribution of the majority of galaxies. 
The wide band and high throughput (\cref{fig:euclid_passband_comparison}) provide a limiting sensitivity 
of $m_{\rm{AB}}=26.7$ ($5\sigma$ point source), so that galaxies 
with a \gls{FWHM} of $0\farcs3$ and 
$m_{\rm{AB}}=25.0$ are detected with $S/N=10$ in a $1\farcs$3 diameter aperture, sufficient for precise shape measurements. At the same time, however, the wide band complicates the shape measurement, because the PSF is chromatic. 
Although the entire VIS optical channel is in reflection, faint off-axis optical ghosts are created by the rear surface of the dichroic, and these must be masked in post-processing. 

The \glspl{CCD} are read out at 73\,kHz to limit readout noise, and require a shutter \citep{Genolet2016} to avoid image trailing from continuous illumination during readout. For reliability, the shutter is a single leaf with dimensions sufficient to cover the $343\,\mathrm{mm} \times 303\,\mathrm{mm}$ \gls{CCD} array. It is momentum-compensated to a fine degree to minimise the disturbance to the \Euclid pointing and hence the recorded \gls{PSF}, which must be modelled to a high level of fidelity for accurate shape measurements. The shutter and a calibration unit for generating flat fields \citep{Cropper2018} are driven by a power and mechanism control unit \citep[PMCU;][]{Cara2018}. This unit and the \gls{CDPU} reside in the \gls{SVM}, while the \gls{FPA} with its electronics, the shutter, and the calibration unit reside in the \gls{PLM}. The full complement of VIS units is shown in \cref{fig:VIS_units}.

\subsection{Near-Infrared Photometer and Spectrometer: NISP}
\label{sec:nisp}

\begin{figure}
    \centering
    \includegraphics[width=1.0\hsize]{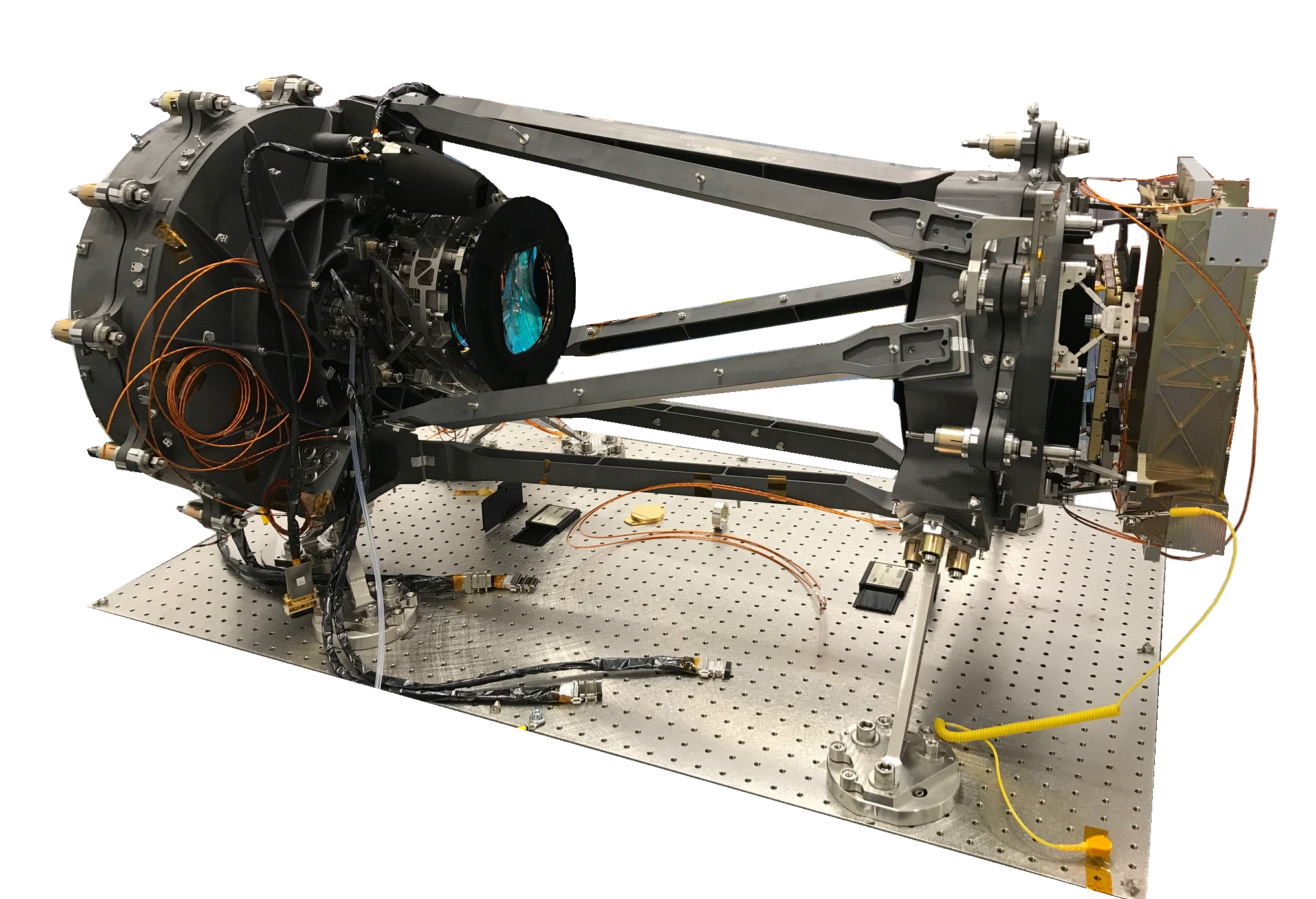}
    \caption{
NISP flight model, before wrapping in light-tight multi-layer insulation. Light enters the filter wheel and grism wheel enclosure (left) through a collimator lens, hidden behind the large round wheel enclosure. A triplet camera lens assembly projects the beam onto the cold detector system at the right end of the structure, with the readout electronics to the very right. The NISP calibration lamp is located to the top left of the camera lens assembly in this picture. See \citet{EuclidSkyNISP} for details.    
    }
    \label{fig:NISP}
\end{figure}

The second instrument on board \Euclid is the Near-Infrared Spectrometer and Photometer (NISP, \cref{fig:NISP}), described in more depth in \citet{Maciaszek22} and \citet{EuclidSkyNISP}.
NISP provides multiband photometry and slitless grism spectroscopy in the wavelength range 920--2020\,nm (\cref{fig:euclid_passband_comparison}), using the light transmitted by the dichroic beamsplitter.

\subsubsection{Hardware overview}
\label{sec:nisp_hardware_overview}

NISP has a common optics and detector system for its photometric and spectroscopic channels, with respective filters and grisms in two wheels.\footnote{It is not permitted to use elements from both the filter and grism wheels at the same time, because both have some optical power. Hence, a simultaneous observation through both elements, for example to produce shorter spectra with smaller overlap fractions, results in strongly defocused images.} A collimator lens provides for each individual source a nearly parallel beam through the filters and grisms. A subsequent camera-lens assembly, together with slight optical power on grisms and filters, focuses the beam in the detector plane. Details and consequences of this design, such as passband variations, are discussed in \citet{Schirmer-EP18}.

The filter wheel includes a dark plate that can be used to block all light from the telescope, but not from the calibration lamp, for specific calibration purposes. The lamp uses five nearly monochromatic \glspl{LED} whose wavelengths span the NISP wavelength range to support a wide spectrum of calibrations (\cref{sec:cal_PVphase,sec:cal_routine,sec:calibration}).

The \gls{FPA} consists of a $4\times4$ grid of Teledyne Hawaii-2RG (H2RG) detectors with $2048\times2048$ pixels. A four-pixel wide border along the detector edges is not light-sensitive, providing baseline reference pixels for detector monitoring.  With a plate scale of \ang{;;0.298}\,pixel$^{-1}$ and including detector gaps, the NISP \gls{FOV} covers a nearly square-shaped 0.57\,deg$^2$ (see \cref{fig:vis_nisp_fov}).

The 16 H2RGs are operated at $T\,{\simeq95}\,$K to optimise detector behaviour, while the main optics is kept at 130--132\,K.
Parallel readout occurs in 32 channels per detector using a \gls{MACC} scheme, with NISP continuously read at a nominal rate of 1.45408\,s per frame. For science and calibration exposures, a group of each 16 successive frames is read non-destructively and averaged by the on-board \gls{DPU}. Between these groups the exposure just continues and photons are collected, but a number of frames are not read (so-called `drops'), either since they would not add up to another group of 16, or due to running into NISP storage limits. At the end, the count rate in each pixel is determined with a linear slope fit to the group values using an iterative algorithm \citep{kubik2016}. 
Standard photometry and spectroscopy exposures for the Euclid Wide and Deep Surveys (\cref{sec:survey}) use four and 15 groups with 16 frames each, and a number of four and 11 dropped frames, respectively; we refer to these modes as MACC(4,16,4) and MACC(15,16,11). The effective integration times for the photo- and spectro-exposures are 87.2\,s and 549.6\,s respectively, while the actual exposure durations are 110.5\,s and 572.9\,s. The latter times are relevant for the survey planning (\cref{sec:survey}), while the former determine the depth of the science data.

NISP uses cold readout electronics at the focal plane, and warm electronics operating at $T\,{\simeq}\,290$\,K located in the \gls{SVM}. The warm electronics contain both the commanding computer and the \gls{DPU}. The latter performs the baseline subtraction using reference pixels, the MACC slope fit, and the data compression. For images the compressed slope-fit image is downlinked together with a 1-bit quality image that encodes whether the slope-fit $\chi^2$ lies above a certain threshold. For dispersed spectra images the full $\chi^2$ information is downlinked in an 8-bit quality image. NISP has full redundancy in its warm electronics and calibration-source \glspl{LED}, as well as the power supply to drive the filter and grism wheels.

\subsubsection{Near-infrared imaging}
\label{sec:nisp-p}

The NISP photometric channel (NISP-P) offers three passbands \YE\ (949.6--1212.3\,nm), \JE\ (1167.6--1567.0\,nm), and \HE\ (1521.5--2021.4\,nm), displayed in \cref{fig:euclid_passband_comparison}. 
The wavelengths refer to the 50\% peak-transmission points near the centre of the \gls{FOV} and are accurate to 0.8\,nm. Passband variations within the \gls{FOV} are characterised to $\lesssim0.1$\,nm. The near-rectangular passband flanks are entirely defined by the 130-mm diameter filters, which carry up to 200 interference coating layers distributed over both filter sides \citep{EuclidSkyNISP}. The total in-band spectral response, including detectors, is close to 80\%. Out-of-band blocking is 10$^{-4}$ or better within 900--2100\,nm, and $10^{-5}$ to $10^{-7}$ outside this range. These excellent blocking capabilities are jointly achieved by the filters, all other coated optical surfaces in the NISP optical path, and the detectors (\cref{fig:chromatic_selection}). Out-of-band contamination is at most 2.0\,mmag for sources with extreme \glspl{SED}, and more typically 0.2\,mmag. More details about this and the NISP photometric system in general are presented in \cite{Schirmer-EP18}. 

The designed $5\,\sigma$ point-source depth of NISP-P for the wide survey is 24.0\,AB\,mag, which we exceed by approximately 0.4\,mag (\cref{sec:preliminaryCommissioning}). The plate scale of \ang{;;0.298}\,pixel$^{-1}$ considerably undersamples the NISP \gls{PSF} that has a typical \gls{FWHM} of 1.10, 1.17, and 1.19\,pixel in \YE, \JE, and $\HE$ when fitting a Moffat profile \citep[see also][]{bernstein2002}. The measured \gls{PSF} size is fully compatible with the on-ground characterisation of the NISP optics \citep{grupp2014}. Details are given in \cref{sec:preliminaryCommissioning}.

\subsubsection{Near-infrared spectroscopy}
\label{sec:nisp-s}

The NISP spectroscopic channel (NISP-S) enables the simultaneous acquisition of slitless spectra for thousands of objects across the \gls{FOV} with uniform quality. The grism wheel houses four different grisms of 140-mm diameter each. The grisms are dispersion gratings combined with a prism whose base is -- just like the filters -- slightly curved for optimal focus. Dielectric coatings improve out-of-band blocking \citep{EuclidSkyNISP}. The total spectral response including all optical surfaces and detectors is shown in \cref{fig:euclid_passband_comparison}.

For the spectroscopic observations of the \gls{EWS} (\cref{sec:wide}) NISP-S uses three red grisms covering the same \RGE passband (1206--1892\,nm; 50\% peak transmission wavelengths), allowing the detection of H$\alpha$ emitters in the range $z=0.84$--$1.88$. These grisms have different dispersion directions of $0^\circ$, $180^\circ$, and $270^\circ$ with respect to the detector columns. By combining the dispersed slitless images of the same field, overlapping spectra from multiple sources can be disentangled (`decontaminated') and clean spectra extracted. The red grisms have a dispersion of $1.372$\,nm\,pixel$^{-1}$ and a resolving power of $\mathcal{R}_{\rm RG}\ga480$ for a source with $\ang{;;0.5}$ diameter. The mission requirement is $\mathcal{R}_{\rm RG}>380$ to achieve a redshift accuracy of $\sigma(z)<0.001\,(1+z)$. 
During the ground tests in 2020 it was discovered that the \ang{270;;} grism does not conform to the specifications and cannot be used for the survey. To achieve spectral decontamination, we use the other two red grisms with additional \ang{4;;} rotational offsets of the grism wheel, providing a total of four different dispersion directions \citep[for details see][]{Scaramella-EP1}. The rotated positions vignette one edge of the detector array by up to 10\%, which is accounted for by our calibration, while still meeting the overall image quality requirements. 

The designed target sensitivity for the wide survey is a $3.5\,\sigma$ detection for an emission line with flux $\SI{2e-19}{W.m^{-2}}$ at 1600\,nm (such as redshifted H$\alpha$), for a source with \ang{;;0.5} diameter. The in-flight performance was not yet available at the time of writing; however, given the excellent spectral image quality and the fact that NISP-P exceeds its designed depth by 0.4\,mag, we are confident that NISP-S also meets its sensitivity requirement.

NISP-S also has a blue grism covering 926--1366\,nm, extending the lower H$\alpha$ redshift limit to $z\,{=}\,0.41$. The blue grism has a resolution of 1.239\,nm\,pixel$^{-1}$ and a resolving power of $\mathcal{R}_{\rm BG}\ga400$. The blue grism is solely used for observations of the \Euclid Deep and Auxiliary fields (\cref{sec:deepsurvey}). Its main purpose is to provide a large reference sample of galaxies with 99\% redshift completeness and 99\% purity required to characterise the typical \Euclid galaxy population, while maximising the legacy value of these fields (\cref{sec:legacy}).


\section{Survey planning}
\label{sec:survey}

\begin{figure*}[t]
\centering
\includegraphics[width=1.0\textwidth]
{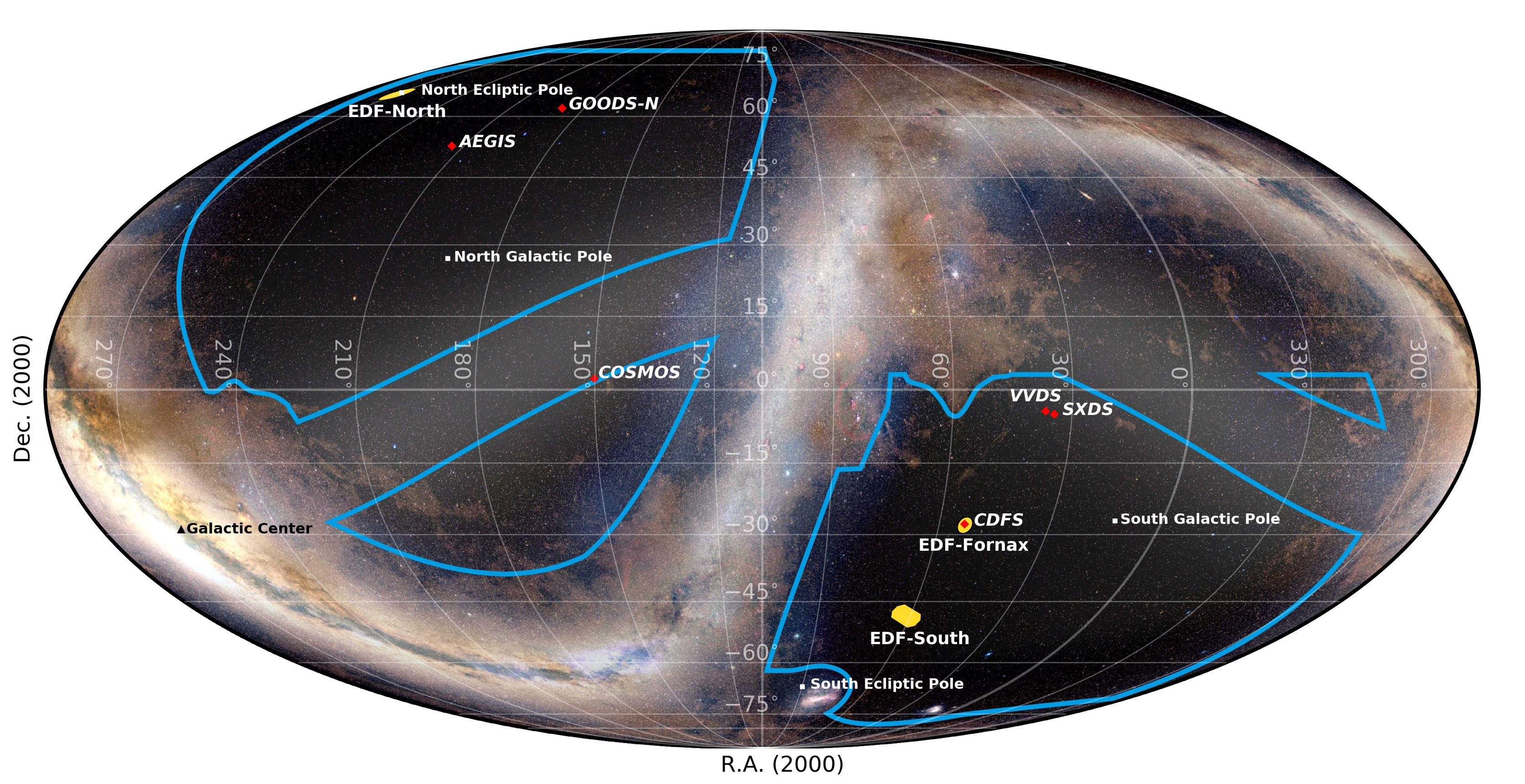}
\caption{\Euclid \gls{ROI} in an all-sky Mollweide projection. The blue borders enclose the 16\,000\,deg$^2$ \gls{ROI} that contains the observed sky of the Euclid Wide Survey. The \gls{ROI} excludes the Galactic and ecliptic planes. The triangular southern `island' near RA = \ang{330;;} is restricted in size since the \gls{LSST} does not extend to more northern latitudes. The Euclid Deep Fields are shown in yellow and the auxiliary fields with red marks (not to scale). }
\label{fig:skysurvey}
\end{figure*}

To achieve its primary cosmology objectives, \Euclid aims to observe a sample of 1.5 billion galaxies for the 3\texttimes2pt analysis, and measure 35 million redshifts for the analysis of the spectroscopic clustering signal. To do so, it needs to cover about 14\,000\,deg$^2$ of extragalactic sky with low zodiacal background and low Galactic extinction over a period of about six years.
In \cref{sec:wide} we summarise the design of the \gls{EWS}. As discussed in \cref{sec:deepsurvey}, these data are complemented by deeper observations over several tens of square degrees, while the performance of the instruments is studied using extensive calibration observations, which are described in \cref{sec:calibration}. Finally, the supporting ground-based observations are summarised in \cref{sec:groundbased} (photometry) and \cref{sec:cog_spec} (spectroscopy).

\begin{table}[t]
\caption{Typical \gls{ROS} data per survey field.} 
\label{table:ROS}
\centering
\begin{tabular}{ll}
\hline\hline
\noalign{\vskip 1pt}
\multicolumn{2}{c}{VIS}\\
{\bf 4 {\boldmath \IE}-band nominal exposures} & 566\,s each\\
{\bf 2 {\boldmath \IE}-band short exposures} & 95\,s each\\
Bias & 2 per day \\
Dark & 4 per day \\
Flat & 6 per day \\
Trap pumping & 6 per day \\
Charge injection & 8 per day \\
\hline
\noalign{\vskip 1pt}
\multicolumn{2}{c}{NISP}\\
{\bf 4 red-grism spectro exposures} & 574\,s each \\
{\bf 4 {\boldmath \YE}-band exposures} & 112\,s each \\
{\bf 4 {\boldmath \JE}-band exposures} & 112\,s each \\
{\bf 4 {\boldmath \HE}-band exposures} & 112\,s each \\
1 Dark & 112\,s \\
\hline
\end{tabular}
 \footnotesize
 \tablefoot{The total duration of an \gls{ROS} is 70.2\,minutes. About 20 fields are observed per day, cycling through different \gls{ROS} configurations. While the science exposures (marked with boldface) remain the same, the inline VIS calibrations vary. \gls{ROS} configurations using the blue grism are used for the \gls{EDF} and \glspl{EAF} only. Most NISP calibrations are taken outside the \gls{ROS} during dedicated calibration blocks.}
\end{table}

\subsection{Euclid Wide Survey}
\label{sec:wide}

\cite{Scaramella-EP1} provides a detailed description of the reference  \gls{EWS}. Here, we summarise its characteristics and highlight the main modifications that were implemented prior to launch. The target area has minimal contamination from the Galaxy, Solar System objects, and the zodiacal background, that is the diffuse sunlight scattered by interplanetary dust in the ecliptic plane. The intersection of the avoided regions around the Galactic and ecliptic planes leaves four separate dark areas on the sky for the \gls{EWS}, which we refer to as the \gls{ROI} of \Euclid, indicated by the blue lines in \cref{fig:skysurvey}.
 These areas are adjusted in size to maximise the overlap with the ground-based surveys providing complementary data (\cref{sec:groundbased}) needed for \gls{PSF} modelling and photometric redshift estimation. In the latest pre-launch configuration,
the \gls{EWS} covers 14\,816\,deg$^2$ of the darkest extragalactic sky, of which 137\,deg$^2$ are lost due to about 800 bright stars  with a magnitude $m_{\rm AB}<4$ in any \Euclid band. As a result, the effective sky area is 14\,679\,deg$^2$. 

As detailed in \cite{Laureijs11}, for the VIS imaging we require a ${\rm S/N}\,{\ge}\,10$ for extended sources with a diameter of $1.2$ times the \gls{FWHM} of the \gls{PSF} at $\IE=24.5$\,AB\,mag. For NISP, we must reach a ${{\rm S/N}}\,{\ge}\,5$ for point sources with $m_{\rm AB}=24.0$ in all three NISP bands, and ${\rm S/N}\,{\ge}\,3.5$ for an H$\alpha$ line flux of $2\times10^{-16}\,\mathrm{erg}\,\mathrm{cm}^{-2}\,\mathrm{s}^{-1}$ at a redshifted wavelength of 1.6\,\micron~in the \RGE spectra. We show in \cref{sec:preliminaryCommissioning} that these depth requirements are met with considerable margin. However, as discussed in more detail in \cref{sec:straylight},  unacceptable levels of stray light were observed for certain spacecraft attitude angles. Avoiding these orientations is possible, but it also means that the effective sky area of 14\,679\,deg$^2$ for the \gls{EWS} can no longer be met within the nominal mission duration.

The \gls{ROS} is the building block of the survey, comprising four dithered exposures that form an `S'-like pattern; the three dithers that follow the first exposure are in ecliptic coordinates\footnote{In this paper we use the Greek letter $\alpha$ for the Galactic longitude to avoid confusion with wavelength $\lambda$.} $(\Delta\alpha,\Delta\beta) = (\ang{;;61},\ang{;;111})$, $(\ang{;;0},\ang{;;111})$ and $(\ang{;;61},\ang{;;111})$, respectively. \Cref{table:ROS} lists the data collected during a typical sequence for a given survey field. In addition to the science data, various calibration exposures are obtained. The \gls{ROS} lasts 70.2\,min, followed by a slew to the next adjacent survey field. Such field slews take 2--4\,minutes. Occasionally, large slews are required to begin a new survey patch on the sky, or to perform specific calibration observations. These last between 7 and 33\,minutes, and their occurrence is kept to a minimum. Hence, about 20 fields are observed per day.

The \gls{EWS} itself consists of about 27\,500 fields, each of which is observed once with the \gls{ROS}. Including \gls{EDS} and calibration observations, the total survey comprises around 49\,000 fields.

\begin{figure}[th]
\centering
\includegraphics[width=0.7\hsize]{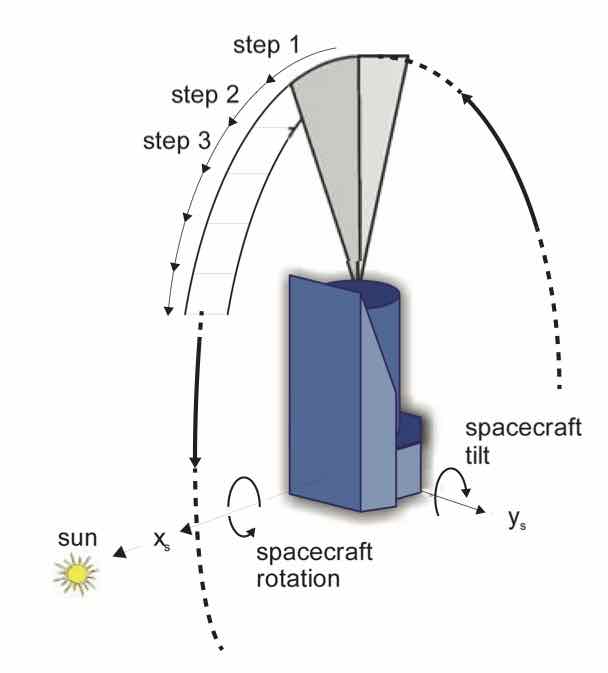}  
    \caption{\Euclid's main step-and-stare observing mode, showing north-south steps along a circle as rotations around the $X$-axis. \Euclid can tilt to another circle by rotating around the $Y$-axis.}
    \label{fig:step_and_stare}
\end{figure}

\subsubsection{Principal survey strategy}

Here, we describe the fundamental principles of the survey design. Significant modifications with respect to the pre-launch strategy are introduced in \cref{sec:newsurveydesign}, because of the need to avoid stray light that was discovered post-launch at certain  spacecraft attitude angles.

For ecliptic latitudes $\ang{-78;;}<\beta<\ang{78;;}$ the \gls{EWS} tessellates the \gls{ROI} with non-overlapping identical tiles aligned with the ecliptic meridians, distributed along parallels of latitude.  This configuration minimises overlaps and maximises survey efficiency.  A tile is a \gls{FOV} placed aligned with the meridian passing through its centre. The size of the tiles is computed from the intersection of the VIS and NISP \glspl{FOV}.  The tiles are observed in a step-and-stare mode by placing the larger \Euclid common \gls{FOV} (\cref{fig:vis_nisp_fov}) on the four dithered positions of each tile. The minimum average overlap between neighbouring fields is 2.2\% in area, occurring when all fields are aligned with their respective tiles. For the polar caps at $|\beta|\geq\ang{78;;}$ a different tiling strategy was chosen to avoid excessive overlaps due to the converging meridians; details are given in \cite{Scaramella-EP1}.

The basic position of the \Euclid spacecraft has its sunshield facing towards the Sun, with the telescope pointing towards the \gls{NEP}. From this position \Euclid can rotate freely around the Sun--spacecraft axis, keeping the sunshield orthogonal towards the direction to the Sun, so that it can point to any field on the transit ecliptic meridian, \ang{90;;} away from the Sun's longitude (\cref{fig:step_and_stare}). The natural observing mode is to step-and-stare along the transit meridian, effectively sweeping the sky with transit meridians, at an approximate rate of \ang{1;;} per day, as \Euclid progresses on its yearly orbit.
The full circle defined by a transit meridian is divided in the `leading side', the half-circle pointing in the direction of the spacecraft orbit, and the `trailing side', the opposite half-circle.  The two half-circles meet at the ecliptic poles.
When observing in the trailing side the \gls{FOV} is rotated \ang{180} in the sky.

The \gls{EWS} is frequently halted to observe calibration fields at specific cadences. Moreover, as discussed in 
\cref{sec:orbitmaintenance}, all observations must stop every 4~weeks for orbital maintenance. During the interruption, the transit meridian moves and in order for the  \gls{EWS} to continue from the same point the spacecraft needs to tilt from its natural position (\cref{fig:step_and_stare}). 
In the latest pre-launch configuration, \Euclid was allowed to point away from the transit meridian by tilting up to \ang{3;;} towards the Sun and up to \ang{20;;} away from the Sun, maximising thermo-optical stability; that is, the \gls{SAA} range was restricted to $[\ang{87;;};\ang{110;;}]$. 

If the interruption is too long, the last observed point of the \gls{EWS} may no longer be visible with the allowed \gls{SAA} range and can only be observed up to 6~months later, when the antipodal meridian is in transit. 

\begin{figure}
\centering
\includegraphics[width=\hsize]{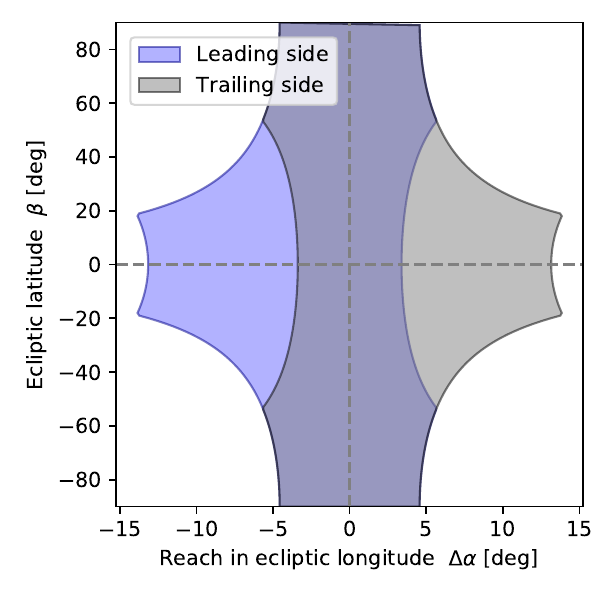}
\caption{Window of visibility. Shown are the reachable ecliptic longitudes around transit as a function of latitude, computed for $\rm{\gls{SAA}}\in[\ang{87;;};\ang{104;;}]$ and 
${\rm \gls{AA}}\in[\ang{-5;;};\ang{5;;}]$. A strict tessellation constraint is imposed, meaning the survey fields are not allowed to rotate with respect to the tessellation.}
\label{fig:windowvisibility1}
\end{figure}

Tilting from the transit meridian to another circle misaligns the \gls{FOV} on the sky with respect to the orientation of the tessellation tiles. A rotation around the $Z$-axis (a change of AA; see \cref{fig:SC}) realigns the field with the tile. In the latest pre-launch configuration, the \gls{AA} range was restricted to $[\ang{-5;;};\ang{5;;}]$ for thermal considerations. The limitations on the range of the two solar angles, \gls{SAA} and \gls{AA}, together with the need to keep the fields aligned (the `tessellation constraint'), defines the `window of visibility' (\cref{fig:windowvisibility1}). The window of visibility shows the span in longitude that \Euclid can reach away from transit, and it is a function of ecliptic latitude $\beta$.  At low latitudes, where tilted observations do not introduce a misalignment of the \glspl{FOV}, the longitude span is identical to the \gls{SAA} range. Since the misalignment for a given tilt increases with latitude, the limited \gls{AA} range available to align the \glspl{FOV} drives the longitude span at high latitudes. Hence the window of visibility broadens towards the ecliptic equator.
We note that without the tessellation constraint the ecliptic poles would have perennial visibility, and the window of visibility would broaden towards high latitudes, with the \gls{AA} range playing no role in its definition. The longitude span is also directly related to how long a tile in the sky remains visible around transit. 
 
Tiles are observed in sequences called `patches', usually covering a latitude-longitude rectangle of the \gls{ROI}.  The viability in scheduling a patch is closely related to the window of visibility.  Tiles must be visible at all latitudes of the tessellated \gls{ROI} ($\ang{10;;}\leq\beta<\ang{78;;}$). A reduced \gls{SAA} range restricted to $[\ang{87;;};\ang{104;;}]$ is used in practice to decrease the asymmetry of the window of visibility between high and low latitudes. Tiles must also be visible for a reasonable span of time; a longer visibility promotes wider patches. 

\subsubsection{Unallocated time\label{sec:unallocatedtime}}

\begin{figure}[t]
\centering
\includegraphics[angle=0,width=1.0\hsize]{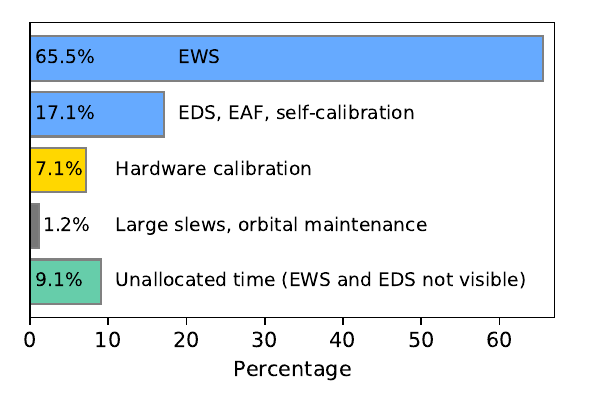}
\caption{Breakdown of activities during routine operations. The blue bars provide on-sky data that are simultaneously valuable for science, target characterisation, and calibration purposes; the instruments take additional calibration data while the data processing units are busy with the science exposures, and while the telescope is slewing. The yellow bar represents pure hardware calibration with little or no astrophysical relevance. Unallocated time arises because the survey runs out of unobserved sky areas (\cref{sec:unallocatedtime}).}
\label{fig:calib_pie_chart_global}
\end{figure}

The \gls{EWS} is built by tiling patches layer by layer, starting from the poles, where the zodiacal background is lower, towards the ecliptic.  
The progression of the \gls{EWS} also aims to observe the areas with complementary ground-based data as early as possible (\cref{sec:groundbased}).
On the other hand, \gls{EWS} observations must be scheduled continuously while the transit meridian scans the sky.
However, longitude-wise, the area of the \gls{ROI} is not uniform.
At some point, the \gls{ROI} area is exhausted where the Galactic plane intersects the ecliptic plane, and no \gls{EWS} fields are available anymore for scheduling. This creates periods of unallocated time, all during the second half of the mission, that  will then reappear regularly with a 6-month cadence, growing in duration. In the latest pre-launch configuration the unallocated time amounts to 9\% of total time during routine operations (\cref{fig:calib_pie_chart_global}).

The periods of unallocated time will be used for different purposes. Foremost, they offer an opportunity for thermal decontamination of the PLM (\cref{sec:photthroughput}), and  to recover survey areas that were lost for example due to intermittent hardware problems, data-transmission losses, and severe space-weather events. Any remaining unallocated time not used for primary \Euclid purposes could become available for targeted observations with the \gls{ROS}, or for mini-surveys, possibly outside the \gls{ROI}. Such opportunities, if any, will be developed and communicated in due time.

\subsection{Euclid Deep Survey}
\label{sec:deepsurvey}
The need to calibrate and monitor the telescope, cameras and electronics requires repeated visits of specific fields that will accumulate substantial depth over time. We also need to characterise the typical \gls{EWS} source population and systematic effects, requiring deep data over a large area. In this section we present the motivation and characteristics of these data.

\begin{figure*}[t]
\centering
\includegraphics[width=0.9\textwidth]
{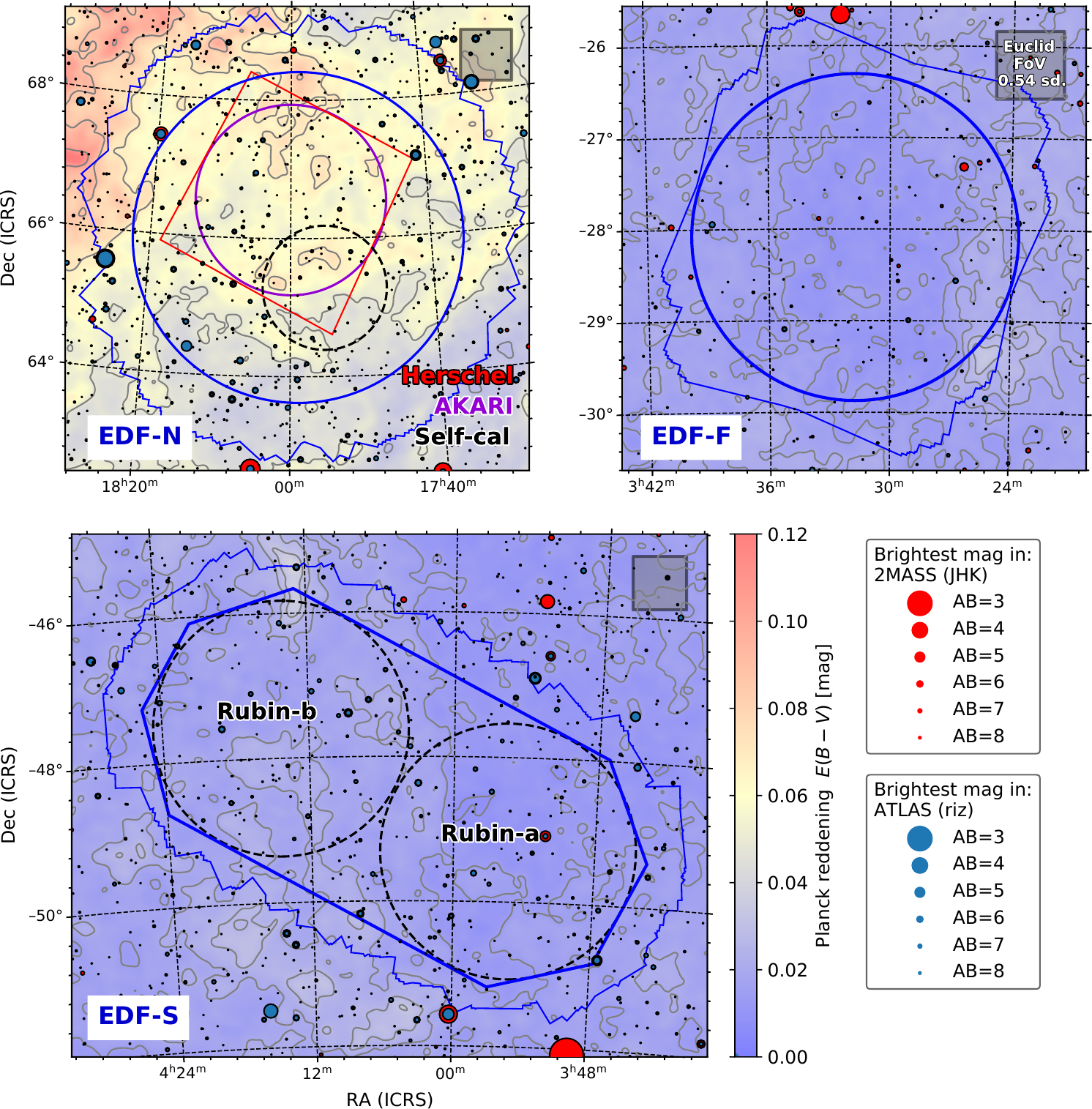}
\caption{Layout of the three Euclid Deep Fields, using coordinates in the \gls{ICRS},
overlaid on top of the reddening map from \cite{Planck2013dust} with bright stars from 2MASS \citep{Skrutskie2006} and ATLAS \citep{tonry2018} indicated. 
The thick blue lines show the areas that will be covered to full depth. The thinner blue lines approximate the wider but shallower extent due to dithering. \textit{Upper-left panel}: The EDF-N contains the \textit{Herschel} \citep{pearson2017} and AKARI NEP-wide surveys \citep{lee2009}, as well as the \Euclid self-calibration field (dashed black circle). \textit{Upper-right panel}: The EDF-F contains the Chandra Deep Field South. \textit{Bottom panel}: The EDF-S will also be observed by two \gls{LSST} deep-drilling fields. All three fields have been fully covered in four \textit{Spitzer} bands \citep{Moneti-EP17}, and are well suited for broad, extragalactic science.}
\label{fig:3fieldsEBV}
\end{figure*}

\subsubsection{Euclid Deep Fields}
\label{sec:deepfields}

About 12\% of \Euclid's on-sky observations are spent on the \gls{EDS}, for which we target a six-fold increase in S/N compared to the \gls{EWS}, or a gain in depth of about 2\,magnitudes. The primary purpose of the \gls{EDS} is an accurate characterisation of the typical \gls{EWS} galaxy population, that is a 99\% complete and 99\% pure spectroscopic sample of at least 120\,000 galaxies, as well as accurate morphologies of galaxies to calibrate systematics in the weak-lensing shape measurement. The EDS also enables numerous legacy purposes, from primeval galaxies, galaxy and \gls{AGN} evolution, and discovery of supernovae, to the structure of our Galaxy. It will uncover numerous targets for follow-up observations, and greatly extends the scientific scope of the mission beyond its core cosmology goals (\cref{sec:legacy}). 

Depending on the zodiacal background, at least 40 repetitions of the \gls{ROS} are required to reach the desired EDS depth. Unlike the \gls{EWS}, the \gls{EDS} includes blue-grism observations with an exposure-time ratio of 5:3 for the blue relative to the red grism. The \gls{EDS} comprises the Euclid Deep Field North (EDF-N), the Euclid Deep Field South (EDF-S), and the Euclid Deep Field Fornax (EDF-F). \Cref{fig:3fieldsEBV} shows the layout of the three fields, which cover a total area of 53\,deg$^2$. The EDF-N is a 20\,deg$^2$ circular field located at the northern ecliptic pole. The EDF-F is a 10\,deg$^2$ circular field including the Chandra Deep Field South (CDFS), which has numerous ground- and space-based ancillary observations. Lastly, the EDF-S is a 23\,deg$^2$ field with an extended shape that encompasses two adjacent \gls{LSST} deep-drilling fields. 

\subsubsection{Euclid Auxiliary Fields}
\label{sec:deepfields_EAF}

The \gls{EDS} is complemented by the \glspl{EAF}. These fields are used for the calibration of photometric redshifts (\cref{sec:photoz}) and to quantify the impact of colour gradients within galaxies on shape measurement in the presence of a chromatic \gls{PSF} \citep[e.g.][]{Semboloni+13}. The \glspl{EAF} include the COSMOS \citep{Scoville07}, AEGIS \citep{Davis07}, SXDS \citep{SXDS}, VVDS \citep{VVDS}, CDFS \citep{CDFS}, and GOODS-North \cite{GOODS} fields (\cref{table:EAF}), which have been extensively observed by ground- and space-based telescopes. The CDFS is included in the EDF-F area, but observed independently for scheduling and technical reasons. The fields are observed up to 4--5 times the depth of the \gls{EWS}.

\subsubsection{Euclid self-calibration field and Ultra-Deep Field}
\label{sec:selfcal_EUDF}

The \Euclid self-calibration field also belongs to the \glspl{EAF}, but we discuss it separately here because of its technical significance. This field was chosen because of its location within \Euclid's northern continuous viewing zone that reaches up to \ang{2.5;;} from the NEP. This \gls{LOS} offers a good stellar density for calibration and system-monitoring purposes, and at the same time gives a view of the extragalactic sky with acceptable reddening (see also \cref{fig:EUDF_cutout}). The \gls{LOS} towards the \gls{SEP} is less favourable due to obstruction by the outskirts of the Large Magellanic Cloud. 

The self-calibration field fits in a radius of \ang{0.9;;} (\cref{fig:3fieldsEBV}), is located within the EDF-N area, and is observed on a monthly basis. We expect to exceed the typical wide-survey exposure time by a factor of 165 after 6 years, resulting in estimated $5\,\sigma$ point-source depths of 29.4\,AB\,mag in \IE, and 27.7\,AB\,mag in \YE, \JE, and \HE, that is about 3.2\,mag deeper than the \gls{EWS}. The central part (\ang{0.5;;} radius) serves as a backup pointing to avoid idling during non-standard operations and maintenance. Thus the self-calibration field will eventually become the \gls{EUDF}. The final depth of the central part is not yet known, given that it will be observed an unknown number of times as a backup field.

\begin{table}[t]
\caption{Basic information about the \Euclid Auxiliary Fields}
\smallskip
\label{table:EAF}
\smallskip
\begin{tabular}{lcccc}
\hline\hline
\noalign{\vskip 1pt}
Field & RA & Dec & Area & Depth\\
\hline
\noalign{\vskip 1pt}
Self-calibration & 268.813 & +65.29 & 2.5\,deg$^2$ & $8\times$\\
AEGIS & 214.827 & +52.82 & 1.0\,deg$^2$ & $4\times$\\
GOODS-North & 189.250 & +62.25 & 0.5\,deg$^2$ & $4\times$\\
COSMOS & 150.119 & +02.21 & 2.0\,deg$^2$ & $5\times$\\
VVDS-Deep & \phantom{0}36.500 & $-$04.50 & 0.5\,deg$^2$ & $5\times$\\
CDFS & \phantom{0}53.117 & $-$27.81 & 0.5\,deg$^2$ & $5\times$\\
SXDS & \phantom{0}34.500 & $-$05.00 & 2.0\,deg$^2$ & $5\times$\\
\hline
\end{tabular}
 \footnotesize
 \tablefoot{The \glspl{EAF} serve multiple calibration and target characterisation purposes. The coordinates are for the J2000.0 epoch. The depth specifies how many times the S/N is expected to improve over the average S/N of the \gls{EWS}.}
\end{table}

\subsection{Calibration observations}
\label{sec:calibration}

\Euclid has tight calibration requirements. The VIS, NISP-P, and NISP-S data must meet respective relative photometric accuracy levels of 1.0, 1.5, and 0.7\% over the full survey area and a 6-year mission duration. 
These requirements enable, respectively, a uniform photometric reference for the ground-based photometry (\cref{sec:groundbased}), accurate photo-$z$ measurements (\cref{sec:photoz}), and a stable selection function for galaxy clustering (\cref{sec:clustering_stats}).
To ensure a sufficiently unbiased weak lensing signal, the uncertainty in the
estimate of the ellipticity of the 
VIS \gls{PSF} model  must be less than $1.5\times10^{-4}$ per ellipticity component, while the relative uncertainty in the area of the model PSF, quantified by its quadrupole moments, must be better than $4.8\times10^{-4}$ \citep{Cropper+13}.

We designed a rigorous calibration programme to monitor the in-flight performance and to counter the effects of space weathering and molecular outgassing \citep{schirmer2023} at any time in the survey. In addition, the galaxy population that \Euclid observes must be characterised in specific fields. Hence our in-flight calibration activities fall into the following three groups.

\subsubsection{Performance verification phase \label{sec:cal_PVphase}}

Commissioning was followed by a 3-month long period for the \gls{PV} phase. During this time
most system calibrations and characterisations for zero-gravity conditions and in-flight temperatures were updated. Not everything, however, could be repeated in-flight, such as measurements of the absolute \gls{QE}. 
A total of 35 observing blocks were executed, the majority of which were based on specialised commanding sequences that were extensively tested on the ground. Whenever possible, on-sky calibrations were taken with the \gls{ROS}, our fundamental survey building block, because it has been well tested and means that the calibration data were taken in the same way as the survey data, sometimes at the expense of increased overheads. Some initial results are highlighted in \cref{sec:preliminaryCommissioning}.

Each calibration block provides data for several calibration products that inform our error budgets. For NISP we have 12 common calibration products that serve NISP-P and NISP-S, such as: baseline map; brighter-fatter effect \citep{plazas2018,hirata2019}; electronic crosstalk; dark current; interpixel capacitance \citep{legraet2022}; lamp flats; nonlinearity; reciprocity failure \citep[count-rate nonlinearity;][]{biesiadzinski2011}; and charge persistence. 

Eight calibration products are for NISP-P, including absolute flux calibration, illumination correction, detector and optical distortions, ghost images, and the \gls{PSF} model. NISP-S comprises eleven calibration products, including a complex calibration chain from astrometric positions on sky to individual wavelengths in the dispersed images, based among other things on observations of the compact planetary nebula SMC-SMP-20 \citep{Paterson-EP32} in the Small Magellanic Cloud. For VIS we have a total of 24 calibration products, the majority of which cover electronic and detector properties such as bias, dark, brighter-fatter effect \citep{antilogus2014}, nonlinearity, crosstalk, \acrlong{CTI}\glsunset{CTI} \citep[CTI;][]{israel2015}, extended pixel-edge response \citep[CTI-EPER;][]{robberto}, and more. The remainder comprises optical aspects such as the shutter map, illumination correction, stray light levels, lamp flats, and absolute flux calibration. Other \gls{PV} activities focused on the telescope's thermal response to solar attitude changes and instrument activities, molecular outgassing, and survey characterisations such as depth, sensitivity, scattered light, and zodiacal background.

\subsubsection{Phase diversity calibration campaign}
\label{sec:cal_PDC}

\Euclid's optics, telescope structure, and baseplate are constructed of SiC that is known for its low thermal expansion coefficient, high thermal conductivity, strength, and stiffness (\cref{sc_sec:PLM}). \Euclid orbits L2 in a thermally stable environment (\cref{sec:orbitmaintenance}). Yet, at the level of our requirements, \Euclid's \gls{PSF} is sensitive to spacecraft attitude changes well below $1^\circ$, causing an initial thermal imbalance and a subsequent heat flow that affects different telescope parts at different times. The survey is therefore designed to minimise attitude changes while stepping from one survey tile to the next \citep{Scaramella-EP1}.

To ensure that an accurate \gls{PSF} model (\cref{sec:psfmodel}) can be derived for all observations in the presence of thermal variations, we must know the wavefront errors that may occur during the survey. Owing in part to the broad VIS passband, the wavefront errors cannot be retrieved from in-focus observations alone. We therefore employ a combination of in- and out-of-focus observations of stellar fields \citep[see e.g.][]{wong2021}. At the centre of this phase diversity calibration are four fields that are observed intra-focal, extra-focal, and in-focus; one of these fields has a high \gls{LOS} polarisation. Six additional fields are observed in-focus only. All fields have known \glspl{SED} from \Gaia and dedicated ground-based observations, and are also observed with NISP to improve the star-galaxy separation for compact sources. 

For each field a different stable thermal state of the \gls{PLM} is prepared prior to the observations, by maintaining the telescope for about seven days at a given solar attitude. Calibration and scientific filler programmes are run during the thermal stabilisation periods. About 60 days are required after the \gls{PV} phase to retrieve the data for the \gls{PSF} model. Routine survey operations began afterwards, in February 2024.

\subsubsection{Routine phase}
\label{sec:cal_routine}

The three deep fields (\cref{sec:deepfields}) will yield at least 200\,000 galaxies with a S/N 6 times that of the \gls{EWS}. This provides improved morphological information that will aid the calibration of the weak lensing signal \citep{Hoekstra+17, Hoekstra+2021}. Blue-grism exposures are needed to provide a spectroscopic subsample of at least 120\,000 galaxies with 99\% redshift purity and 99\% completeness for galaxy-clustering purposes, and to characterise the typical \gls{EWS} galaxy population.

\Euclid also observes the \glspl{EAF} (\cref{sec:deepfields_EAF} and \cref{table:EAF}), for which multiwavelength data by \gls{HST} are available. The spatially resolved colour information from \gls{HST} is needed to quantify shape-measurement biases introduced by colour gradients within a galaxy; these biases are inevitable for a diffraction-limited \gls{PSF} in \Euclid's wide \IE-band \citep{Voigt+12,Semboloni+13,Er+18}. 
Extensive spectroscopic redshift surveys of the \glspl{EAF} make them very suitable for the calibration of photometric redshifts. The \glspl{EAF} are observed with the \gls{ROS} (\cref{sec:wide}), providing full VIS and NISP data sets. In total, 17\% of the time during routine operations is used on the \glspl{EDF}, \glspl{EAF}, and for self-calibration (\cref{fig:calib_pie_chart_global}). 
These data form a scientific cornerstone of the \Euclid mission with substantial legacy value 
(\cref{sec:legacy}) owing to their great depth and large number of revisits.

A central pillar of \Euclid's calibration scheme is the set of monthly visits of the self-calibration field (\cref{sec:selfcal_EUDF}). These last about 23 hours and obtain a large number of images and spectra to recalibrate \Euclid's spectrophotometric response with high accuracy, S/N, and spatial resolution. In this way we counter any adverse effects from space weathering and outgassing \citep{schirmer2023}, and from imperfections in the preceding calibration chain, thus enabling a consistent flux calibration over the full 6-year mission duration. The self-calibration observations also provide data for regular updates of 20 further calibration products. 

The NISP hardware calibration plan foresees monthly lamp flats and nonlinearity calibrations, and a lower cadence of reciprocity failure and wavelength-calibration checks. The VIS calibration plan includes biases, flats, darks, trap-pumping, and charge-injection lines on a daily basis, as these are part of the \gls{ROS}. On a monthly scale we will calibrate the VIS nonlinearity chain, check for radiation damage using trap-pumping, and obtain numerous flat-fields for a high-S/N characterisation of the brighter-fatter effect, and the conversion gain or \gls{PTC}. The VIS \gls{PSF} model is recalibrated every 2--6 weeks, depending on when the survey moves to a new large survey patch \citep{Scaramella-EP1}.  In total, 7\% of the time is used for dedicated hardware calibrations that are not part of the \gls{ROS} (\cref{fig:calib_pie_chart_global}).

\subsection{Complementary ground-based photometry}
\label{sec:groundbased}

\Euclid relies on optical ground-based imaging that complements the VIS and NISP imaging for photometric redshift estimation \citep{Abdalla08} and to assign the correct \gls{SED}-weighted \gls{PSF} to each galaxy in the lensing analysis \citep{Eriksen18}. To this end, 
a large coordinated campaign of ground-based observations with different observatories will provide the necessary multi-band photometry to matching depths across the \gls{EWS} and \gls{EDS} areas. Here, we summarise the characteristics of these data.

The \gls{DES} provides a good starting point for the southern sky. This survey, completed in 2019, covers about 3750\,deg$^2$ of the \gls{EWS}, with achieved depths of $g=24.5$, $r=24.1$, $i=23.6$, and $z=23.4$ ($10\sigma$ for a point source in a $2''$ diameter aperture).
These depths meet the requirements in the $gri$ bands, while it comes close in $z$.
These data are sufficient for the cosmological parameter estimates based on the first data release (see \cref{sec:DR}). To enable more precise 
measurements for the final analyses,
the \gls{DES} data will be superseded by deeper derived data products from the \gls{LSST} \citep{guy2022}, which will overlap with 7534 deg$^2$ of the \gls{EWS}.

The northern sky, however, lacked an equivalent data set: \gls{KiDS} \citep{deJong2015} and the \gls{HSC} survey \citep{Aihara2018, Aihara2022} do not cover sufficient area, and largely target regions of the northern sky that are closer to the ecliptic, whilst the \gls{DESI} Legacy Imaging Survey \citep{Dey2019} is one magnitude too shallow. To address the need for additional complementary imaging data a new collaboration was set up in 2017: the 
\gls{UNIONS}, a wide field $ugriz$ survey of the northern extragalactic sky that is a `collaboration of collaborations'. The Canada-France Imaging Survey collaboration provides $u$- and $r$-band imaging using the \gls{CFHT}, $i$-band and part of the $z$-band data are obtained using the \acrlong{Pan-STARRS}\glsunset{Pan-STARRS} \citep[\gls{Pan-STARRS};][]{Chambers2016}, 
while the \gls{WISHES} team, a collaboration of Japanese scientists, acquires $z$-band imaging with the \gls{HSC} \citep{Miyazaki2018}.
In addition, \gls{HSC} $g$-band data are collected through PI time, via a Canadian Gemini-Subaru exchange and time from the Institute for Astronomy, University of Hawaii: the
\gls{WHIGS}. \gls{UNIONS} is becoming the definitive broadband optical survey of the northern sky and once fully acquired and combined, these will provide the required $ugriz$ coverage of the northern part of the \gls{EWS} over 5711 deg$^2$, joining with the \gls{LSST} at a declination of $+15$ degrees over the north Galactic Cap. For reference, the achieved depths by \gls{UNIONS} are: $u=23.6$, $g=24.5$, $r=24.1$, $i=23.7$, $z=23.4$ ($10\sigma$ for a point source in a $2''$ diameter aperture).
As it remains a challenge to ensure that at every \Euclid data release the footprints covered from space and from the ground overlap as much as possible, the first year of the survey will be prioritised towards the southern sky, while the data collection in the north by \gls{UNIONS} continues and balances out the north and the south for the second \Euclid data release.

\begin{figure*}
    \centering
    \includegraphics[width=0.7\linewidth]{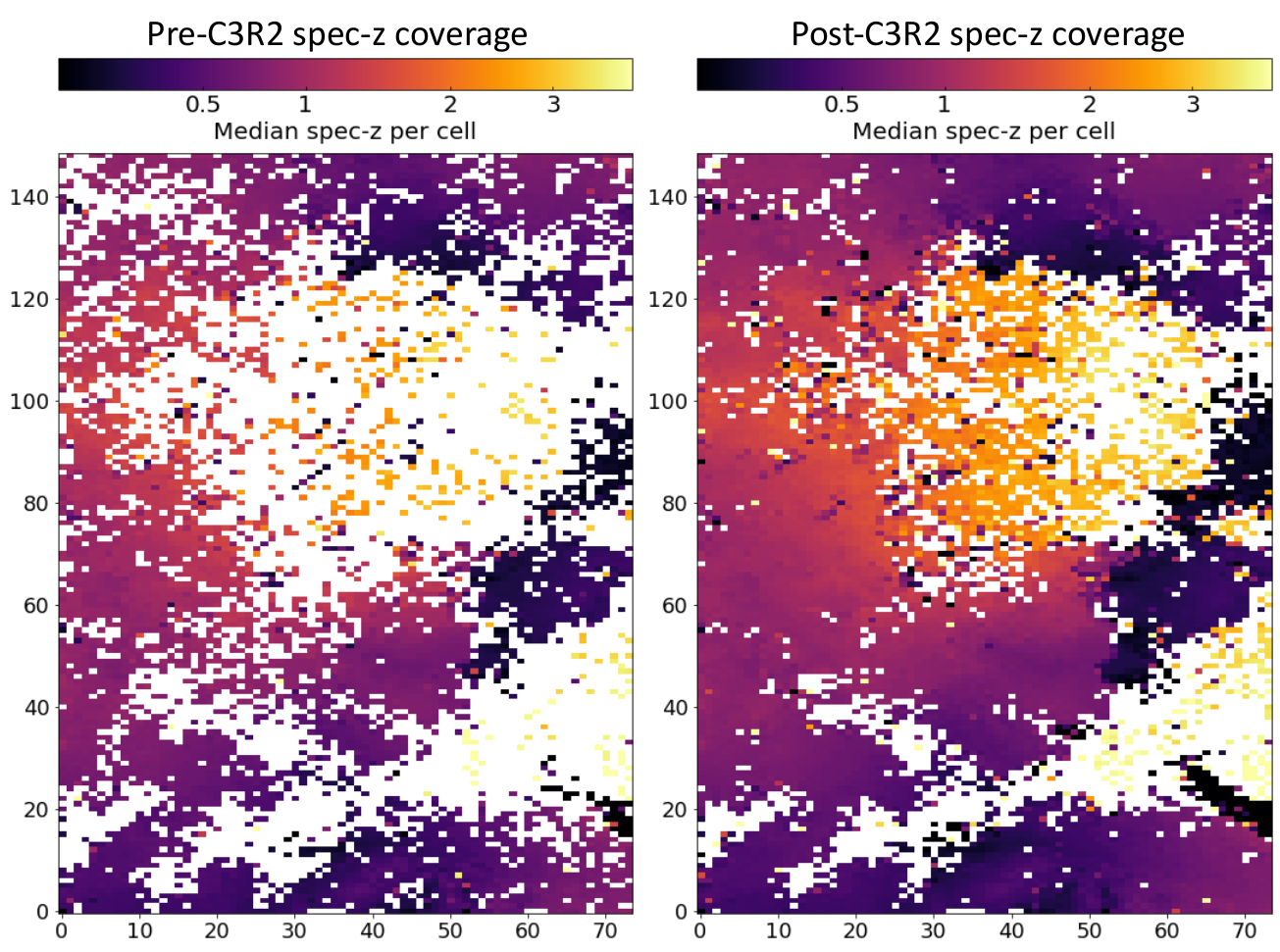}
    \caption{The galaxy multicolour-space to $i=25$\,AB\,mag, encoded in a 2D map with a 150$\times$75 binning using the self-organising map algorithm \citep{Masters15}. On the left is the distribution of spectroscopic coverage of the map prior to the C3R2 effort. The white regions are those parts of galaxy-colour space lacking high-confidence spectroscopic redshifts for calibration. On the right is the current map, after incorporating the $>$5800 C3R2 faint galaxy spectra. The map coverage has increased from about 51\% to $>$90\%, with many colour cells calibrated with multiple galaxies. Spectra to calibrate the remaining empty cells may be obtained as next-generation spectroscopic facilities come online, or they can be addressed with clustering redshift approaches \citep[e.g.][]{Newman08}. We note that the remaining empty regions correspond to lower-density (less occupied) parts of the galaxy-colour space.}
    \label{fig:specfig}
\end{figure*}

The \glspl{EDF} (\cref{sec:deepfields}) and \glspl{EAF} have their own challenges, with limiting-magnitude requirements of around 26\,AB\,mag over 56 square degrees.
The Cosmic Dawn Survey (Euclid Collaboration: McPartland et al., in prep.) is a complementary, UV--IR multiwavelength survey for the EDFs and EAFs that aims to optimise the legacy science returns from these fields, with a primary focus on the 
high-redshift Universe. The DAWN survey combines dedicated and archival observations from CFHT MegaCam, Subaru HSC, \gls{LSST}, \textit{Spitzer} IRAC  and other ancillary data available in the EDFs and EAFs, to depths matching the \Euclid observations. Photometric catalogues of the imaging data were consistently produced using \texttt{The Farmer} \citep{COSMOS2020}. The \textit{Spitzer} observations and data reduction are described in \cite{Moneti-EP17}. A description of the ground-based observations from the Hawaii 20\,deg$^2$ (H20) Survey and the first catalogue data release can be found in Euclid Collaboration: Zalesky et al. (in prep.).

\subsection{Complementary ground-based spectroscopy}
\label{sec:cog_spec}

As discussed in \cref{sec:weaklensing}, the correct interpretation of the photometric clustering and lensing signals depends critically on accurate estimates of their redshift distributions. This, in turn, relies on large samples of robust spectroscopic redshifts.
To calibrate the redshift distributions of the lenses and sources for the weak lensing and photometric clustering measurements, we have collected a substantial amount of complementary deep spectroscopy.

Our baseline approach is outlined by \citet{Masters15} and aims to directly calibrate the relation between galaxy colours and the redshift. Specifically, \cite{Masters15} quantified the expected distribution of galaxy colours using the \gls{SOM} algorithm \citep{Kohonen07}, showing that sizeable regions of the colour space were lacking high-confidence spectroscopic redshifts (see the left panel in \cref{fig:specfig}). They also demonstrated that a targeted campaign of spectroscopic follow-up could obtain the redshifts needed for calibration.

This analysis motivated the Complete Calibration of the Colour-Redshift Relation (C3R2) programme, a coordinated effort between Keck \citep{Masters17, Masters19, Stanford2021}, the Very Large Telescope \citep{Guglielmo-EP8}, and the Large Binocular Telescope \citep{Saglia22}, to measure redshifts for faint galaxies down to $i=25$\,AB\,mag spanning the full galaxy colour space. The Keck programme, using DEIMOS, MOSFIRE, and LRIS, resulted in $>$5100 deep spectra with secure redshifts, while the other observations 
added over 600 more. The improvement in colour-space coverage as a result of the C3R2 efforts is shown in Fig.~\ref{fig:specfig}.

We emphasise that the C3R2 surveys built upon extensive existing spectroscopy to calibrate the colour-redshift relation. Other large-scale deep spectroscopic surveys that contributed substantially to the external redshift calibration sample for \Euclid include DEIMOS 10k \citep{Hasinger18}, DEEP2\&3 \citep{Cooper12}, MOSDEF \citep{Kriek15}, UDSz \citep{Bradshaw13}, VANDELS \citep{McClure18}, VIPERS \citep{Garilli14, Scodeggio18}, VUDS \citep{Tasca17}, VVDS \citep{LeFevre13}, and zCOSMOS \citep{Lilly07}.

All deep spectra were collected in a database for careful source-by-source redshift validation, given the need for high purity in the spectroscopic calibration sample \citep{Ilbert-EP11}. We are currently testing the performance for the planned tomographic redshift binning for 
\Euclid, using the spectroscopic calibration sample together with the Flagship mock galaxy simulation (\cref{sec:flagship}).


\section{Early results from commissioning and PV}
\label{sec:preliminaryCommissioning}

Comprehensive results from the commissioning and \gls{PV} phases will be published once the ongoing data analyses have concluded. The estimated amount of raw data produced by \Euclid during the \gls{PV} phase between August and November 2023 is 21\,TB, compared to 12.1\,TB of raw data created by \gls{HST} up until November 2023.\footnote{File names in the \gls{HST} archive that end in {\tt \_raw.fits} (Matthew Burger, Space Telescope Science Institute, priv. comm.)}
Therefore, in this paper we  present some preliminary results, showing that \Euclid's performance is sufficient for its core scientific goals. More details can be found in the instrument-specific papers for VIS \citep {EuclidSkyVIS} and \gls{NISP} \citep{EuclidSkyNISP}.

\subsection{Photometric throughput and molecular contamination\label{sec:photthroughput}}
For initial estimates of the total system throughput we used the \Gaia \citep{Gaia-DR3}, 2MASS \citep{cutri2003}, and VHS \citep[VISTA Hemispheric Survey;][]{mcmahon2021} magnitudes of field sources. Considerable colour terms exist in the transformations from the four wide \Euclid bands to these external bands (\cref{fig:euclid_passband_comparison}). Our transformations are based on the known passbands and stellar SEDs from our simulations (\cref{sec:sims}); in the case of NISP they are given in \citet{Schirmer-EP18}. 

We find the measured NISP and VIS \glspl{ZP} for sources with a frequency-flat \gls{SED} to be considerably better, by about 0.4\,mag, than required. Accurate \glspl{ZP} and updated \Euclid photometric systems will be based on observations of stable white-dwarf spectrophotometric standards that we already established with \gls{HST}, jointly for \Euclid and \textit{Roman} 
(\href{https://archive.stsci.edu/proposal_search.php?mission=hst&id=16702}{proposal ID 16702}).

Thin layers of water ice formed on optical surfaces due to outgassing, resulting in throughput modulations from interference and scattering. Contamination is expected and typical for spacecraft, and can be countered with thermal decontamination \cite[for details see][]{schirmer2023}.

Throughput monitoring has shown characteristic variations that have been linked to several nanometres of ice on the optics; selective heating of one of the folding mirrors in March 2024 completely restored the transmission to immediate post-launch levels. Because the instrument cavity has little venting area to the outer space, and because outgassing is a continuous process, further decontamination activities are expected over the operation period.

\begin{figure}[t]
\centering
\includegraphics[angle=0,width=1.0\hsize]{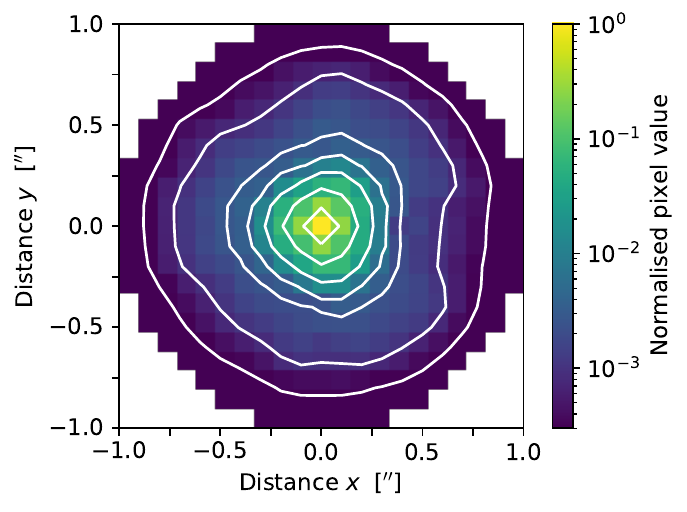}
\caption{VIS image quality. The figure shows a stacked data \gls{PSF} near the centre of the VIS FPA, from an observation of the self-calibration field, averaging over source SEDs. The \gls{FWHM} is approximately \ang{;;0.13} in this data set. 
The effect of trefoil (\cref{sec:pv_psf}) is evident in this log-scale representation.}
\label{fig:VIS_PSF}
\end{figure}

\begin{figure}[t]
\centering
\includegraphics[angle=0,width=1.0\hsize]{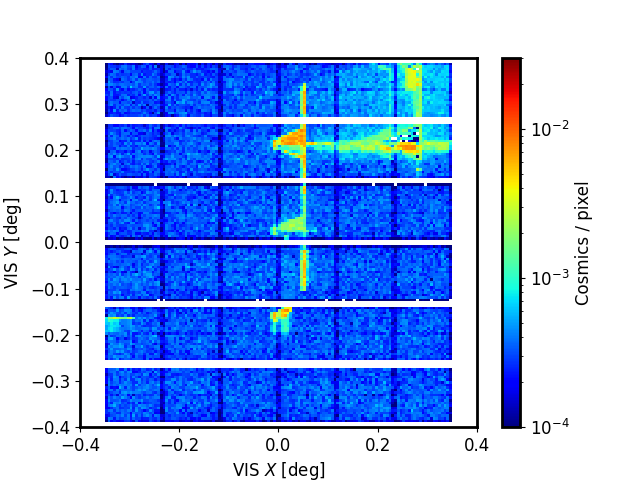}
\caption{Point-like cosmic-ray density in VIS during a low M-class solar flare. The cosmic rays are caused by X-rays impinging onto the detectors after penetrating \Euclid's sunshield in some gaps between the solar cells, causing characteristic geometric patterns. During rare but bright X-class flares, up to 25\% of the VIS detector area must be masked. The location of the pattern and its shape depends strongly on the \gls{LOS} of VIS towards the Sun through the sunshield, and thus on the spacecraft's attitude.}
\label{fig:xrayfig_overview}
\end{figure}

\subsection{VIS point-spread function}
\label{sec:pv_psf}

For the weak lensing measurements we need a detailed and chromatic \gls{PSF} model that we introduce in \cref{sec:psfmodel}. Here we present an initial and coarse evaluation of the PSF. 
With a plate scale of \ang{;;0.1}\,pixel$^{-1}$ the VIS optical PSF is considerably undersampled (\cref{fig:VIS_PSF}). To avoid aliasing in resampled images, the sampling frequency would need to be at least $2.0$--$2.5$\,pixels per \gls{FWHM} \citep{bernstein2002}. 
The requirements on the VIS \gls{PSF} are an ellipticity less than 0.13, and an \gls{FWHM} smaller than \ang{;;0.18} at 800\,nm. The typical ellipticity and \gls{FWHM} we measure in VIS science exposures are 0.04 and \ang{;;0.13}, respectively. This includes jittering effects from the \gls{FGS} for representative guide-star densities and a background of cosmic-ray hits. The jittering is accounted for in the \gls{PSF} model using the time-series of the guiding corrections during every exposure. From an opto-mechanical perspective alone, \Euclid's \gls{PSF} has great stability owing to its SiC components and a thermally stable environment at L2. 

In pre-launch testing, the polished M1 was found to have a small residual amount of astigmatism, which would result in a strong dependence of the \gls{PSF} ellipticity on the telescope's focus. To reduce this astigmatism, a mechanical correction was applied to M1 to compensate the astigmatism prior to launch. It is thought that this resulted in a small, but noticeable, amount of trefoil in the \gls{PSF}, visible as an approximately triangular shape in the \gls{PSF} contours at \ang{;;0.3}--\ang{;;0.5} from the \gls{PSF} core (\cref{fig:VIS_PSF}). Like \gls{PSF} ellipticity, the trefoil is an inherent part of the \gls{PSF} model (\cref{sec:psfmodel}) and thus accounted for in the shape measurements. The trefoil is described --to leading order-- as a spin-3 contribution to the \gls{PSF}. We therefore do not expect that it has a significant impact on the multiplicative and additive shear biases in Eq.~(\ref{eq:mcbias}), as those are described by spin-0 and spin-4 components.

Thanks to the excellent optical design (see \cref{sc_sec:PLM}), most of the light is concentrated in the core of the \gls{PSF} over the full \gls{FOV}, while stray light is suppressed. Thanks to further efforts to minimise scattered light (but see \cref{sec:straylight}), the extended \gls{PSF} 
should ideally decline with distance as $r^{-3}$, the limit posed by diffraction. 
An initial study of the extended \gls{PSF} by \cite{EROData} took advantage of the observation of the bright star HD\,1973. Their results demonstrated \Euclid's exceptional ability for the study diffuse emission around galaxies (see \cref{sec:NearbyGal}), because the extended \gls{PSF} indeed nearly matches a pure diffraction halo.

\begin{figure*}[t]
\centering
\includegraphics[angle=0,width=1.0\hsize]{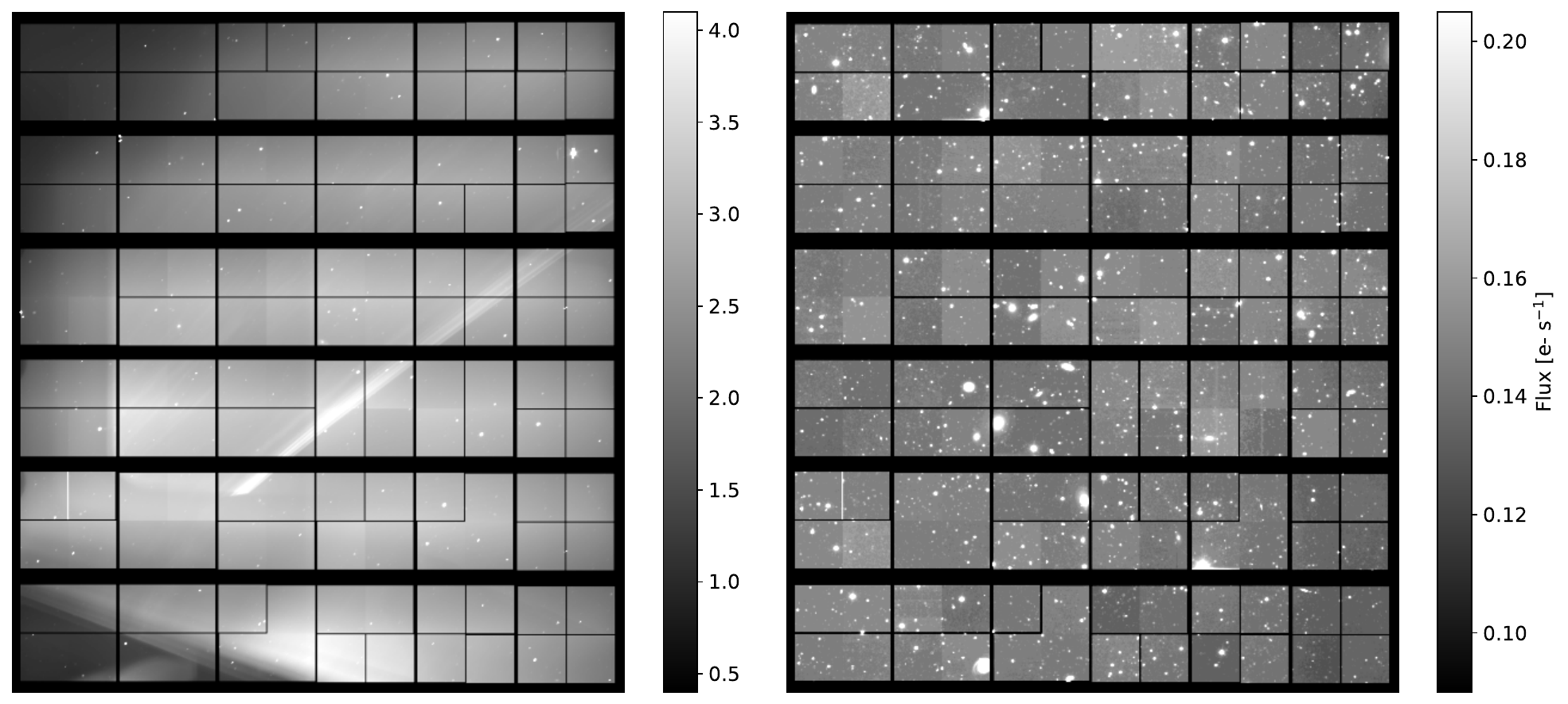}
\caption{Impact of the spacecraft orientation on the VIS background. \textit{Left}: At ${\rm AA}=0$ considerable stray light levels are present that exceed the zodiacal background by more than one order of magnitude. \textit{Right}: For ${\rm AA}<-\ang{2.9;;}$ the stray light is reduced to a few percent of the zodiacal background. It still needs to be modelled for some calibrations and low-surface-brightness science.}
\label{fig:straylight}
\end{figure*}

\subsection{X-ray contamination from solar flares}
\label{sec:xrays}

\Euclid's instruments have radiation shields, and are additionally protected by the spacecraft itself. This also includes protection from X-rays produced during  solar flares, which are absorbed by the silicon in \Euclid's sunshield. However, gaps in the sunshield's solar cells do let X-rays pass, some of which intersect with the VIS \gls{FPA} at an angle-of-incidence of about $60^\circ$. These are then detected as excess cosmic rays in the VIS images (\cref{fig:xrayfig_overview}). NISP is not affected by this.

Contrary to protons that cause displacement damage in the detector's atomic lattice, X-rays harmlessly create electron-hole pairs in the CCD's depletion region, like optical photons. The increased density of cosmic rays renders a fraction of the image unusable for scientific analysis. Using the X-ray sensors \citep[XRS;][]{hanser1996} on board the Geostationary Operational Environmental Satellites (GOES-16 and GOES-18), we find that considerable data loss occurs once the solar X-ray flux approaches about $1\times10^{-5}$\,W\,m$^{-2}$ in the GOES long band (1.5--12.4\,keV).

Typical flares last about 10--60 minutes, so that one or several subsequent VIS images can be affected. As long as particularly active sunspot regions are visible on the Sun, up to 10\% of the VIS images can be substantially affected by flares, judging from the GOES solar activity recorded between March and November 2023. During these periods, we expect to lose data from up to 4\% of all VIS pixels while passing through the solar maximum in 2024--2025. Any area lost could possibly be recovered during periods of currently unallocated time (\cref{sec:unallocatedtime}).

\begin{figure}[t]
\centering
\includegraphics[width=1.0\hsize]
{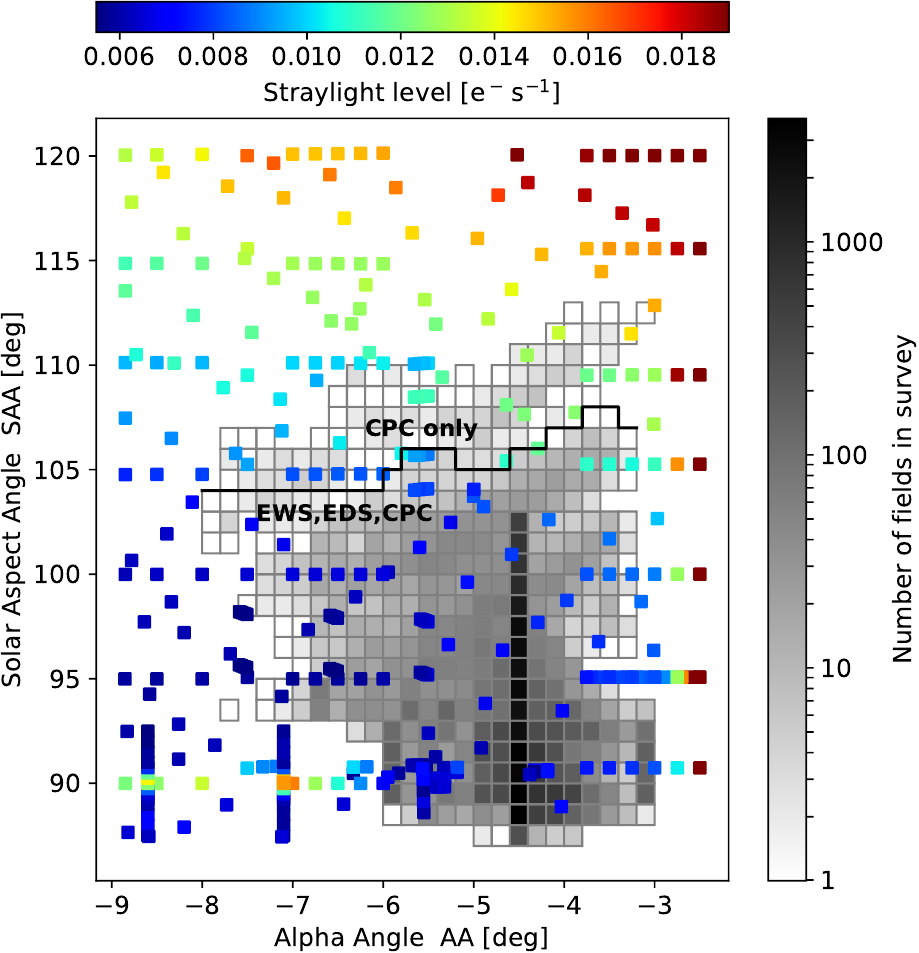}
\caption{Stray light map and survey fields. The coloured squares show the stray light level in VIS dark exposures as a function of spacecraft orientation angles. The log-scaled greyscale map shows the density of fields in the latest survey configuration including calibrations. The survey minimises stray light over the \gls{EWS} and \gls{EDS}, with the majority of the observations to be taken at ${\rm \gls{AA}} = \ang{-4.5;;}$. The \gls{CPC} fields are NISP-specific and include higher \gls{SAA} positions (above the jagged black line); while NISP is not affected by stray light, parallel VIS observations must still be taken.}
\label{fig:straylightmap}
\end{figure}

\subsection{Optical stray light in VIS\label{sec:straylight}}

During commissioning it was found that VIS can be affected by considerable amounts of stray light that exceed the zodiacal background by a factor of 10 or more (\cref{fig:straylight}).
The root cause has not been unambiguously identified, but is thought to be a thruster nozzle that is illuminated by the Sun. The light enters the instrument cavity following a triple scattering process, from the nozzle to the backside of the Sun shield, through a hypothesised opening in the thermal multi-layer insulation, to a mounting leg of the VIS shutter, and from there to the VIS focal plane. Even closed-shutter VIS observations are affected, and several parasitic light paths exist. NISP is not affected, as it is enclosed in black \gls{MLI}, and  parasitic light entering through the dichroic is blocked by baffles.  

Rotating the spacecraft around its $Z$-axis (\cref{fig:SC}) imposing $\rm AA < \ang{-2.9;;}$ moves the nozzle into the shadow, effectively reducing the stray light to levels of a few percent of the zodiacal background (\cref{fig:straylight}). Rotating up to $\rm AA = \ang{-8.5;;}$ is safe following a post-launch evaluation. The reduced \gls{AA} range now available for the survey is  $[\ang{-8.4;;},\ang{-3.0;;}]$ (previously $[\ang{-5;;},\ang{+5;;}]$ ) with a margin of \ang{0.1;;} for orbit uncertainty. The stray light is then negligible for \Euclid's core science. Figure~\ref{fig:straylightmap} shows how the latest survey configuration adapts the spacecraft orientation to the stray light constraint. Low-surface brightness science and some calibrations still require the construction of stray light models from the large number of survey fields and calibration data.

\begin{figure}[t]
\centering
\includegraphics[width=1.0\hsize]
{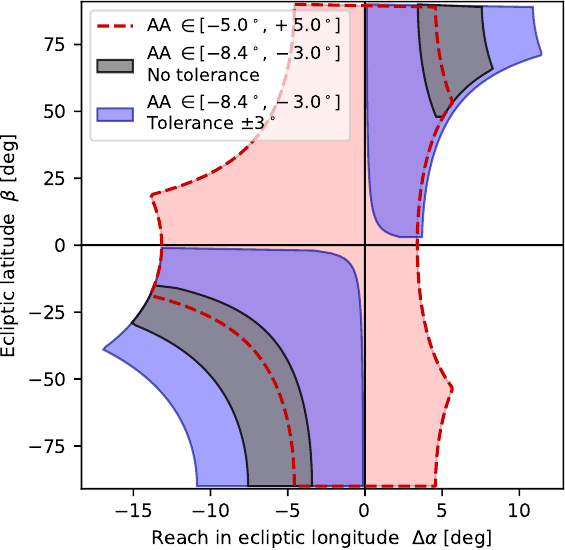}
\caption{Reach in ecliptic longitude around transit for the leading side of the survey, for ${\rm \gls{SAA}}\in[\ang{87;;},\ang{104;;}]$. Shown in red is the reach for the originally planned symmetric \gls{AA} range. To minimise the stray light in VIS, the range was shifted to $\rm{\gls{AA}}\in[\ang{-8.4;;},\ang{-3.0;;}]$ with much reduced visibility (grey) that would not permit the completion of the survey. By allowing the fields to rotate by up to \ang{3;;} with respect to the tessellation, a much larger area of the sky becomes accessible (blue). For observations in the trailing side, the areas must be rotated by \ang{180;;} around the origin.}
\label{fig:windowvisibility2}
\end{figure}

\begin{figure*}[t]
\centering
\includegraphics[width=1.0\textwidth]
{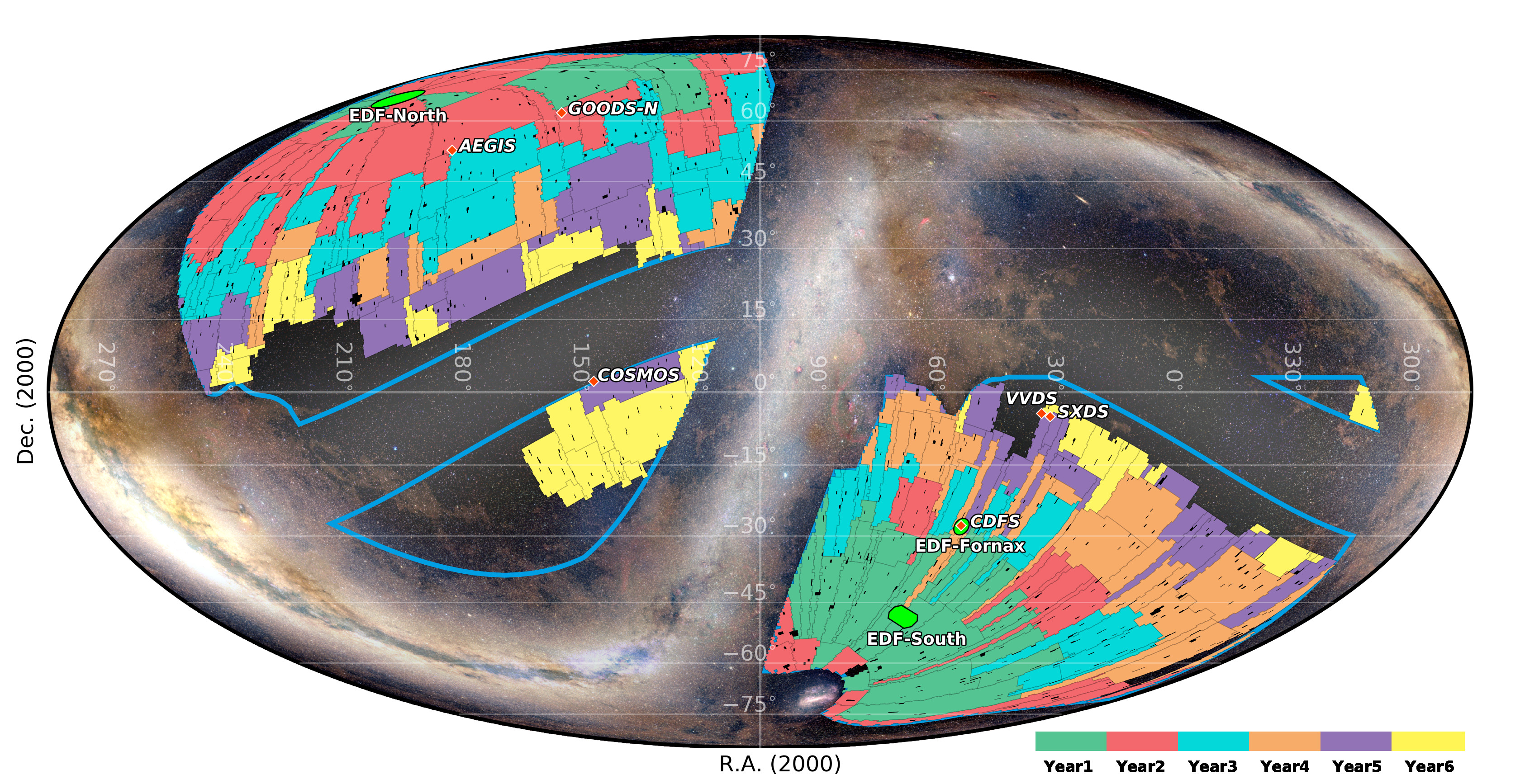}
\caption{EWS coverage and colour-coded yearly progress in an all-sky Mollweide projection. The blue borders enclose the $16\,000\,\deg^2$ \gls{ROI} that contains the $13\,416\,\deg^2$ observed sky of the \gls{EWS}. Small dark regions within the \gls{EWS} are masks for stars brighter than $4$\,AB\,mag.}
\label{fig:skysurveynewdesign}
\end{figure*}

\subsection{Latest survey strategy\label{sec:newsurveydesign}}

The reduction of the \gls{AA} range has the double effect of thinning and skewing the window of visibility, which no longer contains the transit meridian and has a zero reach at some latitudes
(\cref{fig:windowvisibility2}).
This inhibits the scheduling of Wide Survey patches of any substantial size. Furthermore, half the time it would be impossible to schedule fields with $|\beta|\leq\ang{50}$.

Hence, we allowed the FOV to rotate by up to ${\pm}\,$\ang{3;;} with respect to the tessellation tiles, recovering a large fraction of the tiles' original visibility, albeit with a different shape. Visibility was regained at all latitudes. A post-launch check showed that we could also safely increase the maximum \gls{SAA} from \ang{110} to \ang{120}. The new enlarged window of visibility, shown for the leading side in \cref{fig:windowvisibility2}, has two distinct parts in the northern and southern ecliptic hemispheres due to the now asymmetric \gls{AA} range. The visibilities for the leading and trailing sides are east-west and north-south mirrored. In the leading side, the larger reach in longitude favours observations in the southern hemisphere, while the trailing side favours the northern hemisphere.

This relaxed tessellation constraint implies that neighbouring \glspl{FOV} are no longer necessarily aligned. To avoid gaps between \glspl{FOV} we tessellate the sky with smaller tiles, increasing the mean overlap between fields, which reduces the covered sky area. In the latest survey computation, \cref{fig:skysurveynewdesign}, the \gls{EWS} covers 13\,416\,\,deg$^2$ of which 171\,deg$^2$ are lost due to bright stars. The effective sky area is 13\,245\,\,deg$^2$, representing a decrease of 5\% as compared to the target of 14\,000\,deg$^2$ in 6~years (\cref{sec:wide}). The implications of this reduction in survey area for the core science objectives, as well as possible mitigation strategies are being explored.


\section{Simulated data}
\label{sec:sims}

The calibration observations discussed in \cref{sec:calibration} provide important information about the performance of the telescope and the data that are collected. To prepare the pipeline and to interpret the results, simulated data are needed as well. This includes large realistic input universes that can capture survey characteristics, such as the \acrlong{EFS}\glsunset{EFS} (EFS; \cref{sec:flagship}) as well as sophisticated pixel-level instrument simulators (\cref{sec:simdata}). The latter are essential for exploring the sensitivity of the measurements to instrumental effects and test our ability to correct these. 

\subsection{The Euclid Flagship Simulation}
\label{sec:flagship}

The optimal exploitation of the \Euclid data demands the development of large-volume and high-mass resolution numerical simulations that reproduce the large-scale galaxy distribution that the mission will observe with high fidelity. Not only do these help to assess the performance with a realism that cannot be achieved otherwise, but such simulations are also an essential tool for the development of the data processing and science analysis pipelines. The major advance that the \Euclid data will bring implies the need for a dedicated effort. To this end, we developed the \gls{EFS}, which is described in detail in \cite{EuclidSkyFlagship}. Here, we summarise its main characteristics.

The \gls{EFS} features a simulation box of 3600\,$h^{-1}$Mpc on a side with $16\,000^3$ particles, leading to a mass resolution of $m_{\rm p} = 10^9\,h^{-1}M_{\odot}$. This 4 trillion particle simulation is the largest $N$-body simulation performed to date and matches the basic science requirements of the mission, because it allows us to include the faintest galaxies that \Euclid will observe, while sampling a cosmological volume comparable to what the satellite will survey. The simulation was performed using \texttt{PKDGRAV3} \citep{Potter:16} on the Piz Daint supercomputer at the Swiss National Supercomputer Center (CSCS).
The input cosmology differs slightly\footnote{The \gls{EFS} uses the following values for the density parameters: $\Omega_{\rm m} = 0.319$; $\Omega_{\rm b} = 0.049$; and $\Omega_{\Lambda} = 0.681 - \Omega_{\rm rad} - \Omega_{\nu}$, with a radiation density $\Omega_{\rm rad} = 0.00005509$, and a contribution from massive neutrinos $\Omega_{\nu} = 0.00140343$. Additional parameters are: the equation of state of dark-energy $w = -1.0$; the reduced Hubble constant $h = 0.67$; the scalar spectral index of the initial fluctuations $n_{\rm s} = 0.96$; and the scalar power spectrum amplitude $A_{\rm s} = 2.1 \times 10^{-9}$ (corresponding to $\sigma_8 = 0.813$) at $k = 0.05\,{\rm Mpc}^{-1}$.} from the one listed in \cref{tab:fiducials}, but this has no material impact on the applications.

The initial conditions were realised at $z = 99$ with first-order Lagrangian perturbation theory (1LPT) displacements from a uniform particle grid. The transfer functions for the density field and the velocity field were generated at this initial redshift by \texttt{CLASS} \citep{CLASS} and \texttt{CONCEPT} \citep{Dakin22}. As the usual scaling of the linear power spectrum at $z=0$ to the initial redshift of the simulation (known as back-scaling), to generate the initial conditions, was not used, all linear contributions from radiation, massive neutrinos, and metric perturbations \citep[in the $N$-body gauge, see][]{Fidler2015} were included via a lookup table and applied as a small corrective PM (particle-mesh) force at each timestep. This ensures a match to the linear evolution of the matter density field at all redshifts when including these additional linear terms. The main data product was produced on the fly during the simulation and is a continuous full-sky particle light cone out to $z=3$, where each particle was output exactly when the shrinking light surface sweeps by it. This resulting ball of particles contains 31 trillion particle positions and peculiar velocities (700\,TB of data). The 3D particle lightcone data were used to identify roughly 150 billion dark-matter halos using \texttt{Rockstar} \citep{Behroozi:13}, and to create all-sky dark-matter 2D maps in 200 tomographic redshift shells between $z=0$ and $z=99$, with a \texttt{HEALPix} \citep{Gorski2005} tessellation resolution $N_{\rm side} = 8192$, corresponding to \ang{;0.43;} per pixel.

The halo catalogue and the set of 2D dark-matter maps are the main inputs for the Flagship mock galaxy catalogue. A detailed description of the catalogue production is given in \cite{EuclidSkyFlagship}, which we summarise here. Galaxies were generated following a combination of \gls{HOD} and \gls{AM} techniques. Following the \gls{HOD} prescription, halos were populated with central and satellite galaxies. Each halo contains a central galaxy and a number of satellites that depends on the halo mass. 
The halo occupation was chosen to reproduce observational constraints of galaxy clustering in the local Universe \citep{Zehavi2011}. 

\begin{figure*}
	\includegraphics[width=\textwidth]{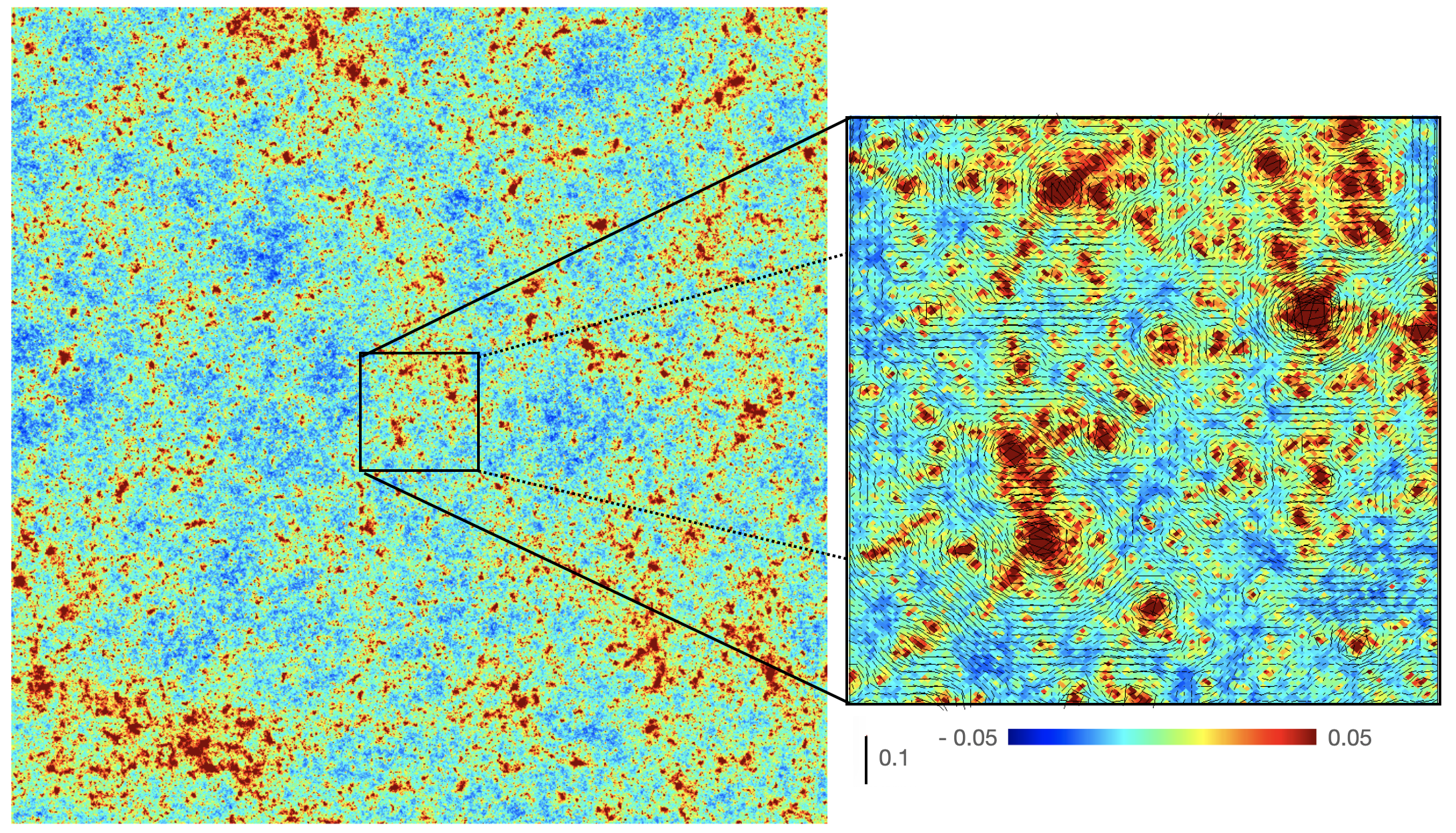}
    \caption{The image on the left shows the lensing convergence for sources with $z_{\rm s}=1$ for a simulated patch of sky covering 50 deg$^2$. A zoom-in of the central square degree is shown on the right, with the sticks indicating the direction and amplitude of the corresponding shear.
    The colour bar of the convergence field displays values within the range $\pm 3\sigma$, where $\sigma$ is the rms value of the full-sky map. The stick at the bottom of the zoom-in image shows a reference amplitude for the shear sticks overlaid on that area of the mass map.}
    \label{fig:lensingmap_z1p0}
\end{figure*}

The luminosities of the central galaxies were assigned by performing abundance matching between the halo mass function of the simulation halo catalogue and the galaxy \gls{LF}. We took as the reference luminosity function a parameterised prescription that tries to fit the observed \gls{LF} throughout the redshift range of the simulation. We then applied a 15\% scatter to the resulting luminosities. The application of scatter is necessary to reproduce the galaxy clustering dependence on luminosity. 
The satellite luminosities were assigned assuming a universal Schechter \gls{LF} for satellites in which the characteristic luminosity depends on the central luminosity in a way that ensures that the global luminosity function agrees with observations. Galaxies were split into three colour types, namely red, green and blue, and the central and satellite galaxies in each group were distributed to match the observed clustering as a function of colour by  \citet{Zehavi2011}. 
The radial positions of the satellites within their halos follow \gls{NFW} profiles \citep{NFW1997} with each colour type having its own concentration: green and blue galaxies are distributed using a concentration that is respectively one-half and one-quarter times that of the red galaxies, consistent what \citet{Collister05} found for blue versus red galaxies in the nearby Universe. This colour segregation is assumed to hold at all redshifts (see \cref{sec:env} for a discussion of this assumption).

To determine the simulated redshifts, we assumed that the central galaxy is at rest in its halo. The satellite velocities were drawn using formula derived by solving the Jeans equation of local dynamical equilibrium for each type, assuming velocity anisotropy profiles consistent with what was measured from the kinematics of low-redshift regular clusters \citep{Mamon19}, with green and especially blue galaxies having more radial orbits around their host haloes. Redshifts of the galaxies were then obtained by projection of these 3D velocities along the \gls{LOS}.

\glspl{SED} were assigned to each galaxy with a procedure that aims to mimic the observed colour distributions as a function of redshift. The resulting SEDs are a linear combination of the ones presented in \citet{Ilbert2009}. The stellar masses were computed from the galaxy luminosities and the mass-to-light ratios of the \glspl{SED}. The star-formation rates were computed from the ultraviolet luminosity of the \gls{SED}. The luminosity of the H$\alpha$ line was computed from the star-formation rate following the Kennicutt recipe \citep{Kennicutt1998}. The galaxy clustering measurements of \Euclid  rely on the detection of this line in the galaxy spectra (see \cref{sec:galaxyclustering}). We therefore want to simulate their distribution to our best current knowledge. Consequently, we then refined the H$\alpha$ luminosities to match the models of \citet{Pozzetti2016}, using abundance-matching techniques.
The luminosities of the other main emission lines were assigned using observed relations, taking the H$\alpha$ line as reference. The shapes and sizes of the galaxies were assigned following relations based on HST observations \citep{Miller+13,Dimauro2018}. The observed fluxes of each galaxy were computed by integrating the \gls{SED} with the filter transmission for the surveys that are expected to be used to obtain photometric redshifts. 

As for galaxy lensing properties, we followed the `onion universe' approach presented in \citet{Fosalba:08} and \citet{Fosalba:15b}, to compute all-sky weak lensing observables (convergence, shear, and deflection) within the Born approximation. The latter agrees within 0.1\% out to a multipole $\ell = 10^4$ with the much more complex and \gls{CPU} time-consuming ray-tracing technique \citep{Hilbert:20}. \Cref{fig:lensingmap_z1p0} shows the convergence (colour-coded) and the overlaid shear field (sticks indicating the amplitude and direction of the shear) for a source redshift $z=1$ in a patch of approximately $50\,\mathrm{deg}^2$ of the \gls{EFS}.

\subsection{Simulated \Euclid data}
\label{sec:simdata}

Pixel image simulations serve as test data for the development of the data-processing pipeline and computing infrastructure, and   to validate the stringent requirements regarding performance and data quality \citep[e.g.][]{Gabarra-EP31}. In this respect, several end-to-end science performance verification (SPV) tests have been performed throughout the mission preparation, which allowed for the reproduction of certain instrumental issues to assess their impact and guide decision making (see \cref{sec:endtoend} for more details). Image simulations have also been essential to develop alternative solutions to critical problems that were discovered during on-ground tests, such as
the non-conformity of one of the three red grisms for the NISP instrument \citep{Scaramella-EP1, EuclidSkyNISP}; or to evaluate the impact of unexpected features discovered in-flight, such as the contamination of VIS images by X-ray photons during solar flares (see \cref{sec:xrays} and \cref{fig:xrayfig_overview}). Finally, image simulations are necessary to calibrate intrinsic biases related to the methods used to measure the shapes of the galaxies in the cosmic shear analysis. A large volume of very accurate and representative simulations is required to determine these biases to the required level of precision \citep[e.g.][]{Hoekstra+17}. 

To ensure a common input for the instruments, the starting point is a `true universe' catalogue that contains all input sources and their corresponding parameters, spectra, and shapes. These parameters are based on the output from the \gls{EFS} (\cref{sec:flagship}),
which provides both spectra (continuum and emission lines) and morphological parameters (bulge and disc models).  The stars are simulated using a hybrid catalogue, using actual stars from 
\Gaia DR3 \citep{Gaia-DR3} at the bright end ($G<18.5$), merged with the fainter end of deep stellar population simulations using the Besan\c{c}on model \citep{Robin03,Czekaj14,Lagarde2021}, including binaries, down to $\IE=26$. All objects are simulated using common coherent libraries to ensure consistent results between the simulated VIS imaging, NISP photometry and spectroscopy, and ground-based imaging. In addition to individual sources, the simulations include zodiacal light, diffuse stray light from stars beyond the simulated pointing, and thermal irradiance caused by the heat of the various elements of the telescope and instruments. The instrument models and the reference survey characteristics are drawn from the central mission database. This database (versioned and controlled by dedicated change control boards) provides the instrument simulators with all the parameters necessary to simulate the numerous instrumental features of each simulation channel.

To create simulated VIS observations and the associated calibration frames, we have developed the  \Euclid VIS simulator (\texttt{ELViS}).  It includes an accurate emulation of the optical response, which is based on the complex \gls{PSF} modelling tools that are described in \cref{sec:psfmodel}. Although challenging, \texttt{ELViS} can capture the complexity of the chromatic \gls{PSF} and its spatial variations.
The sources are projected on the simulated mosaic of 36 \glspl{CCD}, and an extensive list of instrumental signatures can be included, such as bias, pixel response non-uniformity, cosmetic defects, saturation, bleeding, shutter movement, and ghosts from the dichroic plate. Particularly important for \Euclid are simulations of the 
imperfect charge transfer during the readout. The simulated readout electronics include a nonlinear response, saturation of the analogue-to-digital converter, and electric cross-talk, with parameters determined during 
the on-ground test campaigns. 

To capture the dual use of the NISP instrument, two separate codes are used, \texttt{Imagem} and \texttt{TIPS}, to simulate the photometric and spectroscopic channels, respectively. As is done for VIS, both science and calibration images are produced. 
As the photometric and spectroscopic channel share the same optical path and detector units, the NISP-P simulator,
\texttt{Imagen} and NISP-S simulator, \texttt{TIPS} employ common background, \gls{PSF}, optics, detector, and electronics models. Consistency across the two channels is particularly important because the spectroscopic analysis relies on measurements from the imaging data. The simulations start from the same list of astronomical sources that is used by \texttt{ELViS}. The simulated NISP \glspl{PSF} contain the same types of effects that have been included for VIS instrument, but with a simplified module to capture the variation of the \gls{PSF} (given the less strict requirements, we opted to use a tabulated \gls{PSF} as opposed to recomputing it for each object). Each different multiple accumulated sampling readout (\gls{MACC} mode) is simulated for photometric and spectrometric images. At the detector level, variable \gls{QE} is simulated using \gls{QE} estimates obtained from on-ground tests, while the \gls{PRNU} is taken from the on-ground flatfield data. To capture biases introduced during readout, nonlinearity and gain are simulated together with the \gls{MACC} readout modes. On the spectroscopic side (\texttt{TIPS} simulator), the slitless light dispersion is handled by \texttt{aXeSIM}\footnote{\url{http://axe-info.stsci.edu/axesim/}} \citep{Kuemmel2009}, developed by the Space Telescope Science Institute. The simulation of the slitless spectra requires a complete characterisation of the trace dispersion, with sensitivity, diffraction coefficients, grism tilts, vignetting, and \glspl{PSF} at each dispersion order (expressed as a Taylor series expansion). The dispersion order of interest is the first order, where the main spectrum can be recovered. However, the zeroth and second orders of dispersion need to be simulated as well. The readout electronics are shared with the photometric channel and the simulations are therefore also similar (except for the \gls{MACC} modes, which are photometry and spectroscopy specific).

We also produce pixel simulations of the external ground-based surveys that are used to characterise the \glspl{SED}
of the stars and galaxies and to determine photometric redshifts. Here, it is important to capture the key characteristics of the ground-based data, summarised in 
\cref{sec:groundbased}, while providing realistic images that can be used to examine the performance of the various processing steps. With the exception of \gls{LSST}, the surveys have already collected data, so that realistic prescriptions for their main features can be readily implemented. For instance, the simulations include realistic values for the background levels, zero points, filter transmissions, bias levels, flat-field characteristics, detector defects, observed cosmic rays, and \glspl{PSF} for each of the simulated surveys. 
Importantly, the simulated ground-based data use the same input catalogues as the simulations of the \Euclid instruments, so that all the detection and cataloguing steps can be tested, as well as the determination of photometric redshifts. For a more detailed description of the pixel simulations, the reader is referred to \cite{Serrano2024}.


\section{Survey data products}
\label{sec:data}

\Euclid provides high-quality optical and \gls{NIR} imaging, as well as slitless \gls{NIR} spectroscopy over a large fraction of the extragalactic sky.  The processing of these data and supporting ground-based observations is performed by the \gls{SGS}. 
In this section, we present the most salient parts of the pipeline. The key steps in the processing of the \Euclid optical and \gls{NIR} imaging data are highlighted in \cref{sec:ouvis,sec:ounir}, respectively, while the processing of the supporting ground-based imaging is summarised in \cref{sec:ouext}. These data form the basis for the creation of the object catalogues, described in \cref{sec:catalogues}. The processing of the slitless spectroscopy and the subsequent redshift determination are discussed in \cref{sec:gcmeasurement}. 

In \cref{sec:wlmeasurement} we describe how these results are used to derive the key ingredients for the 3\texttimes2pt analysis.
The science-ready data products, summarised in \cref{sec:dataproducts}, include the various two-point statistics, as well as weak lensing convergence maps and catalogues of clusters of galaxies. Although many of these data products are excellent starting points for myriad scientific investigations, several high-level data products will be released as well, with a particular focus on the interpretation of the primary probes. These data, as well as a range of calibrated data products, will be released to the scientific community in a number of data releases (\cref{sec:DR}).  

\subsection{Processing of VIS imaging data} 
\label{sec:ouvis}

\begin{figure*}[t]
\includegraphics[width=0.5\hsize]{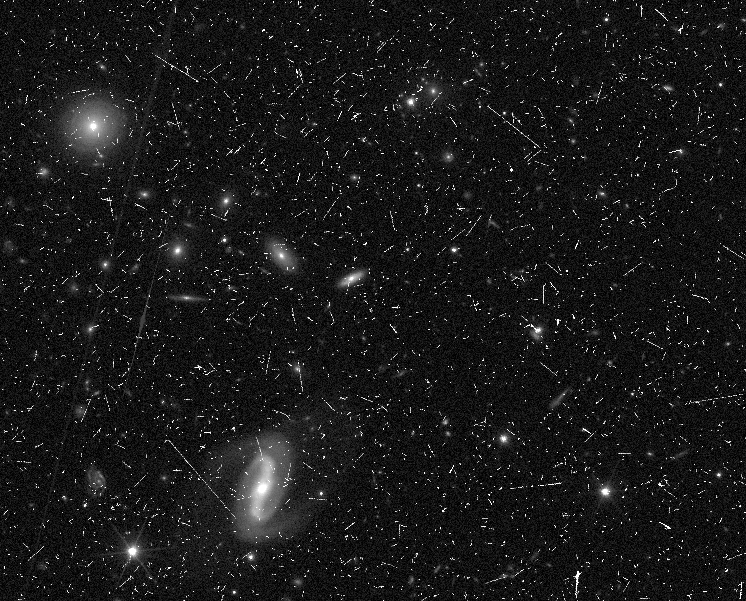}
 \includegraphics[width=0.5\hsize]{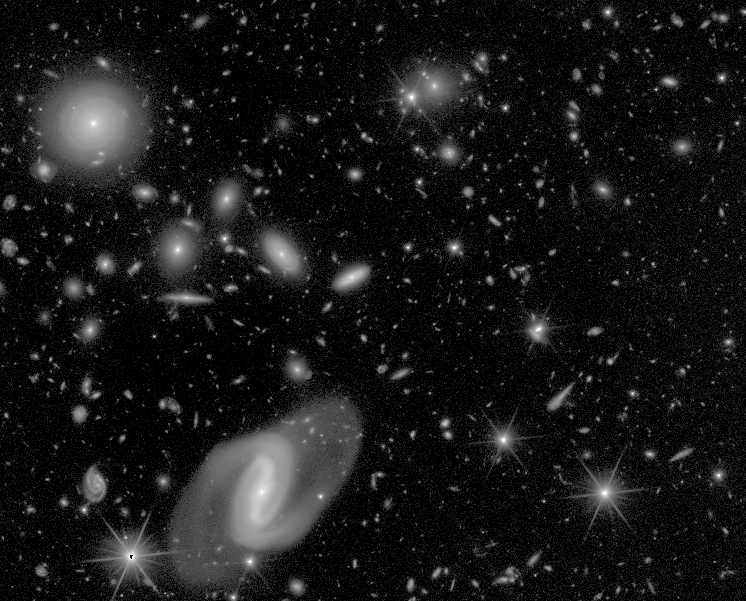}
    \caption{VIS view of a $\ang{;2.5;}\times\ang{;2.0;}$ wide area of \Euclid's self-calibration field (\cref{sec:selfcal_EUDF}) taken during the \gls{PV} phase. \textit{Left}: An unprocessed single exposure, where cosmic rays are clearly visible. \textit{Right}: A VIS-processed stack using 42 exposures, or about 10 times the exposure time of the EWS. The ability of \Euclid to reveal low-surface-brightness features is evident.}
    \label{fig:pv01_vis_cutout}
\end{figure*}

The high resolution imaging data provided by the VIS instrument is the starting point for the weak lensing measurements. To reach the main objectives of \Euclid, the shapes of about 1.5~billion galaxies need to be measured with unprecedented accuracy.
To this end, a range of instrumental effects need to be carefully accounted for \citep{Massey+13, Cropper+13, Paykari-EP6}. This involves fully characterising the performance of the instrument through the processing and analysis of an extensive amount of calibration data. The subsequent processing of the raw VIS data relies on these calibration products to correct all relevant instrumental effects to enable robust shape measurements. 

Before launch, each pipeline processing element was tested and validated using the simulations described in \cref{sec:simdata}. These simulations enabled an assessment of how well instrument models (based on pre-launch observations) could be derived using realistic observing sequences. They also allowed an evaluation of how well measurements (photometry, astrometry, and object shape) could be made after the application of these models. These findings are being updated using the \gls{PV} observations and in-flight calibrations.

At its most basic level, the processing of the VIS data can be divided into three categories: the calibration pipeline that generates or updates the VIS calibration models, which by definition do not alter the input pixels; the science pipeline that alters the input pixels by applying the calibration models; and the validation pipeline designed to assess the performance of the data processing. Because the requirements for weak lensing shape measurement are so strict, the adopted approach is to apply the minimal number of pixel-level corrections, and, if feasible, to provide the information to reverse the correction if necessary (or to provide images for which a given correction has not been applied). 

In  detail, the pipeline that processes the VIS data calibrates and corrects a comprehensive set of pixel-level effects, which would otherwise bias the shear estimation \citep{Cropper+13}.
These include electronic offset (bias), dark current, \gls{PRNU}, detector-chain nonlinearity, brighter-fatter effect \citep{antilogus2014}, charge-transfer inefficiency \citep{israel2015}, illumination correction, and the flagging of cold, hot, and saturated pixels, as well as optical and electronic ghosts and cosmic rays. Moreover, the astrometric solution, required to be better than 30\,mas, and photometric solution are computed, using \Gaia DR3 \citep{Gaia-DR3} as the reference catalogue. Apart from robust shape measurements, these processing steps enable relative photometry measurement with an accuracy better than 1\%. This implies that the collective contribution of residuals from the detection chain, small- and large-scale flat fielding, source extraction, and background and scattered light correction, is smaller than this target. 

The main outputs of this step in the processing are the calibrated individual exposures, as well as stacks and their associated source catalogues. However, important supporting data are also provided, such as the calibration models, the background and flag maps, the distortion model, and a first estimate of the \gls{PSF} model that is used for multi-band photometry during the generation of the main survey catalogue (\cref{sec:catalogues}).  For reference, \cref{fig:pv01_vis_cutout} displays a $\ang{;2.5;}\times\ang{;2.0;}$ view of a single raw frame (left) and the corresponding processed stack (right) of the \Euclid self-calibration field. The left panel demonstrates the need for the robust detection of cosmic rays, whereas the deep image on the right shows the potential of \Euclid to study low-surface-brightness features around galaxies.

\subsection{Processing of NIR imaging data} 
\label{sec:ounir}

The interpretation of the observed weak lensing signal requires accurate estimates of photometric redshifts (\cref{sec:primary}). As shown by \cite{Abdalla08}, the NIR images from the NISP instrument
in the \YE, \JE, and \HE\ bands provide key information to improve the photometric redshift precision (see also \cref{sec:photoz}). To reach this goal, NISP images require an exquisite calibration that takes into account all instrumental effects and a possible time variation of the telescope's throughput (\cref{sec:photthroughput}). Most notably, the relative photometric calibration needs to be better than $1.5\%$ for the entire magnitude range down to 24.0~AB. This requirement has been the main driver for the design of the processing pipeline. 

The generation of the calibration products needed for the reduction of NISP images is carried out through a number of dedicated pipelines running on well-defined blocks of calibration observations. This is the case, for instance, for bad pixel identification, dark current, detector nonlinearity, \gls{PRNU}, large-scale illumination correction, geometric distortions, persistence image creation, and relative and absolute photometric calibration factors. 

The reduction of the \gls{NIR} exposures can be divided into three main steps: the preprocessing, where all detector-related effects common to both NISP photometric and spectroscopic exposures are accounted for; the calibration part, which includes \gls{PSF} estimation, computation of the astrometric solution based on VIS catalogues, application of flat-fielding and photometric calibration based on the precomputed products, and source catalogue extraction from individual dithered observations; and lastly the stacking of frames for each observation sequence along with the generation of their \gls{PSF} and source catalogues. Similarly to the VIS processing, we keep the  amount of pixel-level operations to a minimum, while providing at the same time all relevant information about modelling and corrections, so that they may be reversed if needed or tailored for specific purposes.

For each of the three \gls{NIR} filters, the output of the \gls{NIR} processing consist of calibrated individual dithered observations and stacks with their associated source catalogues. These data products are provided to the subsequent processing step for the creation of the main survey catalogue (\cref{sec:catalogues}), along with all ancillary information, including \gls{RMS}, data quality flags, background, and \gls{PSF} images.

\begin{figure*}[t]
	\includegraphics[width=1.0\hsize]{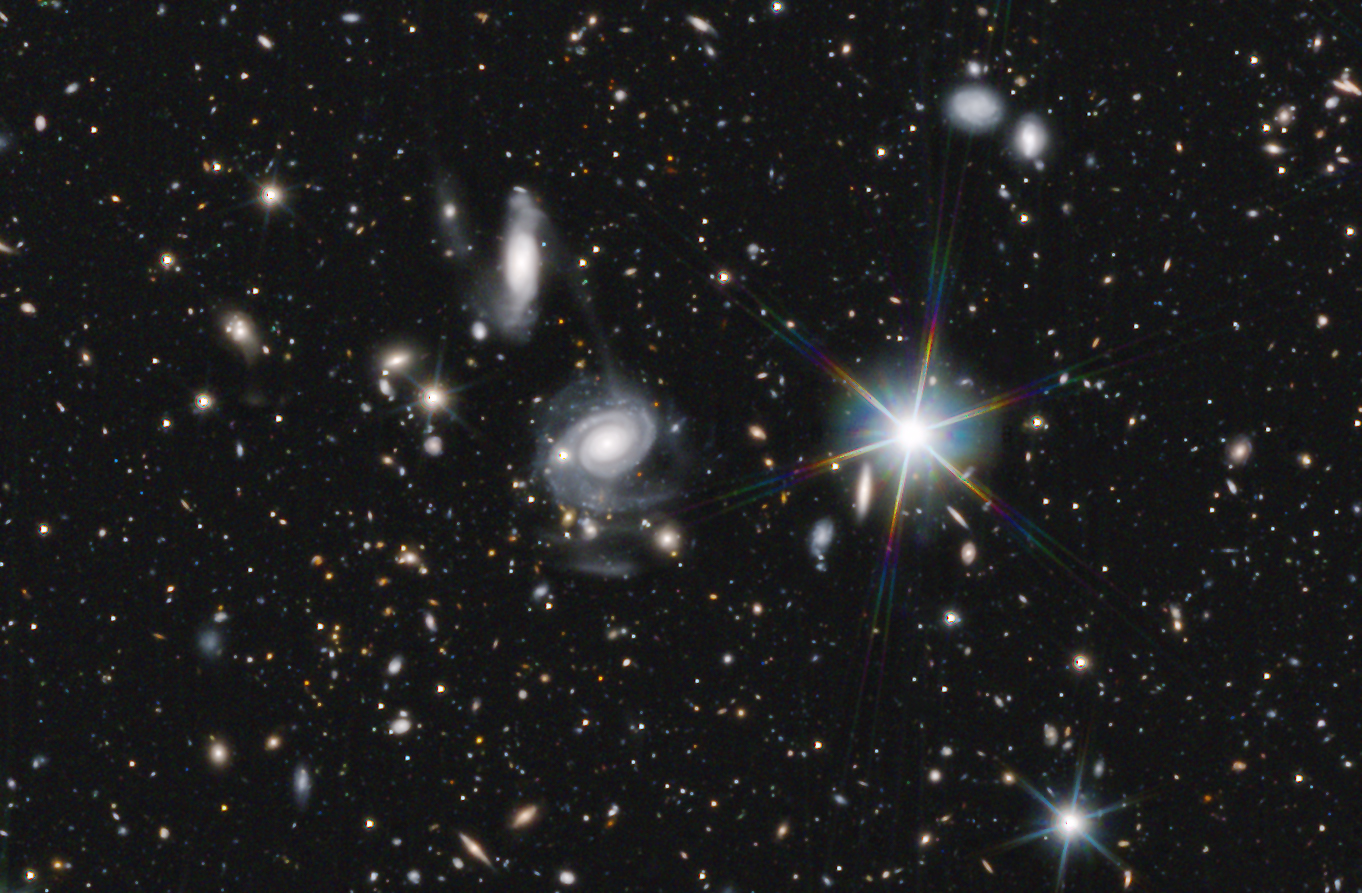}
    \caption{False-colour NISP image of a $\ang{;4.5;}\times\ang{;3.0;}$ area of \Euclid's self-calibration field (\cref{sec:selfcal_EUDF}). Filters \YE, \JE, and \HE~are shown in blue, green, and red, respectively. The depth is that of the \gls{EDS} (\cref{sec:deepsurvey}), about 26.4\,AB\,mag per band. The bright star has 11.5\,AB\,mag, showcasing \Euclid's excellent performance for in-field stray light suppression. Field rotation between observations is evident from the diffraction spikes.}
    \label{fig:EUDF_cutout}
\end{figure*}

Prior to launch, the design and implementation of the various processing steps have been extensively tested through a number of simulation campaigns with increasing realism and coverage of instrumental features, mostly based on modelling from ground calibration campaigns and in some cases by design. Further improvements will be made using the calibrations obtained during the \gls{PV} phase (\cref{sec:cal_PVphase}). \Cref{fig:EUDF_cutout} gives an impression of the quality of data. It shows a small area in the \Euclid self-calibration field (\cref{sec:selfcal_EUDF}).

\subsection{Processing of ground-based imaging data}
\label{sec:ouext}

The \Euclid imaging data are complemented by supporting ground-based observations (\cref{sec:groundbased}) to improve the precision of the photometric redshifts and to determine the \gls{SED}-weighted PSF model for the galaxies used in the weak lensing analysis (\cref{sec:galsed}). The challenge is to uniformly calibrate this vast and heterogeneous external data set with extraordinary accuracy: the \gls{EWS} is composed of three surveys (\gls{UNIONS}, \gls{DES} and the \gls{LSST}) employing five telescopes with differing $u$, $g$, $r$, $i$, and $z$ filters.  The \gls{EDS} and \glspl{EAF} have primarily been observed using \gls{HSC}, but there are additional datasets available featuring, for example intermediate band filters, that are being included to aid in the testing and validation of the photometric redshift estimation and the VIS \gls{PSF} modelling. Hence, the data are complex, with a wide range of single exposure depths and seeing conditions (from roughly \ang{;;0.5} to \ang{;;1.5}), sampled with over 450 different detectors. The resulting data set will be large, ultimately comprising roughly 1.5\,PB of raw (uncompressed) science observations. Crucially, to ensure robust photometric redshifts, the processing of these data should result in 1\% single-filter photometric homogeneity over the entire \gls{EWS}, and subpercent colour homogeneity over the VIS \gls{FOV} \citep{Eriksen18}.

For the cataloguing process described in \cref{sec:catalogues}, the required external data product inputs are coadded ground-based images.  These coadded images are supplemented with associated \gls{PSF} models for all objects that are identified in the VIS- and NIR-based detection process.  These coadded images are built using a single pipeline whose software components were initially developed as part of the \gls{DES} data management system \citep{Mohr2008SPIE.7016E..0LM} and have been tuned and validated to serve our needs and then integrated into the \gls{SGS} code base. The pixel coaddition is based on the widely used {\tt SWarp} code \citep{Bertin2002ASPC..281..228B,Bertin2010ascl.soft10068B}.  We developed a dedicated code to model
the \gls{PSF} to take into account the position-dependent \gls{PSF} models of all input images.

To facilitate the processing of the individual ground-based exposures, a common data model is enforced across all external ground-based data sets.  This common input data product-- termed a \gls{SEF}-- consists of a detrended and astrometrically and photometrically calibrated single CCD image, the associated 
position-dependent \gls{PSF} model and an associated catalogue that includes, at a minimum, the sky positions and PSF-fitted magnitudes of the brighter, unresolved sources.  Typically, these \glspl{SEF} are created using output data products from the external surveys.  For example, the ensemble of $i$-band \glspl{SEF} from \gls{Pan-STARRS} is prepared using the software within the \gls{Pan-STARRS} collaboration \citep{magnier2017.datasystem,waters2016}.  Similarly the $g$- and $z$-band \glspl{SEF} from \gls{WHIGS} and \gls{WISHES}  are produced using output data products from {\tt HSCpipe}  \citep{HSCpipe2018PASJ...70S...5B,LSST2019ASPC..523..521B}, which is software developed by the \gls{HSC} and \gls{LSST} teams.  The $u$- and $r$-band \glspl{SEF} from \gls{CFHT} are created using the {\tt MegaPipe} software developed by the \gls{CFHT} team \citep{Gwyn2008PASP..120..212G}. 

For the \gls{EDS} data, we receive the ground-based data from the Cosmic Dawn team (Euclid Collaboration: McPartland et al., in prep.; \cref{sec:groundbased}) in the case of \gls{EDF-N} and \gls{EDF-F}, while we directly downloaded public data from the \gls{HSC} archive for other \glspl{EAF} such as COSMOS and produce the data ourselves in the case of \gls{EDF-S}.  For the publicly available \gls{DES} data \citep{DESDR12018ApJS..239...18A} and other \gls{DECam} \citep{Flaugher2015AJ....150..150F} data, we produce and calibrate the \glspl{SEF} ourselves using extended versions of pipelines originally developed for \gls{DES} \citep{Mohr2012SPIE.8451E..0DM,Desai2012ApJ...757...83D} that include also image masking \citep{Desai2016A&C....16...67D} and have been reorganised and rewritten to simplify the large-scale processing required. Finally, thanks to a collaboration agreement, the data products from the \gls{LSST} are provided through a joint \Euclid-Rubin working group \citep[see][for details]{guy2022}.
 
To enable an accurate and uniform photometric calibration of the ground-based external data across the full extragalactic sky that will be observed by \Euclid, we leverage \Gaia \citep{Gaia_2016A&A...595A...1G} photometry and spectroscopy that are stable across the sky with a systematic uncertainty of $\sim$2~mmag \citep{Gaia-DR3}.  Initial testing of statistical transformations from \Gaia $G$, $BP$ and $RP$ to each of the external $griz$ bands in \gls{DES} demonstrated a high level of consistency between the \gls{DES} DR2 \citep{DESDR2} calibration and the \Gaia-based predictions \citep{George2020ASPC..527..701G}.  Further tests with the \gls{UNIONS} dataset provided indications that the external photometry predictions from statistical transformations based on \Gaia $G$, $BP$ and $RP$ photometry exhibited improved stability across the sky and improved internal self-consistency in comparison to the original \gls{UNIONS} calibration.  This demonstrated the promise of adopting \Gaia data as a basis for calibrating the heterogeneous ground-based data sets to ensure consistent photometric redshifts and stellar \gls{SED}s across the \Euclid sky.

A challenge in employing the statistical transformation function from \Gaia $G$, $BP$ and $RP$ \citep{George2020ASPC..527..701G} to the external data bands is that systematic changes in the stellar populations over the \Euclid sky could bias the \Gaia-based predictions, introducing errors larger than the advertised $\sim$2~mmag systematic uncertainties in \Gaia photometry.  Therefore, with the availability of the \Gaia DR3 \citep{Gaia-DR3} $BP$ and $RP$ calibrated spectra, we have transitioned to using \Gaia spectra and the associated synthetic magnitudes to calibrate the external data.   We calculated \Gaia synthetic magnitudes using the \Gaia calibrated spectra together with the appropriate bandpass for each external survey band and camera combination.  Under the assumption that the external data bandpasses are robust, these \Gaia synthetic magnitudes are highly accurate, enabling us to meet the \Euclid requirements.  In the case that there are significant residual errors in the external data bandpasses, these errors would enter both the ground-based photometric calibration and the \gls{SED} fitting being performed for photometric redshifts and stellar \gls{SED} constraints.  In addition, we expect to be able to use the \Gaia spectra to aid in characterising bandpass variations across the focal plane in the ensemble of cameras being used to obtain the ground-based data needed for \Euclid.

For the \gls{UNIONS} and \gls{DES} external data sets there are enough \Gaia synthetic magnitude constraints per \gls{SEF} to enable a photometric calibration of individual \glspl{SEF} that approaches the 2\,mmag systematic floor of the \Gaia mission. For the \gls{EDS} and \gls{EAF} data sets, the situation is more complicated, because the typical \gls{HSC} integrations are longer, pushing the saturation limits in these \glspl{SEF} to fainter magnitudes and reducing the overlap between the \Gaia-based synthetic magnitudes and the external data sets.  To address this we incorporate also relative photometric constraints between overlapping pairs of \glspl{SEF}.  This allows us to combine the direct photometric constraints from \Gaia synthetic magnitudes across an ensemble of co-located \glspl{SEF}, delivering a comparable level of photometric calibration accuracy in the \gls{EDS} and \gls{EAF} datasets as we achieve in the \gls{EWS}.

\subsection{Catalogue creation}
\label{sec:catalogues}

The imaging data form the basis of the object catalogue that is used for most of the subsequent analyses. To avoid multiple entries for the same object, the survey area is divided into predefined tiles. Each tile consists of a rectangular extended area and a core area that is defined by a set of \texttt{HEALPix} indices \citep{Gorski2005}.  All data needed for the detection and photometry are generated to cover the extended tile area, which overlaps with neighbouring tiles. Only the objects in the core area, which is unique for each tile, are actually selected for insertion into the object catalogue \citep{Kuemmel2022b}. This procedure avoids multiple detections of identical objects in adjacent tiles, while allowing the proper processing of large, extended objects in the overlap areas between two tiles. 

The pipeline retrieves all imaging data for the tile of interest and creates coadds of the calibrated \Euclid VIS and NISP exposures. As part of this step, the background is subtracted from each exposure and each image is sampled to the native VIS pixel scale of 
\ang{;;0.1}\,pixel$^{-1}$. Information on flagged pixels and areas around bright stars that are affected by diffraction spikes or blooming is propagated, as is the case for the model \gls{PSF} in each exposure.

In principle, the information in all bands could be combined to maximise the prospects for object detection, but this would result in complex selection biases for the primary probes. To avoid this, we opt for a staged process, where we detect and subsequently deblend objects in the VIS and NIR bands separately. This ensures a clean selection of the weak lensing source sample, while recording all objects that are visible in the \Euclid data. For the \gls{NIR} detections we employ a deep image generated from the combined \YE, \JE, and \HE\ data. Source detection is done with {\tt SourceXtractor++}\footnote{\url{https://github.com/astrorama/SourceXtractorPlusPlus}} \citep{Bertin2020}, with detection parameters that are optimised for completeness, while keeping the false detection rate below $1\%$. The {\tt denclue} algorithm \citep{Tramacere2016} is used for the deblending procedure. 
The resulting VIS and \gls{NIR} detections are then combined into a single joint catalogue that is used to perform photometry and to determine a number of other properties. To enable the selection of VIS-detected objects, the combined catalogue contains a flag that indicates whether a source was detected in VIS or in the deep \gls{NIR} image.

Accurate multi-band photometry is essential for the determination of robust photometric redshifts, but optimising the pipeline for this application might not be ideal for other science cases. Therefore, to maximise the usefulness of the catalogue, fluxes are measured in the following ways. 

\begin{itemize}
    \item Total object flux within a Kron aperture on the detection image with \texttt{T-PHOT}\footnote{The transformation of the \gls{PSF} from the VIS band with the highest resolution to the \gls{NIR} and ground-based images, which is necessary for \texttt{A-PHOT} and \texttt{T-PHOT}, is done using convolution kernels as described in \cite{Boucaud2016}.} \citep{Merlin2015}.
    \item Isophotal flux measured by summing the flux of the pixels above the detection threshold in the detection image.
    \item \gls{PSF} flux measured on the VIS image.
    \item Aperture flux measured by \texttt{A-PHOT} \citep{Merlin2019} on images that are \gls{PSF}-matched to the one with the worst resolution (typically a ground-based image). The circular aperture is set to twice the worst FWHM for each object.
    \item Template-fitting photometry computed by \gls{PSF}-convolving the VIS object shape to the different bands and fitting the surface-brightness profiles.
    \item Single-S\'{e}rsic fitting photometry using {\tt SourceXtractor++} \citep{Kuemmel2022b} in all available bands.
\end{itemize}

To increase the scientific value of the catalogues, morphological properties are also determined. A robust separation of point-like and extended objects is achieved by computing the difference between the central surface brightness $\mu_{\rm max}$ and the total brightness. This quantity is translated to a probability of the object being point-like using calibrations based on simulated data (\cref{sec:simdata}). We include 
the non-parametric estimates for  \acrlong{CAS} 
\glsunset{CAS}
 \citep[\gls{CAS};][]{Tohill2021}, as well as the 
Gini index \citep{Lotz2004} for all objects. Moreover, the catalogue includes the best-fit S\'{e}rsic models in all bands obtained using {\tt SourceXtractor++} \citep{Kuemmel2022b}.

\newcommand{\yc}[1]{{\color{orange} #1}}

\subsection{Processing of NIR spectroscopy data}
\label{sec:gcmeasurement}

The first steps in the processing of the dispersed \gls{NIR} images are similar to what is done with the imaging data used for photometry (\cref{sec:ounir}): the same detector-level effects need to be accounted for (bad pixel flagging, nonlinearity correction, persistence masking, dark subtraction, and cosmic-ray rejection). Therefore, these steps are based on a common set of processing elements. The different nature of the images does lead to some modifications, for instance in the cosmic ray rejection. The main difference, however, pertains to the instrument models that are adopted for the two observing modes. Specific calibration products are used for each case. The resulting images are then used to extract the spectra.

\subsubsection{Extraction of spectra} 

To extract the spectra from the preprocessed NISP spectroscopic images, we need to precisely locate the dispersed image of each object (the so-called `spectrogram') in the spectroscopic frame. Specifically, we are interested in the first order of the grism dispersion, where more than 96\% of the object flux is concentrated. To this end, the object catalogues derived from the corresponding direct photometric images are critical. 

First, the two-dimensional first-order spectrograms for each object are located on the full NISP spectroscopic frame, based on its coordinates measured from the corresponding direct image. This involves applying the astrometric solution, which translates sky coordinates into detector reference positions of the different dispersion orders. The precise location of the spectrogram is then traced, accounting for any inclination and curvature. Along the dispersion direction, the spectrum is re-mapped into wavelength steps, using the wavelength solution computed from reference planetary nebulae emission-line spectra \citep[see \cref{sec:cal_PVphase}, and][]{Paterson-EP32}. 

Detector-level pixel-to-pixel variations are corrected from calibrated detector flats, and values for the background over the full focal plane are sampled on detector areas where no spectra are present. These are then averaged per detector, and the values subtracted over the whole frame.

In slitless observations, overlapping spectra from nearby objects represent an important contribution to the noise affecting a given spectrum.  
\Euclid adopts a specific observing sequence to mitigate this, collecting four exposures at varying dispersion directions in each \gls{ROS}, as described in detail in \cref{sec:nisp-s}.
A model of the spurious contribution of each neighbouring source contaminant is built using its spatial extent estimated from the $\JE$-band photometric image, and its intrinsic \gls{SED}. If possible, the latter is estimated from uncontaminated portions of the spectrograms extracted from one of the available exposures taken as part of the \gls{ROS}. If an object happens to be contaminated in all four exposures, 
a power-law \gls{SED} interpolation of the available NIR photometric measurements is used instead. These decontamination operations are performed on each identified 2D spectrogram, ultimately resampled on a rectilinear grid along the dispersion and cross-dispersion directions.

At this stage, 1D spectra can then be extracted from the individual exposures in a \gls{ROS} by properly integrating over the cross-dispersion dimension.  This is performed through optimal extraction \citep{robertson_optimal_1986, horne_optimal_1986}, using a weighting profile derived from the $\JE$-band photometric image of the source. This also includes appropriate rotation, matching of both NISP-P and NISP-S \gls{PSF}s, shear, and resampling, to account, in the case of galaxies, for the inclination of the source with respect to the dispersion direction, so as to minimise effective spectral line-spread function and maximise S/N.  

The spectrum extraction is followed by a chromatic relative flux calibration, using bright point sources in the self-calibration field repeatedly observed over the NISP field of view.  This process normalises to the same relative flux scale all 1D spectra from different observations, detectors, and location in the focal plane.  The absolute flux is then obtained using the overall chromatic sensitivity curves, derived from observations of spectrophotometric standards for each observing setup. 

Finally, the flux-calibrated 1D spectra corresponding to each of the four exposures in a \gls{ROS}\footnote{These may in fact be more or less than four, depending on the specific position of the object on the detector. In the \gls{EWS}, the few cases when more than four sub-exposures are available correspond to objects near the borders of the field, which benefit from the overlaps between pointings.} are averaged into a combined 1D spectrum using inverse-variance weighting. During this operation, statistically outlying pixel values (corresponding, e.g. to cosmic ray or contamination residuals) are identified and discarded. This is this 1D spectrum that is then passed to the next step of the spectral analysis, together with the corresponding statistical variance, bit mask -- that is per-pixel flagging -- and combined effective line-spread function estimate.

\begin{figure}
	\includegraphics[width=\hsize]{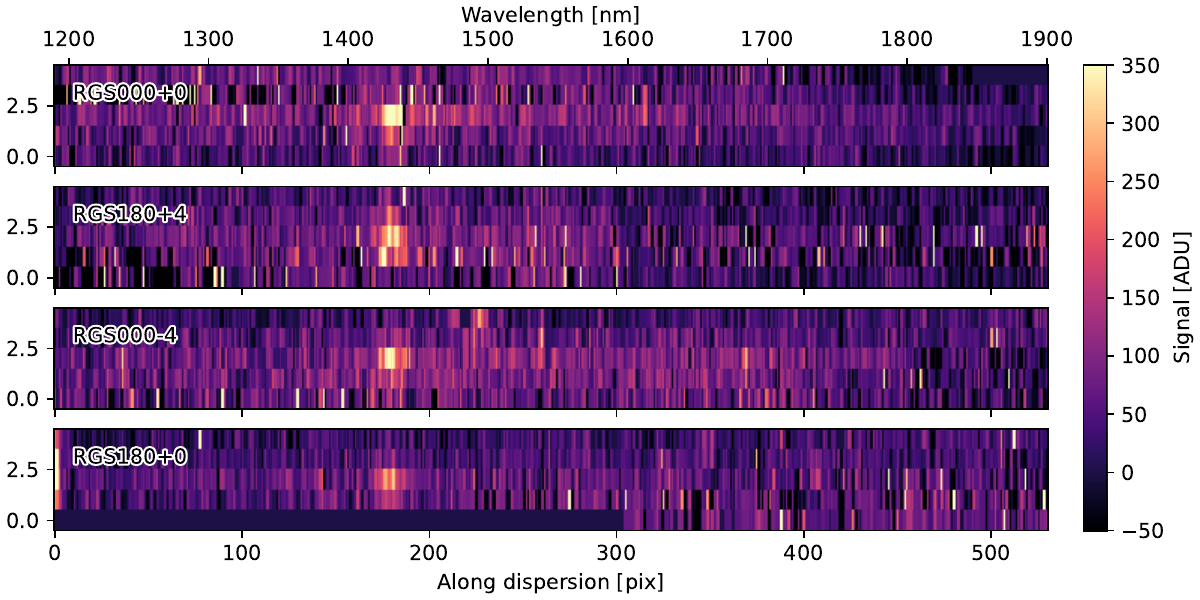}
 \includegraphics[width=\hsize]{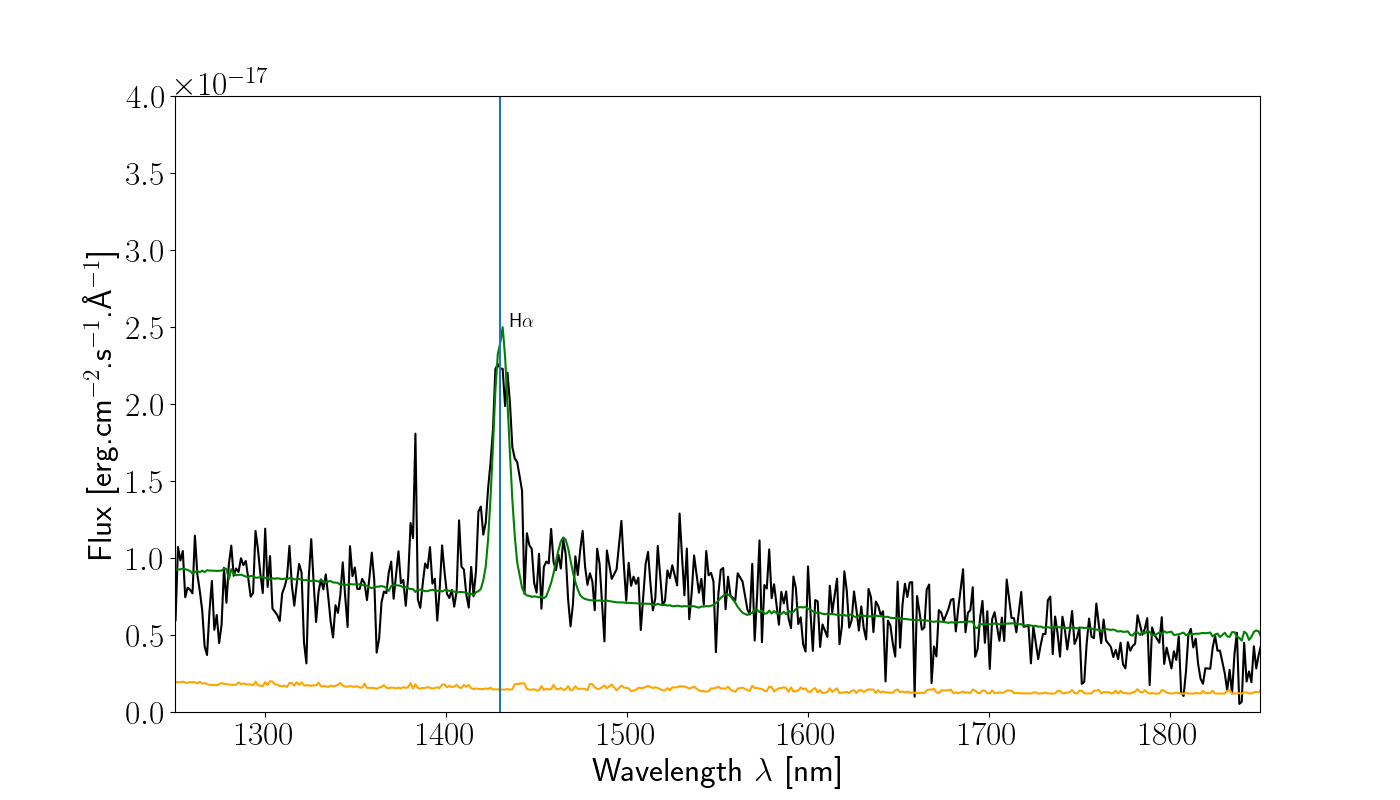}
    \caption{An example of \gls{NISP} spectroscopic data for a galaxy in the COSMOS field with $z=1.1770\pm0.0005$ \citep{Mainieri2007} The top figure shows the four spectrograms, with the H$\alpha$ line clearly visible. The bottom plot shows the corresponding combined and flux-calibrated 1D spectrum (in black) and its associated statistical noise (in orange), while the green line shows the combined continuum and emission line model that fits the data best.  The bright H$\alpha$ line is detected with $\text{S/N}=14$ yielding a redshift of $z=1.1783\pm0.0005$ (vertical blue line), which is in agreement with the previously published value. The flux of the line, $f_{{\rm H}\alpha}=2\times 10^{-15}$\,erg\,cm$^{-2}$\,s$^{-1}$, is approximately ten times higher than the limiting flux for the \gls{EWS}.}
    \label{fig:sample_spectrum}
\end{figure}

\cref{fig:sample_spectrum} shows the results for a galaxy in the COSMOS field with a known redshift of $z=1.1770\pm 0.0005$ \citep{Mainieri2007}. Comparison of the redshift and line flux estimates allows us to quantify the accuracy of the \gls{NISP} spectroscopic calibrations. The top panel shows the four individual spectrograms extracted for this object over the full \gls{RG} domain, after applying the decontamination procedure necessary to remove the signal from nearby objects, and rectilinear resampling.  The bottom panel shows the combined and flux calibrated 1D spectrum for this object, with the H$\alpha$ line clearly detected. We note that this galaxy is not representative for the majority of emission line detections, because its H$\alpha$ flux is approximately ten times brighter than the limiting flux for the \gls{EWS}.

\subsubsection{Redshift measurement}

The next step is to determine redshifts and measure line fluxes from the extracted spectra, and to provide an estimate of the reliability of the measurements. In the case of \gls{EWS} observations, the spectra are obtained with
 the \gls{RG}, while for the 
 \glspl{EDF}, the \gls{BG} spectra are included to enhance the performance of the redshift estimation and to  maximise the purity and completeness. Photometric measurements from VIS and NISP could be optionally included and used for the redshift measurement. 
 
The redshift measurement itself is performed through template fitting over a regular grid of redshifts, spaced logarithmically. 
A first pass uses a coarser logarithmic grid (initial step $\Delta z = 10^{-3}$ logarithmic  grid, after which a \gls{zPDF} is calculated for each model. All those individual \glspl{zPDF} are combined into one so-called `first pass \gls{zPDF}'. 
A first list of redshift solutions is based on the main peaks (up to 10) of this first \gls{zPDF}. The redshift measurement is then refined around each of these solutions by refitting over all the models, but with a finer redshift grid; the \gls{zPDF} is recalculated and the best final solutions (up to five) and corresponding models are then obtained from the strongest peaks. The peaks are ordered following the value of the integral of the \gls{zPDF} under each peak with a $3\,\sigma$ window.

The model includes a set of six distinct continuum models representing various star-formation histories 
from the  \cite{BC03} library. Given the resolution and limited wavelength coverage of NISP, we found that it is not necessary to have more detailed continuum parameters, because it only leads to degeneracies between the different templates. We then add nebular emission lines from templates describing various ratios between the different emission lines. These templates have been built from the more than 35\,000 galaxies of the VVDS programme \citep{LeFevre13}, which provides an unbiased sample of spectra for a wide range of galaxy types with $0.4 < z < 3$, but with magnitude limit of $I_{\rm AB}=24$. Using these templates is more efficient than fitting all possible emission lines independently, since the number of free parameters becomes too high to provide a reliable solution. Finally the model includes intrinsic reddening based on \cite{Calzetti_2000} and \gls{IGM} absorption as described in \cite{Meiksin2005}.

It is also possible to include priors in the calculation of the \gls{zPDF}, 
which can be used to favour solutions where a detected emission line is identified as one of the strongest emission lines usually detected in galaxy spectra (e.g. [\ion{O}{II}], [\ion{O}{III}], H$\alpha$), or just H$\alpha$. Alternatively, one can use the known redshift distribution of H$\alpha$ emitters as a baseline for the \gls{zPDF} instead of a flat prior. 
Tests have shown that  such an empirical prior gives the best results in terms of purity and completeness (also see \cref{sec:clustering_stats}). 

In addition to the galaxy model, the pipeline also provides solutions for quasar and stellar models. The quasar model is built in the same way as was done for galaxies, but including a series of double power-law continuum and Lorentzian broad emission lines. The stellar models are built from a set of 36 templates covering all stellar types \citep{Pickles_1998}. The selected object class corresponds to the highest statistical evidence over the three categories (galaxy, star, or quasar). 

Once the redshift has been calculated, the fluxes of the detected emission lines are measured using both \gls{DI} and a \gls{GF}. In the \gls{DI} method, the spectrum is first continuum-subtracted, using the continuum evaluated from a median-iterate filtering smoothing with a variable window. Each line is then integrated, starting from the position of the peak as provided in the previous step, until the flux remains positive; together with the flux, also the S/N, \gls{EW}, and position of the lines are provided. In the \gls{GF} method, a multi-Gaussian model plus a constant continuum model is considered, with $N$ Gaussians, depending on the line considered: (i) if a blend with adjacent lines is expected (e.g. we assume three Gaussians for the H$\alpha$+[\ion{N}{ii}] doublet complex); or (ii) if the lines are separate but close in wavelength, and their ratios can be linked through physics (e.g. two Gaussians for the [\ion{O}{iii}] and [\ion{S}{ii}] doublets), or if a line is isolated (e.g. one for H$\beta$). In contrast to the \gls{DI} method, this allows for us to deconvolve the contribution of different lines in a complex (e.g. H$\alpha$ from [\ion{N}{ii}]). This model is used to fit the data with a Levenberg--Marquardt algorithm, deriving the flux, S/N, continuum, \gls{FWHM}, \gls{EW}, and wavelength for all the lines, assuming the (up to) five redshift solutions (with their uncertainties) obtained in the step before for each galaxy.

Spectra of objects below redshift 0.9 display very few features in NISP spectra, and hence any artefact in the spectra might be misinterpreted as an emission line. It is therefore essential to identify those spectra that are affected by artefacts: these interlopers could outnumber the targets of the \gls{EWS} ($0.9<z<1.8$) because the redshift distribution peaks around 0.5 for the limiting AB magnitude of $\HE=24.0$ in the \gls{EWS}. To discriminate objects with secure redshift measurements from possible interlopers, the pipeline provides a numeric indicator of the `reliability' of the redshift measurement, which quantifies the quality of the spectrum based on the analysis of the \gls{zPDF} using a deep-learning algorithm. This algorithm will be trained on \gls{EWS}-like observations of the \gls{EDS}, for which a correct redshift at the 99\% confidence level is expected. This training will be applied to the set of \gls{zPDF} of the \gls{EWS} spectra to quantify the reliability of the redshift measurements.

\subsection{Measurements for 3\texttimes2pt statistics}
\label{sec:wlmeasurement}

As discussed in \cref{sec:3x2pt} the shape measurements from \Euclid are combined with photometric redshifts derived from multi-band photometry. The precise photometric redshifts are required to divide the sample into tomographic bins, while their redshift distributions need to be characterised well for a correct interpretation of the clustering and lensing signals. In \cref{sec:photoz} we describe how the photometric redshifts are determined and we plan to calibrate the corresponding redshift distributions to high accuracy.
As described in \cref{sec:classification}, the photometry is also used to classify the objects for further science applications.

The galaxy shape measurements benefit greatly from the sharp diffraction-limited \gls{PSF}. Nonetheless, our objectives require its size and shape to be determined with unprecedented accuracy. This implies that we have exquisite knowledge of the optical properties of the telescope 
and understand how the detectors record the incoming photons. This is, however, not sufficient: the \gls{PSF} is chromatic, which means we need to estimate the appropriate \gls{PSF} for each galaxy based on its observed \gls{SED}. This requires a dedicated procedure that is described in \cref{sec:galsed}.

The requirements for \Euclid were derived by considering the change in the observed unweighted quadrupole moments, $Q^{\rm obs}_{ij}$, when a galaxy image is convolved by the \gls{PSF}. For an object with an observed surface-brightness distribution, $I^{\rm obs}(\mathbf{x})$, and total flux, $F$, its quadrupole moments are defined as 
\begin{equation}
     Q_{ij} = \frac{1}{F}\int\,{\rm d}^2\mathbf{x}\,\;x_i\, x_j\, I^{\rm obs}(\mathbf{x})\;.
\end{equation}
The shape of an object can then be quantified  by combining the quadrupole moments into the complex polarisation (or distortion)\footnote{This definition is related to the third flattening (see \cref{sec:weaklensing}) through $\epsilon=e/(1+\sqrt{1-|e^2|})$.}
, 
$e=e_1+{\rm i}e_2$, where
\begin{equation}
    e_1=\frac{Q_{11}-Q_{22}}{Q_{11}+Q_{22}}\;,\quad {\rm and}\;\;
    e_2=\frac{2\,Q_{12}}{Q_{11}+Q_{22}}\;,
\end{equation}
while the size of an object is captured by
\begin{equation}
R^2=Q_{11}+Q_{22}.
\end{equation}

Although unweighted moments are not practical in the presence of noise and blending, they do provide a convenient framework to quantify the impact of the \gls{PSF} on shape measurement \citep{Paulin-Henriksson09, Massey+13}.
In this case, the observed quadrupole moments of a galaxy are given by the sum of the quadrupole moments of the true galaxy image and those of the \gls{PSF}
\citep{Valdes83}, so that
\begin{equation}
Q_{ij}^{\rm obs}=Q_{ij}^{\rm gal}+Q_{ij}^{\rm PSF}.
\end{equation}

\citet{Cropper+13} used this to allocate tolerances for a wide range of instrumental effects, starting from an acceptable level of bias in cosmological parameters. A limitation of this simple `flow down' is that it did not capture spatial variations of sources of bias, resulting in conservative estimates for the impact of residual systematics on the observed lensing signal \citep{Kitching2019}. 
To explore a more realistic scenario, \citet{Paykari-EP6} considered a `flow-up' from perturbations to the defocus of the \gls{PSF} model to the bias in cosmological parameters. They found 
that \gls{PSF} variations at the level of the requirements induced biases of about 4\% of the expected statistical uncertainty in dark energy parameters, well within requirements.
Further tests, using simulated data, confirmed that the requirements derived by \citet{Cropper+13} are adequate and conservative and thus remain the basis for the calibration of VIS data (\cref{sec:calibration}) and the modelling of the \Euclid \gls{PSF}, which is described in detail in \cref{sec:psfmodel}. The challenging task of accurate shear estimation from the \Euclid images is discussed in \cref{sec:shapemeasurement}.

\subsubsection{Photometric redshift estimation} 
\label{sec:photoz}

As shown in \cref{eq:limber}, the amplitude of the lensing signal depends on the redshift distribution of the sources, in addition to the cosmological parameters. Moreover, being able to separate the overall galaxy sample into subsamples that are separated in distance enhances the amount of information that can be extracted from a weak lensing, or 
3\texttimes2pt analysis. The current baseline configuration uses 13 evenly populated bins  in the redshift range $0.2<z<2.5$. 

The mean redshift of these subsamples is of particular importance for cosmic shear, since a bias in the estimated mean redshift directly translates into a bias in cosmological parameter inference. 
Given the objectives of \Euclid, this implies that the mean redshift must be known to $\sigma_{\langle z\rangle}<0.002\,(1+z)$ per bin \citep{Ma06, Amara07}. 
Importantly, the modelling of the photometric clustering signal in the 3\texttimes2pt analysis relies not only on accurate knowledge of the mean redshift, but also of the width of the redshift distributions of the different subsamples \citep[e.g.][]{Tutusaus20,Porredon22}. 
The need for precise photometric redshifts is therefore two-fold: to place objects in narrow tomographic redshift intervals, minimising the overlap between subsamples; and to ensure that the distribution in redshift of the sources in each subsample is accurately known. These two needs are addressed separately in the photometric redshift pipeline, since the former requires that we optimise the photo-$z$ for precision, while the latter demands very high accuracy.

\begin{figure}[ht]
\centering
\includegraphics[width=0.495\textwidth]{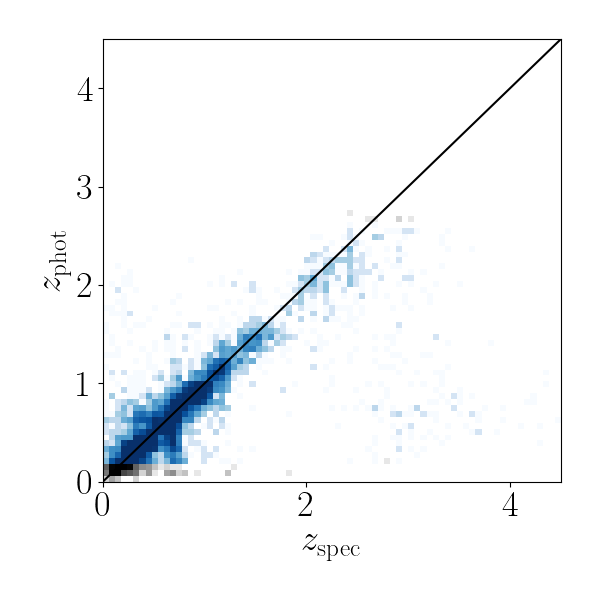}
\caption{Photometric redshift performance of the mode of individual probability distributions using \texttt{NNPZ}, taken from \cite{Desprez-EP10} who used simulated \gls{DES} and \Euclid \gls{NIR} data.
Regions of photometric redshift space that will be excluded from the weak lensing analyses are shown in grey.}
\label{fig:zphotzspec}
\end{figure}

For individual galaxies, the need to place objects in tight redshift intervals, with $\sigma_z<0.05\,(1+z)$ being a key requirement, leads to two considerations: the effective dispersion of the \gls{zPDF} around its assumed true redshift (`photo-$z$ scatter'); and the probability that our assumed redshift for an object is far from its true redshift (`outlier rate'). 
Further requirements are placed on the choice of the photo-$z$ algorithm by the nature of the \gls{EWS}. It must be able to run efficiently for the billions of galaxies we will observe and to be able to account for effects that vary on an object-by-object basis, namely the impact of Galactic reddening and the variation of filter transmission curves across the \gls{FOV} of an instrument \citep{Paltani-EP31}. 
These additional concerns led to the development of the \texttt{NNPZ} (Nearest-Neighbour Photometric Redshifts) algorithm; its performance on a sample of spectroscopic redshifts taken from the COSMOS field is shown in \cref{fig:zphotzspec}. These results were taken from the \Euclid photo-$z$ data challenge of \citet{Desprez-EP10}, which compared different methods using simulated \gls{DES} (optical) and \Euclid (NIR) data.
The photometric redshift quantity that is used to assign tomographic redshift bins is the mode of the \texttt{NNPZ} \gls{zPDF}, and for the sample shown in \cref{fig:zphotzspec} the scatter ($\sigma_z=0.059$) and outlier fraction ($\eta=11.1\%$) of this photo-$z$ with respect to the spectroscopic redshifts lie slightly outside of our requirements ($<0.05$ and $10\%$, respectively). This is adequate for the \acrlong{DR1}
\glsunset{DR1} (\gls{DR1}; see \cref{sec:DR}), while we expect to achieve a better performance for later releases, once much deeper \gls{LSST} data become available.

Machine-learning algorithms such as \texttt{NNPZ} rely heavily on how representative the reference training data are of the target data set. Complete and representative spectroscopic samples are impractical to assemble for the \gls{EWS}, and so instead we will build a sample of galaxies with accurate and precise photometric redshift \glspl{zPDF} from a carefully selected and curated set of deep reference fields, the \glspl{EAF}. These reference field photo-$z$s will be computed using a new custom \gls{SED}-fitting package called \texttt{Phosphoros} \citep[see Sect. 3.3 in][for a brief description]{Desprez-EP10}. Through tagging the \glspl{SED} with properties, such as a mass-to-light ratio or metallicity, \texttt{Phosphoros} is able to additionally produce sampled posterior distributions of physical quantities (e.g. stellar masses) for each reference galaxy. The \gls{zPDF} and multi-dimensional physical-property posterior information can then be propagated from these reference field galaxies to the \gls{EWS} galaxies with \texttt{NNPZ}. In this sense, \texttt{NNPZ} operates as an accelerator for the template-fitting approach, and its lack of an explicit training phase allows us to account for effects that vary on an object-by-object basis \citep{Paltani-EP31}. The \glspl{zPDF} we output at this stage are still expected to be subject to biases, due to our imperfect knowledge of galaxy \glspl{SED} and the priors we use, and are therefore used only for tomographic redshift binning in the weak lensing analysis. The galaxy physical properties inferred through this process are not used in the weak gravitational lensing analyses, but will be used for a vast range of non-cosmological applications (see \cref{sec:legacy}).

Once galaxies have been assigned to their tomographic redshift samples, we must reconstruct their collective distribution in redshift, $n(z)$, for cosmological inference of the weak lensing signal. The strategy to achieve the target accuracy of $\sigma_{\langle z\rangle}< 0.002\,(1+z)$ per tomographic bin was outlined in \citet{Masters15} and has evolved only a little since then. Briefly, the 8- or 9-band photometric space of the target galaxy sample is quantised by way of a \gls{SOM} \citep{Kohonen07}, resulting in a 2-dimensional array of vertices or `cells' (also see \cref{fig:specfig}). Each cell of the \gls{SOM} is represented by a vector of values, where each value corresponds to the ratio of flux in a photometric band with respect to the $\IE$ value, and thus each cell represents a possible galaxy \gls{SED} with free amplitude. Galaxies in the weak lensing sample are each assigned to their closest \gls{SOM} cell  
in the flux-ratio space used to construct the \gls{SOM}. The \gls{SOM} algorithm preserves locality of the input space, in the sense that similar vectors of flux ratios with respect to $\IE$ will be nearby to one another in the 2-dimensional \gls{SOM} map, and therefore each cell has a finite size in the input parameter space. The \Euclid survey is supported by ground-based observations with different instruments and filters for the northern and southern regions (see \cref{sec:groundbased}). To place all galaxies correctly in the \gls{SOM} we perform a band-standardisation step whereby a per-galaxy colour term is computed and applied to the observed photometry in each band. Corrections for Galactic reddening and bandpass variation are also carried out during this step.

Construction of the $n(z)$ is based on the spectroscopic sample built up through the C3R2 programme (see \cref{sec:cog_spec}). The sample is designed to cover the diversity of galaxy \glspl{SED} in the shear sample, but the measured \gls{SED} of an object depends on the noise properties of the observations and other survey characteristics. In order to be able to account for varying survey depth, the spectroscopic sample is drawn from the \glspl{EAF} (\cref{sec:deepfields_EAF}), where the ${\rm S/N}$ of the data are five times higher than the \gls{EWS}. We can then draw multiple realisations of the \gls{EWS} noise properties and apply them to these objects to form a calibration sample that covers both the range in galaxy \glspl{SED} and photometric scatter of the shear sample. 

With the calibration sample in hand we perform the $n(z)$ estimation for each tomographic redshift subsample separately. Each galaxy in a subsample is assigned to its best-matching \gls{SOM} cell and the shear weights of objects (see \cref{sec:shapemeasurement}) are summed within each \gls{SOM} cell. These sums of weights represent the statistical power of each \gls{SOM} cell, with an analogous quantity determined by each objects' probability to pass the selection flux cut in the case of photometric clustering. The calibration objects are treated in the same way, with the exception that their `shear weights' are always unity, that is, their \glspl{zPDF} are measured via \texttt{NNPZ} and their artificially noised flux measurements are dereddened and band-standardised in the same way as for the shear sample. The calibration objects are then assigned to tomographic bins on the basis of the mode of the \gls{zPDF}, exactly as for the shear sample objects. They are also rendered into the \gls{SOM} and the number of them per \gls{SOM} cell, within the given tomographic bin, is computed. The redshift distribution of a tomographic bin is then just a weighted histogram of the spectroscopic redshifts, where the weight of a spectroscopic object is the sum of shear weights in its cell, divided by the number of spectroscopic objects in that cell. In this way, all trusted redshifts per cell are used, but the relative sampling difference between the shear sample and the selection of objects for which we have spectroscopic redshifts is accounted for. 

There are some \gls{SOM} cells containing galaxies that can be used for the weak lensing analysis, but that lack spectroscopic measurements. Those galaxies are flagged and removed from the analysis because we are unable to represent them in the recovered redshift distribution. 
With this strategy, the $n(z)$ distribution is subject to sample noise, but \citet{Masters15} showed that we will be able to meet our target requirement of $\sigma_{\langle z\rangle}<0.002\,(1+z)$ on the mean redshift per tomographic bin, provided that each tomographic subsample is represented by a large (${>}\,600$) number of cells.

As a further validation of the redshift distributions estimated with the SOM method, we will use clustering redshifts \citep{Newman08}. This method employs angular cross-correlation measurements of the positions of the sources in the tomographic bins and spectroscopic calibration samples that overlap on the sky. Crucially, this drops the assumption of the calibration sample fully covering the colour and magnitude range of the source sample. As such, different calibration samples that are typically brighter and cover larger areas are used, which makes this approach highly complementary to the SOM method. Measuring the angular cross-correlation amplitude between the tomographic source samples and the spectroscopic calibration samples finely binned in redshift can result in an accurate reconstruction of the tomographic redshift distributions. As shown in \citet{Naidoo23}, an overlap area of only a few hundred square degrees between the \textit{Euclid} source sample and spectroscopic surveys like BOSS, DESI, and the \textit{Euclid} NISP-S sample is sufficient to reach the requirement on the accuracy of the mean redshift of the tomographic bins.

\subsubsection{Classification}
\label{sec:classification}

The objects detected in \Euclid images comprise galaxies, stars, QSOs, globular clusters, Solar System objects, and a diverse array of contaminants and artefacts. In order to avoid introducing selection effects in the weak gravitational lensing analysis, each detected object with a full complement of photometric information available is treated as a possible galaxy at the photometric redshift determination stage. Contaminants and stars are determined during the shape measurement process (see \cref{sec:shapemeasurement}) by their low shear weights. However, to facilitate the modelling of the \gls{PSF} we must identify a very pure sample of stars that spans the range of stellar colours. Using only morphological information, such as an object's apparent compactness, risks introducing biases into the PSF model, and so this step is ideally performed using only photometric information.

Our classification uses a set of three pretrained \glspl{pRF}, each a binary classifier for a type of object, such as star versus not a star. The \gls{pRF} returns a probability that an object is of that type and thus each object receives three probabilities, one for each of the considered types, namely star, galaxy, and QSO. An object is assigned a given class if its probability to be that class exceeds a predetermined threshold, and a single object can be given multiple classes. Only those objects classed as stars and not any other additional classes are considered suitable for \gls{PSF} modelling. Class probabilities are retained for use in non-cosmological legacy science processing (see \cref{sec:legacy}).

\subsubsection{SED modelling}
\label{sec:galsed}

The convolution of galaxy images by the \gls{PSF} is the dominant source of bias for weak lensing studies, and an accurate estimate of the \gls{PSF} is required to obtain unbiased shape measurements. Here, we focus on the challenges that arise from the fact that the 
\Euclid telescope is diffraction limited and the VIS passband is very broad \citep[see][for more background]{Cypriano10, Eriksen18}. In the analysis we need to take into account that each galaxy is convolved by an effective \gls{PSF} that depends on its \gls{SED}. Moreover, to determine the underlying optical \gls{PSF} model successfully, we must know the \gls{SED} of each of the stars that is used in this process.

For bright stars ($G\,{\la}\,16.5$) \Gaia spectra can be used directly, but for fainter objects we must estimate their \glspl{SED} at high accuracy from their broad-band photometry \citep{Eriksen18}. To perform this task we re-use \texttt{NNPZ}, the nearest-neighbour method that was developed for the main photometric redshift pipeline, but choose as target quantities the fluxes at different wavelengths. Similar to what is done to determine the redshift \glspl{PDF}, we construct a reference data set of objects that is in principle representative of the objects for which we need to recover the \gls{SED} information. Specifically, we use the \Gaia spectra, assuming that across the \Euclid footprint we have examples of all relevant stellar types and metallicities. An analysis of the impact of differential Galactic reddening between brighter and fainter stars of the same types (and thus different radial distances) is ongoing. We integrate the \Gaia spectra through a series of 55 synthetic narrow-band filters of width 10\,nm, filling the range $450\,{\rm nm}\,{<}\,\lambda\,{<}\,1000\,{\rm nm}$, and recover the weighted mean flux from the $30$ closest neighbours. The weight is a pseudo-likelihood computed from the $\chi^2$ distance in flux space between the target object and a reference object, with a free \gls{SED} amplitude parameter.

The measurement of galaxy \glspl{SED} follows a similar procedure, but in this case we lack an equivalent to the \Gaia spectro-photometric data set. Moreover, existing sets of galaxy \gls{SED} templates are either incomplete or biased and would thus introduce biases in the \gls{PSF} construction that would propagate to cosmological parameter estimation \citep[see][]{Eriksen18}. 

The strategy to create the \texttt{NNPZ} reference sample for galaxy \glspl{SED} is summarised as follows. Instead of using spectra, we begin with broad- and intermediate-band photometry from the COSMOS and CDFS fields to provide a coarse sampling of galaxy \glspl{SED} through much of the \IE\ bandpass range. We then apply a combination of \gls{GP} interpolation and template-guided filter colour terms to achieve the finer 55 narrow-band sampling that we also use for the stellar \glspl{SED}. To reduce 
the impact of sample variance we plan to collect additional medium band data across the \glspl{EAF}.

\subsubsection{VIS PSF model}  
\label{sec:psfmodel}

As discussed in detail in \citet{Cropper+13}, accurate measurement of weak lensing shear imposes stringent requirements on model accuracy for the \Euclid VIS \gls{PSF}. In the case of \Euclid, this means that the \gls{PSF} model must be known throughout the mission lifetime with a residual temporal-spatial model uncertainty of $\sigma(e) < 2\times10^{-4}$ per ellipticity component, while the size needs to be known such that $\sigma(\Delta R_{\rm PSF}^2)/R_{\rm PSF}^2 < 10^{-3}$. 
Although the \Euclid VIS system possesses a highly stable, diffraction-limited \gls{PSF}, meeting these requirements is nonetheless challenging because of a number of design choices.

First, to reach the required depth, VIS uses a broad bandpass, but this also results in a strong chromatic dependence for the \gls{PSF}. Hence, as already discussed in \cref{sec:galsed}, the \gls{PSF} varies between stars and galaxies according to their colour or \gls{SED}, and within galaxies due to local changes in stellar populations \citep[i.e. colour gradients within galaxies;][]{Semboloni+13, Er+18}.
Moreover, \Euclid utilises a dichroic to split the visible and \gls{NIR} components of the beam, which was designed to produce a hard bandpass edge with minimal out-of-band light (in conjunction with coatings also applied to two of the fold mirrors). This is achieved using a complex multi-layer dielectric coating, which induces wavelength- and polarisation-dependent phase errors on the wavefront and consequent effects in the \gls{PSF} model \citep{GasparVenancio2016}. This is further influenced by chromatic, polarised reflection from the silver optical surfaces and fold mirror coatings, and transmission at the detector. Finally, the pixel size of the \Euclid VIS \glspl{CCD} is such that the \gls{PSF} is undersampled at all wavelengths in the VIS bandpass.

Hence, in addition to capturing the 
 variation across the \gls{FOV}, the 
VIS \gls{PSF} model also needs to allow for the propagation of the model between stars and galaxies with differing \glspl{SED}. Together with the need to produce an oversampled model, to address pixel undersampling, this motivates utilising a \gls{PSF} forward-modelling approach. The Fraunhofer condition links the image-plane \gls{PSF} to the wavefront error at the exit pupil of the telescope \citep[e.g.][]{hopkins1970}. Under this condition, one may forward model the \gls{PSF} on the detector \gls{FPA} using a combination of chromatic wavefront modelling for the optical contributions and both chromatic and achromatic convolutional kernels for the detector and guiding error contributions
\citep[also see][]{Ma08}.
When modelling the \gls{PSF}, it is essential to analyse image data that have had all linear and nonlinear detector effects corrected, including detector nonlinearity and the \gls{BFE}, the latter currently being corrected by a modified version of the algorithm due to \citet{coulton2018}.

An important benefit of forward modelling is that a wide range of effects can be included in a consistent fashion.
In the case of the VIS \gls{PSF} model, these include the following.

\begin{itemize}
    \item A model for vignetting of the pupil by the secondary optics structure, constructed from industry-supplied CODE~V\footnote{CODE~V\textsuperscript{\textregistered} is a commercial optical design tool.} model inputs and tested against CODE V model outputs.
    \item Optical path differences due to optical layout, modelled locally on the pupil plane as a weighted sum of Zernike polynomials. Zernike polynomials are a natural basis set for this contribution, since they are defined to be orthogonal on a circular pupil, although the  vignetting breaks the orthogonality of the polynomial set in this application. Variation across the \gls{FPA} is included as a polynomial fit across the \gls{FOV} for each Zernike polynomial.
    \item The \gls{SFE} contribution to the wavefront error, resulting from imperfections of each reflecting surface, are included by propagating a model for the beam footprint on pre-launch measured \gls{SFE} maps for each optical surface.
    \item Telescope jitter, or guiding error, of the telescope during an observation induces a field-dependent, achromatic convolutional effect \citep{Ma08}. This convolutional kernel is calculated from three-axis measurements of the \gls{RPE} from the \gls{FGS}, Wiener filtered to remove noise.
    \item Detector charge diffusion is modelled as in \citet{Niemi2015}, as a chromatic convolutional effect.
    \item Chromatic dependencies, including polarisation-dependent telescope throughput, detector quantum efficiency, and source \glspl{SED}, are included as spectral weightings in the broadband, chromatic \gls{PSF} model, which is computed by spline interpolation of monochromatic calculations at multiple wavelengths, followed by integration across the VIS passband and out-of-band optical wavelength region.
    \item The chromatic response to polarisation from mirror coatings, most notably the layered dielectric coatings of the dichroic, the silver coating on FOM\,3, the coatings on FOM\,1 and 2 at high angles of incident light, and the hafnium oxide coating on the detector. In particular, the layered dichroic coating induces a strong chromatic and \gls{FOV} dependence, which must be accounted for in the model.
    To characterise this dependence and to provide inputs for the PSF modelling, we have commenced an extensive test campaign on the spare model of the dichroic \citep{Baron2022, Baron2023, Baron.phd}.
    The polarisation dependence of the VIS optical system is modelled following \citet[][chapters 11,16]{chipman2019-PolarisationPSF}. An \gls{FOV}- and wavelength-dependent set of four `Jones pupils' is produced from the CODE~V model of the optical surfaces and layout. The Jones pupils describe the system's polarised response to orthogonal input polarisation states across the pupil. They are included in the construction of the complex electric field at the exit pupil, and transformed to the monochromatic \gls{PSF} at a given wavelength for a given set of Stokes parameters, via the construction of an `amplitude response matrix' and Mueller point-spread matrix. We emphasise that this effect must be included even for unpolarised incident light, which should be considered as a combination of two incoherent orthogonal polarisation states.
    \item Detector undersampling. The model \gls{PSF} is produced on oversampled pixels at the \gls{FPA}, ensuring that the exit pupil is Nyquist sampled. This oversampled image may be downsampled to the \Euclid VIS detector pixel sampling for comparison with data.
\end{itemize}

The wavefront model for the \Euclid VIS \gls{PSF} allows for a fully flexible and modular parameterisation, including chromatic, spatial, and temporal variations. The \gls{SFE}, dichroic, and detector contributions, with the exception of \gls{CTI}, are expected to be invariant across the mission lifetime and therefore the models for these are fixed using laboratory measurements. The telescope pointing stability is measured during observations using the \gls{FGS}, and propagated to the model for any given exposure. 
However, the optical model is expected to vary throughout the mission, due to thermal variations arising from changing telescope orientation with respect to the Sun and due to variations in spacecraft heat inputs. Distortions to the telescope structure under gravity and perturbations resulting from the launch process mean that laboratory measurements of the optical alignment cannot predict wavefront errors, and instead these must be inferred from in-orbit data. 

The \gls{PSF} model is therefore calibrated across the lifetime of the survey using a hierarchical calibration process consisting of three steps.
First,  the initial state of the \gls{PSF} model is inferred from \gls{PV} (\cref{sec:cal_PVphase}) and \gls{ESOP} data across a range of spacecraft attitudes with respect to the Sun vector. This is used to produce a basis set of spatially dependent \gls{PSF} variations, which can be fit to further observations at other attitudes. Since \gls{PSF} optical modes are expected to be degenerate with detector modes (e.g. the charge diffusion kernel), this process is conducted iteratively, alongside fits for detector modes, until requirements on validation metrics are met. Because detector modes are not expected to vary across \gls{PV} and \gls{ESOP} calibration data, they are fit in a meta-analysis of the full calibration data set. This basis set is then updated from continuous observations with a monthly cadence, and any perturbations to the basis set modes are fed through to further optimisation. The monthly calibrations are taken on a limited set of defined fields, to simplify the interpretation of \gls{PSF} variations between observations. This process should capture long-timescale variations.
Finally, the resulting calibrated basis set is fit to each science observation, producing short-timescale variation on the timescale of science observation cadence. This enables an accurate \gls{PSF} model to be produced for shear measurement in each science observation.

\begin{figure*}[!ht]
\begin{center}
	\includegraphics[angle=270,origin=c,width=0.65\hsize,trim={80 0 92 0},clip]{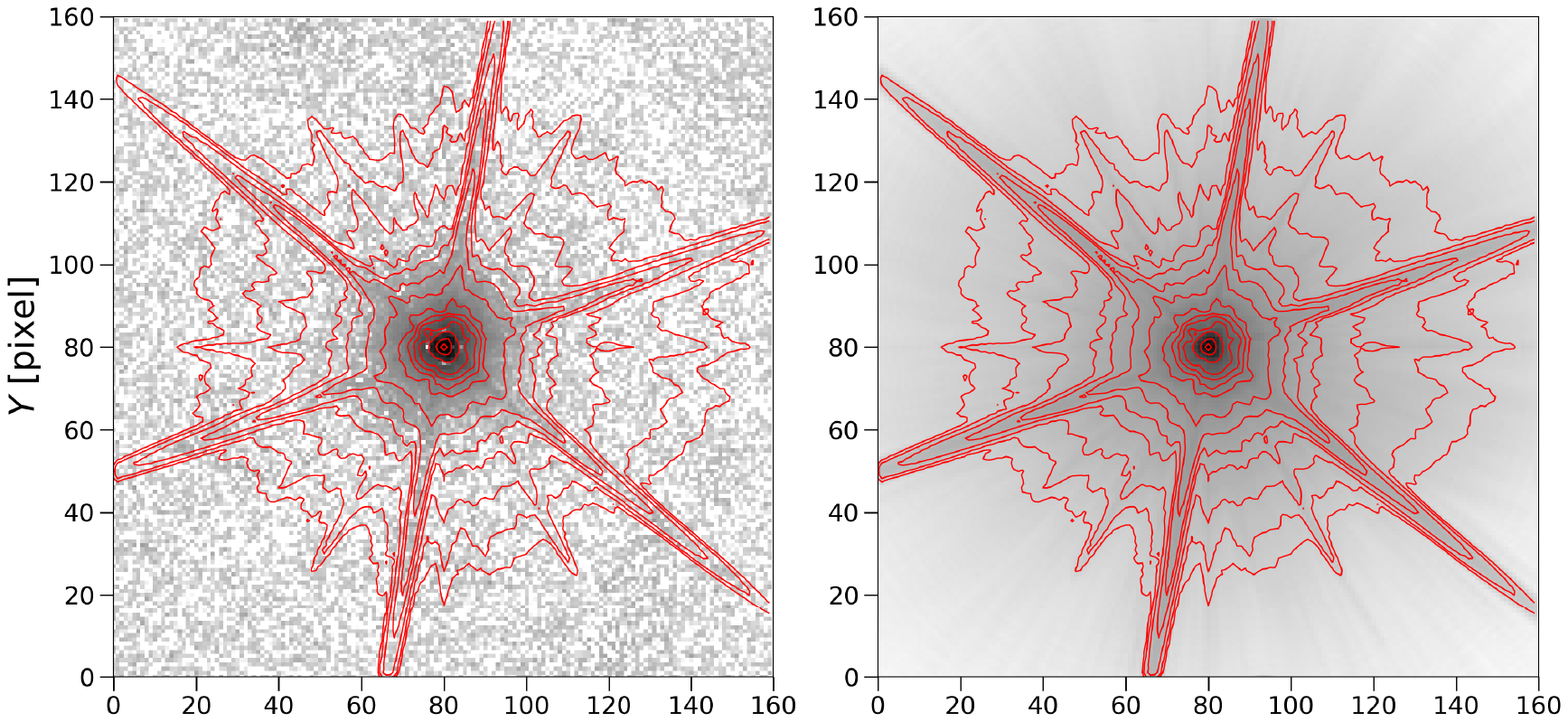}
 
    \vspace{-3cm}

    \includegraphics[angle=270,origin=c,width=0.65\hsize,trim={80 0 80 0},clip]{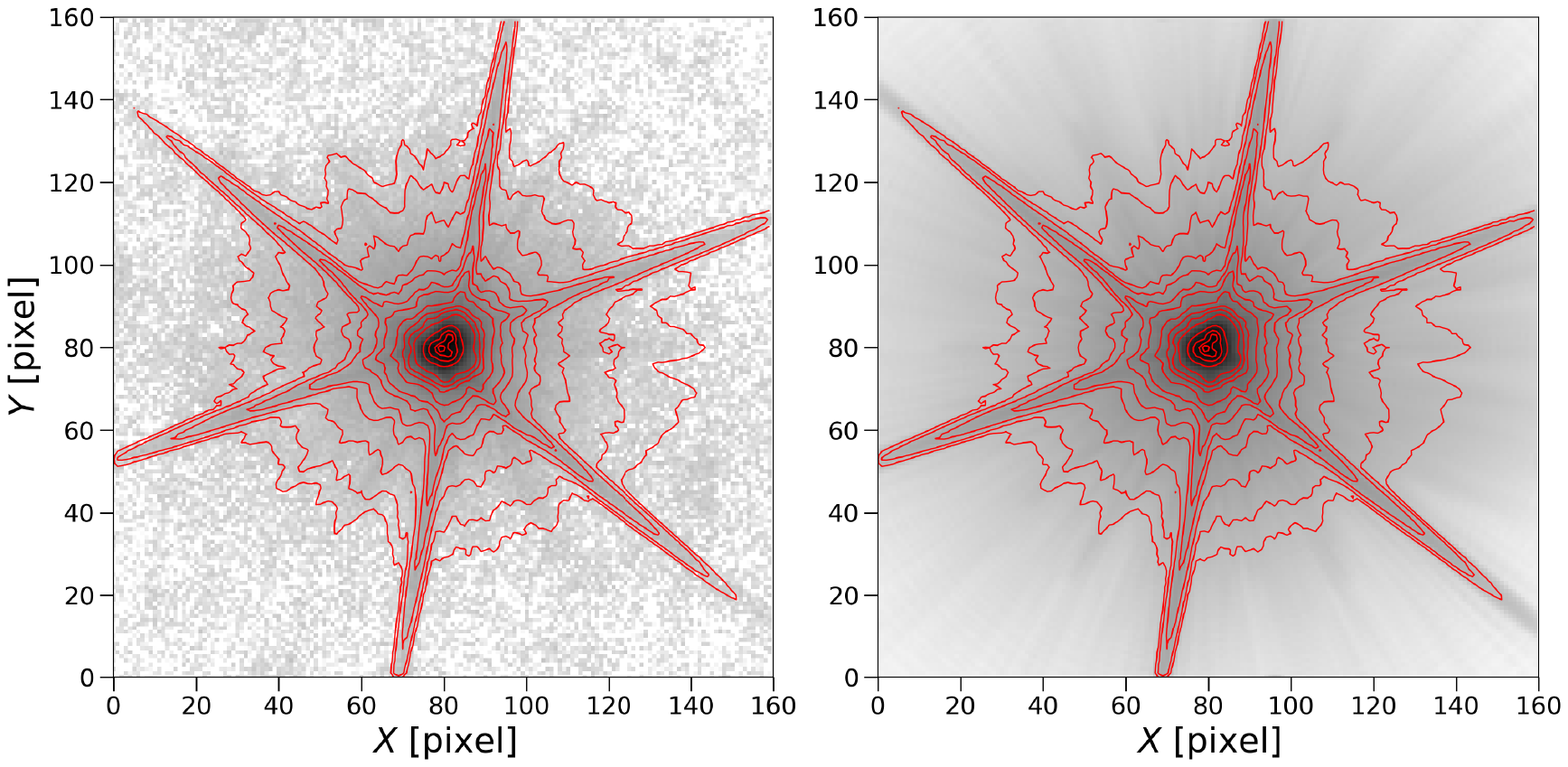}

    \vspace{-2.5cm}
    
    \caption{Comparison between the mean profile of stars and the VIS \gls{PSF} model during the \gls{PV} \gls{PDC} test data. The left panels show stacked images of stars selected from \gls{CCD} 4-5, while the right panels show the corresponding stacked model \glspl{PSF}. The top panels show in-focus data and models, while the bottom panels show defocused data and models, obtained with M2 movement of $-18\,\micron$. All images are produced as the mean of flux-selected stars, applying 2\,$\sigma$ clipping on pixels to remove contaminating objects. Red contours are isophotes of the model \gls{PSF}, superimposed on both the model and data images. The model is a pre-calibration one, fit to a small \gls{PDC} test data set.}
    \label{fig:mini-pdc-psf-comparison}
    \end{center}
\end{figure*}

In the calibration, the initial telescope optical state is assumed to be maximally unknown, whereas the \gls{PSF} model contains degenerate effects, both between optical Zernike modes and between detector and guiding error effects.
Together, this can make it difficult for initial \gls{PSF} calibration to robustly determine the underlying optical parameters that are used to infer the \gls{PSF} across the mission lifetime.

We therefore utilise phase-diverse observations to calibrate the \gls{PSF} model during the observation of \gls{PV} and \gls{ESOP} calibration images (see section \ref{sec:cal_PDC}). The phase diversity technique \citep[e.g.][]{gonsalves1982, fienup1982} adds known aberration to the observations, and for the \Euclid mission we add defocus by deliberate movement of the M2 mirror along the beam propagation direction. Adding a known defocus serves two purposes: first, it causes optical modes to become distinct from detector modes, due to dispersion of the \gls{PSF} across a wider scale of pixels; and second, perturbations in optical modes induce differing impacts on the \gls{PSF} image for each defocus, allowing a joint fit that lifts degeneracies between optical modes. Together, the phase-diverse calibration allows for the model parameter optimisation to robustly find the true best-fit model parameters, by simplifying the posterior that is optimised, through lifting degeneracy and multi-modality. These phase-diverse data are analysed with a model-fitting approach, to account for the wide VIS passband and the image undersampling \citep[cf.][]{fienup1999}.

\Cref{fig:mini-pdc-psf-comparison} shows the pre-calibration VIS \gls{PSF} model, fit to a \gls{PDC} test data set taken during the \gls{PV} phase. This comprised three sets of 10 exposures, taken at nominal in-focus and with two defocus offsets, applied as a movement of the M2 mirror by $\pm 18\,\mu$m, corresponding to about 0.8 waves of defocus. Larger defocus amounts are prohibited by the requirement to operate the \gls{FGS}. Also shown are stacked profiles of stars in that data set, applying pixel-level sigma-clipping to remove contaminating sources. This stacking procedure was carried out on both the stars and the model images fit to those stars. Red contours show isophotes for the model \gls{PSF}. We see that the features of the \gls{PSF}, as observed in the stars selected in the data, are well described by the \gls{PSF} model. There is good agreement on all scales, including in the wings of the \gls{PSF}, with good alignment of the diffraction spikes caused by the spiders holding the M2 mirror, including lateral translation of those spikes in the defocused images, and the halo caused by \gls{SFE} on M1. We note that these models were fit to a set of phase-diverse test data with approximately 15 times fewer exposures than the \gls{PV} and \gls{ESOP} \gls{PDC} calibration sets, and the models shown are not yet expected to satisfy the model accuracy requirements discussed above.

\subsubsection{Shape measurement} 
\label{sec:shapemeasurement}
 
Although \Euclid is designed to provide sharp images of distant galaxies, the observed shapes are biased because of the convolution with the PSF, noise in the images, blending with other galaxies, the presence of cosmic rays, as well as a variety of detector-related sources of bias. These amount to biases that dwarf the lensing signal itself, and thus all need to be carefully accounted for.

Over the years much effort has been spent on developing and testing methods that aim to provide accurate ellipticities for individual galaxies \citep[see e.g.][for the results from several community-wide efforts]{Heymans+06, mm:Massey2007,Bridle2010,great10g,great10s,2012arXiv1204.4096K,great3}. The sensitivity of algorithms to certain systematics will differ, and hence it is advantageous to consider several approaches. A well established method is to fit parameterised models to the observed surface brightness. To this end, we have developed \texttt{LensMC} \citep{Congedo2024}, which is based on forward modelling galaxies with fast \gls{MCMC} sampling and marginalisation over nuisance parameters. It measures the properties of objects from the multiple exposures jointly, while mitigating the bias due to detected neighbours by measuring objects in groups. The robustness has been proven on realistic simulations of \gls{EWS} images, using galaxies drawn from the \gls{EFS} as input. The development of \texttt{LensMC} has
 benefited from the knowledge gained from the ground with \texttt{{\it lens}fit} \citep{Miller+07,2008MNRAS.390..149K, Miller+13},
which has been successfully applied to the various data releases of \gls{KiDS} \citep[e.g.][]{FenechConti+17,Kannawadi+19,SSLi23}.

An alternative route uses the observed moments of the surface brightness. For instance, \texttt{MomentsML} \citep{Tewes+19} is a machine-learning algorithm that trains a neural network to estimate the shape parameters from the moments of the surface-brightness profiles of galaxy images. In principle, 
this can reach the required accuracy, but it does rely critically on the realism of the simulated training data. This is a general concern for shape measurement
\citep[e.g.][]{Hoekstra+17, Kannawadi+19}, also because biases are already introduced during the object detection stage \citep[e.g.][]{FenechConti+17, Hoekstra+2021}. To (partially) circumvent these concerns, \texttt{MetaCal} \citep{Sheldon+17,Huff+17} measures the shear response of an object for any shape measurement method by directly distorting its observed images. It has already been extensively applied to the \gls{DES} lensing analysis \citep{Zuntz+18,Gatti+21}. Moreover, tests on realistic simulated images have shown its potential 
for \Euclid \citep{Hoekstra2021,Kannawadi2021, Hoekstra+2021}. Hence, work is ongoing to develop a \texttt{MetaCal} setup for the analysis of \Euclid data.

Simulated images are also needed to calibrate the biases introduced by other complications, such as the blending of galaxies at different redshifts \citep{MacCrann22, SSLi23} or specific detector effects. The latter includes the non-uniformity of the pixel response over the \gls{CCD}, bleeding due to charge overflow in a pixel, and \GLS{CTI} in the readout process. Other effects that need to be quantified are the various sources of background noise, variation in the star density, cosmic rays, galaxy blending, and contribution of unresolved galaxies to the background \citep{EuclidIV}. Many of these effects only become relevant because of the precision that \Euclid can achieve. Some are very specific to space-based observations where radiation damage leads to \gls{CTI}. The impact on the VIS images can be modelled following \citet{Massey+14}, 
All these effects must be accurately included in the simulation pipeline described in \cref{sec:simdata} to reach a successful calibration \citep[e.g.][]{Hoekstra+15,Hoekstra+17}. 

As already discussed in \cref{sec:weaklensing}, the shear biases can be split into a multiplicative and additive contribution (see \cref{eq:mcbias}), assuming the applied shear in the simulations is sufficiently small, or the algorithm sufficiently linear. In some cases it can be useful to isolate \gls{PSF} leakage from the additive bias, adding the term $p_{\rm leak}\,\epsilon^{\rm PSF}$ to \cref{eq:mcbias}, where 
$p_{\rm leak}$ quantifies the imperfect correction for 
the \gls{PSF} ellipticity $\epsilon^{\rm PSF}$. 

\citet{Massey+13} showed that to
reach our scientific goals (also see \cref{sec:primary}), the 
uncertainties in the multiplicative and additive shear bias must remain below $2\times10^{-3}$ and $2\times10^{-4}$, respectively, for each component of the shear.
These numbers represent the total error budget, which then needs to  be divided among several contributions
\citep{Cropper+13}. Reaching such accuracy on shear measurement requires calibrating on realistic image simulations over a wide enough area so that measurement errors on these biases are negligible compared to the target accuracy. The required number of simulated galaxies, $N_{\rm g}$, to reach a given uncertainty in the total multiplicative shear bias, $\sigma_m$, is given by \citep{FenechConti+17,Congedo2024}
\begin{equation}
    N_{\rm g} = \left( \frac{\sigma_\epsilon}{\sigma_m \, |\gamma|} \right)^2 \;,
\end{equation}
where $|\gamma|$ is the modulus of the shear applied to the simulated galaxies,\footnote{The simulations may use a different shear for each galaxy to capture the impact of the blending of  galaxies at different redshifts \citep{MacCrann22,SSLi23} or a constant shear across the entire scene \citep[e.g.]{Kannawadi+19, SSLi23}} and $\sigma_\epsilon$ is the dispersion of intrinsic ellipticities, taken to be $\sigma_\epsilon=0.26$ in the case of \Euclid \citep{EuclidIV}. However, this number can be lowered by shape \citep{mm:Massey2007} and pixel \citep{EuclidIV} noise cancellation. In the former, pairs of galaxies with $90^\circ$ rotation are considered, so that the mean intrinsic ellipticity reaches $0$ in the absence of noise, and in the latter an extra identical pair with noise of opposite sign is included to lower the impact of shot noise in the calibration process. Together, these cancellations improve the runtime of the calibration simulations by a factor of 7 in the case of \Euclid \citep{Jansen24}.

Since galaxy morphologies depend on redshift, the calibration must be performed for each tomographic bin to avoid any undesired selection effects \citep[e.g.][]{Kannawadi+19}. This process multiplies the number of nuisance parameters related to the calibration that need to be passed to the likelihood by the chosen number of tomographic slices. It also means that the shear and photometric redshift calibration must be performed jointly and that the galaxy morphology and photometry dependence on redshift must be accurately reproduced in the calibration simulations \citep{SSLi23}. Finally, the chromaticity of the VIS \gls{PSF}, combined with the \gls{SED} variation across galaxy bulges and disks, introduces a non-negligible residual bias in shear measurements. This effect is referred to as `colour gradient' bias \citep{Voigt+12}.  \citet{Semboloni+13} showed that it is possible to correct for it on average using estimates for the bias from \gls{HST} imaging in two narrow filters (F606W and F814W), which has been confirmed to be sufficient for \Euclid by \citet{Er+18}.

\subsection{Science-ready data products}
\label{sec:dataproducts}

The carefully calibrated data products that have been described in the previous sections can be used for scientific exploitation. The enhanced galaxy catalogues contain information about the redshifts and stellar masses of galaxies, as well as detailed morphological information and extensive photometry in all bands. To enable the various cosmological analyses, however, the catalogues are processed further. The cosmological information contained in the catalogues is compressed into the various two-point statistics that have been introduced in \cref{sec:primary}. In \cref{sec:clustering_stats} we summarise the steps in the calculation of the clustering statistics of the spectroscopic clustering sample, while \cref{sec:weaklensingstatistics} describes the summary statistics needed for the 3\texttimes2pt analysis. Moreover, high-resolution maps of the projected mass distribution (\cref{sec:massmap}) and a catalogue of clusters of galaxies with a well defined selection function (\cref{sec:clusterfinding}) are provided.  

\subsubsection{Spectroscopic clustering statistics} 
\label{sec:clustering_stats}

As discussed in \cref{sec:galaxyclustering}, the statistical properties of the galaxy distribution can be quantified in terms of a set of clustering moments, which compress the cosmological and astrophysical information from galaxy surveys. The most fundamental measures of galaxy clustering are two-point statistics, which fully characterise the fluctuations in a Gaussian density field. 

The starting point is the spectroscopic galaxy catalogue, described in \cref{sec:gcmeasurement}. Since the spectroscopic measurements are made with a slitless spectrograph, it is necessary to apply a selection process to identify the galaxy sample with confident redshift measurements. The \gls{EWS} will be primarily sensitive to luminous emission line galaxies in the redshift range $0.9 < z < 1.8$ with an H$\alpha$ line flux above the nominal limit of $\num{2e-16}\,\mathrm{erg}\,\mathrm{cm}^{-2}\,\mathrm{s}^{-1}$. The catalogue will be selected to maximise the galaxy number density, while keeping the fraction of spurious redshift measurements under control. It is important to account for potential systematic effects that can modulate the detection limit and remove genuine galaxies from the sample, or lead to errors in the redshift measurement that add interloper galaxies and reduce the purity. These inaccuracies in the selection have a direct impact on clustering statistics. 

The redshift purity and completeness of the sample will be evaluated using the \glspl{EDF} (\cref{sec:deepfields}), which enable a detailed assessment of the spectroscopic selection function across the \gls{EWS}. The selection function will be characterised by a random catalogue of mock, unclustered objects that is constructed to closely trace the mean density of galaxies in the \gls{EWS} and exhibit the same angular and radial selection effects. Constructing this random catalogue involves a forward-modelling process that relies on the noise level estimated in the NISP exposures and approximate end-to-end simulations of the spectroscopic measurement pipeline. 

The resulting selected galaxy catalogue, and its corresponding random catalogue, are provided as data products.  Their `reconstructed' versions are also provided; they are obtained by applying a nonlinear transformation to the observed positions of the galaxies aimed at reducing the effect of the nonlinear evolution of cosmic structures and to enhance signal-to-noise on mildly nonlinear scales.
Reconstructions can be performed either using the standard Zeldovich approximation approach 
\citep[e.g.][]{Padmanabhan2012} or with an efficient implementation of the cosmological least-action method \citep{sarpa2021}, both of which provide excellent results at the \gls{BAO} scale.

Original and reconstructed catalogues are used to compute the clustering statistics. The clustering moments can either be measured in configuration or Fourier space, and while they are theoretically equivalent, in practice their estimation can lead to different types of uncertainties and they can be differently impacted by systematic effects. To ensure accuracy, it is important to measure them in both spaces.
In configuration space, we measure the spatial \gls{2PCF} of galaxies, which quantifies the excess probability of finding two objects at a given separation in a discrete sample of mass tracers with respect to a random Poisson sample. Estimating this quantity involves counting pairs, which can be computationally intensive. Therefore, we prioritise minimising memory allocation and maximising computational efficiency in our implementation of the estimator, while ensuring accuracy within sub-percent requirements.

Apparent deviations from statistical isotropy in clustering measurements depend on the geometry of the Universe and the growth rate of cosmic structures, breaking the circular symmetry of iso-correlation contours. To quantify these effects accurately, our estimator accounts for deviations by decomposing pair separation vectors into polar or Cartesian coordinate systems, which define the transverse and parallel components to the \gls{LOS} within a local plane-parallel approximation.
To achieve computational efficiency and minimise redundancy, we implement two parallel pair-counting algorithms, namely `chain-mesh' and `kd-tree' techniques, both exact in counting all object pairs within a predefined separation range and with varying performance levels in \gls{CPU} time and memory usage at different scales \citep{marulli2016}. Our estimator also incorporates the random-splitting technique to enhance computational speed \citep{keihanen2019} in the validation tests. This technique involves splitting the random sample, which typically contributes significantly to the computational load, into smaller subsets. The number of subsets can be determined by the user, finding a balance between computational efficiency and accuracy needs.
The outputs of our estimator include the first five Legendre multipole moments,
see \cref{eq:xi_l}, of the \gls{2PCF}, the anisotropic, 2-dimensional \gls{2PCF} in polar and Cartesian coordinates, the projected \gls{2PCF}, and data and random pair counts. 
Depending on the option selected by the user, the estimator computes the auto-correlation function of the objects in the catalogue or the cross-correlation of the objects in two different catalogues. Finally, the \gls{2PCF}s of the original and of the Zeldovich reconstructed catalogues will be computed.

Various effects render the observed distribution of galaxies non-Gaussian and thus transfer information to higher clustering moments. In order to enable extraction of that information, \Euclid will additionally produce measurements of three-point statistics. In particular, we have developed an efficient estimator for the \gls{3PCF}, which employs both the local spherical harmonics approximation technique proposed by \citet{slepian2015} and the brute-force triplet-counting approach, which is mainly used for verification purposes. The three-point estimator takes the same input as the two-point estimator, but produces different outputs, namely the connected and reduced 3PCFs for all triangle configurations, as well as the corresponding data and random triplet counts.

The Fourier-space analogue of the two- and three-point correlation functions are the power spectrum and bispectrum, respectively. Their implementation within \Euclid follows the methodology of the standard \acrlong{FKP}\glsunset{FKP} \citep[FKP;][]{FelKaiPea9405} estimator, which measures correlations of Fourier modes from a weighted fluctuation field. This field is constructed from the difference between the galaxy and random catalogues (see paragraph above), multiplied by weights, which are chosen such that they minimise the variance of the estimator and depend on the number density of the selected galaxy sample. The fluctuation field is then smoothed onto a regular grid that is sufficiently large to encompass the full galaxy catalogue, which enables the use of \glspl{FFT}.

In the presence of \glspl{RSD}, the clustering moments acquire a dependence on the \gls{LOS} towards a given pair or triplet of galaxies. Measurements of the power spectrum and bispectrum must appropriately account for the variation of the \gls{LOS} over the extent of the survey footprint, which led \citet{YamNakMas0602} to derive an extension of the \gls{FKP} estimator \citep[see also][]{SzaMatLan9805,YooSel1502,Sco1510}. However, since the Yamamoto estimator presents a computational challenge for data sets as large as those produced by \Euclid, in practice one approximates the \gls{LOS} by that of one of the galaxies in the pair or triplet. With this approximation it is still possible to extract the anisotropic signal using \glspl{FFT} \citep{Sco1510,BiaGilRug1510}, making the computation efficient, while consequences of this choice of \gls{LOS} can be addressed at the level of the theoretical models \citep[e.g.][]{BeuCasZha1903}. The implementation for \Euclid allows us to measure the first five Legendre multipoles of the power spectrum, 
\cref{eq:Pelldef}, in addition to the monopole, quadrupole, and hexadecapole of the bispectrum.

The clustering two-point statistics moments estimated in Fourier space carry the imprint of the survey selection function (also called the survey window function), which must be accounted for when fitting theoretical models to the measurements. The required correction of the theoretical templates can be computed from the power spectrum of the random catalogue, which is therefore a separate data product of the \Euclid pipeline.

\subsubsection{Photometric 3\texttimes2pt statistics} \label{sec:weaklensingstatistics}

\begin{figure*}%
    \centering%
    \includegraphics[width=0.9\textwidth]{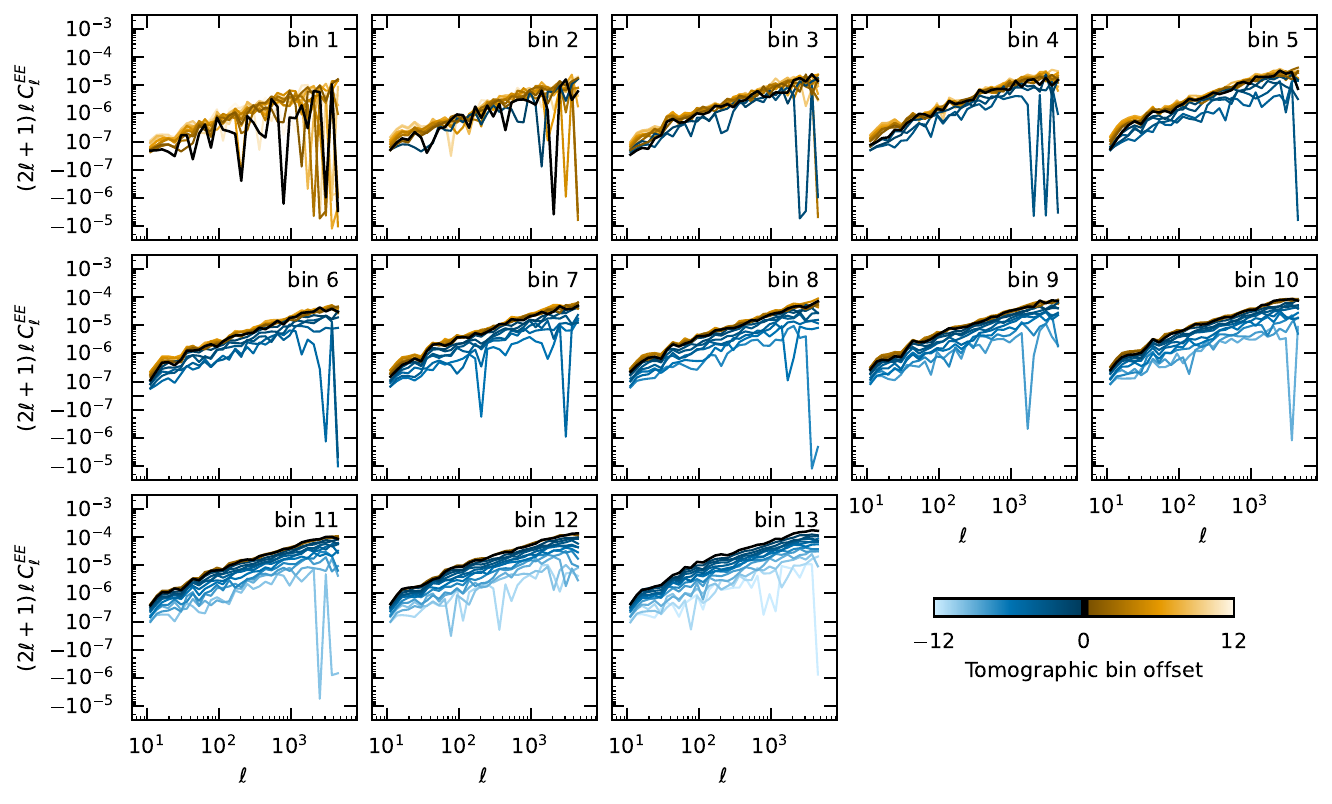}%
    \caption{%
        Measured $E$-mode angular power spectra for cosmic shear from the galaxy
        ellipticity in the Flagship simulation, after applying the
        expected survey footprint of the northern part of \Euclid's first data
        release (DR1).  Shown for each tomographic redshift bin (numbered
        panels) are the cosmic shear signal of that bin (black) and the
        cross-correlations with both lower-numbered (blue) and
        higher-numbered (orange) bins.  The shading of the colour
        indicates the difference between the two bin numbers.
        Galaxy ellipticities have intrinsic ellipticity variations (`shape
        noise'), but no shape measurement error has been added here.  Spectra
        are binned into 32 logarithmic bins.  The $y$-axis changes to linear
        when crossing zero.
    }%
    \label{fig:le3-pk-wl_shearshear}%
\end{figure*}

\begin{figure*}%
    \centering%
    \includegraphics[width=0.9\textwidth]{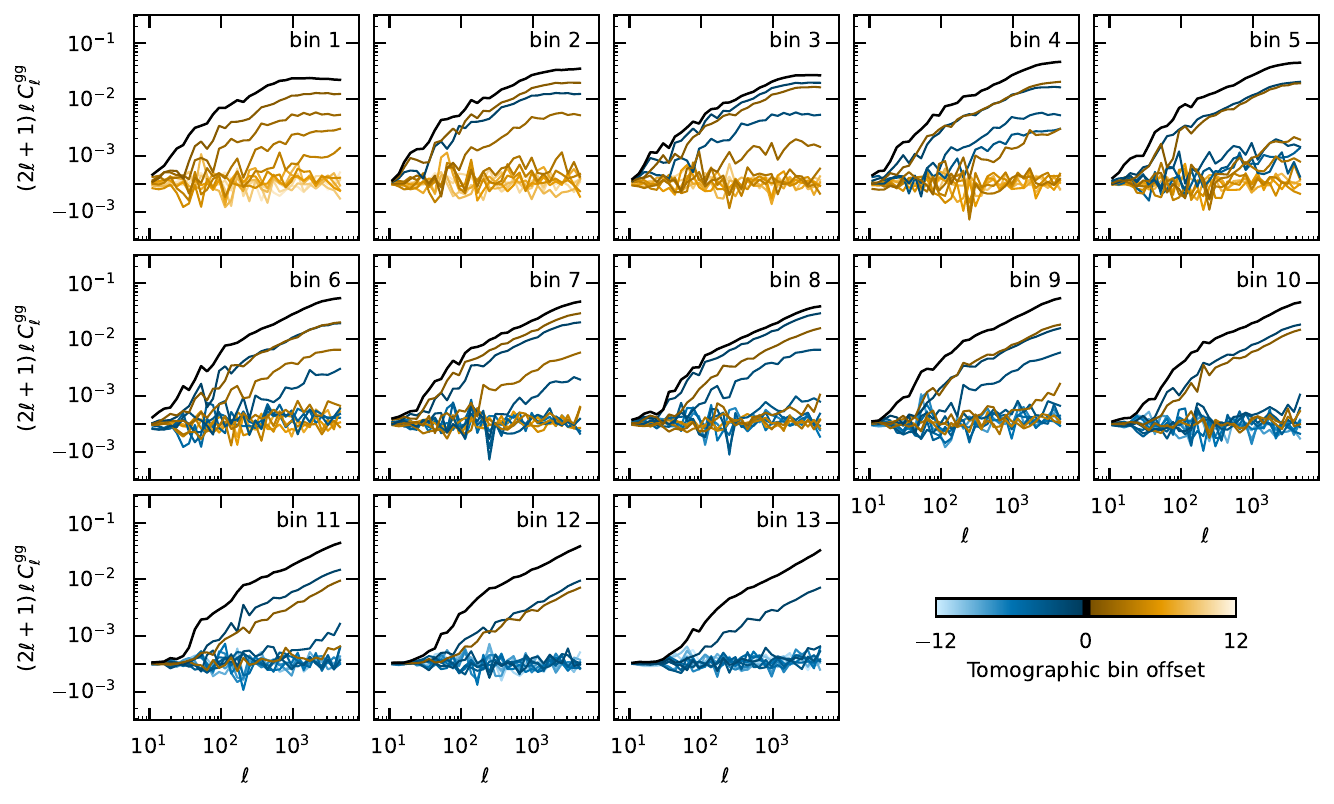}%
    \caption{%
        Similar to \cref{fig:le3-pk-wl_shearshear}, but for angular clustering
        from galaxy positions.
    }%
    \label{fig:le3-pk-wl_pospos}%
\end{figure*}

\begin{figure*}%
    \centering%
    \includegraphics[width=0.9\textwidth]{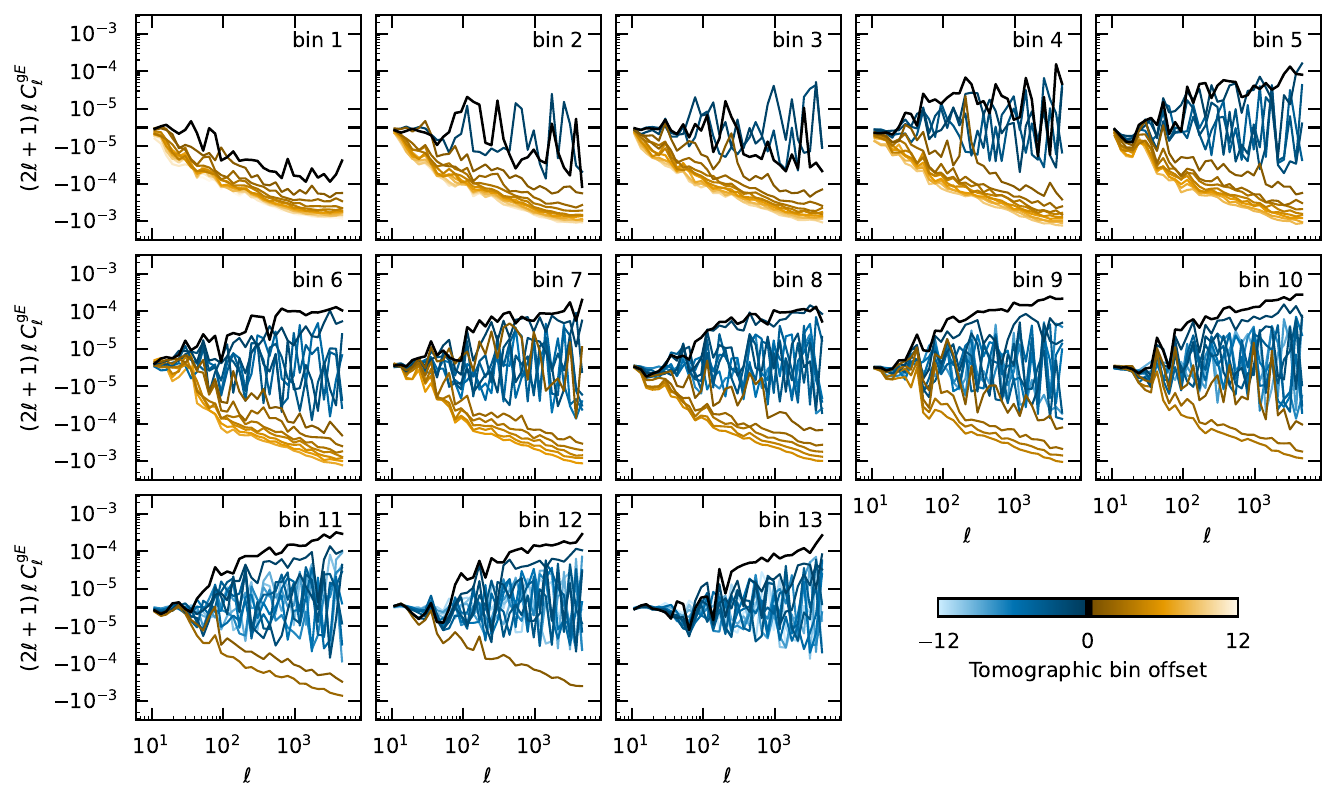}%
    \caption{%
        Similar to \cref{fig:le3-pk-wl_shearshear}, but for galaxy--galaxy
        lensing from the positions of galaxies and their ellipticity $E$-mode.  Here,
        cross-correlations in each panel are shown for positions in that bin
        and foreground or background ellipticities.  In harmonic space, the
        galaxy-galaxy lensing signal is negative; the apparent positive signal
        at higher redshifts is due to the intrinsic alignment of galaxies.
    }%
    \label{fig:le3-pk-wl_posshear}%
\end{figure*}

\textit{Euclid}’s photometric galaxy samples are determined by their detection in the \IE\ filter. Nominally, this includes all galaxies with $\IE<24.5$
($10\,\sigma$, extended source) and with valid photometry in all \gls{NIR} and
all available ground-based bands. The resulting samples and their key
properties, such as number density and redshift distributions, are further
modulated by statistical weights, which typically correlate strongly with S/N
in \IE. Data processing enables the full galaxy sample to be split into up to
13 tomographic bins \citep{Pocino-EP12} via point estimates for the photometric
redshifts. The bin boundaries will be placed such that the statistical
constraining power is optimised \citep[see e.g.][]{Zuntz2021,Pocino-EP12},
while also keeping systematic trends and cross-correlation in the redshift
distribution calibration to a minimum.

The galaxy samples used for shear measurement and for photometric galaxy
clustering will generally differ, due to different statistical weights and
the number and placement of tomographic bins, as well as potentially brighter
cuts in \IE\ to reduce spurious clustering patterns caused by spatially varying
selection effects.  These selection effects are not only caused by systematic
variations in the \IE\ detection efficiency, such as  background noise
level, foreground stellar density, or Galactic extinction, but also by trends
in all bands entering the photometric redshift estimation, including those from
ground-based surveys. This makes it challenging to predict or simulate the full
complement of selection effects. By default, we will therefore apply a
data-driven approach that learns systematic galaxy density variations from the
observed data, closely following methods applied to current imaging surveys
\citep{Johnston2021,RodriguesMonroy2022}. Thus, we will construct visibility
maps and, equivalently, random catalogues for each tomographic bin in the
photometric galaxy clustering sample.

As discussed in \cref{sec:weaklensing,sec:3x2pt}, we will extract cosmological
information from the 3\texttimes2pt statistics of the photometric galaxy
catalogue.  To do so, we measure the two-point correlations of:
\begin{enumerate}
    \item galaxy ellipticities, tracing cosmic shear;
    \item galaxy positions, tracing angular galaxy clustering; and
    \item cross-correlations between galaxy positions and ellipticities,
        tracing galaxy-galaxy lensing.
\end{enumerate}
The 3\texttimes2pt statistics are angular statistics, which are measured in
projection on the sphere. To obtain information from the cosmological evolution
of the galaxy sample, the 3\texttimes2pt statistics are computed for all auto-
and cross-correlations of the full set of tomographic redshift bins.

We can measure angular statistics on the sphere either in real space, as
observed, or in harmonic space, after a spherical harmonic transform of the
observations.  To maximise the scientific return of \Euclid, we will measure
the 3\texttimes2pt statistics in both real and harmonic space.  While there is
a mathematical relation such as \cref{eq: model xipm} for each angular
correlation function and its power spectrum, the transformation between real
and harmonic space requires information on all angular scales, from the full
sky to the infinitesimally small.  In practice, we therefore cannot exactly
transform one measured statistic into the other, even if both ultimately probe
the same information, and we will hence measure them separately.

The main real-space measurement consists of the following estimates of the
3\texttimes2pt angular correlation functions:
\begin{enumerate}
    \item the cosmic shear estimator $\hat{\xi}_{\pm}(\theta)$ is a weighted
        average of the observed shears \citep{Schneider2002};
    \item angular clustering is measured separately using
        estimators $\hat{w}(\theta)$ of \citet{LandySzalay1993} and
        \citet{Hamilton1993}, together with catalogues of random positions
        tracing the survey footprint and systematics;
    \item the galaxy-galaxy lensing estimators $\hat{\gamma}_{\rm t}(\theta)$
        and $\hat{\gamma}_{\times}(\theta)$ correlate galaxy positions of a `lens' sample with shear estimates from a `source' sample (see \citealt{Joachimi2021,Prat2022} for recent discussions of these estimators).
\end{enumerate}
Each real-space estimate is computed in linear or logarithmic bins of angular
separation $\theta$, and for all combinations of tomographic redshift bins.  A
secondary set of derived two-point statistics is also measured, consisting of
band-power spectra \citep{Schneider2002} and \acrlong{COSEBIs} \glsunset{COSEBIs} \citep[\gls{COSEBIs};][]{Schneider2010},
which are different linear combinations of the primary correlation function
measurements.

The harmonic-space measurement consists of the following angular power spectra:
\begin{enumerate}
    \item the cosmic shear angular power spectrum from pairs of galaxy
        ellipticity maps;
    \item the angular clustering power spectrum from pairs of galaxy number
        density maps, together with visibility maps accounting for the survey
        footprint and systematics;
    \item the clustering-shear cross-correlation power spectrum from
        combinations of galaxy ellipticity and number density maps.
\end{enumerate}
These spectra are measured from partial-sky data that are observed only within
the \Euclid survey footprint, which imprints the respective angular selection
function of each tomographic bin on the measurements
\citep{2005MNRAS.360.1262B}.  To account for this effect, every harmonic-space
2-point measurement comes with a so-called mixing matrix, which imprints the
same angular selection of the survey on the full-sky angular power spectra
obtained from theory.

Error estimates for the measured two-point statistics, in both real and harmonic
space, are provided by a delete-one jackknife computation, where the available
survey area is divided into a set of smaller regions, and all measurements are
repeated while leaving out each region in turn.  The resulting sample is then
used to estimate the covariance of each set of 3\texttimes2pt statistics.
Finally, all two-point measurements are repeated for all shape measurement
methods (\cref{sec:shapemeasurement}) that are used in a particular data
release.

To illustrate the wealth of data that \Euclid's 3\texttimes2pt statistics will
deliver, we measure a synthetic harmonic-space data vector from the \gls{EFS}.
Galaxies from the simulation are selected using an approximate footprint for
the northern part of \Euclid's first data release (DR1), and binned into 13
photometric redshift bins with equal number density.  The results are shown in
\cref{fig:le3-pk-wl_shearshear,fig:le3-pk-wl_posshear,fig:le3-pk-wl_pospos}.

\subsubsection{Weak lensing convergence maps}  
\label{sec:massmap}

The spatial correlations in the galaxy shapes provide direct information on the projected mass distribution along the \gls{LOS}: in the weak lensing regime, the observed shear as a function of position can be used to reconstruct the corresponding convergence field.
The result can be compared to the distribution of luminous matter, but the main use of these maps is to compute a range of statistics that provide additional cosmological information, beyond the two-point statistics discussed so far. These include the one-point probability distribution function, peak counts, and Minkowski functionals \citep[for an extensive overview of possible estimators, see][]{Ajani-EP29}.
Hence, convergence or mass maps have become a standard product of weak lensing surveys
\citep[e.g][]{mm:Massey2007, mm:VanWaerbeke2013, mm:shan2014, mm:Jeffrey2018,mm:Oguri2018}, and \Euclid is no exception.

Given the substantial survey area of nearly 14\,000\,deg$^2$, mass mapping\footnote{We refer to `mass mapping' when reconstructing the convergence maps from the observed reduced-shear field.} can be performed either in the plane by dividing the \Euclid survey into small fields, so that the flat-sky approximation remains valid, or directly on the curved sky. The latter, also known as spherical mass mapping, allows for large scales to be probed and is particularly suited for cross-correlations with other observables, such as \gls{CMB} measurements, the distribution of galaxies, or all-sky cluster catalogues.
However, the resolution of the spherical convergence maps and the complexity of the algorithm are limited by the computation time and memory required. As a consequence, planar mass mapping remains important to reconstruct convergence maps with a good resolution and precision, needed in particular to probe the non-Gaussian features of the weak lensing field, for example for higher-order statistics or to study the complex mass distribution in merging clusters \citep[e.g.][]{Clowe06, Jee12}. Since the two methods are complementary, \Euclid will provide convergence maps over the full survey area based on both approaches, as well as high-resolution maps of smaller patches around the most massive clusters of galaxies. The angular resolution of these maps will depend on the galaxy density and should not exceed 2$\arcmin$ for the wide-field and 1$\arcmin$ for the small-field maps. 

Irrespective of the approach, the reconstruction of the convergence maps from the shear is a difficult task because of shape noise, irregular sampling, complex survey geometry, and the fact that the shear is not a direct observable. 
Taking these considerations into account, the \Euclid data releases will include mass reconstructions based on two different planar inversion algorithms. The first method is the standard method proposed by \cite{mm:KaiserSquires1993}.  
This \gls{KS} method has several shortcomings, but it is nevertheless commonly used. We therefore include it to allow cross-checks with previous measurements. The second algorithm employs the nonlinear inversion method KS+, described in \cite{mm:Pires2020}. This aims at reconstructing the convergence with minimal information loss, while controlling systematic effects. Details of the algorithms and an assessment of their performance are given in \cite{mm:Pires2020}.  Extensions of these two methods to the curved sky have been implemented
 \citep[e.g.][]{Vanshika2023}, and the results of those algorithms 
are also included in the \Euclid data releases. A key application of the results is the study of higher-order statistics, which complement the constraints from the two-point correlation functions (see \cref{sec:additional}). For instance, \citet{Ajani-EP29} showed that various higher-order weak lensing statistics can tighten constraints on $\Omega_\mathrm{m}$ and $\sigma_8$ by a factor of about 2, with prospects for further improvement.

\subsubsection{Catalogue of clusters of galaxies} 
\label{sec:clusterfinding}

Another way to study the peaks in the matter distribution is to identify overdensities of galaxies. Such clusters of galaxies represent an extreme environment, affecting the star-formation histories and morphological properties of galaxies \citep[e.g.][]{Dressler80}, and as such can be used to test 
models of galaxy formation. Moreover, their abundance as a function of mass and redshift is sensitive to the underlying cosmological model \citep[e.g.][also see \cref{sec:clusterofgalaxies}]{Allen2011,Kravtsov2012}. \Euclid probes a large cosmological volume, and thanks to the deep \gls{NIR} imaging it will extend the redshift range for cluster studies considerably, thus complementing large surveys that use the Sunyaev-Zeldovich effect \citep{planck2014-a36,Bocquet2019,Hilton2021} and X-ray emission \citep[e.g.][]{Liu2022,Bulbul2024} to detect and study them.

The investigation of individual clusters, such as the interacting cluster 1E0657$-$558 \citep[also known as the Bullet Cluster;][]{Clowe06}, can provide useful insights, but most applications involve large samples that are used in statistical analyses. To correctly interpret the results, understanding how the sample is established is paramount. Moreover, the efficacy of detecting clusters in a multi-band imaging survey depends on a wide number of parameters in the algorithms employed.

\citet{Adam-EP3} explored this issue in detail by comparing the performance of various cluster-finding algorithms using synthetic \Euclid data. The main criteria were their performance in terms of sample purity and completeness. This resulted in the selection of two algorithms that will be used to generate the catalogues of galaxy clusters. We describe their main features below. Importantly, the two codes employ rather different approaches to detect clusters, thus enabling a useful internal cross-check on any potential systematics associated with cluster detection. 

The Adaptive Matched Identifier of Clustered Objects 
\citep[\texttt{AMICO};][]{maturi2005,bellagamba2018,maturi2019} employs an optimised matched-filter algorithm that can be trained directly upon the survey data. The baseline cluster model incorporates a \gls{NFW} radial density profile \citep[][]{NFW1997} and a Schechter \gls{LF} for the cluster members, although different models can be used as well. Clusters are iteratively identified within a 3D (angular position and redshift) significance map. This algorithm has already been used to find clusters in 
KiDS \citep{maturi2019} and the Javalambre-Physics of the Accelerating Universe Astrophysical Survey \citep[miniJ-PAS;][]{gonzalezDelago2022,maturi23}.

The second algorithm, \texttt{PZWav} makes minimal assumptions about expected cluster properties. This code uses a difference-of-Gaussians smoothing kernel to detect galaxy overdensities on the physical scale of galaxy clusters. Clusters are identified as statistically significant overdensities within a 3D data cube, with angular pixels with sizes of roughly $12^{\prime\prime}$  and bin widths in redshift that are appropriate for the photometric redshift \glspl{PDF}. Versions of this code have been used for multiple surveys, including the 
\acrlong{ISCS}\glsunset{ISCS} \citep[\gls{ISCS};][]{eisenhardt2008}, and a search in the S-PLUS fields \citep{werner2023}.

Both codes have been validated and
tests on simulated data indicate that they will yield samples with  high purity and completeness \citep[e.g.][]{Adam-EP3}. We recently used the
Flagship simulation (version 2.1.10; \cref{sec:flagship}), processed with \texttt{Phosphoros}, plus \texttt{AMICO} and \texttt{PZWav}. 
The analysis accounted for the uncertainties associated with the photometry, the photometric redshifts, the efficiency of the detection algorithms, and the intrinsic scatter of the cluster astrophysical properties. Based on these findings, we expect that \Euclid will detect
about $10^6$ clusters within the redshift range $0.1<z<2.0$, of which 30\% are expected to lie at $z\geq1$.
The final sample selection will be based on the purity level estimated with tailored numerical simulations and the data-driven approach implemented in \texttt{SinFoniA} \citep{maturi2019}. These results are in agreement with the earlier forecasts presented in \cite{Sartoris2016}.

The pipeline for cluster finding will yield a merged cluster catalogue containing detections from both the \texttt{AMICO} and \texttt{PZWav} algorithms, which will be matched using both geometrical and membership-based procedures. Information from the detection (e.g. cluster position and significance) will be augmented with additional information from subsequent analyses for a detailed optical characterisation of the sample. This includes refined photometric and spectroscopic redshift estimates, as well as richness and membership estimates based upon the approaches of \citet{Castignani2016} and \citet{Andreon2016}. For a subset of clusters for which measurements are possible, the catalogue will also include weak lensing masses, cluster density profiles, luminosity functions, and velocity dispersions. 
Additionally, the size and the sensitivity of the \gls{EWS} will allow the construction of a shear-selected cluster catalogue \citep[see, e.g.][]{Schneider96,Wittman01,Schirmer07} with more than $10^4$ clusters.

\subsection{Data releases}
\label{sec:DR}
 
The \gls{SGS} processes \Euclid data as soon as they become available. The data acquired on board are downloaded to Earth once a day 
(\cref{sec:downlink}) and received at the \gls{MOC}, located at \gls{ESOC} in Darmstadt, Germany. It is then transferred to the \gls{SOC} at \gls{ESAC} in Spain, where the raw instrument files, together with the telemetry of the spacecraft are turned into processable \gls{FITS} files, which are then ingested in the central \gls{SGS} processing archive. The subsequent processing of the data is distributed among the 9 \glspl{SDC} which all run the same processing pipeline, but on different parts of the data. 

This daily `on-the-fly' processing uses the latest version of the pipeline and calibration products, and is used to monitor the health of the instrument and the quality of the data product produced by each of the processing steps. Once validated, these processed data are published in the \href{https://eas.esac.esa.int/sas/}{ESA Science Archive}. From that moment, these data are available to all members of the \gls{EC}. In principle, these data can be used for scientific investigation, but special care has to be taken, because 
the calibrations and processing functions continue to improve. Although such upgrades are integrated into the pipeline in a controlled manner, over time the data become increasing inhomogeneous. 
This may not pose a problem for many applications, but it can prevent robust high-level tests that are needed to validate the data for further cosmological analysis. To this end, about 500\,deg$^2$ of \gls{EWS} data are regularly reprocessed to ensure homogeneity and all data products are made available without blinding the summary statistics (see \cref{sec:blinding} about our blinding strategy). 

\begin{figure}[ht]
\centering
\includegraphics[width=0.45\textwidth]{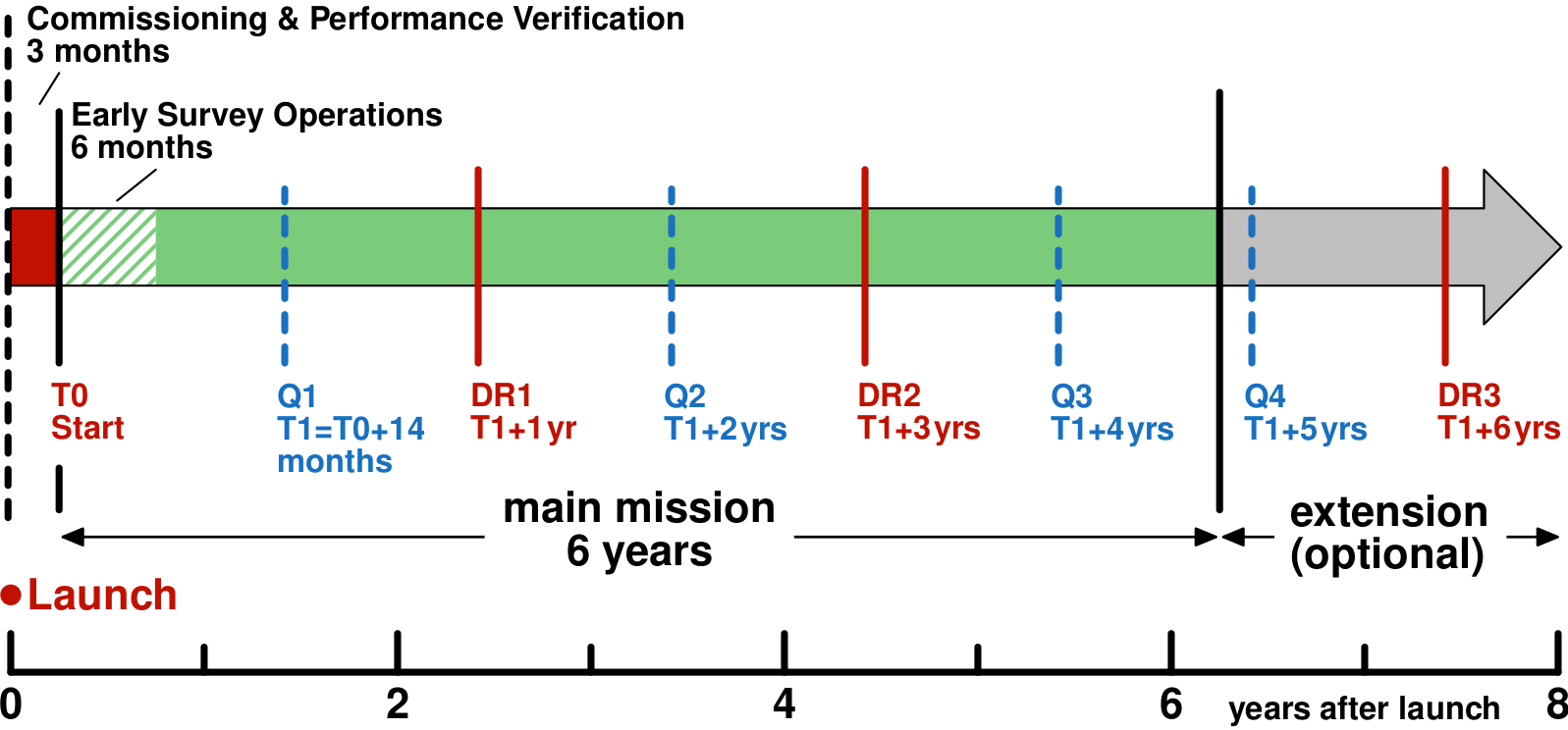}
\caption{Tentative timeline for public data releases, indicating the three main \glspl{DR} as well as four smaller quick releases (Q1--Q4). The moment of release is linked to the start of early survey operations, but unforeseen changes to the mission operation may lead to some changes to this nominal schedule.}
\label{fig:timelineDR}
\end{figure}

The data processing is also homogenised for each of the three major public \glspl{DR}. All \glspl{DR} will occur through the \href{https://eas.esac.esa.int/sas/}{ESA Science Archive}, which is already on-line and currently gives access to the Early Release Observations data, presented in accompanying papers.
A tentative release schedule, relative to the start of early survey operations, is shown in \cref{fig:timelineDR}. The three \glspl{DR} will contain the complete \Euclid data set for roughly the first year of the survey, the first three years of the survey, and the complete survey, respectively. Each \gls{DR} will comprise all the \gls{EWS} data products detailed in the preceding sections, similar products obtained on the \gls{EDS} during the period covered by the \gls{DR}, as well as the associated calibration products.
The \gls{SGS} software version used for the \gls{DR} will also be made public at the release time.

As \cref{fig:timelineDR} shows, the \glspl{DR} are interspersed with so-called quick releases (Q) of smaller volume. At the time of writing, only the contents of Q1 have been planned. This release comprises data for a single
visit over the \glspl{EDF}: 20\,deg$^{2}$ of the EDF-N, 10\,deg$^{2}$ of EDF-F, and 23\,deg$^{2}$ of the EDF-S. We aim to release the imaging and spectroscopic data, as well as catalogues and photometric redshifts. 


\section{Constraining cosmology with the primary probes}
\label{sec:cosmology}

A major development in cosmology since the publication of \cite{Laureijs11} and the selection of \Euclid has been the release of the results from the analysis of \Planck data, resulting in cosmological parameter constraints with high precision \citep{PlanckParams2013,PlanckParams2015,PlanckParams2018}. Moreover, results from other cosmological probes have also continued to improve. As a result, the parameters that describe the concordance $\Lambda$CDM model are now well constrained, even though the physical nature of the main ingredients is yet to be explained. 
Interestingly, in recent years, some inconsistencies between probes have been claimed. 
For instance, local measurements of the Hubble constant 
\citep{Riess21} differ from the preferred value reported by \cite{PlanckParams2018}. Similarly, weak lensing studies find a lower amplitude for the lumpiness of matter, quantified by $S_8$ \citep[e.g.][]{Asgari2021}. Whatever the origin of these differences, it is clear that the much smaller statistical uncertainties and reduced systematic effects of the \Euclid results will have a significant impact in resolving the current debate and defining a new standard model of cosmology.

To exploit the impressive statistical power of \Euclid, it is essential that all sources of systematic errors, whether of instrumental, astrophysical, or theoretical origin, are properly identified and their impact fully assessed. This includes assessing the choice of observables, but also the determination of the likelihood and exploring options for the modelling of various systematic effects. Doing so requires a comprehensive analysis of the performance of the mission, from the pixel-level data to the cosmological inference.

In \cref{sec:endtoend} we present the general methodology that has been adopted to assess the impact of systematic errors and describe some of the tools that have been developed to ensure that cosmological constraints derived from \Euclid are robust. These efforts are supported by extensive sanity and consistency checks of the data, which in turn drive some of the calibration needs. In \cref{sec:forecast} we summarise the work done to date to ensure robust parameter estimation, and provide updated predictions for the primary probes. In \cref{sec:beyondlcdm} we explore the prospects of constraining models beyond standard $\Lambda$CDM using \Euclid.

\subsection{Science performance verification}
\label{sec:endtoend}

As discussed in  \cref{sec:primary}, the science objectives that drove the design of the project are to derive precise constraints on the parameters $w_0$ and $w_a$, which allow us to probe the dynamical nature of dark energy, and to derive constraints on the growth-related $\gamma_\mathrm{g}$ parameter. The performance of \Euclid for the former is quantified by the \gls{FOM} defined by \cref{eq:FOM}. The choices for observables and requirements on residual systematic effects all derive from these core objectives. 

It is often not clear how  experimental design choices are related to changes in the precision and accuracy with which cosmological parameters can be determined. A first attempt was presented in \cite{Laureijs11} in order to establish the feasibility of the mission. Since then, new insights have led to additional sources of bias that need to be accounted for, while others may now be better understood. Also, as design choices are made, requirements should be updated, as values derived from generic considerations tend to be conservative \citep[for example, see the discussion in][for the weak lensing case]{Paykari-EP6}.

To quantify how the inferred values of cosmological parameters from a set of observables depend on the adopted underlying cosmological model, uncertainties in our understanding of astrophysical sources of bias, or imperfections in the calibration of the instruments, we have developed tools that enable an end-to-end analysis.
This allows us to link changes in the mission design to biases in the cosmological parameters of interest, which in turn helps to prioritise their importance, and to derive requirements for specific steps in the analysis pipelines. This is the main objective of the \gls{SPV} exercise.

The basis for this \gls{SPV} are the \glspl{EFS}, which provide a sample of realistic galaxies in the redshift range of interest, for both the \gls{EWS} and \gls{EDS}. As described in \cref{sec:flagship}, there are multiple purposes of such simulations. In the context of \gls{SPV}, they are first used to define the set of fiducial values of the reference \Euclid model, resulting in a common input catalogue for realistic synthetic observations. As discussed in \cref{sec:simdata}, such simulated data are used to validate the performance of the elements of the pipeline in terms of purity, redshift precision, and photometric performance, as well as evaluating residuals in shape measurements. Such \gls{SPV} studies also allow us to take global systematic effects into account. For instance, the star density and Galactic extinction varies across the survey. Similarly, radiation damage or ice build-up introduce large-scale variations that need to be quantified. 
 
The performance of individual pipeline elements can be used to approximate the full end-to-end by a series of catalogue-level operations. Although this may not capture all co-dependencies, it provides a fast way to produce catalogues for large areas, while capturing the various sources of bias in a realistic fashion. As an example, biases in shape measurement are introduced by instrumental effects on the scale of the galaxy image. Hence, it is not necessary to create simulated images for a full \Euclid survey. Instead, determining the biases as a function of relevant properties for a representative sample is sufficient. These dependencies can then be used to propagate biases at the catalogue level. For instance, \cite{Paykari-EP6} used this approach to explore the impact of spatially varying \gls{PSF} and \gls{CTI} residuals for the \Euclid weak lensing measurements.

\subsubsection{Reference observables and nuisance parameters}\label{sec:spv3}

\begin{table*}[ht!]
\caption{Reference values and prior probability distributions for the cosmological and nuisance parameters of the $w_0 w_{\rm a}$CDM and $\Lambda$CDM + $\gamma_{\rm g}$ models (adopting a flat geometry). 
}
\centering
\begin{tabularx}{\textwidth}{llXX}
\multicolumn{2}{l}{Parameters} & Fiducial value & Prior \\
\hline
\hline
\multicolumn{4}{c}{Cosmology} \\
\hline
Dimensionless Hubble constant & $h$ & $0.6737$ & $\mathcal U(0.55, 0.91)$ \\
Present-day physical baryon density & $\Omega_\mathrm{b} h^2$ &  $0.0227$ & $\mathcal N(0.0227, 0.00038)$\\
Present-day physical cold dark matter density & $\Omega_\mathrm{c} h^2$ & $0.1219$ & $\mathcal U(0.01, 0.37)$\\
Dark energy equation-of-state parameters & $\{w_0,w_a\}$ & $\{-1,0\}$ & $\{\mathcal U(-3.0, -0.5),$ $\mathcal U(-3.0, 3.0)\}$\\
Slope primordial curvature power spectrum  & $n_\mathrm{s}$ & $0.966$ & $\mathcal U(0.87, 1.07)$ \\
Amplitude of the primordial curvature power spectrum  & $\ln{10^{10}\,A_{\rm s}}$ & 3.04 & $\mathcal U(1.6, 3.9)$ \\
Growth index & $\gamma_\mathrm{g}$ & $0.545$ & $\mathcal U(0.01, 1.1)$ \\
Baryonic feedback efficiency factor of the \texttt{HMCode} emulator & $\log_{10}(T_{\rm AGN}/\mathrm{K})$ & $7.75$ & $\mathcal N(7.75, 0.17825)$ \\
\hline
\hline
\multicolumn{4}{c}{Photometric sample} \\
\hline
Amplitude of intrinsic alignments & $A_{\rm IA}$&  $0.16$ & $\mathcal U(-2, 2)$\\
Power-law slope of intrinsic alignment redshift evolution & $\eta_{\rm IA}$&  $1.66$ & $\mathcal U(0.0, 3.0)$\\
Coefficients of cubic polynomial for clustering bias & $b_{\mathrm{gal}, i=0\ldots3}$ & $\{1.33291,$ $-0.72414,$ $1.01830,$ $-0.14913\}$ & $\mathcal U(-3, 3)$\\
Coefficients of cubic polynomial for magnification bias & $b_{\mathrm{mag}, i=0\ldots3}$ & $\{-1.50685,$ $1.35034,$ $0.08321,$ $0.04279\}$ & $\mathcal U(-3, 3)$\\
Per-bin shear multiplicative bias$^\ddag$ & $m_{i=1\ldots13}$ & $0.0$ & $\mathcal N(0.0, 0.0005)$\\
Per-bin mean redshift shift  & $\Delta z_{i=1\ldots13}$ & $\{-0.025749,$
 $0.022716,$
 $-0.026032,$
 $0.012594,$
 $0.019285,$
 $0.008326,$
 $0.038207,$
 $0.002732,$
 $0.034066,$
 $0.049479,$
 $0.066490,$
 $0.000815,$
 $0.049070\}$ & $\mathcal N\left[z_i^{\rm fid}, 0.002\,(1+z_i^{\rm fid})\right]$\\
\hline
\hline
\multicolumn{4}{c}{Spectroscopic sample} \\
\hline
Per-bin linear bias & $b_{1,i=1\ldots4}$& $\{1.412,$ $1.769,$ $2.039,$ $2.496\}$ & $\mathcal U(1.0, 3.0)$\\
Per-bin second-order bias & $b_{2,i=1\ldots4}$& $\{0.695,$ $0.870,$ $1.162,$ $2.010\}$ & $\mathcal U(-5.0, 5.0)$\\
Poissonian shot noise for extra-stochastic parameters & $\alpha_{P,i=1\ldots4}$ & $\{0.056,$ $0152,$ $0.144,$ $0.309\}$ & $\mathcal U(-1.0, 2.0)$\\
Per-bin counter term for Legendre monopole & $c_{0,i=1\ldots4}$ & $\{11.603,$ $14.475,$ $15.667,$ $26.413\}$ & Fixed\\
Per-bin counter term for Legendre quadrupole & $c_{2,i=1\ldots4}$ & $\{35.986,$ $44.914,$ $43.819,$ $62.353\}$ & Fixed \\
Per-bin counter term for Legendre hexadecapole & $c_{4,i=1\ldots4}$ & $\{56.943,$ $55.443,$ $44.214,$ $42.89\}$ & Fixed \\
Per-bin purity factor (assuming Poisson distributed interlopers) & $f_{i=1\ldots4}$ & $\{0.195,$ $0.204,$ $0.306,$ $0.121\}$ & $\mathcal N\left(f_i^{\rm fid}, 0.01\right)$ \\
\hline
\end{tabularx}
\footnotesize
\tablefoot{These fiducial values are used to compute the self-generated synthetic data used in \cref{sec:constraints}. For the photometric nuisance parameters, we use a polynomial fitting formula for both the galaxy $b_{\mathrm{gal}}$ and magnification $b_{\mathrm{mag}}$ biases, whose coefficients run from $i = (0, 3)$, whereas we have a constant multiplicative bias $m_i$ and shifts in the bin redshift means $\Delta z_i$ per each of the 13 bins (fiducial values measured from EFS). For the spectroscopic nuisance parameters, we have one per each of the 4 redshift bins. The prior probability distributions are either uniform $\mathcal{U}$ or Gaussian $\mathcal{N}$.
\\
$^\ddag$ We assume a constant nuisance parameter for the multiplicative bias in each photometric redshift bin \(i=1\ldots13\), which corresponds to the \(m_0^{\rm bias}\) parameter appearing in \cref{eq:mcbias}. The other components (i.e. \(m_4^{\rm bias}\) ) are assumed to be negligible.}
\label{tab:fiducial_model}
\end{table*}

As a preparation for the scientific exploitation of \Euclid, an evaluation of its performance for the core science goals has been carried out that incorporates our current best knowledge of the mission. Here, we focus on the primary probes
(described in \cref{sec:primary}) and limit the evaluation using the observables defined in harmonic space. 
\begin{figure*}[htbp!]
\centering
\includegraphics[width=1.0\textwidth]{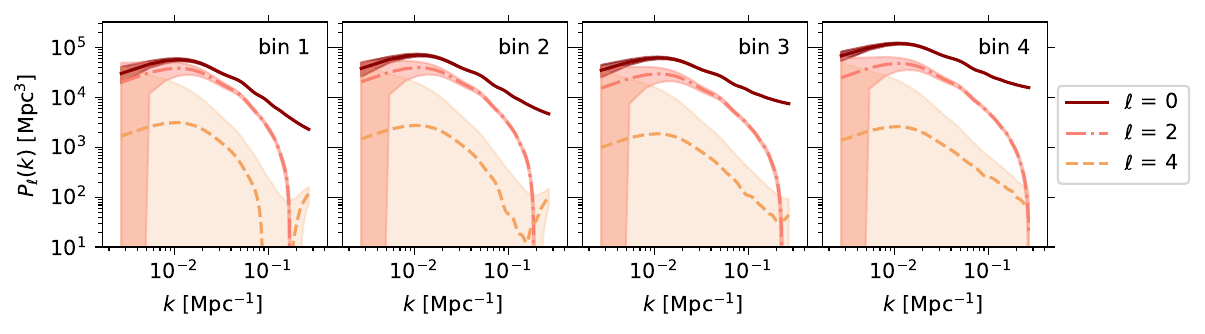}
\caption{Legendre multipoles of the redshift-space power spectrum of galaxy clustering, $P_{\ell}(k)$, as expected from the spectroscopic survey data within four redshift bins (respectively, $0.9<z<1.1$, $1.1<z<1.3$, $1.3<z<1.5$ and $1.5<z<1.8$, where the $P_{\ell}(k)$ are evaluated at the mean of the redshift intervals). The plots show the monopole ($\ell = 0$, solid line), quadrupole ($\ell = 2$, dashed-dotted line), and hexadecapole ($\ell = 4$, dashed line), together with their error corridors (shaded regions). The latter simply connect the 1-$\sigma$ errors from the diagonal values of the analytical covariance matrix, computed for narrow bins of $\Delta k = 0.0017\, h\,\mathrm{Mpc}^{-1}$. As a result of this fine binning, the shaded areas do not fully reflect the actual constraining power of the measurements.
}
\label{fig:CLOE_euclid_probes_spectro}
\end{figure*}

For the spectroscopic clustering measurements, we used the first three even multipoles of the galaxy clustering spectra $P_\ell(k)$, defined in \cref{eq:Pelldef}. As was done in \citet{Blanchard-EP7}, we used four redshift bins, with mean redshift values $z \in \{1, 1.2, 1.4, 1.65\}$ and we used the same binning in $k$.
\Cref{fig:CLOE_euclid_probes_spectro} shows the predicted signals for the four redshift bins, where the shaded indicate the uncertainties for the \gls{EWS} based on the analytical covariance matrix (\cref{sec:covmat}). Scale cuts in $k$, which need to be applied for the cosmological inference, were chosen to alleviate so-called projection effects \citep[also known as prior volume effects that shift marginalised posterior distributions;][]{moretti2023}. We used a maximum value of 
 $k_{\rm max}=0.3$$\,h\,\mathrm{Mpc}^{-1}$ in the analysis. 

\begin{figure}[ht]
\centering
\includegraphics[width=0.485\textwidth]{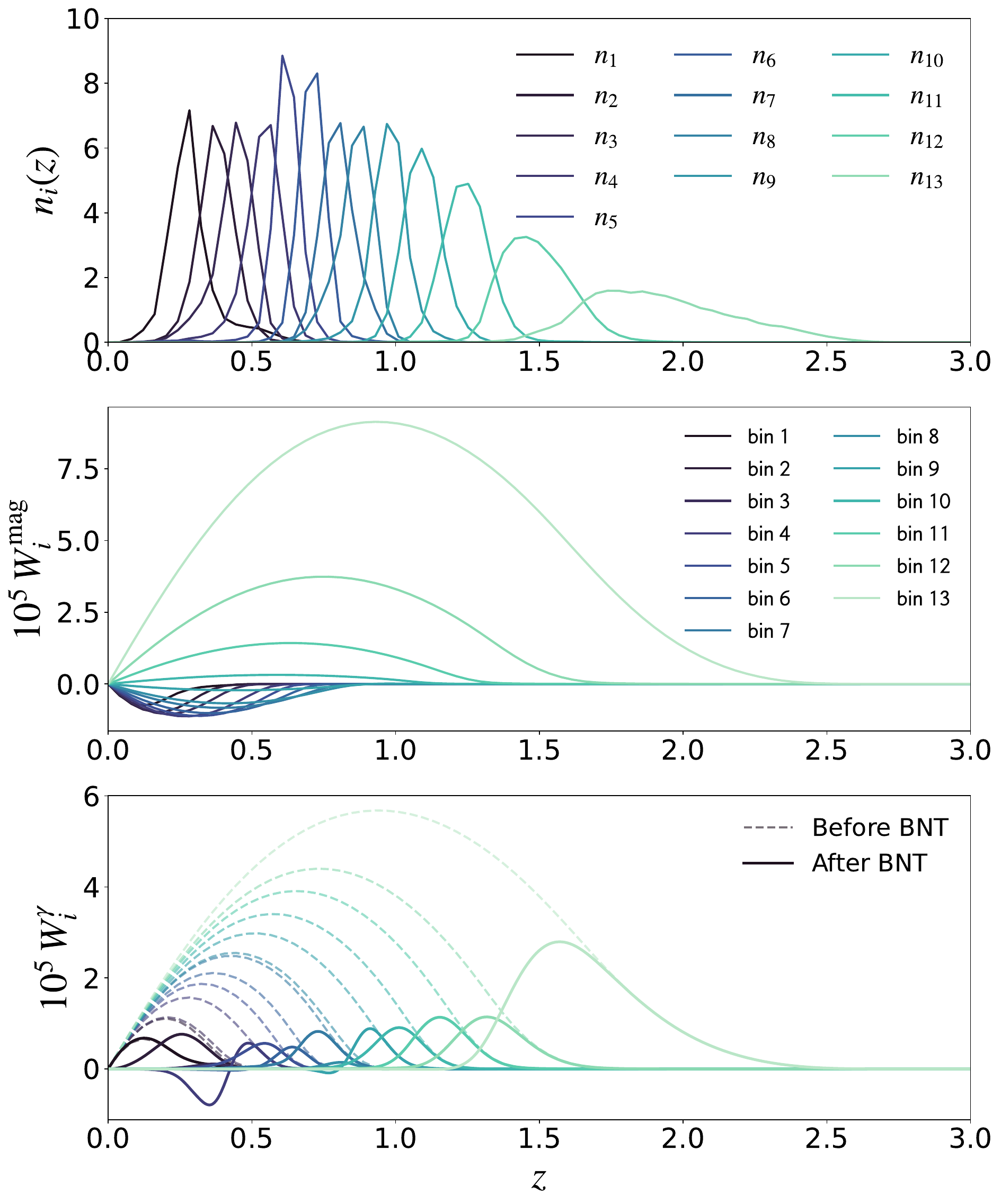}
\caption{\textit{Top}: Normalised redshift distributions $n(z)$, measured from the \gls{EFS}, for the 13 equi-populated bins that were used for the 3\texttimes2pt analysis for the \gls{SPV}. \textit{Middle}: Resulting photometric magnification kernels for the 13 redshift bins shown above.
\textit{Bottom}: Corresponding shear kernels before (dashed) and after (solid lines) BNT transformation. The latter case gives a better grasp of the tomographic information that can be inferred from \gls{WL} observations.}
\label{fig:nzbins}
\end{figure}

The 3\texttimes2pt analysis combines the auto- and cross-angular spectra $C_\ell$ for the weak-lensing and photometric galaxy-clustering probes. We use 13 equi-populated redshift bins, which are presented in \cref{fig:nzbins}. These are based on photometric redshifts derived from the \gls{EFS} assuming \gls{LSST}-like external photometric data (\cref{sec:photoz}). As such, these represent the best-case scenario, because the initial analysis will be based on shallower ground-based data. In the following, we also assume that the samples used for weak lensing measurements and photometric clustering are the same. For a more exact treatment, an estimation of the joint errors in photometric redshift and multiplicative bias would be needed.

\begin{figure*}[htbp!]
\centering
\includegraphics[width=0.95\textwidth]{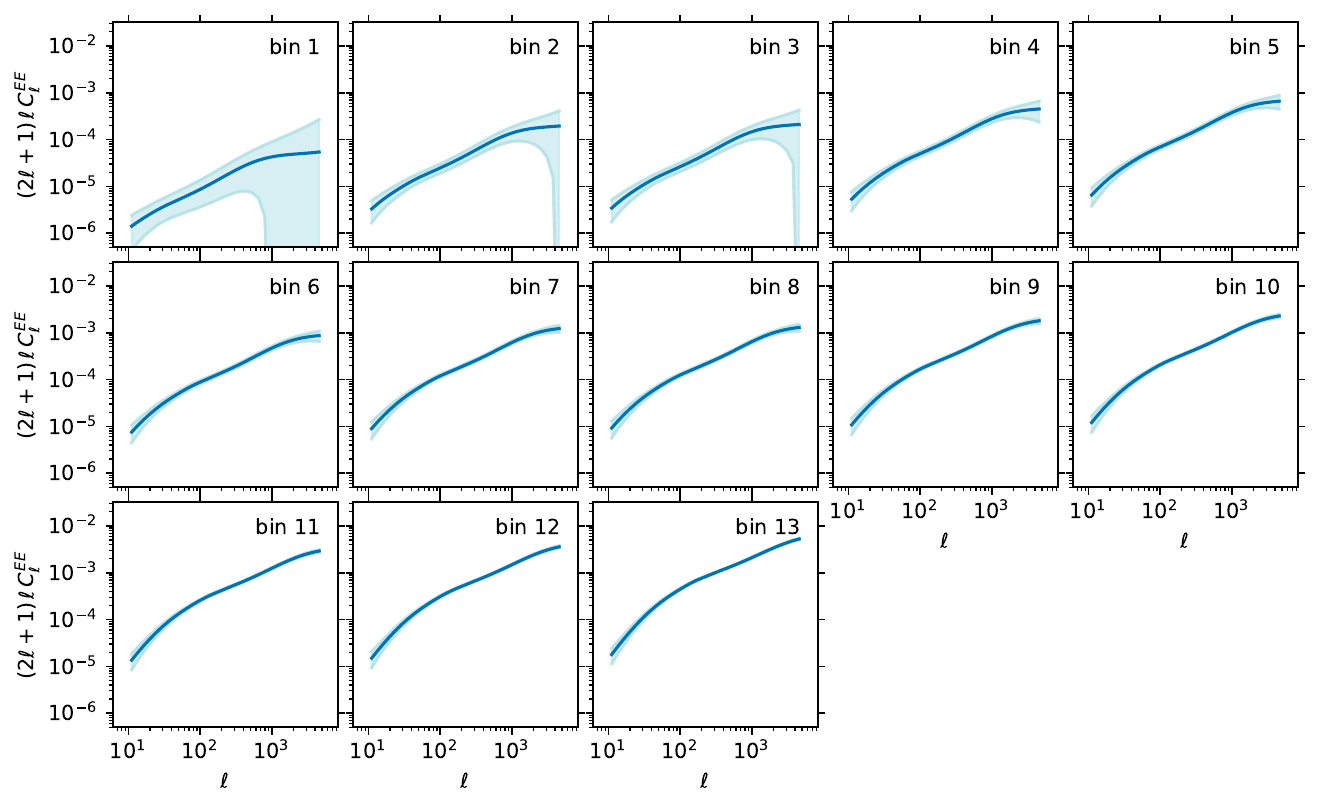}
\caption{Synthetic angular power spectra $C_\ell$ for weak lensing ($EE$) for the auto-correlation between the 13 photometric redshift bins shown in \cref{fig:nzbins}. The shaded light blue area shows the corresponding uncertainty given by the corresponding analytical covariance matrix, including the super-sample covariance (SSC) term.}
\label{fig:CLOE_euclid_probes_WL}
\end{figure*}

\begin{figure*}[htbp!]
\centering
\includegraphics[width=0.95\textwidth]{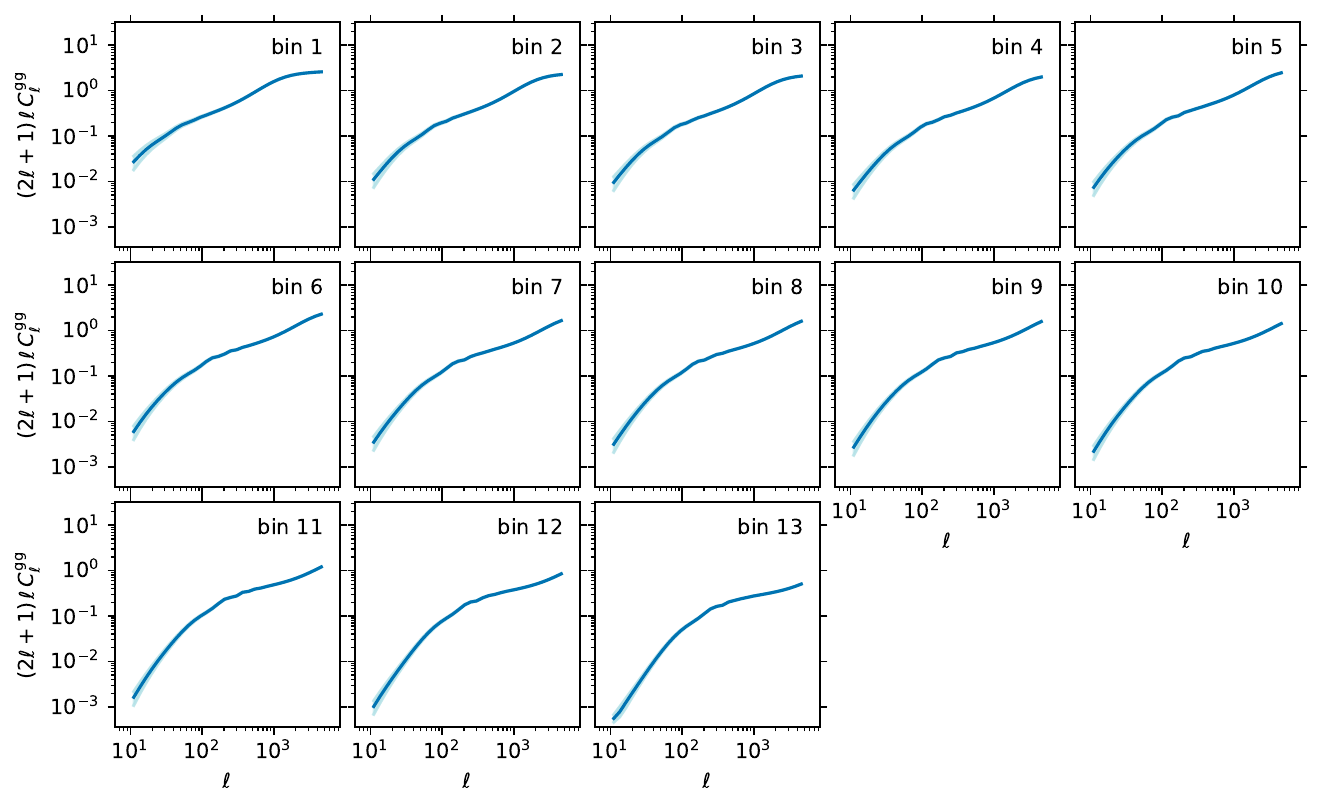}
\caption{Similar to \cref{fig:CLOE_euclid_probes_WL}, but for the photometric galaxy clustering (gg) for the auto-correlation between the 13 photometric redshift bins shown in \cref{fig:nzbins}.}
\label{fig:CLOE_euclid_probes_GC}
\end{figure*}

Although magnification does not improve cosmological constraints, we do need to account for it to avoid biased parameter estimates \citep[e.g.][]{Duncan22,Mahony22}. The middle panel of \cref{fig:nzbins} shows the magnification kernels, which highlight that  magnification is particularly important for the highest redshift bins. The bottom panel shows the shear kernels. We also show results after applying the BNT transformation \citep{
2014MNRAS.445.1526B,2018PhRvD..98h3514T}, which allows us to control the mixing of scales from \gls{LOS} projections.
\Cref{fig:CLOE_euclid_probes_WL} shows the 
synthetic angular power spectra for the weak lensing auto-correlations between the redshift bins, while \cref{fig:CLOE_euclid_probes_GC} shows the corresponding photometric clustering signals. We do not show the cross-correlations, but they are included in the \gls{SPV} analysis.
For the binning in $\ell$-space we adopted a 
set of 32 log-spaced bins, ranging from $[10, 5000]$, for the weak lensing and photometric galaxy clustering measurements.
This ensures a manageable size of the data vector, but these choices may be adjusted for the actual data analysis. Cuts in $\ell$ are implemented on the observables in harmonic space after \gls{BNT} transformation, so that $k>k_{\rm max}$ accounts for less than 20\% of each $C(\ell)$. For the 3\texttimes2pt analysis, we chose $\ell_{\rm max}=5000$ for cosmic shear and $\ell_{\rm max}=3000$ for angular clustering and galaxy-galaxy lensing.

The \gls{SPV} aims to capture the impact of biases in the data or errors in the modelling itself. In the cosmological inference our ignorance is quantified by nuisance parameters that need to be used consistently, in particular when combining probes. The nuisance parameters can describe astrophysical quantities, such as galaxy biases, or aggregated quantities, such as uncertainties in the purity factor, the mean redshift, or residual multiplicative biases.  The nuisance parameters used here are listed and described in \cref{tab:fiducial_model}.
Furthermore, careful modelling needs to take place at nonlinear scales (see \cref{sec:nonlinear} for details), to ensure that the maximum information encoded at large $k$-values is used to exploit the full \Euclid cosmological constraining power.
Consequently, a number of key assumptions have been made, regarding the impact of baryonic feedback processes and the modelling of galaxy bias, magnification bias, interlopers, and intrinsic alignments, which all will have to be revisited once data are acquired. 

We review the model predictions in \cref{sec:nonlinear}, but briefly discuss our choice for the \gls{IA} model here. For this study case, we adopt the so-called
\gls{NLA} model \citep{Bridle2007}
to describe the scale dependence of the signal, while we assume that the amplitude of the signal scales as 
\begin{equation}
    f_{\rm IA}(z)=-A_{\rm IA}\,\mathcal{C}_{\rm IA}\,\frac{\Omega_{\rm m }}{g_+(z)} 
    (1 + z)^{\eta_{\rm IA}}\label{eq: IA_function}
\end{equation}
where $g_+(z)$ is the nonlinear growth rate.
We assume that the redshift dependence is described by a power law with slope $\eta_{\rm IA}$. The amplitude is quantified by the dimensionless parameter $A_{\rm IA}$, which is scaled by the constant 
$\mathcal{C}_{\rm IA}=5\times 10^{-14}\,h^{-2}\,{M}_\odot\,{\rm Mpc}^{-3}$, whose value is determined by the amplitude measured by \cite{Brown2002} in the low-redshift SuperCOSMOS survey \citep{Hambly2001}.

\subsubsection{Data covariances}
\label{sec:covmat}

The \gls{SPV} relies on an accurate estimate of the uncertainties. Hence, attention needs to be paid to the computation of covariance matrices. 
These are also essential to capture the correlations between probes, redshift bins, and angular scales. For the results presented in \cref{sec:likelihood}, we made a number of assumptions. First of all, we adopted a Gaussian distribution for the data vector, so that its statistical properties can be entirely characterised by the covariance matrix.
Although it might be difficult to go beyond this assumption, we plan to assess its validity in future work. Second, we consider the spectroscopic clustering results to be independent of the 3\texttimes2pt measurements. This has been shown to be a reasonable approach for a \Euclid-like mission \citep[see][]{TaylorMarkovic2022}. 
Finally, we evaluated the covariance matrix only for the fiducial model. As shown in \citet{Carron13}, this is required in the Gaussian likelihood approximation to avoid the introduction of spurious parameter information. We assumed that the true model is not too far from the fiducial model (see, e.g. \citealt{Harnois19} for a discussion of the impact of this assumption on cosmic shear constraints). 

The computation of the covariance matrix of the observables is a challenging task. From a theoretical perspective, the computation of its Gaussian part is well understood. The difficulties arise from the fact that the observed modes are not statistically independent, but are coupled. One dominant source of mode coupling is the nonlinear growth of gravitational instabilities. The effect is particularly important on small scales for the observables in the 3\texttimes2pt analysis. 
The mode coupling through super-sample effects is another significant contributor \citep[see e.g.][]{Sciotti23,Beauchamps:2021fhb}. For the \gls{SPV} analysis this term alone is included since we have found that this should give a realistic estimate.
Another source of mode coupling is the impact of masks and visibility functions. When finite volumes and masks have to be taken into account, the harmonic components become coupled \citep{2005MNRAS.360.1262B}. This depends on the details of the survey, as well as the masking procedures. For now, only the size of the footprint has been taken into account in the \gls{SPV} exercise -- via a rescaling of the covariance by the sky fraction $f_{\rm sky} = \Omega_{\rm S}/4\pi$, with $\Omega_{\rm S}$ the solid angle subtended by the survey, in steradians. A more comprehensive treatment is left for future analysis.

The multi-probe Gaussian covariance is given by \citep{Blanchard-EP7}:
\begin{align}\label{eq: covgauss}
\tens{C}_{\rm G} & \left[C_{ij}^{AB}(\ell), C_{kl}^{CD}(\ell^{\prime})\right]=
\left[(2\,\ell + 1)\,f_{\rm sky} \, \Delta \ell \right]^{-1}
\delta_{\ell \ell^{\prime}}^{\rm K}  \nonumber \\
 & \times \Bigg\{ \left[C_{ik}^{AC}(\ell) + {N}_{ik}^{AC}(\ell)\right]
\left[C_{jl}^{BD}(\ell^\prime) + N_{jl}^{BD}(\ell^\prime) \right]  \nonumber \\
& + \left[C_{il}^{AD}(\ell) + N_{il}^{AD}(\ell) \right]
\left[C_{jk}^{BC}(\ell^\prime) + N_{jk}^{BC}(\ell^\prime) \right] 
\Bigg\}  \; .
\end{align}
In the above equation, the Kronecker delta $\delta_{\ell \ell^{\prime}}^{\rm K}$ enforces the aforementioned independence of the different $\ell$ modes in the absence of convolution with the mask. The noise terms $ N_{ij}^{AB}(\ell)$ are non-zero only for the auto-correlations between probes and tomographic bins:
\begin{equation}
N_{ij}^{AB}(\ell) = \left \{
\begin{array}{ll}
\displaystyle{\delta_{ij}^{\rm K}\,\sigma_{\epsilon}^2/\bar{n}_{i}^{\rm S}} & \displaystyle{A = B = {\rm L}} \;\; \\
 & \\
\displaystyle{0} & \displaystyle{A \neq B} \\
 & \\
\displaystyle{\delta_{ij}^{\rm K}/\bar{n}_{i}^{\rm L}} & \displaystyle{A = B = {\rm G}} \; , \\
\end{array}
\right . 
\label{eq: noiseps}
\end{equation}
where $\sigma_{\epsilon}^2$ is the variance of the total intrinsic ellipticity of the sources. Finally, $\bar{n}^{X}_i(z)$ are the number densities of sources ($X={\rm S}$) and lenses ($X={\rm L}$), relevant for cosmic shear and photometric galaxy clustering respectively.

\subsection{Expected cosmological parameter constraints}
\label{sec:forecast}

Comparing the \Euclid measurements to model predictions is not straightforward, owing to the small statistical uncertainties, the need to marginalise over a large number of nuisance parameters (that are needed to quantify residual systematic effects), our limited knowledge of the nonlinear evolution of structure, and the impact of astrophysical processes on the matter distribution. Moreover, an accurate covariance matrix is needed so that all correlations between measurements can be correctly accounted for.
Here, we provide an overview of the tools and procedures that have been developed in order to derive cosmological parameter estimates for \Euclid.  This updates some previously published forecasts presented in \cite{Blanchard-EP7}. 
 
\subsubsection{\texttt{CLOE}: the Cosmology Likelihood for Observables in Euclid}
\label{sec:likelihood}
\glsadd{CLOE}

Given the unprecedented precision of the \Euclid data, it is important that the comparison with theoretical predictions uses codes that have been tested rigorously. To this end, we have developed the Cosmology Likelihood for Observables in \Euclid (\texttt{CLOE}), a highly flexible modular analysis pipeline written in {\tt python3}. To ensure the fidelity of the results, the development of \texttt{CLOE} has combined the practices of continuous integration and delivery, enforcing automation in its construction, with careful unit testing and deployment of the code against similar pipelines (Euclid Consortium: Martinelli et al., in prep.).

As a baseline, \texttt{CLOE} provides the theoretical predictions for \Euclid's primary cosmological probes\footnote{The development of \texttt{CLOE} is open to the whole Euclid Consortium, allowing the merging of additional cosmological probes, such as cross-correlations with the CMB or cluster of galaxies.} for a given set of cosmological and nuisance parameters. It computes the corresponding likelihood given the measurements (\cref{sec:dataproducts}) and outputs the posterior probability distributions for the cosmological and nuisance parameters. It relies on the publicly available Boltzmann solvers {\tt CAMB} \citep{CAMB} and {\tt CLASS} \citep{CLASS} to compute the theoretical background parameters that are the foundation of the calculations of the primary observables. \texttt{CLOE} computes the predictions for the primary cosmological probes in both harmonic and real space.

A Bayesian approach is used to determine constraints on a given set of cosmological parameters, $\theta$, given the \Euclid data vector, ${\vec d}$. According to Bayes' theorem, the key ingredient in the estimation of the posterior distribution of the parameters, $P({\theta}|{\vec d}, M)$, is the likelihood function $\mathcal{L}({\vec d}|{\theta}, M)$, which describes the plausibility of a certain parameter value $\theta$, given a model $M$, after observing a particular outcome.
To sample the full posterior distributions of the cosmological parameters of interest, {\tt CLOE} can be linked to the Bayesian analysis frameworks {\tt Cobaya} \citep{Cobaya} and {\tt CosmoSIS} \citep{CosmoSIS} as external likelihoods. As a result, {\tt CLOE} can employ a large number of different \gls{MCMC} sampling algorithms, such as Metropolis-Hastings, classic nested sampling \citep[e.g. \texttt{PolyChord};][]{Polychord, Polychord2} or advanced nested samplers \citep[e.g. \texttt{Nautilus};][]{nautilus}. 

Generally, we assume that the likelihood probability distribution $\mathcal{L}(\vec{d}|{\theta}, M)$ of these measurements $\vec{d}$ and the underlying physical model $M$, given the \Euclid primary observables $\vec{t}(\theta)$, is Gaussian,\footnote{The impact of non-Gaussian terms in the likelihood has been thoroughly studied, and found to depend on the range of scales employed. For our choice of scales for the angular power spectra of photometric probes, non-Gaussian terms have been demonstrated to be negligible for \Euclid \citep{2021MNRAS.503.1999U, 2022PhRvD.105l3527H}.} with a covariance matrix $\tens{C}$ that does not depend on cosmology, so that, up to an additive constant,
\begin{equation}\label{eq:likelihood}
 -2\log {\cal{L}}({\vec{d}}|{\theta}, M) \simeq \left[{\vec{d}}-{\vec{t}}({\theta})\right]^{\sf T}\;\tens{C}^{-1}\;\left[{\vec{d}}-{\vec{t}}({\theta})\right]\;,
\end{equation}  
where $\tens{C}^{-1}$ is the inverse of the covariance matrix and $\vec{t}(\theta)$ is the theory vector constructed with the predictions for the \Euclid primary observables assuming a cosmological model $M$. \texttt{CLOE} can compute non-Gaussian terms of the likelihood distribution $\mathcal{L}({\vec{d}}|{\theta}, M)$ if the type of covariance matrix $\tens{C}$ selected for the statistical analysis is numerical, according to \citet{2016MNRAS.456L.132S} and \citet{2022MNRAS.510.3207P}. 

Specifically, {\tt CLOE} consists of a series of semi-autonomous {\tt python} modules that interface with a Bayesian statistical framework tool, such as {\tt Cobaya}, to read the relevant data vectors and covariance matrices, to compute the theory vectors (see \cref{fig:CLOE_euclid_probes_spectro,fig:CLOE_euclid_probes_WL,fig:CLOE_euclid_probes_GC}), and to calculate the likelihood. The modules that contain the relevant recipes that are needed to compute the theoretical predictions interface with another module that includes the modifications arising from nonlinear structure formation (\cref{sec:nonlinear}). In an exercise of \textit{Open Science}, {\tt CLOE} participates in the pilot study case of \href{https://datalabs.esa.int}{\texttt{ESA datalabs}\footnote{\url{https://datalabs.esa.int}}} \citep{Navarro2024} as one the selected \Euclid software pipelines to directly interface with the \Euclid science archive system in the near future. As a demonstration of the capabilities, all the \texttt{CLOE}-related figures in this paper have been computed using  \texttt{ESA datalabs} as the reference analysis framework.

\subsubsection{Nonlinear structure formation}
\label{sec:nonlinear}

A major strength of CMB measurements is that the signal for a given combination of cosmological parameters can be computed directly, because the primary CMB fluctuations are in the linear regime. This is no longer the case for \Euclid, unless the cosmological interpretation is restricted to rather large scales. To exploit the information contained in the smaller scales, nonlinear structure formation and the complexities of galaxy formation need to be taken into account. The challenge is to find an appropriate balance between the desire to minimise the statistical uncertainties, whilst ensuring that the predictions are robust. This involves defining the smallest scales that can reliably be used for a particular observable \citep[e.g.][]{Martinelli21a}. The problem is not limited to the signal itself, but also involves robustly quantifying the covariance between measurements at different scales and the combination of observables that probe common structures.
As a minimum, this implies a coherent description of the primary probes of \Euclid, ensuring that theoretical and astrophysical sources of bias are adequately accounted for. These efforts combine analytical and numerical calculations. Below, we summarise the different approaches that have been adopted to model the various summary statistics from \Euclid. We do expect further developments, especially in the implementation of emulators \citep[e.g.][]{Eggemeier:2022anw, Pellejero-Ibanez:2021tbe}. 
As a first step, efforts have focused on analytical prescriptions and emulators that have been extensively validated with simulations and used in some of the most recent data analyses \citep[e.g.][]{Bose:2020wch, Arnold:2021xtm, carrilho2023, Piga:2022mge}.

For the spectroscopic galaxy clustering probe, we need to model the nonlinear galaxy power spectrum in redshift space. We will use a 1-loop perturbation theory model, with counter-terms computed using the \gls{EFT} of \gls{LSS}
\citep{Ivanov:2019pdj, damico2020}, which allows us to predict the clustering of matter in the mildly nonlinear regime. 
To this end, we have developed a new code \citep{moretti2023} using the \texttt{FAST-PT} algorithm for the loop-integral evaluations \citep{mcewen2016, Fang:2016wcf}. The code has been validated against $N$-body simulations \citep[][also see \cref{fig:Pk-prediction}]{oddo2021, tsedrik2023} and has been used for the re-analysis of \gls{BOSS} data \citep{moretti2023, carrilho2023}. 

For the analytical covariance, the prescription of \citet{Wadekar:2019rdu} is used, including contributions from nonlinear effects, the window function, super-survey modes, and the integral constraint. As part of the analysis of \gls{BOSS} DR12, this prescription has been shown to be in excellent agreement with the covariance estimated from over 2000 numerical mocks up to $k=0.6\,h\,{\rm Mpc}^{-1}$, leading to negligible differences in cosmological parameters using either approach \citep{Wadekar:2020hax}. 

\begin{figure}
\centering
\includegraphics[angle=0,width=1.0\hsize]{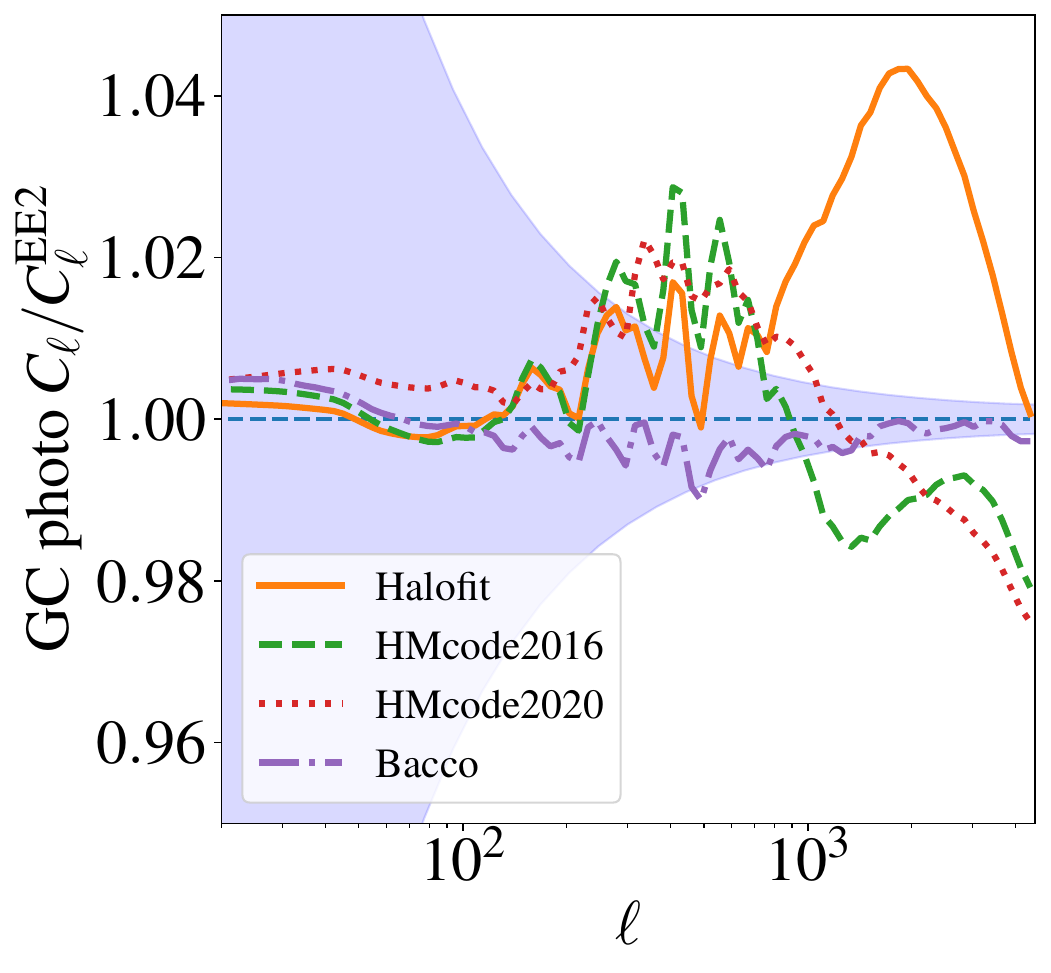}
\caption{Ratio of photometric galaxy clustering $C_\ell$ between different nonlinear models and the result for {\tt Euclid Emulator 2}, for the auto-correlation of the redshift bin centred at $z=0.83446$. Also shown is the expected \Euclid error bar, including the contribution from super-sample covariance.}
\label{fig:GCphot-nonlinear}
\end{figure}

The modelling of the 3\texttimes2pt signals presents its own challenges, because of the desire to probe the matter power spectrum on nonlinear scales, and the need to account for baryonic feedback for the shear-shear and galaxy-shear correlations \citep{Semboloni2011}. In this case, two emulators, based on $N$-body simulations, have been developed and validated. The resulting \texttt{Euclid Emulator} \citep{EE2} and the \texttt{bacco} emulator \citep{Angulo:2020vky} can predict the nonlinear power spectrum with an accuracy of about $1\%$ out to $k = 10\,h\,{\rm Mpc}^{-1}$ in the redshift range $0 < z < 3$, and include predictions for massive neutrinos and $w_0 w_a$CDM. Ongoing development of these emulators 
aims to ensure that their accuracy is sufficient for the analysis of the \Euclid data. For baryonic feedback effects, we have implemented the \texttt{BCEmu} \citep{Giri:2021qin} and \texttt{bacco} \citep{Arico21} emulators. Other popular prescriptions like \texttt{halofit} \citep{Smith:2002dz, Takahashi:2012em} and \texttt{HMCode} \citep{Mead2016, Mead2021} are also available (see \cref{fig:GCphot-nonlinear}). Ongoing simulated data challenges will determine the minimum number of baryonic feedback parameters needed for unbiased parameter inference and the associated scale cuts required, while we also plan to implement nonlinear and non-local bias models for photometric galaxy clustering with prescriptions based on either perturbation theory \citep{DES:2020yyz} or hybrid emulators for biased tracers 
\citep{Zennaro:2021bwy}.

As alluded to in \cref{sec:weaklensing}, we need to account for intrinsic alignments of galaxies \citep{Joachimi2015, Troxel15}.
To capture a wide range of possible alignments, we have implemented the tidal alignment and tidal torque (\texttt{TATT}) model \citep{Blazek2019}, used in the \gls{DES} cosmic shear analyses \citep[see e.g.][]{DES:2017qwj}. Progress will come from linking \gls{IA} models to observations, including those made by \Euclid, so that the dependence of the \gls{IA} signal on galaxy properties can be used to reduce the number of nuisance parameters \citep{Fortuna:2020vsz}.  

Finally, to consistently model the correlations between scales, tomographic bins, and the different probes, we need an accurate covariance matrix for the 3\texttimes2pt measurements. Currently, several independent codes have been studied and validated, including the \texttt{PySSC} module \citep{Lacasa:2018hqp}. Thus far, the focus has been on quantifying the impact of super-sample covariance, which is the largest of the expected non-Gaussian contributions  \citep{Barreira:2017fjz, Upham22, Beauchamps:2021fhb, Sciotti23}.

We also want to explore models that include complexity beyond the baseline $\Lambda$CDM model (see \cref{sec:beyondlcdm}). Although the range of possibilities is vast, we have focused the development on the main science objectives of \Euclid. Hence, we have included nonlinear modelling prescriptions for $w$CDM, $\gamma_\mathrm{g}$, and massive-neutrino cosmologies. Ongoing efforts include developing and implementing nonlinear models for a suite of exotic dark energy and modified gravity cases. 

\subsubsection{Parameter estimation}
\label{sec:constraints}

Since the estimates for the performance of \Euclid were presented in \cite{Laureijs11}, the fidelity of the predictions has steadily improved. In particular, \citet{Blanchard-EP7} presented results from the first collaborative analysis to verify forecasting tools. This study focused on forecasts based on Fisher matrix techniques applied to both primary probes. A key aspect involved the comparison of different numerical implementations. The results showed optimistic and pessimistic scenarios for several cosmological models (flat and non-flat, and different cuts of the nonlinear scales), highlighting the role of the cross-correlations, especially for models beyond a cosmological constant, potentially increasing the dark energy \gls{FOM} by at least threefold.  

In this section, we update the  forecasts from \citet{Blanchard-EP7} using the primary probes only (see \cref{sec:spv3} for details about our setup). We limit the discussion to the baseline models of interest: a spatially flat cosmological model, using \cref{eq:w0wa} to describe the dark energy equation of state ($w_0 w_a$CDM); and a model where the parameter $\gamma_\mathrm{g}$ is left free ($\Lambda$CDM + $\gamma_{\rm g}$). 
 
We generated synthetic noiseless data vectors by running \texttt{CLOE} v2.0.2 (Euclid Consortium: Joudaki et al., in prep.) using the fiducial values for the cosmological model parameters presented in \cref{tab:fiducial_model}. For this analysis we used the analytical super-sample covariance matrix for the 3\texttimes2pt observable \citep[see][]{Sciotti23}, and a Gaussian covariance matrix for the spectroscopic probe (GCsp), following the specifications of the \gls{SPV} exercise for the survey area and other experimental systematics. For the correction of the matter power spectrum at nonlinear scales, we used the latest version of \texttt{HMcode} \citep{Mead2021} for the photometric probes (allowing the baryonic feedback efficiency factor of the code emulator, $\log_{10}(T_{\rm AGN}/\mathrm{K})$, to be free) and a 1-loop perturbation theory model based on \texttt{FAST-PT} for the spectroscopic measurements, as described in \cref{sec:nonlinear}.

\begin{figure*}
\centering
\includegraphics[width=0.9\textwidth]{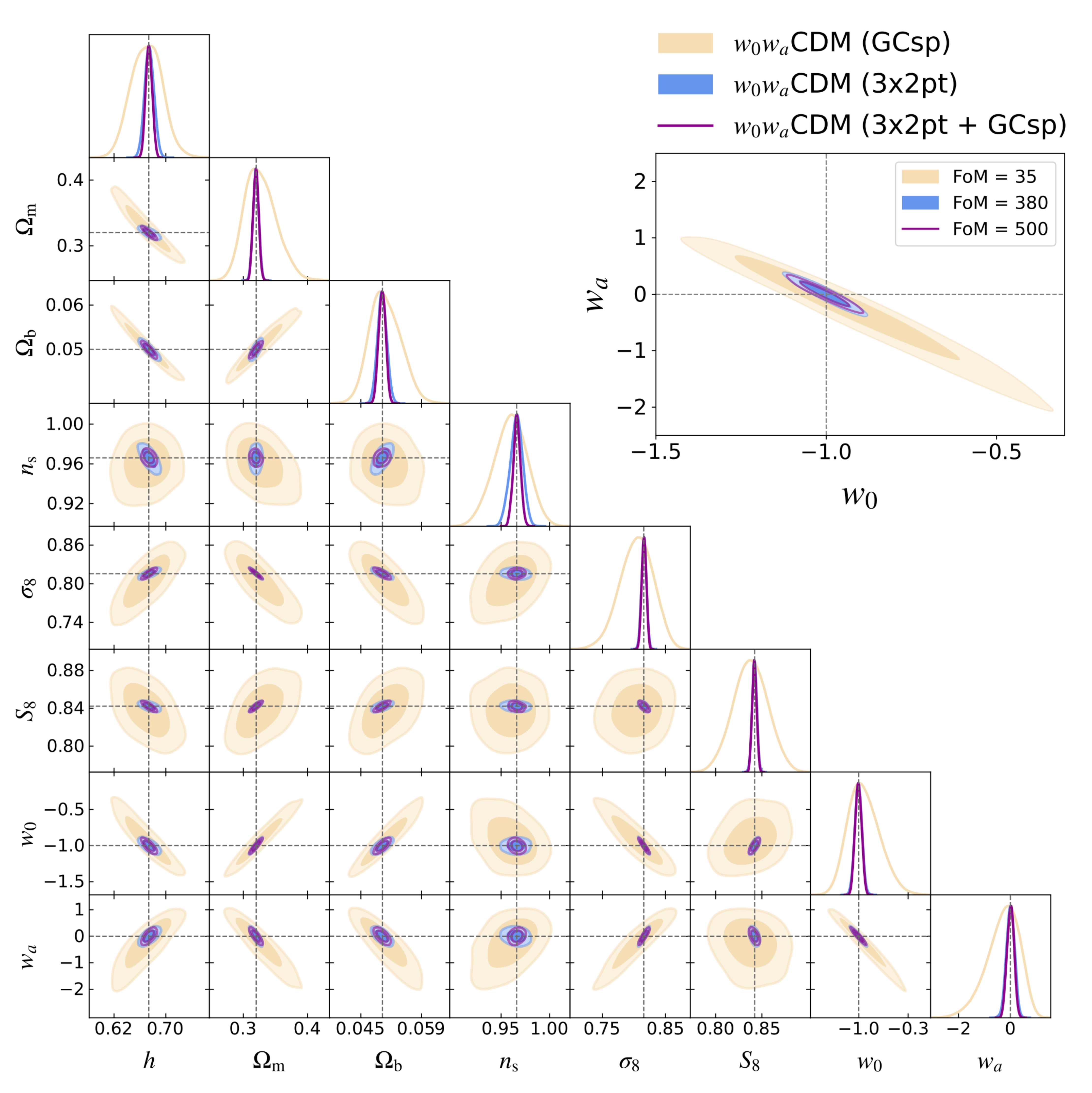}
\caption{Forecast of the constraints for the $w_0$$w_a$CDM cosmological model (adopting a flat geometry) using only the \Euclid primary probes, as described in \cref{sec:spv3}. The sampled parameter space also included the cosmological parameters ($\Omega_{\rm b} h^2$, $\Omega_{\rm c} h^2$, $H_0$, $n_{\rm s}$, $A_{\rm s}$, $w_0$ and $w_a$) and several nuisance parameters listed in \cref{tab:fiducial_model}. The grey dashed lines show the fiducial values of the parameters, that are also listed in \cref{tab:fiducial_model}. The posterior distributions were obtained using \texttt{CLOE} v2.0.2 and the sampler \texttt{Nautilus}, with 4000 live points and 16 neural networks. For the photometric probes, we used $\ell_{\rm max} = 5000$ for cosmic shear and $\ell_{\rm max} = 3000$ for photometric angular clustering, and galaxy-galaxy-lensing, while for the spectroscopic probe we used, $k_{\rm max} = 0.3\,h\,{\rm Mpc}^{-1}$. We show the 2D-posterior distribution for the parameters $w_0$ and $w_a$ in detail, citing the corresponding \gls{FOM} obtained for each probe as well as for the combination of both.}
\label{fig:triangle_plot_w0waCDM}
\end{figure*}

The corresponding theoretical predictions and the calculation of the \Euclid likelihood (see \cref{eq:likelihood}) were computed using \texttt{CLOE} v2.0.2. The sampling of the posterior distributions were obtained using the novel nested sampler \texttt{Nautilus}, interfaced with \texttt{Cobaya}, imposing the priors for all the free cosmological and nuisance parameters presented in \cref{tab:fiducial_model}. For forecasting purposes, and to speed up the sampling process, we have used \gls{BBN} information as a Gaussian prior for the baryon density parameter $\Omega_{\rm b} h^2$ \citep{Cooke18}. In this section, we show the results corresponding to three different forecasting cases: 3\texttimes2pt, GCsp and  3\texttimes2pt + GCsp (see Euclid Consortium: Ca\~nas-Herrera et al., in prep, for a more complete discussion of the forecasts with \texttt{CLOE}). For the GCsp case, we need to sample a total of 23 free parameters, while we need 44 free parameters for the 3\texttimes2pt case, and 60 for the joint one. Moreover, we track several derived parameters on the fly, and simultaneously fit 364 (3\texttimes2pt) and 12 (GCsp) different spectra. During the analysis, we have fixed the per-bin counter terms (see \cref{tab:fiducial_model}) in the 1-loop perturbation theory model used for the GCsp probe, in an effort to mitigate the so-called projection effects. We thus implicitly assume that in future analyses we will be able to impose tighter priors on these model nuisance parameters. 
To achieve convergence, the parameter space is explored by calculating the Bayesian evidence using 4000 live points\footnote{In nested sampling, a \textit{live point} is a sample of the likelihood distribution as given by the prior, that is later used to construct the evidence.} in each iteration. To obtain the results presented here, we used approximately 0.2 million CPU hours in total.

\begin{figure*}
\centering
\includegraphics[width=0.9\textwidth]{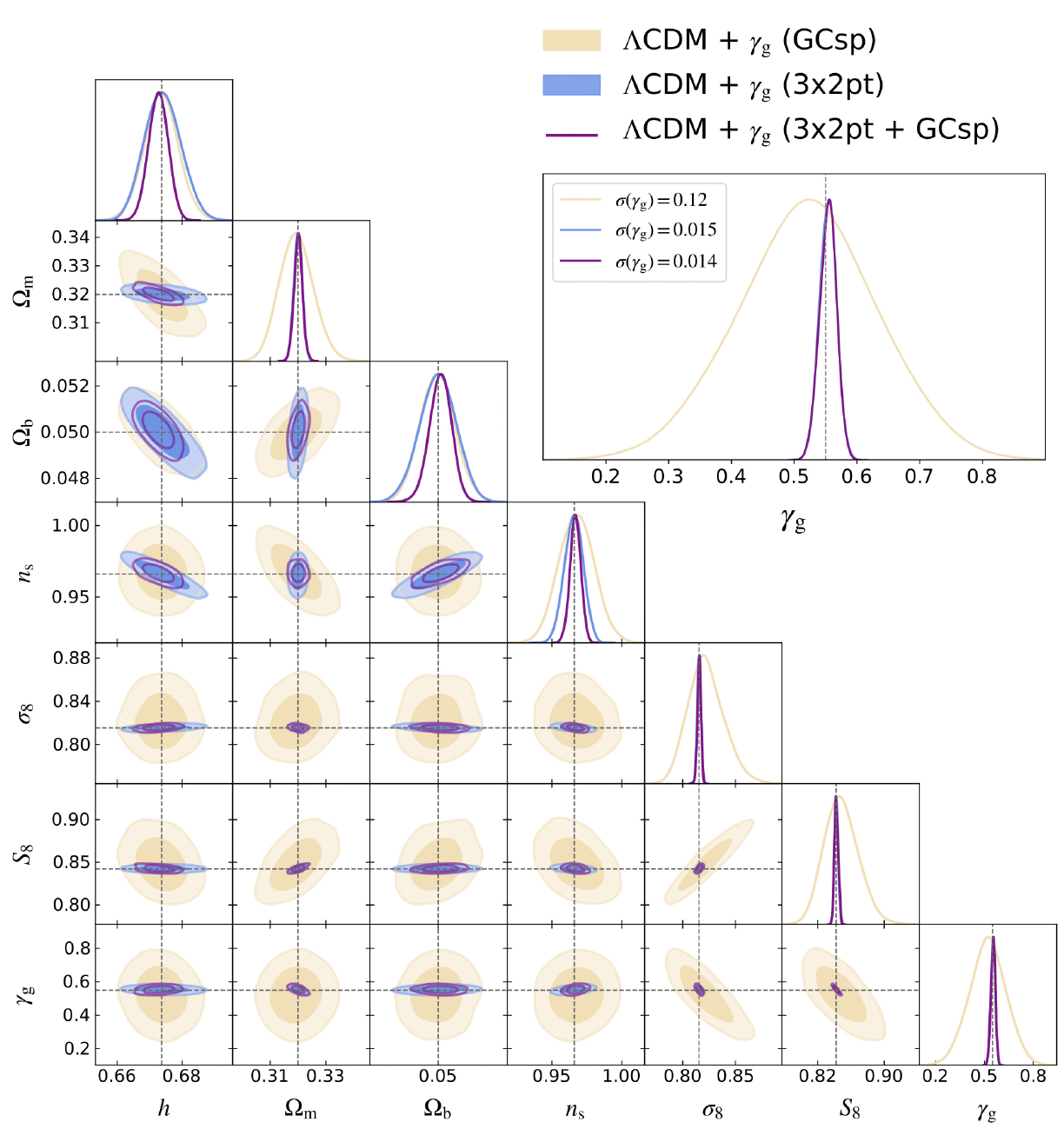}
\caption{Similar to \cref{fig:triangle_plot_w0waCDM}, but for the $\Lambda$CDM + $\gamma_{\rm g}$ model (adopting a flat geometry). We show the 1D-posterior distribution for the $\gamma_{\rm g}$ parameter in detail, citing the corresponding 1-sigma uncertainty associated with each probe as well as for the combination of both.}
\label{fig:triangle_plot_LCDM_gamma}
\end{figure*}

\Cref{fig:triangle_plot_w0waCDM} shows constraints on the dark energy parameters $w_0$ and $w_a$, as well as the corresponding constraints on the other cosmological parameters, while \cref{fig:triangle_plot_LCDM_gamma} shows constraints on the $\gamma_{\rm g}$ parameter. Regardless of the cosmological models, similar converged distributions are obtained for the nuisance parameters. For all the cases, the fiducial values are recovered for all the sampled parameters, with associated uncertainties that improve by one order of magnitude compared to current surveys \citep{Abbott2022}. The values obtain for the \gls{FOM} of the dark energy parameters $w_0$ and $w_a$ are consistent with those of \citet{Blanchard-EP7}, and they are obtained by marginalising over all the sampled parameters.

\subsubsection{Blinding strategy}
\label{sec:blinding}

Cognitive bias of the scientists undertaking an experiment can lead to priors on data analyses linking together otherwise independent measurements. Blinding strategies are designed to broaden these priors by separating the influence of the scientists' predictions for the measurements from the results themselves. 
We aim to avoid such biases introduced by the way data are processed, selected, or modelled. This requires some care, because the tendency to either consciously or unconsciously select, process, or interpret data in such a way as to confirm prior beliefs often leads to biased results \citep{Nickerson98}. The quest to determine cosmological parameters is no exception \citep{Croft11} because certain values for cosmological parameters may be preferred based on theoretical grounds. 

Many recent cosmological analyses have taken steps to avoid cognitive bias by adopting a so-called `blinding' strategy. This can take many forms, for example, by shifting values or theoretical models in plots \citep[e.g.][]{Wong20}, modifying the data vectors \citep[e.g.][]{Muir2020}, or adjusting the covariance matrix \citep{Sellentin20}. The \Euclid data are required to pass stringent validation tests, while the software that is employed to process the data is under strict control. This distinguishes it from most previous cosmological experiments, but there is still the potential to introduce biases during the scientific analysis. 

The blinding strategy we propose for \Euclid starts with the extensive use of synthetic data, such as those described in \cref{sec:simdata}, to build and test pipeline elements. We will then allow the analysis of the first 500\,deg$^2$ observed at the start of the project without any blinding. This represents $1/30$ of the expected final sample. Assuming that variance scales with the inverse of the volume, the results from the early data should have an error 5.5 times larger than that from the final data and should be comparable to current constraints. Thus we can consider that analysing this sample is equivalent to blinding the signal at the 5.5$\,\sigma$ level. As the pipeline is developed and refined before the \acrlong{DR1}\glsunset{DR1} (DR1; \cref{sec:DR}), development will concentrate on this sample, and it will be the only sample upon which revisions in the pipeline are retrospectively applied until the DR1 sample is constructed.

As described in \cref{sec:calibration}, the \Euclid mission has defined a careful and comprehensive calibration strategy. The observations taken to facilitate the calibration tend to use the instrument in a different way to the \gls{EWS}, so that potential systematic problems can be identified and quantified. Crucially, the calibration measurements do not strongly depend on cosmology, and hence can be analysed without additional blinding.

For the core cosmology measurements to come from DR1, we adopt a strategy similar to that adopted by the \gls{DES} \citep{Muir2020}, but without forcing the shift to match between galaxy clustering and weak lensing measurements. Independent shifts mean that differences between statistics are blind: while we cannot then use them to test our analysis methodology, the impact of \Euclid on any tensions will be blinded. We choose to introduce blinding at the level of derived statistics rather than raw data to facilitate calibration and validation tests and to ensure we do not impact other analyses of these data.

Specifically, we will shift the two-point data products (e.g. the correlation function and power spectrum multipole moments) by the expected difference between two cosmological models. This includes shifts in the effects of all of the key cosmological processes to be measured. The model offset applied will be between the best-fit flat $\Lambda$CDM model of \citet{PlanckParams2018}, and one randomly chosen within a 3$\,\sigma$ interval.

\subsection{Beyond \texorpdfstring{$\Lambda$CDM}{LCDM} models}
\label{sec:beyondlcdm}

As shown above, the primary probes of \Euclid will place tight constraints on the parameters of the $\Lambda$CDM model. However, as already highlighted in \cref{sec:primary}, the data also greatly advance our ability to explore various extensions to the standard cosmological model, potentially shedding light on the dark constituents of the Universe and the underlying theory of gravity. Here, we highlight some of these cases, where we note that further improvements can be achieved when combining the \Euclid results with
complementary cosmological probes, such as the ones summarised in \cref{sec:additional}.

The combination of the primary probes over the \gls{EWS} area offers the unique opportunity to test gravitational physics on cosmological scales, placing constraints on several modifications of \gls{GR} and models of dark energy \citep{Amendola18}. In addition to testing specific models, more agnostic descriptions of modified gravity and dark energy can be explored. For instance, the phenomenological functions $\mu_{\rm mg}(k,z)$ and $\Sigma_{\rm mg}(k,z)$  parameterise, respectively, the relation between the matter density contrast and the Newtonian and lensing potentials in Fourier space \citep[e.g.][]{Pogosian:2010tj}, while the \gls{EFT} of dark energy \citep{Frusciante:2019xia} provides a framework to explore deviations from \gls{GR} consistently. Provided the observables can be modelled accurately on nonlinear scales, \Euclid will provide outstanding constraints on extensions of $\Lambda$CDM, especially when cross-correlations with \gls{CMB} measurements are also included \citep[see \cref{sec:cmbx};][]{Casas23a,Frusciante23}.

Massive neutrinos suppress the matter power spectrum on small scales. Hence, a measurement of this subtle signature allows the sum of neutrino masses, $\sum m_\nu$, to be constrained using cosmological data. Current CMB and LSS measurements provide stringent upper bounds \citep[$\sum m_\nu \lesssim 0.1\,{\rm eV}$, 95\% confidence;][]{PlanckParams2018, Alam2021, Palanque-Delabrouille:2019iyz}, well below the limits of current laboratory experiments studying $\beta$ decay \citep[upper limit on effective electron anti-neutrino mass $m_{\nu}<0.8\,{\rm eV}$, 90\% confidence, ][]{KATRIN:2021uub}. The absolute neutrino mass sum is still unknown, but this situation will change thanks to \Euclid. 
In the minimal $\Lambda$CDM+$\sum m_\nu$ model, \Euclid's primary probes alone can constrain the neutrino mass with a $1\,\sigma$ error $\sigma(\sum m_\nu)=0.05\,{\rm eV}$ for a fiducial neutrino mass of $0.06\,{\rm eV}$.
In combination with \Planck, a precision of $\sigma(\sum m_\nu)=0.02\,{\rm eV}$ can be reached, implying a $3\,\sigma$ detection of a non-zero neutrino mass \citep{KP_nu}. If the true neutrino mass is below $0.08\,{\rm eV}$, these constraints will also provide evidence in favour of the normal neutrino mass ordering.\footnote{The minimum mass allowed by neutrino oscillation experiments in normal (inverted) ordering is $0.058\,{\rm eV}$ ($0.100\,{\rm eV}$).}
We note that these estimates are conservative because they assumed the pessimistic scenario of \citet{Blanchard-EP7}. Ongoing efforts to improve the modelling of nonlinear structure formation (\cref{sec:nonlinear}) should ultimately result in smaller uncertainties.

\Euclid will also improve the constraints on the effective number of relativistic degrees of freedom $N_{\rm eff}$, which accounts both for the number of standard model neutrinos \citep[$N_{\rm eff}^{\rm SM}=3.044$,][]{Froustey:2020mcq,Bennett:2020zkv} and for additional light particles, dubbed `dark radiation'. The sensitivity of \Euclid, in combination with current and future CMB surveys, to $N_{\rm eff}$ will potentially exclude several theoretically well-motivated particles beyond the standard model \citep{KP_nu}. Therefore, \Euclid will also shed light on dark matter models predicting a deviation of $N_{\rm eff}$ from the standard model value, such as models involving interactions between dark matter and dark radiation. Moreover, \Euclid will inform the models for alternative dark matter scenarios beyond the cold dark matter paradigm, by improving the constraints on warm dark matter, and decaying dark matter 
(Euclid Collaboration: Lesgourges et al., in prep.).

Current observations are consistent with initial conditions that correspond to a flat Universe with nearly Gaussian adiabatic perturbations, whose spectrum is described by a simple power law \citep{Planck:2018nkj,Alam2021}. Thanks to the different sensitivity of the primary probes to the expansion of the Universe and to the growth of structure, \Euclid provides an invaluable snapshot of the initial conditions at low redshift. The measurements will reduce the uncertainty on the spatial curvature $\Omega_K$ an order of magnitude below the current constraints from galaxy surveys \citep{Alam2021}. A similar improvement is expected for the uncertainty in the value of
the scalar spectral index $n_{\rm s}$ and its running 
$\alpha_{\rm s}= \diff n_{\rm s} /\diff \ln k$. \Euclid will also improve the constraints on features in the primordial power spectrum as forecast in \cite{Ballardini23}.

\Euclid will test the statistics of primordial fluctuations beyond the power spectrum. The spectroscopic
survey is expected to improve constraints on $f_{\rm NL}^{\rm local}$, the local shape of primordial
non-Gaussianity, by approximately a factor 8 over current results
\citep{Mueller:2021tqa}, reaching an uncertainty of about $3.4$ when combining power spectrum and bispectrum information and assuming universality for the halo mass function.  These uncertainties on initial conditions are comparable to those obtained by \Planck \citep{Planck:2018jri,Planck:2019kim}, but target a markedly different range in redshift and scale.


\section{Additional cosmological probes}
\label{sec:additional}

\Euclid is designed with the primary probes in mind, but the data enable a wide range of additional measurements that can improve cosmological parameter constraints \citep{Laureijs11}. For instance, the cosmological information is not limited to the two-point statistics that we have focused on so far. In \cref{sec:hos} we discuss how 
higher-order clustering and lensing statistics can be used to improve 
cosmological parameter constraints.
In \cref{sec:clusterfinding} we already highlighted the large number of clusters that \Euclid will discover.
Their use to improve cosmological parameter constraints is reviewed in \cref{sec:clusterofgalaxies}. As discussed in \cref{sec:stronglensing}, the sharp imaging data are ideal for the discovery of strong gravitational lenses, which enable a unique study of the distribution of dark matter on small scales, as well as additional tests of cosmology. The wide area galaxy and matter maps (see \cref{sec:massmap}) can be cross-correlated with measurements of the \gls{CMB}, enabling new probes that are presented in \cref{sec:cmbx}. Finally, high-redshift quasars with X-ray data can complement low-redshift cosmological probes in the determination of the cosmological parameters as discussed in \cref{sec:qso}, while \cref{sec:cosmic_chrono} explores the use of passive galaxies as chronometers to provide an independent constraint on the expansion history.

\subsection{Higher-order statistics}
\label{sec:hos}

While the two-point statistics would capture all cosmological information in the \gls{LSS} if it were Gaussian, nonlinear structure formation has introduced non-Gaussian features into the cosmic matter distribution. The full information content can, therefore, only be unlocked with \gls{HOS}. A wide variety of observables that capture the higher-order information have been proposed, which can be roughly grouped into two categories: those that consider $N$-point correlation functions and $N$th-order moments of the density distribution; and those that use topological information of the density distribution. Examples for the first category are higher-order moments \citep[e.g.][]{Gatti2022, Porth2021}, higher-order correlation functions \citep[e.g.][]{Heydenreich2023, Burger2024} and one-point probability distributions \citep[e.g.][]{Barthelemy2020, Boyle2021}. The second category includes peak statistics \citep[e.g.][]{Martinet2018, Harnois-Deraps2021}, Minkowski functionals and persistent homology \citep[e.g.][]{Parroni2020, Heydenreich2022}, and scattering transforms \citep[e.g.][]{Cheng2020, Cheng2021}. 
Many estimates can be inferred from \gls{WL} convergence maps (see \cref{sec:massmap}), while some, for example, higher-order correlation functions and aperture mass moments, can be directly measured from shear catalogues \citep{Jarvis2004, Secco2022, Porth2023}. 

When combined with two-point statistics, \gls{HOS} enhance cosmological constraints by (partially) resolving parameter degeneracies \citep[e.g.][]{Kayo2013, Heydenreich2023}. \citet{Ajani-EP29} found that combining each of ten different \gls{WL} \gls{HOS} with two-point statistics results in a twofold improvement in constraining $\Omega_\mathrm{m}$ and $\sigma_8$ compared to relying solely on two-point statistics. Combining all \gls{HOS} leads to a factor of about 4.5 improvement.

\begin{figure}
    \centering
     \includegraphics[width=\linewidth]{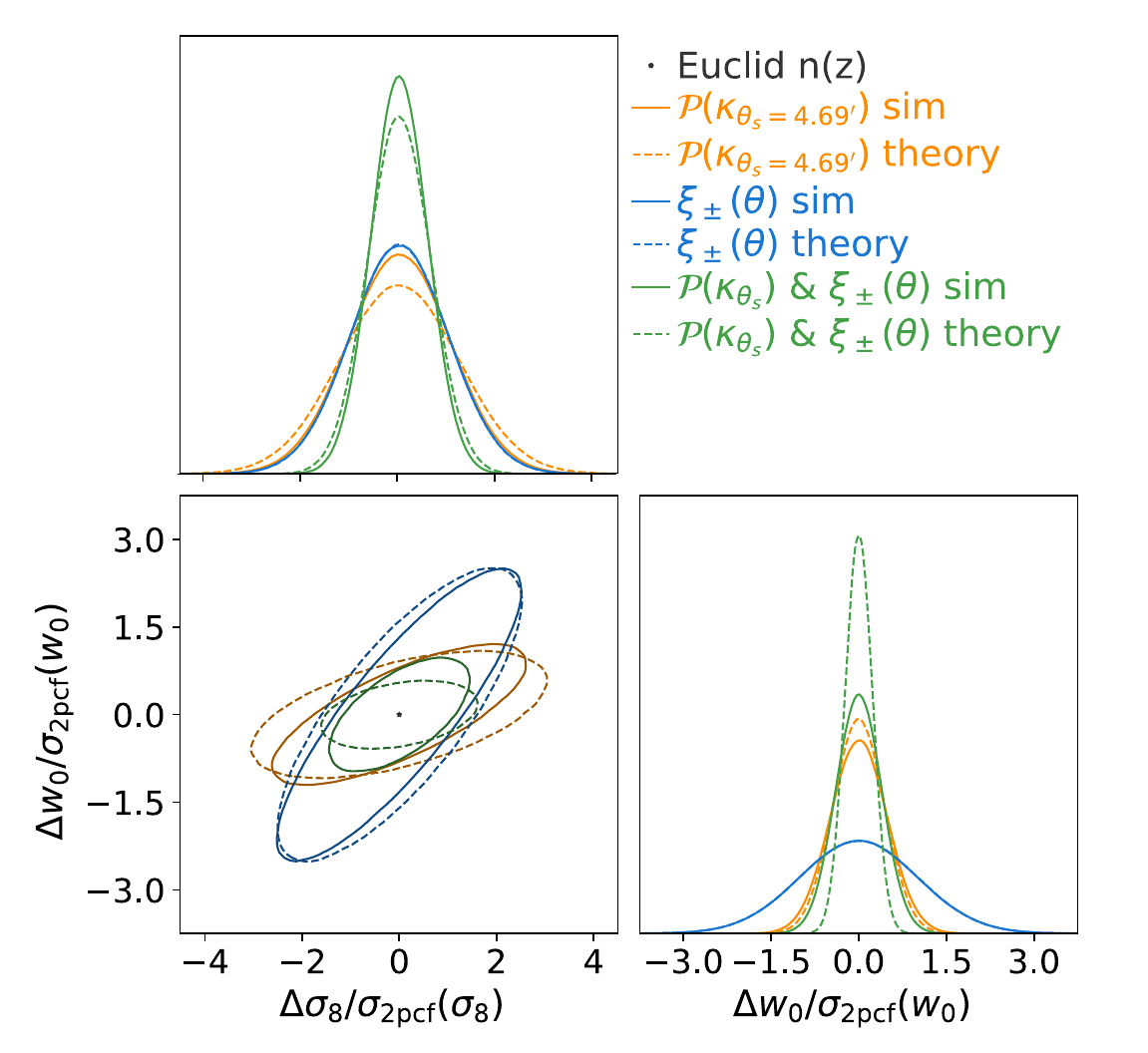}
    \caption{Constraints on $\sigma_8$ and $w_0$ from a Fisher analysis of $\xi_\pm$ and the convergence \gls{PDF}, when keeping all other cosmological parameters fixed, normalised by the constraints of second-order statistics alone. We assumed a \Euclid-like source redshift distribution to derive the results.
    The $\xi_+$ and $\xi_-$ values were taken in the range of \ang{;1.65;} to \ang{;201;}. The \gls{PDF} was measured for convergence fields smoothed by a tophat filter of radius \ang{;4.69;}. Covariances were estimated from the \gls{SLICS} \citep{Harnois-Deraps2018}, derivatives were either modelled analytically (dashed lines) or estimated from the DUSTGRAIN-pathfinder simulations \citep[][solid lines]{Giocoli2018}.}
    \label{fig:HOS}
\end{figure}

The power of \gls{HOS} is illustrated in Fig.~\ref{fig:HOS}, which shows expected constraints from a Fisher forecast analysis on $\sigma_8$ and $w_0$ with all other cosmological parameters fixed, using the shear correlation functions $\xi_\pm$, the convergence  \gls{PDF}, or the combination of both. The covariance for the analysis is estimated from the \gls{SLICS} $N$-body simulations \citep{Harnois-Deraps2018} and the derivatives of the data vectors are taken either from theoretical predictions (see \citealp{Boyle2021} for details on the \gls{PDF} modelling) or from the DUSTGRAIN-pathfinder simulations \citep{Giocoli2018}, where for the simulations we used the Kaiser--Squires mass reconstruction scheme (see Sect.~\ref{sec:massmap}) The \gls{PDF} shows a different degeneracy direction between $w_0$ and $\sigma_8$ than the second-order statistics, illustrated by the tilted ellipse in the lower-left corner of Fig.~\ref{fig:HOS}. This change in the degeneracy leads to a tightening of the constraints on $\sigma_8$ and $w_0$ when $\xi_\pm$ and the \gls{PDF} are combined. Consequently, the HOS carry additional cosmological information, which needs to be included to unlock all of \Euclid's potential.

\gls{HOS} can also test for residual systematics and constrain astrophysical effects such as intrinsic alignment \citep{Pyne2021}, baryonic feedback \citep{Semboloni13b},  or galaxy bias \citep{Huterer2006}. Since the \gls{HOS} react differently to these effects than two-point statistics, combined analyses allow us to simultaneously constrain cosmological and nuisance parameters without additional data sets.

\subsection{Clusters of galaxies}
\label{sec:clusterofgalaxies}
 
Galaxy clusters have long proven to be a valuable cosmological tool \citep[e.g.][]{Bahcall1998,Borgani2001,Haiman2001,Weller2002,Reiprich2002,Vikhlinin2009,Mantz2015,PlanckSZ2016, Bocquet2019,Costanzi19, desy1cl, Lesci22, Chiu23,Ghirardini2024}.  Arising from the highest density peaks of the initial matter density field, the abundance and spatial distribution of clusters contains information on the growth of structures and expansion history of the Universe \citep{Allen2011, Kravtsov2012}. More specifically, the cluster abundance is a sensitive probe of the parameter $S_8$. The evolution of cluster counts, which effectively measures the growth rate of cosmic structure, constrains dark energy and modified gravity models \citep[e.g.][]{Mantz2015, Cataneo2015, Bocquet2019}. The galaxy cluster correlation function, probing the same matter field traced by galaxies, is sensitive to the same cosmological effects and parameters detailed in \cref{sec:galaxyclustering}. The relative lower \gls{S/N} of the cluster clustering measurement, due to the sparser nature of the cluster sample, is mostly compensated by a theoretically predictable halo bias \cite[e.g.][]{Sartoris2012}.
The combination of cluster counts and clustering has the potential to deliver independent, competitive, and complementary cosmological constraints to those provided by the primary \Euclid probes, but depends sensitively on our ability to accurately calibrate their masses \citep{Sartoris2016,Fumagalli24}. Indeed, while it is possible to predict with percent level accuracy the abundance of dark matter halos as a function of mass and redshift in an arbitrary cosmology \citep[e.g.][]{Tinker2008, Castro21}, halo masses themselves are not directly observable. In cluster cosmology studies, it is hence crucial to identify and calibrate observational proxies -- such as the number of member galaxies (richness), X-ray luminosity, or the \gls{SZ} signal -- against mass.  At present, the calibration of these scaling relations represents the main limiting factor for cluster cosmology studies at all wavelengths \citep[e.g][]{PlanckSZ2016,desy1cl}.

The combination of \Euclid\/’s wide sky coverage and high-quality optical data will allow the detection of an order of $10^6$ clusters above $\sim 10^{14}\,M_\odot$ out to redshift 2, increasing the number of detected systems by more than an order of magnitude compared to current surveys \citep{Sartoris2016}. In particular, \Euclid will perform a census of the cluster population above $z \sim 1$ for the first time in the optical-\gls{NIR} wavelength regime, a critical stage of the Universe's evolution for studying dark energy.
The unprecedented large statistics, along with the large volume probed by the survey, will allow \Euclid to beat down shot noise and sample variance, enabling a statistically significant measurement of the cluster correlation function at $z\ga0.3$. Along with the exquisite imaging data provided by \Euclid’s space observations, this will enable a weak lensing calibration of the observable-mass relation out to $z \simeq 1$ \citep{Koehlinger15}, while the sparse spectroscopic data for cluster-member galaxies at $0.9 \la z \la 1.8$ will provide a valuable mass proxy for high-redshift systems \citep{Sartoris2016}. 

The large number of clusters and the precision of the lensing measurements demand stricter control over systematic effects compared to current cluster surveys. This is especially true for a photometric cluster survey such as \Euclid's, capable of detecting systems down to group mass scales and becoming mass-complete above $M > \times 10^{14}\,M_\odot$ (see \cref{sec:clusterfinding}). These systems, which outnumber their more massive descendants by orders of magnitude, potentially encode valuable cosmological information, but prove to be difficult to include in abundance studies \citep{desy1cl}; the limited resolution that a photometric cluster-finder algorithm can achieve along the \gls{LOS} leads to unavoidable uncertainties and biases in the richness estimate, which become more severe in the low-\gls{S/N} regime. The correlation of these systematics with others affecting the lensing measurements, or dynamical mass proxies, further hampers the characterisation of these systems \citep{Sunayama20, Wu22}. For \Euclid, the calibration of such selection effects, as well as the determination of the threshold for the minimum cluster richness, will be tackled using a combination of simulated and multi-wavelength data analyses \citep[e.g.][]{Costanzi19,Costanzi21,2021MNRAS.504.1253G}, along with the inclusion of the clustering of clusters statistics. The latter, thanks to the mass dependence of the halo bias, will enable us to break the degeneracy between cosmological and scaling-relation parameters when combined with the other cluster observables \citep[e.g.][]{To21}.
In summary, from the combination of cluster counts, cluster clustering, and \Euclid’s mass-proxies, we expect to increase the precision on the estimation of $S_8$ by an order of magnitude compared to current galaxy cluster studies based on imaging surveys, such as \gls{DES} and \gls{KiDS}, or \gls{SZ} surveys, such as \gls{ACT} and \gls{SPT}. Moreover, \Euclid's ability to sample the cluster population at $z>1$ will ensure an improvement of a factor of 2 on the DE \gls{FOM}, compared to cluster surveys at lower redshift \citep[e.g.][]{Sartoris2016,Bocquet2019}.

Furthermore, the inclusion of  
\Euclid  weak lensing data 
is expected to dramatically improve the cosmological constraints derived from intra-cluster-medium-selected cluster samples, such as the ones provided by the \acrlong{eROSITA}
\glsunset{eROSITA}
\citep[\gls{eROSITA};][]{Bulbul2024}, or the high-resolution \gls{SZ}
surveys conducted by \gls{SPT} 
\citep[e.g.][]{2015ApJS..216...27B,spt3g} 
and \gls{ACT} \citep{Hilton2021}.

\begin{figure}
    \centering
    \includegraphics[width=\linewidth]{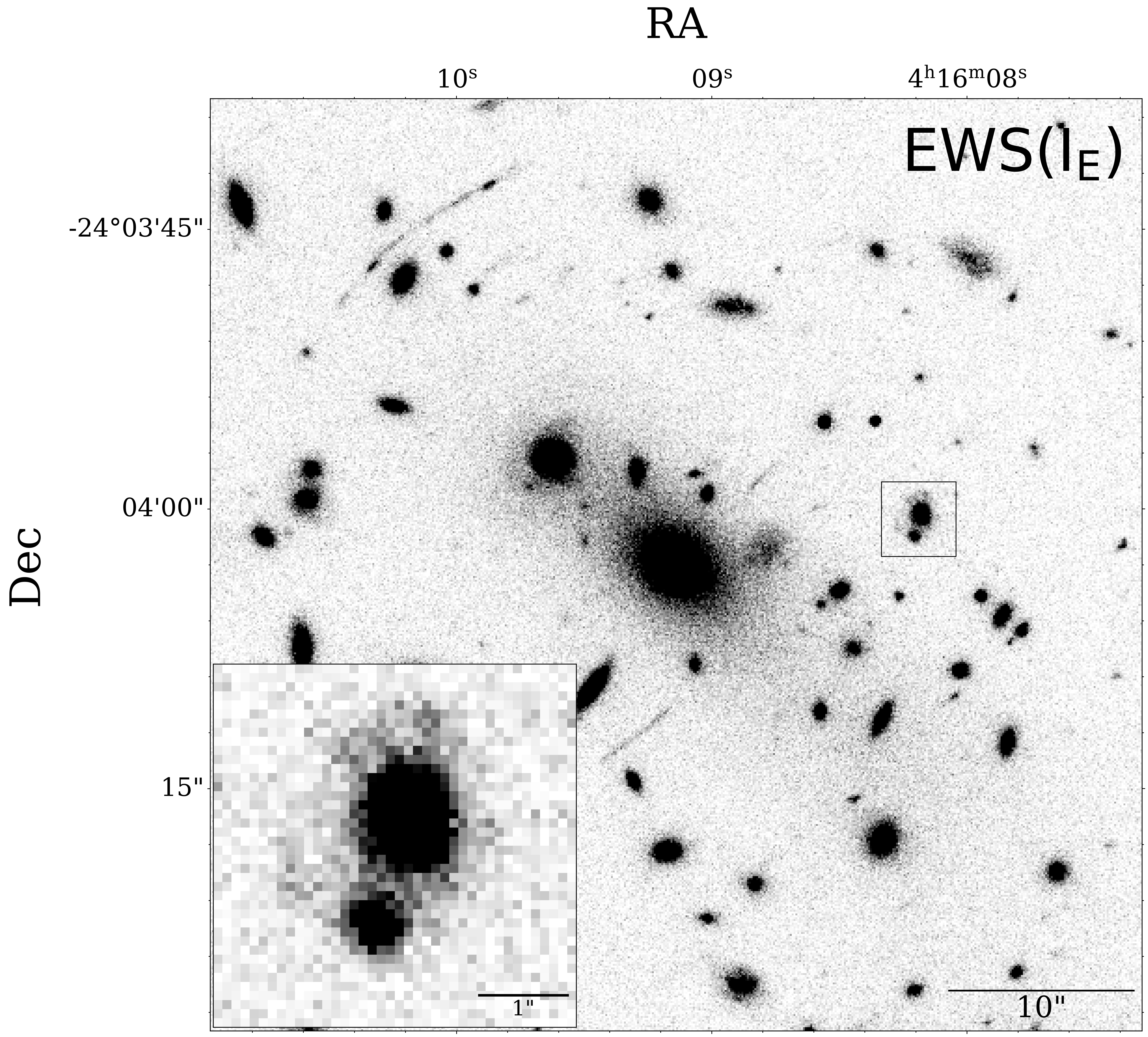}
    \caption{Simulated \Euclid observation in the \IE\ band of the central region of the strong lensing galaxy cluster MACSJ0416.1$-$2403 \citep[$z=0.397$,][]{2016ApJS..224...33B}. The image was obtained with the code \texttt{Hst2Euclid} (Bergamini et al., in prep.), using \gls{HST} observations taken as part of the Hubble Frontier Fields Survey \citep{2017ApJ...837...97L}. The image reproduces the depth of the \gls{EWS} and several giant arcs are clearly visible. The inset shows a zoom into a known galaxy-galaxy strong lensing system, where the lens is a cluster member and the source a background galaxy at redshift $z=3.222$ \citep[ID14,][]{2017ApJ...842...47V}.}
    \label{fig:m0416_hst2euclid}
\end{figure}

\subsection{Strong gravitational lensing}
\label{sec:stronglensing}

Multiple images of a distant source can be produced if light rays pass sufficiently close to a massive structure along the \gls{LOS}. Such favourable configurations are rare, and typically are not resolved in ground-based observations. Thanks to the sharp images provided by \Euclid, vast numbers of strong lenses will be discovered, and we highlight some of the main applications here.
Strong lensing studies with \Euclid will cover a wide range of mass, from galaxies to groups and clusters of galaxies (see e.g. \cref{fig:m0416_hst2euclid}), while probing the distribution of dark matter on relatively small scales.

We expect to detect approximately 200\,000 galaxy-scale strong lenses, which thus make up the largest fraction of systems. These typically feature an $M^\ast$ lens at a median redshift $z_{\rm d}=0.6$ within the \gls{EWS}. The sources in these systems are expected to have redshifts ranging from $1 < z_{\rm s} < 2$ \citep{collett2015,2019A&A...625A.119M,2024A&A...681A..68E}, resulting in an average Einstein radius of $\theta_{\rm E} = \ang{;;0.5}$, which aligns well with the capabilities of \Euclid, but is too small for ground-based surveys \citep[e.g.][]{Petrillo+19_CNN}. In the \gls{EDS}, approximately 3500 lenses are expected, compared to 500 in the equivalent area of the \gls{EWS}. On average, the unlensed \IE\ magnitude of the sources is 2 magnitudes fainter than that of the lensed ones, highlighting the significant magnifying effect of lensing \citep{2017ApJ...837...97L}. This magnification enables the search for ultra-high redshift (lensed) sources up to $z = 8$ in the \gls{EDS}, down to an unlensed \IE\ magnitude of 28. This approach is critical for investigating questions about re-ionisation by distant protogalaxies. As \Euclid targets the Ly\,$\alpha$ line and benefits from magnifications up to $\mu = 30$~\citep{2015ApJ...813...21M,2015ApJ...814...69A,2017MNRAS.467.4304V}, we anticipate the discovery of 35 to 75 such galaxies in the \gls{EDS}, while brighter and rarer events may be discovered in the \gls{EWS}.
Moreover, in spite of the lower \gls{NISP} resolution, in the \gls{EWS} we expect to find tens of thousands of bright lensed \gls{NIR} sources in the \gls{NISP} photometric data that are too faint to be detected in the \IE~band \citep{pearson24}. Most of these will be sub-millimetre galaxies (at $z\sim2$--4) or evolved red galaxies.

We expect the identification of up to 2300 lensed quasars, with approximately 16\% predicted to be quadruply imaged \citep{oguri2010}. The discovery of lensed supernovae  \citep[e.g.][]{Pascale2024} will complement the lensing of quasars; due to the EDS observations being divided into roughly 40 independent epochs over the six-year mission duration, the detection of several lensed type-Ia supernovae is expected. Both lensed quasars and supernovae will facilitate time-delay cosmography \citep{Treu2022}, allowing for measurements of the Hubble constant~\citep{1964MNRAS.128..307R,2015Sci...347.1123K,2018ApJ...860...94G,Wong20}. This process requires adding the time dimension to the precise \Euclid data, demonstrating a significant synergy with \gls{LSST}.

Moreover, microlensing in lensed quasars can be utilised to estimate the fraction of dark matter in compact form in galaxies~\citep{1998MNRAS.295..587M,2003MNRAS.339..607M,2005MNRAS.363.1136K,2009MNRAS.400.1583V,2012Natur.481..341V,2016ApJ...823...37H,2017MNRAS.471.2224N,2020MNRAS.491.6077G,2023ApJ...942...75W,2023MNRAS.524L..84P}. Analysing the flux ratios between lensed images or their distribution across the sample of lensed quasars provides insights into the quantity and distribution of low-mass dark matter halos in lensing galaxies and along the line of sight (\gls{LOS}). High spatial resolution follow-up observations, using very-long-baseline interferometry or adaptive optics at large ground-based optical telescopes, will facilitate the detection of such halos through gravitational imaging. 

Furthermore, compound lenses, where multiple sources at different redshifts are lensed by the same foreground galaxy, are instrumental in overcoming the mass-sheet degeneracy inherent in lensing. This enables measurements of both the mass slope in lensing galaxies and the distance ratios between lenses and sources. Compound lenses serve as a robust tool for both mass profile measurement and cosmography, with additional sensitivity to dark energy parameters~\citep{2008ApJ...677.1046G,2016MNRAS.456.2210C,2023JCAP...04..001S}, provided that the multi-lens-plane mass-sheet degeneracy is broken \citep{Schneider2014}.

\Euclid is set to observe strong lensing features, such as families of multiple images and giant arcs, in thousands of galaxy clusters within the redshift range of $0.2 \lesssim z \lesssim 1.0$ \citep{2012MNRAS.427.3134B,2016MNRAS.457.2738B}. These observations will allow for the creation of detailed mass models of clusters' inner regions \citep[e.g.][]{Kneib_1993, Bradac2005, Diego2005, lie06, Coe08, Jullo_2007, Zitrin_2009, Oguri_2010, Zitrin_2013, Lam2014}. The resulting constraints will test the predictions of $\Lambda$CDM and alternative dark matter models, such as self-interacting dark matter, on a cluster scale~\citep{2001MNRAS.325..435M,2002ApJ...564...60M,2020Sci...369.1347M,2013MNRAS.430...81R,2013MNRAS.430..105P,2022A&A...668A.188M,2023A&A...678L...2M,2023A&A...679A.124G}.

In addition, akin to the aforementioned compound lenses, the simultaneous observation of numerous sources at varying redshifts, all lensed by the same clusters, will provide constraints on cosmological parameters such as $\Omega_{\rm m}$ and $w$ through the lensing sensitivity to angular-diameter distances~\citep{2009MNRAS.396..354G,2010Sci...329..924J,Moresco2022,2022A&A...657A..83C,2023A&A...680L...9A,2024arXiv240104767B}.

Lastly, \Euclid's unique ability to integrate both strong and weak lensing measurements will enable the determination of mass profiles of galaxy clusters from kiloparsec to megaparsec scales~\citep{1996ApJ...464L.115B,Bradac2005,2016ApJ...821..116U}. These measurements are essential for accurately determining the total cluster mass and are crucial for constraining the shape and redshift evolution of the cluster concentration-mass relation. Hydrodynamical simulations indicate that the concentration of dark matter halos correlates with the universe's density at the time of their collapse~\citep{NFW1997,2008MNRAS.387..536G,2013MNRAS.432.1103L}. Therefore, measuring this key relation is vital for validating the $\Lambda$CDM cosmological framework~\citep{2011A&A...530A..17M,2014ApJ...797...34M,2015ApJ...806....4M}. 

\subsection{Cross-correlation with CMB observables}
\label{sec:cmbx}

During their journey towards us the CMB photons interact with the large-scale structures of the Universe as they are forming. These structures leave their imprint on the CMB through gravitational lensing \citep{lensing-review} and via the scattering of CMB photons with electrons having significant thermal and bulk velocities, called the \gls{tSZ} and \gls{kSZ} effects, respectively \citep{tsz-review}. Additionally, the decay of gravitational potentials caused by the accelerated expansion in the late Universe generates new anisotropies in the \gls{CMB} temperature at large angular scales \citep[\gls{ISW} effect;][]{SachsWolfe}.  Maps of the \gls{CMB} lensing convergence ($\kappa$ hereafter) and of the strength of \gls{tSZ} (parameterised through the Compton $y$ parameter) can be extracted from high-resolution multi-frequency observations of \gls{CMB} anisotropies. They are sensitive to the total integrated matter or pressure distribution along the \gls{LOS} between us and the surface of last scattering, respectively. The \Euclid survey overlaps on the sky with the major existing \gls{CMB} data sets, such as \Planck, \gls{SPT}-3G and \gls{ACT} \citep{planck-lensing,spt3g,spt-lensing,act-lensing}, as well as future ground-based \gls{CMB} experiments such as the \acrlong{SO}\glsunset{SO} \citep[SO;][]{so} and CMB-Stage~4 \citep{cmbs4} or space-based experiments such as LiteBIRD \citep{litebird}.

The cross-correlation power spectra between CMB lensing and \Euclid galaxy clustering ($C_\ell^{\kappa\mathrm{g}}$) and weak lensing maps ($C_\ell^{\kappa \gamma}$), together with the CMB lensing auto-correlation $C_\ell^{\kappa\kappa}$, will provide additional observables that are sensitive to cosmological parameters affecting the angular-diameter distances and the growth of the matter perturbations, and as such they can tighten the statistical uncertainties \citep{sailer2021}.  Moreover, they are also free from additive systematic biases \citep{vallinotto2012,schaan2017}. They will thus allow us to break degeneracies and minimise the impact of systematic effects and theoretical uncertainties that might affect the \Euclid observables when analysed on their own, or provide new estimators that are less sensitive to systematic effects \citep{giannantonio2016,bianchini2018}. These \gls{CMB}-\Euclid cross-correlations will therefore not only add statistical power, but they will also allow us to marginalise over parameters describing systematics with minimal loss of constraining power. Adding these observables on top of the 3\texttimes2pt analysis in a combined 6\texttimes2pt analysis has become a standard in the field for current surveys \citep{des6x2pt2023,robertson2021,marques2020}. Considering also the correlation of galaxy clustering with \gls{CMB} temperature $C_\ell^{\kappa\mathrm{g}}$ in a full 7\texttimes2pt analysis with multiple tomographic redshift bins, the constraining power of cosmological parameters for the joint analysis of \Euclid with SO and CMB-S4 data can reduce the statistical uncertainty by a factor of 2--3 and in some cases even more than 10, in particular for generalised cosmological models including curvature or modifications of gravity \citep{Ilic-EP15}.

Further improvements are expected if higher-order statistics, involving mixed bispectra are used, correlating one or more \gls{CMB} lensing fields with \Euclid probes \citep{chen2021,farren2023}. Recent studies have also shown that cross-correlations between $\kappa$ and biased density tracers (such as galaxies or QSOs) can be used to set competitive constraints on local primordial non-Gaussianity through the scale-dependent galaxy bias, Cross-correlations do this in a more robust way compared to what is achievable with the auto-correlation analysis of the tracers, which is plagued by large-scale survey systematic effects, such as inhomogeneous depth, air mass, or selection effects \citep{rezaie2023,krolewski2023}. \Euclid's space-based observations will allow us to carry out these measurements with exquisite precision, as has recently been demonstrated in the context of the \Gaia mission \citep{alonso2023,quaia}. The AGNs and QSOs sample detected by \Euclid (see \cref{sec:qso,section:agn}) in particular will probe primordial non-Gaussianity in a redshift range never surveyed before.

\Euclid probes can also be cross-correlated with SZ maps. Cross-correlation with the \gls{tSZ} $y$ maps ($C_\ell^{y\mathrm{g}}, C_\ell^{y\gamma},C_\ell^{y\gamma}$), on top of being interesting probes in their own right to probe properties of the hot gas in the Universe, are highly sensitive to the physics of baryons.  This can be used to constrain (and marginalise) models of feedback or other baryonic effects in 3\texttimes2pt analyses of the \Euclid probes \citep{pandey2022,troster2022,osato2020,koukoufilippas2020,Kou2023} using a full combination 10\texttimes 2pt analysis with tSZ and CMB lensing \citep{fang2023}. Unlike for the \gls{tSZ} effect, the \gls{kSZ} signal cannot be separated from \gls{CMB} temperature maps, since it has the same frequency dependency as the \gls{CMB} itself and hence can only be seen in combination with an external tracer of the \gls{LSS}, either in cross-correlation with 2D matter tracers or through velocity-weighted stacking techniques using \Euclid spectroscopic sample \citep{hill2016,schaan2021}.  The cross-correlation between the squared \gls{CMB} temperature of SO or CMB-S4 maps and \Euclid galaxy clustering and weak lensing data will enable measurements of the \gls{kSZ} effect with an overall \gls{S/N} of about 20, providing statistical constraints on the parameters describing the shape of the gas radial density profile in halos (and thus on the underlying physical mechanisms) at the 10-20\% precision level \citep{bolliet2023}. Stacking techniques will be extended for the first time to higher redshifts, thanks to the capabilities of the NISP instrument. The spectroscopic power of \Euclid will also allow us to detect the \gls{kSZ} effect, and more generally the cosmological information encoded in the velocity field, through  approaches like the pairwise momenta \citep{Hand2012,PlanckkSZ2016}, velocity field reconstructions \citep{Dedeo2005,PlanckkSZ2016,ACT2016}, or cross-correlation to maps of the so-called `angular redshift fluctuations', which are sensitive to galaxy radial motions \citep{CHM2021,JChaves-Montero2021}. Combined with angular galaxy clustering,  the latter are expected to improve constraints on the dark energy equation of state by almost an order of magnitude, compared to angular galaxy clustering alone \citep{legrand2021}.

\subsection{Cosmology with high-redshift quasars}
\label{sec:qso}

Quasars have long been known to obey a nonlinear relation between the rest frame 2500\,\AA\ ($L_{\rm UV}$) and the rest frame 2\,keV ($L_{\rm X}$) emission \citep[e.g.][]{avnitananbaum79,zamorani81}, parameterised as $L_{\rm X}\propto L_{\rm UV}^\gamma$, with $\gamma\simeq0.6$. Recently, this relation has been employed to provide an independent measurement of quasar distances, thus turning these objects into {\it standardisable\/} candles and extending the distance modulus--redshift relation (the so-called Hubble diagram) of Type~Ia supernovae to a redshift range that is still poorly explored ($z>2$; \citealt{rl15}, see also \citealt{Moresco2022}). 

The applicability of this technique is based upon two main factors. First, the understanding that most of the observed dispersion in the $L_{\rm X}$-$L_{\rm UV}$ relation is not intrinsic to the relation itself, but due to observational issues (e.g. X-ray absorption by gas, UV extinction by dust, calibration uncertainties in the X-rays, variability, and selection biases associated with the flux limits of the different samples). Once corrected for and with an optimal selection of {\it clean\/} sources (i.e. where the {\it intrinsic\/} UV and X-ray quasar emission is measured), the dispersion in the relation becomes rather small, namely ${\simeq}\,0.2$\,dex \citep{lr16}. Second, the realisation that the slope of the $L_{\rm X}$-$L_{\rm UV}$ relation does not evolve with redshift up to $z\simeq4$, which is the highest redshift where the source statistics are currently sufficient to verify any possible dependence of the slope with distance. Before \Euclid, the largest quasar sample that can be used for cosmological analysis is composed of approximately 2400 sources out to $z\simeq7.5$ \citep[][with about 500 quasars beyond redshift 2]{lusso2020}. As of today, the precision achieved by the combined use of quasars and Type~Ia supernovae is on the order of 25\% on $\Omega_{\rm m}$ and 20\% on $w_0$, assuming an evolving dark energy equation-of-state parameter as in \cref{eq:w0wa} -- see table~1 in \cite{bargiacchi2022}. 

We expect a quasar sample of around $2\times10^6$ sources in the \gls{EWS} out to $z\simeq5$ detected in all four \Euclid bands ($5\,\sigma$ detection following the colour-colour \gls{AGN} selection described in \cref{section:agn}), with X-ray emission above the \gls{eROSITA} limiting depth (at a 2--10\,keV flux limit higher than $5.2\times10^{-18}\,\mathrm{W}\,\mathrm{m}^{-2}$; see Selwood et al., in prep.\ for details).\footnote{\gls{eROSITA} sources detected in the 0.2--2.3\,keV energy range with a detection threshold ${>}\,6$, corresponding to a point source flux limit of $6.5\times10^{-18}\,\mathrm{W}\,\mathrm{m}^{-2}$, assuming a power law with photon index $\Gamma=1.9$ \citep{brunner2022}.} We forecast that roughly 20\% of \Euclid quasars will fulfil the selection criteria described in \cite{lusso2020}, that is, unobscured at both UV and X-ray energies, radio quiet, with negligible contamination from the host galaxy, with several thousands of objects at $z>2$ surviving the selection cuts. Approximately, 10\% of this sample will have spectroscopic redshifts from the NISP instrument (see also \citealt{lusso2023arXiv}). 
The \Euclid quasar sample, complemented with \gls{eROSITA} data and additional X-ray data available in the archives (e.g. XMM-{\it Newton} and {\it Chandra}), will provide measurements of the cosmological parameters with a precision of 2\% on $\Omega_{\rm m}$ and 5\% on $w_0$, assuming an equation of state given by \cref{eq:w0wa}. The results from the \Euclid quasar Hubble diagram will thus be highly complementary, in both physical and observational terms, to all the other cosmological tests that \Euclid will enable.

\subsection{Cosmology with cosmic chronometers}
\label{sec:cosmic_chrono}

Passive galaxies are not only a powerful resource to set constraints on galaxy formation and evolution (\cref{sec:passiveGalaxies}), but can also be used to provide cosmological constraints when used as cosmic chronometers. As discussed in \cite{Moresco2022}, the measurement of the differential age evolution within an interval d$t$ of the Universe using a redshift bin of width d$z$ can provide direct and cosmology-independent constraints on the expansion rate of the Universe, since by only assuming a FLRW metric it is possible to derive that $H(z)=-1/(1+z){\rm d}z/{\rm d}t$. Very massive and passively evolving galaxies represent the ideal chronometers in the Universe, since many observational pieces of evidence indicate that they represent the oldest objects in the Universe at a given redshift, they experienced a synchronised formation, and they are a homogeneous population in terms of their physical properties. Hence, by measuring their differential ages as a function of redshift, it is possible to obtain an independent and complementary measurement of the Hubble parameter \citep[for a detailed review on the cosmic chronometer method, see][]{Moresco2022}.

\Euclid will detect thousands of passive galaxies at $1.5\,{<}\,z\,{<}\,3.2$ in the \gls{EDS} (see \cref{sec:passiveGalaxies}), and from their spectra it will be possible to detect features that have been demonstrated to provide robust tracers of their differential age (in particular the feature at 4000 \AA, the D4000 break, also see Fig.~\ref{fig:SFR-mass_spec}). A test study using quiescent galaxies with strong D4000 breaks as cosmic chronometers to derive cosmological constraints was presented in \cite{Moresco2015}. In this work, only 29 high-redshift ($1.4\,{<}\,z\,{<}\,2.2$) massive and passive galaxies ($19.7\,{\la}\,H_\mathrm{AB}\,{\la}\,22.2$) were analysed, showing the potential of this method to constrain the expansion history of the Universe in a cosmology-independent way up to $z\,{\sim}\,2.5$. Moreover, in \cite{Moresco2022} forecasts were presented showing how this method, applied to \Euclid data, will be able to constrain the Hubble constant and the dark energy equation-of-state parameter with a precision of 4\% and 30\%, respectively. While these constraints are less precise than the ones obtained with the main cosmological probes in \Euclid, they provide useful complementary information, that combined with the other approaches can contribute to maximise the scientific harvesting of \Euclid data.

\section{Non-cosmological science with \Euclid}
\label{sec:legacy}

\Euclid's combination of data of unrivalled fidelity and volume will have a significant impact in other areas of astronomy, especially once combined with various complementary data. In this section we highlight some of the other science cases where we expect \Euclid to have a major impact.  This discussion updates and extends the relevant discussion in \cite{Laureijs11} and highlights where the main challenges are. We refer to these as `legacy' science cases to emphasise their expected long-lasting value over the coming decades. 

For instance, the high spatial resolution, \gls{PSF} stability, and photometric depth of \Euclid offer unprecedented opportunities to study resolved stellar populations in the Milky Way and nearby galaxies out to beyond 5\,Mpc.  The ability to detect and characterise faint cool stars, either dwarf stars nearby in the Galactic disc or luminous evolved stars throughout the Local Volume, along with the contiguous \glspl{FOV} of regions of the sky is expected to be transformative for a variety of science cases. Moreover, the \Euclid surveys will provide imaging and spectroscopy of very large samples of galaxies. Indeed, the majority of sources in the final catalogues will be galaxies and this will enable a wide variety of extragalactic studies over a vast range in redshift. These include spatially resolved and integrated measures of star formation in galaxies, detailed morphologies of galaxies, and the detection and characterisation of transient phenomena, as well as 
distant galaxies in the \gls{EOR}. Closer to home, \Euclid will also enable the study of objects in the Solar System.

\subsection{The Milky Way and the Local Volume}
\label{sec:MWRSP}

Starting with the Milky Way, major advances are expected in studies of low-mass stars and star clusters. For example \glspl{UCD} are the lowest-mass, coldest and faintest products of star formation. Defined as objects with spectral types M7 and later \citep{2005ARA&A..43..195K}, they have masses of $M \leq 0.1\,M_{\odot}$, and effective temperatures $\leq 2700\,$K. They encompass the stellar-substellar limits, including the lowest mass stars (late M and early L) as well as brown dwarfs and planetary mass objects across the whole L-, T-, and Y-dwarf sequence. The \gls{EWS} is poised to dramatically increase the census of UCDs in the solar vicinity 
\citep{2021MNRAS.501..281S,2021A&A...655L...3M}, and spectra from the
NISP instrument will allow these objects to be classified into subtypes \citep{Zhang2024}. High-redshift quasars are contaminants for this kind of study (cf.~\cref{section:highz}) but these can be distinguished from \glspl{UCD} through use of VIS photometry and/or adoption of statistical modelling techniques \citep[e.g.][]{Barnett-EP5}. With large complete samples of \glspl{UCD}, studies of the oldest Milky Way populations, the structure of the Galactic disc, and the form of the substellar initial mass function will be possible. The sharp VIS \gls{PSF} offers the opportunity to resolve \gls{UCD} binaries and study the binary properties for an unprecedented number of systems.
Moreover, \Euclid will provide a particularly exciting window on some of the rarest low-mass objects currently known -- the coolest low-mass objects, the so-called Y dwarfs \citep[e.g.][]{2011ApJ...743...50C}, and young planetary mass objects, both free-floating and in wide
binaries \citep[e.g.][]{2013ApJ...777L..20L}. A showcase of \Euclid's capabilities to detect planetary
mass objects in star-forming regions is provided by the \gls{ERO} programme in Orion \citep{EROOrion}.  

\Euclid will observe many star clusters and star-forming regions throughout the local disc and halo. Of special interest are the roughly 25 \glspl{GC} that will lie within the \gls{EWS} footprint. \Euclid will enable a variety of studies, such as multiple population signatures at the end of the main sequence, the present-day mass function and \gls{GC} ages using the near-IR \gls{CMD} `knee' -- such studies have thus far been possible for only a small handful of \glspl{GC} \citep[e.g.][]{2016A&A...586A..51M, 2022ApJ...927..207D}. The wide-area coverage will also facilitate unmatched studies of the peripheral structures of \glspl{GC}, including the search for very sparse tidal features as demonstrated by the \gls{ERO} data for NGC~6254 and NGC~6397 \citep{EROGalGCs}. Even more detailed analysis will be possible for the halo \gls{GC} AM-1 which falls within the \gls{EDS} South. AM-1 lies at a Galactocentric radius of 120\,kpc and is one of the most distant \glspl{GC} currently known in the Milky Way. The multi-epoch imagery will permit an extremely deep \gls{CMD} study of AM-1, as well as a search for RR Lyrae variable stars that can be used to accurately measure its distance, and hence refine the measurement of its age \citep[e.g.][]{2008AJ....136.1407D}.

Our understanding of the assembly history of the Milky Way has undergone a transformation in the last few years thanks to the ESA \Gaia mission.  Results include the discovery of a significant accretion event early in the history of our Galaxy \citep{2018Natur.563...85H, 2018MNRAS.478..611B} and a large number of new nearby stellar streams \citep[e.g.][]{2019ApJ...872..152I}.  However, one of the most fascinating and poorly understood components of the Milky Way lies beyond \Gaia's reach but will be accessible with \Euclid\/ -- the outer stellar halo.  \Euclid will detect main sequence stars to Galactocentric radii of $\ga100$\,kpc, providing the first detailed window of the outer stellar halo with its repository of diverse dwarf satellites, ancient GCs and copious tidal debris from past accretion events \citep[e.g.][]{2008ApJ...689..936J}. The outer halo also provides an excellent laboratory for hunting for clues about the nature of dark matter. The existence of gaps, spurs, and peaks in cold tidal streams could signify impacts with dark matter sub-halos \citep[e.g.][]{2019ApJ...880...38B}.  The existence of many other possible perturbing sources (e.g. the rotating bar, giant molecular clouds, and spiral arms) in the inner halo of the Galaxy has complicated work of this nature to date, but the outer halo offers a much cleaner environment in which to characterise and interpret the origins of these density variations.
 
Results from the \Gaia mission have also brought to the fore the importance of understanding how representative our Milky Way is of the disk galaxy population at large.  Indeed, the archaeological record in our nearest large neighbour, M31, suggests a much more active accretion history than that experienced by the Milky Way \citep[e.g.][]{2019Natur.574...69M}, raising the possibility that our home galaxy may be unusual.  \Euclid is poised to have an enormous impact by enabling studies of resolved stellar populations in the halos of galaxies throughout the Local Volume.  Thanks to the long dynamical timescales, these parts are expected to contain the richest and best-preserved fossil record of the accretion history of a galaxy \citep[e.g.][]{2008ApJ...689..936J}. The resolved star approach is extremely powerful, having sensitivity to surface-brightness levels well below $\sim 30\,\mathrm{mag}\,\mathrm{arcsec}^{-2}$, but has proved a challenge from the ground due to star-galaxy separation at faint magnitudes \citep[e.g.][]{2023MNRAS.518.2497Z}.  \Euclid's high-resolution imagery and stable \gls{PSF} allows it to resolve luminous evolved stars, such as red giant and age-sensitive asymptotic giant branch stars, in the low surface-brightness peripheries of galaxies to distances of 5--7\,Mpc \citep{ERONearbyGals}. This volume encompasses several hundred systems, ranging from the smallest dwarf galaxies to large spirals like the Milky Way.  Systematic studies will be possible of tidal streams and stellar halos across the galaxy mass spectrum and in environments ranging from the field to small groups. 
The detection and characterisation of new dwarf satellites and halo \glspl{GC} around these galaxies will be achievable, as well as a search for free-floating \glspl{GC} across roughly a third of the sky \citep[e.g.][]{2016MNRAS.460L.114M}. 

\subsection{Nearby galaxies and diffuse structures}
\label{sec:NearbyGal}

The superb ability of \Euclid to detect \gls{LSB} features of galaxies \citep{Scaramella-EP1, Borlaff-EP16} makes it an ideal facility to study galaxy evolution, as exemplified by the first applications presented in \citet{EROPerseusOverview}, \citet{EROPerseusDGs}, \citet{EROPerseusICL}, and \cite{ERONearbyGals}. In the hierarchical paradigm of structure assembly, massive galaxies and their host dark matter halos are assembled from smaller ones, leaving observable signatures such as \gls{LSB} stellar streams, shells, and tidal remnants around galaxies. As shown in a number of observational \citep{Duc2015,Trujillo2016,Spavone2017,Buitrago2017,Martinez-Delgado2023} and theoretical \citep{Cooper2010,Martin2019,Perez-Montano2022} works, the diffuse \gls{LSB} light in the outskirts of galaxies contains tidal streams, tails, shells, and extended stellar halos; these features encode information about the past merging history of galaxies and helps to reconstruct their mass assembly through major or minor mergers \citep{Conselice2003,raj2020,spavone2020}.
 
\Euclid's unrivalled combination of area, resolution, low background, wavelength coverage, and \gls{PSF} stability has the potential to revolutionise these fields of research. \Euclid will reach a photometric depth of $\IE=29.5\,\mathrm{mag}\, \mathrm{arcsec}^{-2}$ (measured as 3$\,\sigma$ fluctuations in $10'' \times10''$ boxes) in the \gls{EWS}. This is equivalent to the deep surveys done so far from the ground over much smaller areas: hundreds of square degrees versus many thousands of square degrees for the \gls{EWS}. Furthermore in the \gls{EDS}, a gain of 2\,magnitudes will in principle be achievable \citep{Scaramella-EP1,Borlaff-EP16}. In the NIR, a regime for which ground-based \gls{LSB} studies are almost impossible, \Euclid has no competitor, with expected 1$\sigma$ \gls{EWS} depths of $\YE=28.2$, $\JE=28.4$, and $\HE=28.4\,\mathrm{mag}\,\mathrm{arcsec}^{-2}$ \citep{Scaramella-EP1}.
These surface-brightness limits have been confirmed observationally by \citet{EROData} and \citet{ERONearbyGals}.
 
 \Euclid will also reveal a population of low-surface-brightness and ultra-diffuse dwarf galaxies, both as satellites around massive hosts and as isolated field galaxies \citep[e.g.][]{vandokkum2015,vanderBurg16,marleau2021,venhola2022}. These populations are the most abundant galaxies by number at any redshift, and tend to be missed by large-scale surveys. They contribute to the faint end of the galaxy \gls{LF}, which is poorly known in the environments that \Euclid will probe. However, the detection of these \gls{LSB} structures, some located towards foreground Galactic cirrus, is challenging and will require non-standard data reduction procedures. 
 \Euclid will provide the crucial NIR regime that, combined with deep multi-wavelength imaging from future synoptic facilities such as the Rubin Observatory or \textit{Roman} telescopes, will constrain stellar populations and enable the characterisation of \gls{LSB} emission in and around dwarfs and more massive spirals and early-type galaxies.
 This is demonstrated by the census of dwarf galaxy satellites in the Perseus cluster of galaxies \citep{EROPerseusDGs} and around a nearby spiral galaxy \citep{ERONearbyGals}.

 Detailed studies of semi-resolved stellar populations in nearby galaxies will be possible with the superb spatial resolution of \Euclid's VIS and NISP imagers. The \gls{NIR} wavelengths are crucial because they trace the bulk of the stellar mass in galaxies by directly sampling the peak of the \glspl{SED} of the cool, low-mass stars that dominate stellar populations. 
 Spectral modelling of independent pixels in galaxy images will spatially resolve stellar mass densities, ages, metallicities, dust extinction, and other properties \citep[e.g.][]{Abdurrouf2022a, Abdurrouf2022b}, as well as their variations with environment and galaxy type. It may also help ameliorate the well-known degeneracies among age, metallicity, and dust extinction because of additional constraints from neighbouring pixels.
 With the \gls{EWS} and \gls{EDS}, exploiting the important \gls{NIR} regime, it will be possible to construct a census of resolved galaxy demographics on a statistical basis never before possible.
 
\Euclid's spatial resolution, sensitivity, and \gls{PSF} stability also provide a new, photometrically uniform, view of
\acrlong{EGC}s \citep[\acrshort{EGC}s; e.g.][]{powalka2017}. \Euclid VIS spatial resolution in combination with VIS/NISP colours help to distinguish 
EGCs from foreground stars and background high-$z$ galaxies \citep{munoz2014,Cantiello2020,saifollahi2021b, EROFornaxGCs}, and identify \acrshort{EGC}s around galaxies spanning a wide range of mass and environment, in particular \acrshort{EGC}s belonging to \gls{LSB} dwarf galaxies \citep{georgiev2009,lim2018,muller2021,saifollahi2022}. Accumulating the statistics of \acrshort{EGC}s around galaxies in the volume where \Euclid\ can detect these faint point sources provides strong constraints on the dark matter halo mass assembled through hierarchical merging \citep[e.g.][]{Zaritsky2022,Burkert2020}. The \gls{EWS} and \gls{EDS} offer unprecedented statistical constraints on any systematic variations of \acrshort{EGC} demographics with environment and galaxy mass concentration, as well as the populations of nuclear star clusters \citep{voggel2016,carlsten2022}. Furthermore, within the Local Universe, \Euclid VIS images also resolve ultra-compact dwarf galaxies, thus helping to complete our current understanding of these systems in a high-density environment \citep[e.g.][]{voggel2020,wang2023}.

Finally, \Euclid has the potential to furnish definitive estimates of galaxy distances, in particular through \glspl{SBF}. \gls{SBF} distance measurements have to be calibrated with respect to the galaxy stellar populations, that is, their age and metallicity, and have been typically applied to massive galaxies \citep[e.g.][]{tonry2000,mei2007,blakeslee2009,cantiello2018}. It would be particularly important to apply the \gls{SBF} methodology for distance estimates to dwarf galaxies, because they tend to be too faint for an emission line analysis that would provide a redshift. Dwarf galaxies make up the bulk of galaxy populations, but detecting them without being able to measure their distances makes demographic studies impossible. \Euclid's sensitivity to the abundant dwarf galaxy population enables extending the \gls{SBF} method to dwarf galaxies and massive galaxies in different environments, with statistics that will only be possible through the \gls{EWS}.
Although the \gls{PSF} is somewhat undersampled in \Euclid images, preliminary results show that the \gls{SBF} signal can be detected even as far as the Perseus cluster (Cantiello et al., in prep.).

\subsection{Galaxy structure and morphology}
\label{sec:morphology}

\begin{figure}[!ht]
    \centering
    \includegraphics[width=\linewidth]{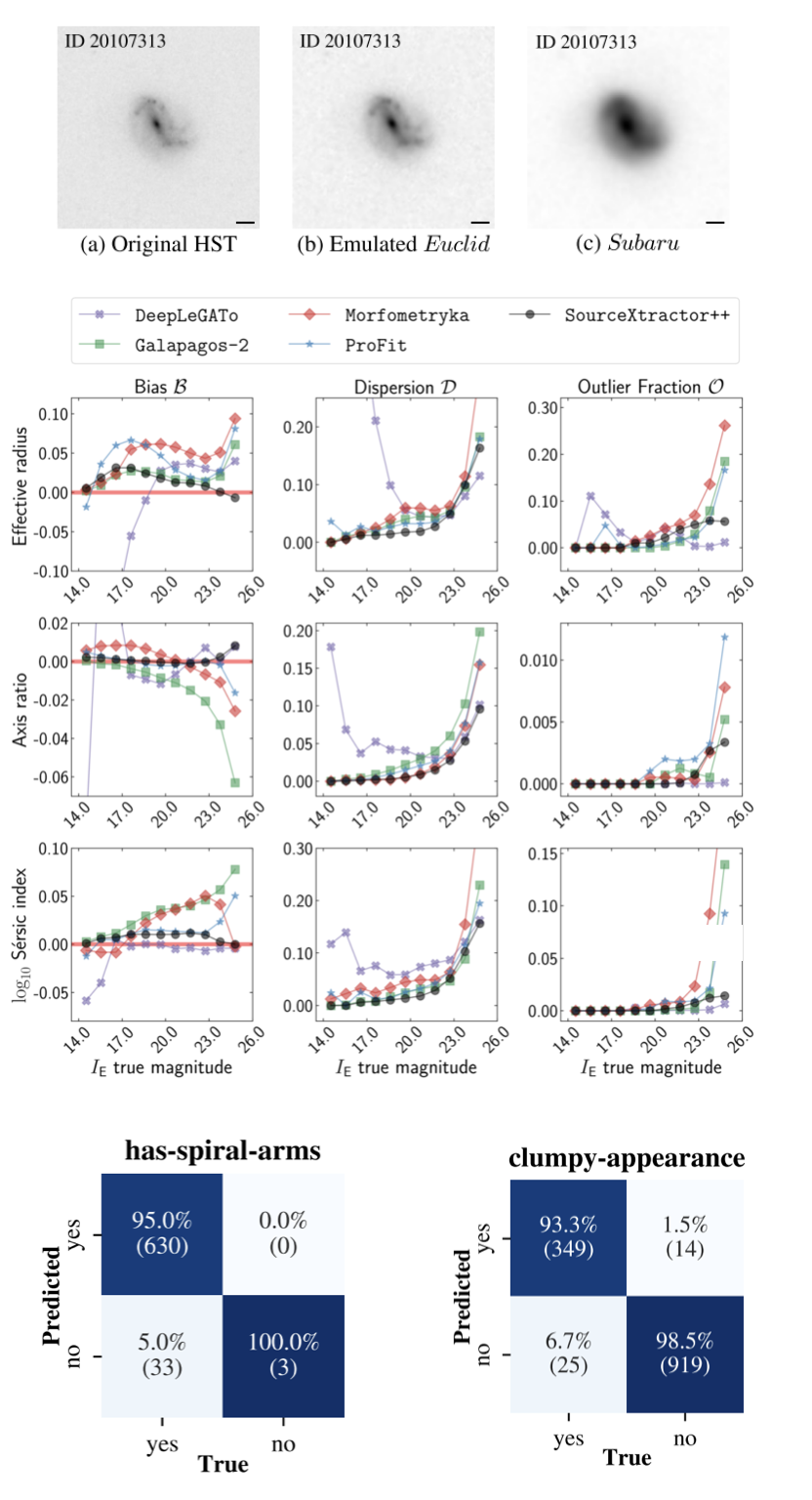}
    \caption{Illustration of \Euclid's capabilities to measure galaxy morphologies. {\it Top panels:} Example of a simulated galaxy observed with VIS as compared to \gls{HST} and Subaru/HSC. The horizontal black line indicates a $1^{"}$ length. {\it Middle panels:} Comparison of the bias (left column), dispersion (middle column) and outlier fraction (right column) of the effective radii (top row), axis ratio (middle row) and \Sersic~index (bottom row) for the best-fit \Sersic~profiles obtained with different state-of-the art surface brightness fitting codes applied to simulated \Euclid galaxies as a function of \IE. \Sersic~parameters can be obtained with errors smaller than $\sim10\%$  down to a \IE=24. {\it Bottom panels:} Accuracy of deep learning based morphological classifications on simulated \Euclid observations of galaxies trained on human based labels.  The confusion matrices show the accuracy for identifying spiral arms (left) and clumpy galaxies (right). Figure adapted from~\cite{2024arXiv240210187E} and~\cite{Bretonniere-EP26}
}
    \label{fig:galaxymorphology}
\end{figure}

Unlike \gls{LSST} or completed large ground-based imaging surveys, such as \gls{SDSS}, \gls{KiDS}, and \gls{DES}, \Euclid can reveal features in the surface brightness distribution for a 
considerable fraction of the galaxies it will image. With a \gls{PSF} \gls{FWHM} of \ang{;;0.13} (see~\cref{sec:pv_psf}) \Euclid's resolution will be similar to that of \gls{HST}(\cref{fig:galaxymorphology}). However, the \gls{EWS} will cover an area that is over 1000 times larger than what \gls{HST} has imaged since 1990, as well as covering several deeper fields.  This will allow investigations into the structures and morphologies of galaxies, which can then be correlated and applied to determine the physical drivers of galaxy evolution with redshift, and
to establish how the environment affects galaxy properties.

We know mostly from \gls{HST} that galaxies in the early Universe were more morphologically peculiar, compact in size, and undergoing more star formation, compared with galaxies at $z=0$ \citep[][]{Conselice2005, Conselice2014, 2016MNRAS.462.4495H}. \gls{JWST} is transforming our view of this topic, with the discovery that disk galaxies are much more common at high redshifts than we had previously thought \citep[][]{2022ApJ...938L...2F, 2023arXiv230207277V,2023arXiv230502478H}.  \Euclid will make a unique contribution to this research area by providing orders of magnitude more resolved galaxy structures in the distant Universe than what we could ever obtain with \gls{HST} and \gls{JWST}. This includes parametric and non-parametric morphological investigations of galaxy structure and how these properties evolve to the highest redshifts where \Euclid can resolve galaxies.  The benefit of \Euclid is that the rarer, more massive galaxies can be studied in detail. Hence, we will be able to examine the structures of the most massive and largest systems up to $z \sim 6$. 

\def\nSer{n_\sfont{Ser}}

There are several ways in which the morphological properties of galaxies measured by \Euclid will be investigated. One of the main methods for characterising galaxy morphology is to use the \Sersic\ fitting method, whereby a predetermined profile is fit to galaxy light distributions. As discussed in \cref{sec:catalogues}, such fits are performed by the main pipeline. The output of this gives the size of each galaxy, as well as its \Sersic\ index $\nSer$ \citep{Sersic1963}, with $\nSer = 1$ being an exponential disk and $\nSer = 4$ a de Vaucouleurs profile. 
The \Sersic\ modelling can be made more complex by adding further components to model the surface-brightness profiles of both bulges and disks, which is usually referred as bulge/disk decomposition. \Euclid will be able to measure single \Sersic\ parameters with around 10\% accuracy down to an apparent magnitude of $\IE<23$, which roughly corresponds to 450 million galaxies with a median redshift of $z\,{\sim}\,1.5$. For bulge-disc decomposition, the same 10\% accuracy is reached for galaxies with $\IE<21$ \citep{2023A&A...671A.101E,Bretonniere-EP26}. 
These measurements will be critical to examine how scaling laws, such as the size-mass and size-environment relations, evolve with redshift, with unprecedented statistics.  With \Euclid we can determine accurate sizes from both \Sersic\ fits and Petrosian radii \citep{Petrosian1976}, and use these to examine in great detail how galaxy sizes have changed with time, over a wide range in stellar mass.

The analysis pipeline will also provide non-parametric descriptions of galaxy structure, without the recourse to a predetermined parametric model, which is used in the \Sersic\ fitting. As discussed in \cref{sec:catalogues}, this includes the \gls{CAS} parameters \citep{2003ApJS..147....1C}, as well as the Gini and M20 parameters \citep{Lotz2004} as part of the standard \Euclid pipeline. These parameters can then be used to examine the formation histories of galaxies, as well as the morphological properties of galaxies as a function of redshift \citep[e.g.][]{Conselice2003, Conselice2014}.  These non-parametric measurements will allow us to measure structural evolution in a quantitative way, as opposed to simple visual estimates. Moreover, it allows us to find galaxies that are undergoing mergers to trace the merger history of galaxies, and thereby measure the role of merging in galaxy formation and evolution.

In recent years, deep learning has also been extensively used to provide morphological classifications of galaxies \citep[see][for a review]{2023PASA...40....1H}. This is particularly useful for providing detailed morphological descriptions of the internal structure of galaxies (e.g. clumps, bars) for samples that are too large to be visually inspected.
\cite{2022A&A...657A..90E} estimated that the \gls{EWS} will be able to resolve the internal morphological structure of galaxies down to a surface brightness of $22.5\,\mathrm{mag\,arcsec}^{-2}$, and the \gls{EDS} down to $24.9\,\mathrm{mag\,arcsec}^{-2}$, which roughly corresponds to 250 million galaxies at the end of the mission. This magnitude limit is typically brighter than for the S\'ersic fits because internal features need to be detected. The \Euclid photometric pipeline will provide neural network-based morphologies for this sample, first using existing labels from several Galaxy Zoo projects \citep[e.g.][]{Lintott08} and subsequently complemented with classifications done on \Euclid images~(\citealp{2024arXiv240210187E}; see also \cref{fig:galaxymorphology}). These detailed morphologies will enable a large variety of scientific analyses aiming at constraining the physical processes that drive the structural evolution of galaxies, for instance by comparing to predictions from simulations.

\subsection{Active galaxies across redshift}
\label{section:agn}
 
Active galaxies have compact regions at their centres with characteristics indicating that their luminosity is not produced by stars but is the result of the accretion of matter onto a \gls{SMBH} at the centre of its host galaxy. 
While all local massive galaxies show some level of \gls{AGN} activity \citep{Sabater2019}, phenomenology caused by a high accretion rate onto the central black hole is seen in less than 10\% of massive galaxies, and is thought to be short-lived (500 Myr to a few Gyr).

\begin{figure}
    \centering
    \includegraphics[width=\linewidth]{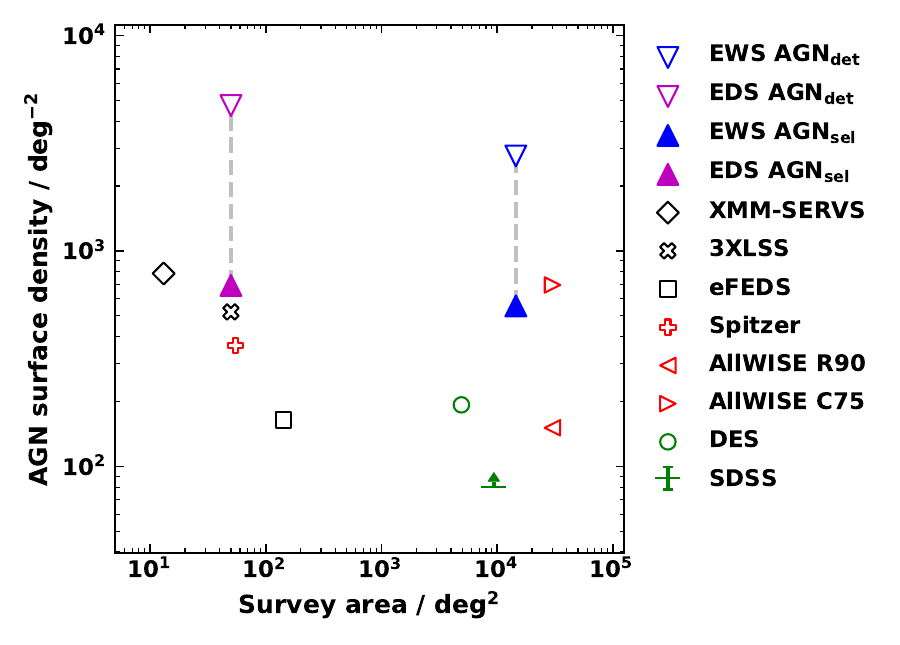}
    \caption{\gls{AGN} surface density (deg$^{-2}$) versus survey area (deg$^2$) for \gls{EWS} and \gls{EDS} compared with wide field and medium area surveys in
different wavebands (according to the legend). Unfilled downwards triangles show the surface density of \gls{AGN} detected in at least one \Euclid band (at 5$\,\sigma$), while filled upwards triangles represent the surface density of \gls{AGN} selected by using a simple colour criterion with \Euclid and \gls{LSST} colours, in both \gls{EWS} and \gls{EDS}.
}
    \label{fig:AGNNumber}
\end{figure}

Type~1, or unobscured \gls{AGN}, typically show broad emission lines (${\rm FWHM}>1000\,\si{\kms}$) and power-law continuum emission originating from the accretion disk, while Type~2, or obscured \gls{AGN}, typically show extreme emission line ratios compared to the normal galaxy population and continuum emission with host galaxy features. In orientation-based unification models \citep[e.g.][]{antonucci93,urry95, netzer15}, Type~2 \gls{AGN} have been described as obscured Type~1 \gls{AGN} with the broad-line emitting region and accretion disk being hidden behind a partially opaque torus. According to this simple scheme, Type~1 and 2 \gls{AGN} should have similar distributions in terms of redshift, luminosity, host galaxy properties, and black hole mass. By contrast, in \gls{AGN} evolutionary scenarios \citep[e.g.][]{2008ApJS..175..356H, 2009ApJ...696..891H}, obscured Type~2 \gls{AGN} may represent an earlier evolutionary phase compared to unobscured systems, and thus have different physical properties. The transition from obscured to unobscured accretion in \gls{AGN} is postulated to occur through an outflow phase, during which energetic feedback is deposited in the host galaxy, impacting the formation of new stars \citep[for a review see][]{2018NatAs...2..198H}. Given their transient nature, comprehensive studies necessitate very large sample sizes to decipher their evolutionary paths, link to their host galaxies, clustering tendencies, and large-scale environments. 

The unique combination of spatial resolution, depth and wide area coverage of \Euclid will allow us to explore the \gls{AGN} population like never before using both photometric and spectroscopic selection criteria. \Euclid's spatial resolution will provide critical observational constraints on \gls{AGN} morphology and merger rate of their host galaxies. The depth of the \gls{NIR} observations will allow for the first time the detailed study of Type~1 and Type~2 \gls{AGN} sub-populations and their co-evolution with galaxies during the so-called `Cosmic Noon' ($1<z<3$) and beyond. Finally, due to the depth and wide area coverage, rare and extreme states, such as red quasars, will benefit from detailed analysis of their morphologies and large-scale environment. 

The most luminous Type~1 \gls{AGN}, namely quasars, can be identified using \Euclid photometry alone or in combination with multi-band coverage from optical surveys, such as \gls{LSST}. We estimate that $4\times10^7$ ($2.4\times10^5$) \gls{AGN} will be detectable (at 5$\,\sigma$) in at least one \Euclid filter in the \gls{EWS} (\gls{EDS}), corresponding to a surface density of $3-5\times10^5$ AGN per square degree (\ref{fig:AGNNumber} open downwards triangles; Selwood et al., in prep.). This large sample will include about 30\% Type~1 and 70\% Type~2 \gls{AGN}, based on population studies in the X-rays \citep{2016A&A...587A.142F,Merloni2014MNRAS.437.3550M}. A colour selection using \textit{u} band in combination with the \textit{i}, \textit{r}, or \textit{z} filter, reaches completeness and purity $\sim$  81\% (77\%) and 92\% (91\%) for the \gls{EWS} (\gls{EDS}), respectively (Bisigello et al., in  prep.). As shown in \cref{fig:AGNNumber}, this corresponds to a total of $8.1\times10^6$ ($3.5\times10^4$) \gls{AGN} in the \gls{EWS} (\gls{EDS}) identified as \gls{AGN}, by using \Euclid and \gls{LSST} colours (filled triangles). This simple colour selection will provide quasar surface densities that are better or en par with current AGN surveys, from X-rays \citep[e.g. XMM-SERVS, ][]{chen2018}, optical \citep[e.g. \gls{DES}][]{Yang2023ApJS..264....9Y} and mid-infrared \citep[e.g. WISE 75\% completeness -- C75, ][]{Assef2018ApJS..234...23A}.
On the other hand, the identification of Type~2 \gls{AGN} is challenging with \Euclid and optical filters alone. Longer wavelength observations, in combination with more sophisticated methods, are therefore necessary. 
 
The identification of \gls{AGN} in \Euclid will also rely on spectroscopic data.
Type~1 \gls{AGN} will be classified through broad emission-line detection, while Type~2 \gls{AGN} will be identified by using the narrow emission-line \nii{6584}/\ha\ versus \oiii{5007}/\hb\ diagnostic diagram, called the
\gls{BPT} diagram \citep{1981PASP...93....5B}. A notable challenge for the \gls{BPT} \gls{AGN} identification using \Euclid spectroscopy is the limited spectral resolution ($R\simeq450$), leading to the blending of key emission lines (\ha\ and \nii{6548,6584}) and a restricted availability of diagnostic narrow emission lines in specific redshift ranges within the \gls{EWS} and \gls{EDS} \citep{lusso2023arXiv}.
 
With \Euclid we will use BPT diagrams to classify and characterise Type~2 \gls{AGN} within the narrow redshift range $1.5 < z < 1.8$ ($1 < z < 1.8$) in the EWS (EDS),  which corresponds approximately to the peak of star-formation activity. Additionally, bright Type~2 \gls{AGN} will be identified through the detection of high-ionisation emission lines, such as \nev{3426} \citep{2013A&A...556A..29M} at $1.7 < z < 4.4$ and \ion{C}{iv} at $z > 4.9$ in the EDS \citep[e.g.][]{2019A&A...626A...9M}. 
Furthermore, spectroscopic redshifts will be available for millions of \Euclid \gls{AGN}.  We expect to determine spectroscopic redshifts for around 90\% of the Type~2 \gls{AGN} in the redshift range $0.9 < z < 1.8$  down to an emission line flux of 
about $3\times10^{-16}\,\mathrm{erg}\,\mathrm{cm}^{-2}\,\mathrm{s}^{-1}$ for the integrated \ion{H}{$\alpha$}$+$[\ion{N}{ii}], whilst the same redshift completeness percentage occurs for an emission line flux more than a factor of two higher for Type~1 \gls{AGN}, namely $8.5\times10^{-16}\,\mathrm{erg}\,\mathrm{cm}^{-2}\,\mathrm{s}^{-1}$ \citep{lusso2023arXiv}.

\Euclid's exceptionally extensive \gls{AGN} data set presents a unique opportunity to constrain the \gls{LF} of \gls{AGN} in the near-infrared, over larger magnitude and redshift ranges, and larger area than previously employed to generate \gls{NIR} galaxy \glspl{LF} \citep{bell2003,cirasuolo2007}.
The \gls{AGN} \gls{LF} and its evolution with time are key observational quantities for understanding the origin of \glspl{SMBH} and accretion onto them \citep{aird2013,shankar&2013,schulze2015}.
Additionally, the measurement of the \gls{SMBH} mass serves as a key parameter for studies aimed at establishing scaling relations between black holes and host galaxy properties. These scaling relations are vital for testing black hole feedback mechanisms \citep[e.g.][]{2015MNRAS.448.1504S} and cosmological hydrodynamical simulations of structure formation that investigate the relationship between galaxy and black hole growth \citep[e.g.][]{2015MNRAS.454..913D}. 
The \gls{AGN} \gls{LF} and the \gls{SMBH} demography \citep{lusso2023arXiv} will be important observational products of \Euclid, along with the \gls{AGN} clustering, which links the evolution of growing \glspl{SMBH} and the large-scale cosmic structure \citep{allevato2019A&A...632A..88A}, and the study of \gls{AGN} close pairs and galaxy mergers.

\subsection{Galaxies and quasars in the epoch of reionisation}
\label{section:highz}

\Euclid's combination of optical and \gls{NIR} instruments makes it ideally suited for identifying galaxies and quasars at high redshifts, taken here to be $z \geq 7$.  Whilst other telescopes can reach these distant redshifts, most notably \gls{HST} and \gls{JWST}, \Euclid has the advantage that it covers a large area of the sky and therefore can find rarer but brighter sources. The data from \Euclid will be sufficiently deep such that it will probe into the \gls{EOR} whereby all sources at these redshifts will exhibit Ly$\alpha$ breaks \citep{1965ApJ...142.1633G}, redshifted to an observed wavelength of $\lambda_\alpha \simeq \{0.97 + 0.12 \, (z - 7)\}$\,\micron. Such sources would be detectable in \Euclid's NISP images and grism spectra but will be VIS dropouts, so they could potentially be identifiable using \Euclid data in isolation.  
The difficulty is that the vast majority of optical dropouts are not high-redshift sources, so the reliable selection of these sources represents a significant data analysis challenge (even with access to external data sets); however, the potential \Euclid science return from high-redshift galaxies and quasars justifies this effort.

\Euclid will cover 50\,deg$^2$ to depths of about 26.4 AB magnitudes ($5\sigma$ point source limit) in the \gls{NIR} filters in the \gls{EDS} fields, which means that it probes a unique parameter space in the selection of $z > 7$ galaxies. Previously, space-based \gls{NIR} telescopes have been limited in their ability to select the brightest and hence rarest sources due to their small \glspl{FOV}: the \gls{WFC3} on \gls{HST} has an \gls{FOV} of 4.5\,arcmin$^2$ and the \gls{NIRCAM} on \gls{JWST} has an \gls{FOV} of 9.7\,arcmin$^2$. Even with large efforts to produce mosaics, these cover $\la 1\,{\rm deg}^2$.  Hence, while the study of ultra-high-redshift galaxies was undoubtedly revolutionised with the deep \gls{NIR} surveys of \gls{HST} and \gls{JWST}, which allow galaxy candidates to be identified up to $z \simeq 11$--13, the identified galaxies are typically limited to $L < L^\ast$ (with $L^\ast$ being the knee in the rest-frame UV \gls{LF}; see review by \citealp{Stark2016}).  In the traditional `dropout' formalism, galaxies will be selected with \Euclid via the Lyman-break as \IE dropouts at $z \simeq 7$, \YE-dropouts at $z \simeq 8.5$, and \JE-dropouts at $z \simeq 11.5$, and \Euclid can in principle detect sources up to $z \simeq 15$ because the \HE\ band extends up to 2\,\micron.  
Importantly, the higher resolution of the \Euclid imaging will aid in the identification of the main contaminants, brown dwarfs (cf.~\cref{sec:MWRSP}), since the galaxies are expected to be resolved at these bright magnitudes. 
Moreover, contamination fractions from $z < 6$ galaxies can be reduced from up to 40\% to less than 5\% with the inclusion of deep optical photometry, such as \gls{LSST} \citep{vanMierlo-EP21}.

These luminous galaxies provide signposts to the most ionised regions of the neutral \gls{IGM} at $z > 7$, while also representing a challenge to theoretical models and being key candidates for multi-tracer follow-up, for example with \gls{ALMA} and \gls{JWST}.  Recent results from \gls{JWST} \citep[e.g.][]{Naidu2022, Donnan2023, Finkelstein2023} have revealed a surprising abundance of luminous sources, of which several have shown unusual spectral features (e.g., GNz11; \citealp{Oesch2016, Bunker2023}).  These results highlight the huge potential of \Euclid-selected high-$z$ sources in understanding the earliest stages of galaxy formation and reionisation.

Much of this work in the deep fields will be aided by observations from the Cosmic Dawn Survey (\cref{sec:groundbased}), which provides matching depth  $u$-band from \gls{CFHT}, $griz$-bands from \gls{HSC} on Subaru, and 3.6\micron\ and 4.5\micron\ from Spitzer/IRAC  as described in more detail in \cref{sec:groundbased}.  The science goals of this survey are primarily focused on the high-redshift Universe, but will also facilitate many aspects of legacy science for the \Euclid mission. Leveraging the deep wide area data from the Cosmic Dawn Survey will provide robust measurements of the galaxy stellar mass function to $z=8$ and the UV luminosity function to $z=10$. Other science goals include mapping the topology of reionisation, studying the formation of large-scale structure, quantifying the prevalence of high-redshift protoclusters, and characterising the first quenched galaxies. The survey data have immediate value for studying galaxy formation and evolution, and that legacy value will only increase as more data are collected.
 
Previous work on smaller scales shows the potential of \Euclid discoveries. Degree-scale ground-based surveys from UKIRT and VISTA have provided the first view of the \gls{LF} at $L > L^\ast$ for $z = 7$--$10$, showing an excess of sources over the predictions of \gls{HST} surveys and revealing an apparent lack of evolution in the number density of the brightest sources \citep[e.g.][]{Bowler2020, Harikane2022}. With the unique combination of space-based, wide-area, deep \gls{NIR} imaging provided by \Euclid, the number of rare luminous galaxies will be increased by factors of 100--1000 over current samples, providing key constraints on the bright end of the \gls{LF}. Integrating the current best-fit \glspl{LF} leads to predictions of thousands of sources from the deep fields, down to an absolute UV magnitude of $M_{\rm UV} = -21$. This allows the shape at the bright end -- which depends sensitively on astrophysical effects such as dust, scattering, and lensing -- to be unambiguously determined.
If the slope of the bright end continues as a power law beyond observational limits, there is the possibility of detecting a substantial number of extremely bright galaxies with NISP  $m_{\rm AB}\lesssim24.0$ within the \gls{EWS} (approximately 2000 versus 50 for a double-power law or Schechter function, respectively, at $z = 7$; \citealp{Bowler2017}).  These additional objects will not only be useful for deriving the \gls{LF}, but can be used to study the structures, stellar masses, and star-forming properties of these early systems and to compare them to lower redshift systems, as well as with galaxies at lower masses studied with \gls{HST} and \gls{JWST}.

High-redshift quasars are much more luminous than comparably distant galaxies (and, unlike $\gamma$-ray bursts, are non-transient) and so are among the most useful probes of the high-redshift universe.
As described in the review by \cite{Fan_etal_2023}, spectroscopic observations of high-redshift quasars probe the growth of the first 
\glspl{SMBH}, provide several unique constraints on cosmological reionisation, and also give a record of the evolution of elemental chemical abundances.  
Luminous high-redshift quasars are expected to be extremely rare: while the redshift range $6 \leq z \leq 7$ is reasonably well explored, with approximately 300 quasars known, just eight bright quasars have so far been spectroscopically confirmed to be at $z > 7$ \citep{Fan_etal_2023}, along with some more speculative fainter detections at even higher redshifts using JWST \citep{2024Natur.627...59M,2024NatAs...8..126B}.  It is in this regime that \Euclid should be able to make a transformational contribution, thanks primarily to the large area coverage of the \gls{EWS} (\cref{sec:wide}). The headline prediction given in \cite{Laureijs11} was that \Euclid imaging data would be expected to include 55 $z > 8.1$ quasars brighter than $\JE = 22.5$, sufficient to completely revolutionise this field. \cite{Barnett-EP5} then presented a more realistic simulation, which explored the impact of a range of contaminants (M, L, and T dwarfs, as well as early-type galaxies at $1 \la z \la 2$), different selection methods, the availability of external optical data (from \gls{LSST}), and uncertainty about the evolution of the quasar \gls{LF}.  

While these simulations showed that the \Euclid data should be able to produce complete high-redshift quasar samples to a greater depth of $\JE \simeq 23$ (particularly if it is possible to cross-match to \gls{LSST} optical data, this is outweighed by the steeper decrease in quasar numbers with redshift found by \cite{Jiang_etal_2016}. So the realistic predictions from \cite{Barnett-EP5}, given in full in Table~3 of that paper, are broadly that the \gls{EWS} photometry\footnote{It is also possible that the brightest high-redshift quasars could be identified from the \Euclid grism spectra directly. Unfortunately, the simulations described by \cite{Roche_etal_2012} are no longer relevant because of the subsequent reduction in the capabilities of the grisms, and no updated studies of this possibility have been published to date.} will yield approximately 100 robust quasar candidates with $7.0 \leq z \leq 7.5$, around 25 quasars beyond the current record of $z \simeq 7.5$ \citep{Banados_etal_2018, Yang_etal_2020, Wang_etal_2021}, and perhaps 10 quasars at $z \geq 8.0$.

While the \Euclid photometry will be necessary to identify (candidate) $z > 7$ quasars, confirmation and characterisation will (in contrast to most other projects described in this paper) come primarily from external follow-up observations. Most important will be \gls{NIR} spectroscopy; hence there is a particular utility to finding the brightest sources with $M_{1450} \la -26$, corresponding to $\JE \la 21$. Even one such detection at $z \ga 8$ (plausible in DR1) would represent major progress in this field; the full \gls{EWS} should yield a well-characterised population of quasars out to at least $z > 8$, with the observational frontier possibly pushed back as far as $z \simeq 9$.

\subsection{Galaxy evolution and environment}
\label{sec:env}

While it is largely agreed that the cosmic environment in which a galaxy evolves strongly correlates with its measured properties \citep[e.g.][]{2017ApJ...837...16D}, we still need to understand all the physical processes that drive its evolution as well as their relative importance. \cite{Dressler80} showed that, at least in the local Universe, galaxies in denser environments are more likely to be elliptical galaxies, whilst systems found within regions of lower density are generally galaxies with spiral and irregular morphologies. Furthermore, in the local Universe, with the exception of radio-bright galaxies that preferentially reside at the centres of clusters \citep[see][for a review]{maglio22}, \gls{AGN} are observed to avoid massive structures \citep[e.g.][]{popesso06}. 

A wide diversity of other properties are observed to correlate with various measures of environment as well. For instance, on average, high-density regions are mostly populated by redder, brighter, more metal rich, and less star-forming galaxies, while the opposite is true for low-density regions (e.g. \citealt{kauffmann04}; \citealt{blanton06}; \citealt{2021MNRAS.505.4920W}). Additionally, `green valley' galaxies, likely to be transitioning from star-forming to quiescent phases, have a lower specific star-formation rate in groups and clusters than in the field \citep[e.g.][]{2020ApJ...894..125J}. Another example is the positive (negative) correlation between the star-formation rate (\hi\ gas deficiency) of galaxies, at fixed mass, and their distance to the closest filamentary structure
(\citealp{2017MNRAS.470.1274M,2018MNRAS.474.5437L,2018MNRAS.474..547K,2020A&A...635A.195G}, \citealp{2018ApJ...852..142C}), or the recent evidence that galaxies that are strongly connected to the cosmic web form fewer stars and are more pressure supported than those that are  weakly connected. (\citealp{2020MNRAS.491.4294K}). 
Furthermore, a bimodality in the central surface brightness of disks is found for galaxies in voids, filaments and knots, but not in sheets (\citealp{2016MNRAS.455.2644S}). All these findings seem to hold at least out to $z\sim 1$ \citep[e.g.][]{cucciati,2020ApJ...894..125J}.

At higher redshifts the situation is less clear, although there seems to be increasing evidence for a reversal of at least some of the scaling relations mentioned above. As an example, it has been found that at $z\sim 1.5$--2 more \gls{AGN} reside within dense regions than in the field \citep[e.g.][]{martini}. However, the exact details of the processes leading to such observed correlations at $z>1$ are still unknown. Evolutionary paths followed by galaxies are extremely complex, especially because several processes are at play in an intertwined way. For instance, \citet{2009MNRAS.400..937M,2012MNRAS.423.1277D,2012MNRAS.421.1949V} and \citet{2019A&A...632A..49S} showed that galaxies are probably preprocessed in groups (filaments) before falling into clusters where they finish their transition to passive states.

Given its extremely wide area coverage, \Euclid will probe all environments in which galaxies are found, even the most extreme ones, over the widest redshift range to date. Indeed, it will be able to measure properties (such as redshifts, stellar masses, star-formation rates, presence and relevance of a central AGN) for billions of galaxies in regions of varying density out to $z \sim 2$ and, in the three deep fields, even out to $z\sim 3$. 

In order to exploit the exquisite quality of \Euclid data for environmental studies, we have developed several codes for the reconstruction and characterisation of the observed density field traced by \Euclid galaxies at all scales. These all deal with the search of overdense structures such as high-$z$ clusters and protoclusters. We also optimised existing codes for the detection of cosmic filaments \citep[e.g.][]{disperse} in order to work with both photometric and spectroscopic samples. 
These codes will complement those already used for the detection of voids (e.g. \texttt{VIDE}, \citealt{VIDE} and \citealt{Contarini2022}) and clusters (\texttt{AMICO} and \texttt{PZWav}, \citealt{Adam-EP3}), allowing the reconstruction and investigation of the full density field at all scales. This will permit us to shed light on the link between galaxy and \gls{AGN} formation and evolution and their environments out to $z\sim 3$.

Furthermore, the codes developed for the investigation of the clustering properties of \Euclid galaxies both in configuration space (two- and three-point correlation functions) and in Fourier space (power spectra and bispectra, cf.~\cref{sec:clustering_stats}) will also be used to relate the properties (stellar mass, colour, luminosity, star-formation rate, metallicity, star-formation history, black hole type, accretion, luminosity, etc.) of the galaxies and \gls{AGN} observed by \textit{Euclid} with the \gls{LSS} they trace. This will permit the establishment of a direct connection with the dark halos inhabited by these sources, via the so-called bias function \citep[e.g.][]{mo,scoccimarro}.

\subsection{Star-forming galaxies across time}

The combination of photometry and spectroscopy will enable detailed investigations into the physical properties of star-forming galaxies, such as stellar populations, star-formation rates, and dust attenuation, as well as their evolutionary processes and the interplay between star formation and black hole accretion at the peak of \gls{AGN} activity and star-formation history, during Cosmic Noon.

\begin{figure*}[!ht]
\centering
\includegraphics[width=0.995\textwidth]{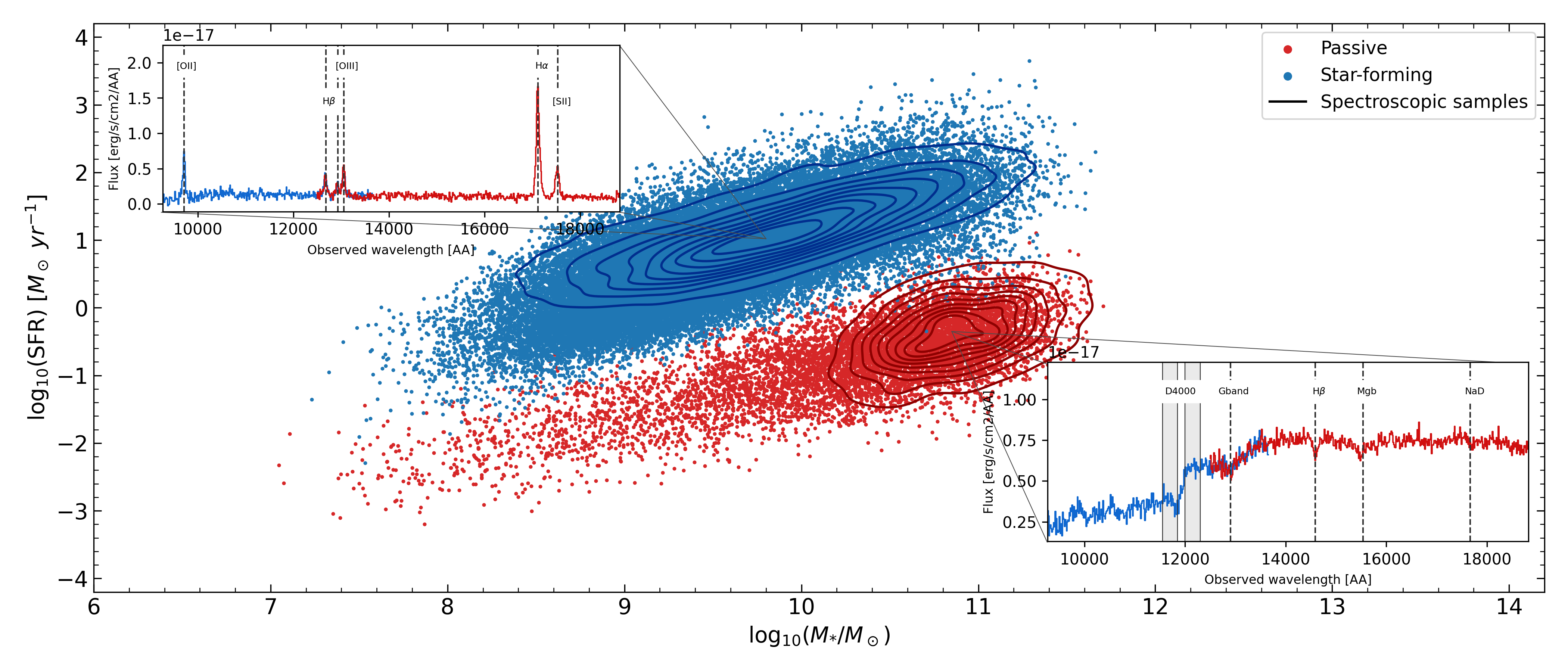}
\caption{\gls{SFR}-stellar mass diagram. The points represent the photometric sample, divided into star-forming (in blue) and passive galaxies (in red). Galaxy type was assigned as a function of mass and redshift from the stellar mass function (SMF) by \cite{2010ApJ...721..193P} and \cite{Ilbert2013}. The coloured contours highlight the spectroscopic sample (for the \gls{EWS} in the case of star-forming galaxies in blue, and for the \gls{EDS} in the case of passive galaxies, in red). The two insets show two examples of star-forming and passive galaxies as observed in the \gls{EDS}, comprising both the blue and the red grisms, simulated from the \gls{MAMBO} mock catalogue \citep{Girelli.phd}  taking into account all instrumental and observational effects. }
\label{fig:SFR-mass_spec}
\end{figure*}

In the \gls{EWS} \Euclid will map 2000--4800 H$\alpha$ emitters per deg$^{2}$ in the redshift range $0.9 < z < 1.8$ at a flux limit of $2\times10^{-16}\,\mathrm{erg}\,\mathrm{cm}^{-2}\,\mathrm{s}^{-1}$, totalling 28--67 million sources in the \gls{EWS}, while in the \gls{EDS} we will amass about 32\,000--48\,000 H$\alpha$ emitters per deg$^{2}$ (i.e. a total of 1.6--2.4\,million) at a flux limit of $5\times 10^{-17}\,\mathrm{erg}\,\mathrm{cm}^{-2}\,\mathrm{s}^{-1}$, in the broader redshift range $0.4 < z < 1.8$ enabled by the addition of the blue grism (\citealt{Pozzetti2016}). For these galaxies it is not only possible to obtain robust spectroscopic redshifts, but we will also be able to study a wide range of properties. First of all, the \gls{SFR}, which can be estimated from the intensity of the \ha\ emission line \citep[e.g.][]{1998ApJ...498..541K}, corrected for dust attenuation using \ha/\hb, when \hb\ is available or statistically otherwise: with such a sample of tens of millions of \ha\ emitters we will be able to analyse in detail the tight correlation between \gls{SFR} and stellar mass, the so-called galaxy \gls{MS}, down to a few $M_\odot\,\mathrm{yr}^{-1}$ at $z\sim 1$ in the \gls{EWS}, as shown in \cref{fig:SFR-mass_spec}. 
In particular, with the \gls{EWS} sample we will be able to probe the most massive part ($\logten M_\ast/M_\odot > 10$) of the \gls{MS} with unprecedented statistics at these redshifts, and help provide the physical explanation for the observed turnover in the \gls{MS} shape and its evolution \citep[e.g.][]{2023MNRAS.519.1526P}.  Additionally, the \gls{EDS} will allow us to explore the fainter end of \gls{SFR}-stellar mass space. Moreover, thanks to the wider wavelength range due to the availability of the blue grism, multiple spectral lines will be available for individual galaxies (e.g. \hb, \oiii{5007}, and \oii{3727}), enabling us to recover dust attenuation and gas-phase metallicities, and to put solid constraints on a variety of scaling relations \citep[e.g. the fundamental metallicity relation and mass-attenuation][]{2019A&ARv..27....3M}. 
In the \gls{EWS} we expect to find about 440 [OIII]5007
emitters per deg$^{2}$ (resulting in a total of about $6\times 10^6$ over the entire \gls{EWS} area) in the range $1.5<z<2.3$ \citep{2020ApJ...897...98B}. 
Finally, a complementary sample of \Pb\ emitters would also be mapped at low redshift ($z<0.48$), similar to \cite{Cleri22} with HST grism G141. 

\Euclid photometric information, complemented by ground-based data, will also be used to derive the physical properties of galaxies \citep{Bisigello-EP23}. In particular we will be able to classify galaxies from their photometric SED \citep{Bisigello20} and compute the \gls{SMF} of galaxies of different classes to derive its differential evolution and infer the history of cosmic mass assembly. The enormous \gls{EWS} area will enable the study of the evolution of the number density of rare populations, like massive star-forming galaxies, in order to relate them to the mass growth and buildup of passive galaxies, while assessing the contribution due to mergers, thanks to the exquisite morphological information that the \Euclid high resolution will provide (cf.~\cref{sec:morphology}).  By comparing the \glspl{SMF} of galaxies classified with different methods (i.e. using their morphology or colours), it will also be possible to understand the timescale of the morphological transformation and stellar population ageing \citep[e.g.][]{2010A&A...523A..13P} that lead to the present-day galaxies and to derive constraints on the quenching mechanisms \citep{2010ApJ...721..193P} responsible for turning off star formation. 

The combination of physical and morphological properties will provide constraints on the size-mass relation of blue star-forming galaxies and insights on the connection between the compactness of galaxies, their light profiles, and the efficiency of the star-formation process, leading to an understanding of the range of properties of the blue cloud galaxies quenched by internal mechanisms \citep{2022MNRAS.512.1262H}. 
To explore  the parameter space sampled by the \gls{EWS} and \gls{EDS} even further, we will
consider samples of galaxies sharing similar properties, for instance by reducing the high-dimensional photometric space of colours and fluxes with machine-learning algorithms (e.g. \glspl{SOM}, as in \citealt{2022A&A...665A..34D}, see also \cref{sec:photoz}).
The spectra of these groups of galaxies can then be analysed with stacking techniques to improve the signal and eventually study the \gls{MS} at 0.9$<z<$1.82 at \glspl{SFR} as low as $0.1\,M_\odot\,\mathrm{yr}^{-1}$ in the \gls{EWS}, and $\sim 0.01 \,M_\odot\,\mathrm{yr}^{-1}$ in the \gls{EDS}.
Similarly, SED fitting will be performed on the photometric composite SEDs with high \gls{S/N} to analyse the physical properties and their evolution especially of faint galaxies. Taking advantage of the 3-dimensional distribution of galaxies (see \cref{sec:env}), we will study how the physical properties of galaxies depend on the environments they inhabit.

The source-subtracted \gls{CIB} fluctuations uncovered in deep {\it Spitzer} data  \citep[e.g.][]{Kashlinsky2005,Kashlinsky2012} are coherent with the unresolved \acrlong{CXB} \glsunset{CXB}\citep[\gls{CXB};][]{Cappelluti2013,Mitchell-Wynne2016,Cappelluti2017,Li2018}, which could be caused by a new population \citep{Helgason2014} containing a fraction of black holes in excess of what appears in the known galaxy populations. As discussed in 
\cite{Kashlinsky2018}, \Euclid will play an important role in directly resolving the nature of the \gls{CIB} at near-IR wavelengths, because the source-subtracted \gls{CIB} measurements in the three \Euclid near-IR bands over the \gls{EWS} can be cross-correlated with the unresolved \gls{CXB} from eROSITA  to constrain model predictions \citep{Kashlinsky2019}.
  
In the far-IR, a statistical characterisation of star-formation processes across cosmic epochs can be obtained from a joint analysis of \Euclid data and maps of the far-IR \gls{CIB}, which probes the unresolved emission of dusty star-forming galaxies since the epoch of reionisation \citep{dole2006}. 
Maps of the far-IR \gls{CIB} can be extracted from multi-frequency \gls{CMB} observations through component separation or from direct observation at relevant frequencies, such as
those of {\it Herschel}, which is, however, hard to extract due to residual Galactic foreground contamination in far-IR \gls{CIB} maps \citep{planck2013-p13,maniyar2019,mccarthy2023}.
Instead, the cross-correlation of \gls{CIB} maps of {\it Planck}, {\it Herschel}, or future ground-based instruments (such as the Cerro Chajnantor Atacama Telescope) with the \Euclid observables (galaxy or QSOs clustering and cosmic shear) can be used to constrain models of star-formation history, its efficiency across time, and the connection to host halo mass and environment \citep{jego2023a,jego2023b}. 
\Euclid's \gls{NIR} images from both the \gls{EWS} and \gls{EDS} can be used to create background maps (after masking the detected sources) and directly cross-correlating with \gls{CIB} maps from the same external sources to extract additional complementary information \citep{Lim2023}.

\subsection{Passive galaxies and galaxy quenching}
\label{sec:passiveGalaxies}

Massive ($M_\ast > 10^{11}M_{\odot}$), quiescent galaxies are key systems for understanding galaxy formation, but they are rare and it is therefore challenging to assemble large samples. \Euclid spectroscopy can play a unique role in studying these systems in the \gls{EDS} at high redshifts. Indeed, due to the rapid decline of their number densities at $z>1.5$ \citep{Weaver23}, finding and observing over the widest possible redshift range is a crucial yet difficult task.

Much work has been done out to $z\,{\sim}\,1$, but the peak of star-formation activity and the assembly of passive galaxies is at $z\,{>}\,1$ \citep{Ilbert2010, Ilbert2013, Madau2014}. The study of the number densities of massive passive galaxies at high redshift -- that is the galaxy \gls{SMF} -- is crucial for understanding how the evolution proceeds from star-forming to passive galaxies, that is when galaxies are quenched, and when and how they assemble their mass, which is still a challenge for models of galaxy formation. The \gls{EDS} with the \gls{BG} will be truly unique for finding and spectroscopically identifying, at the continuum limit $\HE<21$ ($22$), the rarest and most massive quiescent galaxies at $z>1.4$, with $\logten(M_\ast/M_{\odot})>11.3$ (11). The evolution of the luminosity and mass functions of passive galaxy types can be followed in different environments \citep[e.g.][show the possibility to measure galaxy environments in the \gls{EDS} over the redshift range $0.9\,{<}\,z\,{<}\,1.7$, see also \cref{sec:env}]{Cucciati2016}. Therefore, the \Euclid deep \gls{BG} data can be used to probe the assembly of the red sequence out to $z\,{\sim}\,2$.

The \gls{BG} provides a fundamental complement to the \gls{RG} for the identification of a large sample of quiescent galaxies based on the identification of the Balmer or D4000 break.
We consider for the \gls{BG} the wavelength range 0.926--1.366\,\micron, and 1.206--1.892\,\micron~for the \gls{RG} (50\% peak transmission wavelengths) 
and a reference \gls{EDS} area of 50\,deg$^2$. The blue limit of \BGE, around 0.926\,\micron, allows us to detect the $D4000$ break starting at $z\sim 1.5$ and up to $z\simeq 2.2$, while the \gls{RG} allows its detection only in the redshift range $2.2\,{<}\,z\,{<}\,3.2$, where the expected number of passive galaxies is significantly smaller.

To obtain quantitative estimates of the expected numbers of passive galaxies, we have used the COSMOS2020 catalogue \citep{COSMOS2020}  
to derive the number density of passive galaxies expected for different \HE\ limits. We have selected passive/quiescent galaxies using the colour-colour selection $(\mathrm{NUV}-r)$ versus $(r-J)$ and derived the redshift distributions $(\diff N/\diff z)$ for different \HE\ limits. At the \BGE limit, about 5000 passive galaxies will be detectable at $z>1.5$ within the sensitivity limits of NISP continuum spectroscopy over 50 square degrees at \HE=21 in the \gls{EDS}, and over 30000 at \HE=22.

The red grism extends the possibility of detecting passive galaxies, through the D4000 detection at $z\,{>}\,2.2$, but only at $\HE<23$. A typical example is shown in \cref{fig:SFR-mass_spec}.
Keeping in mind the drop in the number of passive galaxies above $z=1.5$, the importance of having a blue grism is clear, since it allows for the detection of more than $5000$ massive quiescent galaxies with $\HE < 21$ and approximately $4\times10^4$ ($10^5$) galaxies with  $\HE<22$ (23) in 50\,deg$^2$. These numbers are to be contrasted with the several hundreds of spectroscopically confirmed objects in the \gls{RG}. 

\Euclid will not only provide a huge sample of passive/quiescent galaxies compared to what has been available before, but will also be fundamental for finding and spectroscopically identifying the rarest and most massive passive galaxies (${>}\,10^{11}\,M_\odot$) with unparalleled statistics.  This is because their number density (10--100\,deg$^{-2}$) makes their identification difficult with \gls{NIR} instruments with a small \gls{FOV} from space (e.g. \gls{JWST} or \gls{HST}), while from the ground strong sky residuals limit the sensitivity (e.g. VLT+MOONS or SUBARU+PFS). Only \gls{JWST} can compete with \Euclid in this area, but will provide much smaller samples.

At high redshift ($z>1.5$), which is the most active period of galaxy assembly, but where ground-based
spectroscopy is inefficient, \Euclid will truly revolutionise the field and will be a unique facility for galaxy evolution analysis. The large number of passive galaxies provided in the \gls{EDS} with the \gls{BG} plus \gls{RG} can be used to follow in detail the growth of this fundamental class of rare galaxies that remain a challenge for models of galaxy formation.

The detection of passive galaxy spectroscopic pairs could be used to evaluate the dry galaxy merger contribution to the assembly of massive passive galaxies. Galaxy mergers are, indeed, an essential part of the evolution of galaxies in any hierarchical cosmological model, but current observational constraints on the merger rate cannot distinguish between models due to the small existing samples \citep{Duncan2019,Conselice2022}. 

\Euclid's \gls{BG} plus \gls{RG} spectra in the \gls{EDS} will allow us to study, using spectral fitting techniques, the evolution of massive and passive galaxies in terms of the physical properties of their stellar populations (such as ages, metallicities, dust content, and velocity dispersions), with extraordinary statistics compared with present and future ground-based data sets. High-\gls{S/N} ($\mathrm{S/N}>10$) spectra are needed to reconstruct physical properties \citep[see, e.g.][]{Citro2016} with high accuracy; we can perform such a study on the brightest subsample or on stacked spectra.

\subsection{The galaxy-halo connection from gravitational lensing}

\begin{figure}[hbtp!]
    \centering
    \includegraphics[width=0.495\textwidth]{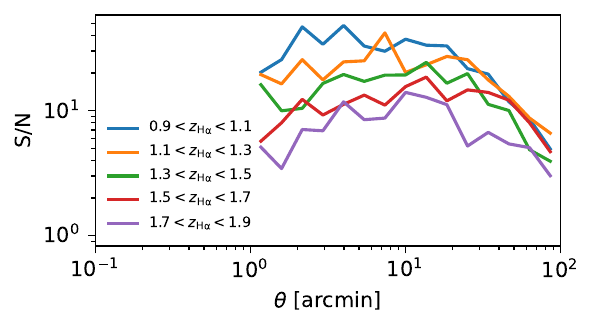}
    \caption{Predicted signal-to-noise ratio of the weak gravitational lensing signal (the tangential shear) per angular bin produced by NISP-detected \ha\ emitters selected in five redshifts bins. Even at such high redshifts, the combination of \Euclid image quality, depth, and area results in a strong detection.
    }
    \label{fig:ggl_halpha_z_snr}
\end{figure}

Science with \gls{GGL} focuses on the relation between galaxies, their baryonic properties (luminosity, stellar mass, etc.), and their dark matter halo properties (mass, density profile, shape, and environment), that is, the galaxy-halo connection. 
We will exploit the precise lensing signal around foreground galaxies with the aim to constrain the halo mass as a function of luminosity and/or stellar mass and to distinguish between central and satellite galaxies. The lens samples will be split by various properties, such as rest-frame colour, redshift, and size to explore the variation in the \gls{SHMR} as a function of galaxy type and to quantify the evolution of this relation up to $z < 1.9$. The \Euclid data should be able to constrain this from the \gls{GGL} signal alone, but the analysis may need to include additional priors, for instance from the luminosity or stellar mass function.
For this, we will use lens samples obtained from \Euclid photometric redshifts, from large ground-based spectroscopic surveys such as \gls{DESI} \citep{DESI2023}, as well as from \Euclid spectroscopy. 
In the particular case of NISP emission-line galaxies detected in spectroscopy with \Euclid, we will model their \glspl{HOD}, and use this information to build mock galaxy catalogues for the cosmological analysis. As shown in Fig~\ref{fig:ggl_halpha_z_snr}, although at quite high redshift, we expect a \gls{GGL} signal of $\mathrm{S/N} > 10$ in five redshift bins in the range $0.9 < z < 1.9$ and down to 1\,arcmin scale.

We also anticipate \gls{HOD} constraints with \acrlong{G3L}\glsunset{G3L} \citep[G3L;][]{Schneider2005, Simon2008}, which encompasses the correlation between the positions of pairs of lens galaxies with the shear of a source galaxy (shear-lens-lens G3L), and the correlation between the shears of pairs of sources with the position of individual lenses (shear-shear-lens \gls{G3L}). \gls{G3L} depends on \gls{HOD} parameters in a way that is different than for the second-order statistic \gls{GGL}. Hence, similarly to \glspl{HOD} for cosmological inference (see \cref{sec:additional}), combining \gls{GGL} and \gls{G3L} can reduce parameter degeneracies in the \gls{HOD} that define the occupancy of central and satellite galaxies (such as minimal halo masses for galaxy formation and satellite fractions) and improve constraining power. Moreover, \gls{G3L} can probe the correlations between different lens samples, for example, red-sequence and blue-cloud galaxies \citep{Linke2022}. 

While \gls{G3L} has been detected in smaller and shallower ground-based surveys \citep{Simon2013, Linke2020}, \Euclid will provide an immense improvement in the \gls{S/N} due to its larger area and higher galaxy number density. This is particularly important for the shear-shear-lens \gls{G3L}, whose signal is about two orders of magnitude lower than shear-lens-lens \gls{G3L} \citep{Simon2013}.

Measuring the radial mass density profiles of galaxies is important because baryonic cooling and feedback effects likely alter this from the standard dark matter-only halo density profile \citep{SonnenfeldLeauthaudAuger2018}. Empirically determining the profile is critical for both understanding galaxy formation and for its impact on cosmology.  This is possible through careful measurement and modelling of the excess surface density, derived from the \gls{GGL} tangential shear.  Moreover, by combining weak and strong lensing, we can obtain tight constraints over the whole profile, provided that the selection function for strong lenses can be quantified.

Measuring the flattening of the dark matter halo is also possible with \gls{GGL}. We will use the quadrupole moment of the tangential shear referenced to the major axis of the light, correcting for spurious signal caused by cosmic shear, large-scale alignments between galaxy shapes, and the tidal field of the matter, as well as additive \gls{PSF} systematics. The measurements will be conducted for lens galaxies split, for example, according to colour or luminosity/stellar mass, to facilitate the comparison to predictions from simulations.  
With \gls{GGL} alone, it is not possible to measure the ellipticity of a halo independently from the misalignment between the orientation of the halo and its central galaxy \citep[e.g.][]{Bett2012}. To break this degeneracy and so constrain the halo ellipticities, we will combine \gls{GGL} with shear-shear-lens \gls{G3L}, which is sensitive to the overall halo ellipticity \citep{Simon2012, Adhikari2015, Shirasaki2018}.

Finally, we will use \gls{GGL} to assess the link between galaxy content, galaxy halos, and their large-scale environments, extending the work described in \cref{sec:env}. For example, we expect that satellite galaxies will have tidally stripped halos \citep[e.g.][]{Sifon18}, which will become increasingly truncated as they spiral into their host group or cluster. Moreover, we will explore larger-scale structures such as filaments between galaxies \citep{EppsHudson2017}, a form of \gls{G3L} (lens-lens-shear), and weak lensing by cosmic voids and the galaxies within them.

\subsection{Clusters as testbeds for astrophysical processes}

Clusters of galaxies are powerful testbeds for various astrophysical phenomena, from \gls{AGN} heating to tidal stripping; this is because almost all of the matter in them can be probed. The dark matter is so densely concentrated that it can be traced by strong and weak gravitational lensing \citep[e.g.][]{Bradac2005, 2016ApJ...821..116U}. The intracluster medium is visible through X-ray and sub-millimetre observations and carries information about the thermal and chemical enrichment history \citep[e.g.][]{Voit2005RvMP}. Member galaxies, although representing a minor contribution to the total and baryonic mass budget, trace the dynamics of the system and retain memory of the formation and evolutionary path in their morphologies and colours.
Additionally, the diffuse glow of intracluster light, situated between the cluster galaxies, enables the tracking of stars that have been stripped from previous galaxy mergers and interactions \citep{Montes2022NatAs}. However, due to the rarity and size of clusters, our knowledge of their physical properties is limited to relatively small samples, primarily focused on core regions and predominantly at $z\la1$.

\Euclid will be transformational, since it only needs to be combined with a probe of the gas (e.g. from {eROSITA}, XMM-{\it Newton}, {\it Chandra}, {\it Planck}, \gls{ACT}, or \gls{SPT}) to provide a comprehensive view of the state of the clusters.  The high-resolution VIS images enable the mass distribution to be mapped using weak lensing, the combination of the spectra and galaxy positions will reveal the dynamical state of the cluster, and the low sky-background in the \gls{NIR} means intracluster light can be observed out to $z\sim2$.

\Euclid will be able to measure the weak lensing masses of nearly  \num{3000} massive clusters with a
relative uncertainty of less than 30\%
\citep[][also see \citealt{Koehlinger15}]{EP-Sereno}. Weak lensing analyses of individual clusters or stacked samples will accurately probe the mass distribution from the inner and virial region, with accurate measurements of the inner slope and concentration \citep{2016JCAP...01..042S,ECGIOCOLI24}, to the cluster boundaries and beyond, constraining the splashback radius (the radius at which accreted matter reaches its first orbital apocentre after turnaround) and the infalling region \citep{2017ApJ...836..231U, Contigiani19, 2024arXiv240206717G}, and the correlated matter in the cluster environment \citep{2018NatAs...2..744S,GIOCOLI21, 2022MNRAS.511.1484I}.

With the redshifts, the masses, and the dynamical states of the clusters provided by \Euclid observations, the properties of the gas and the baryon budget will be readily predicted from theories of gravitational collapse \citep{Voit2005RvMP}. Any deviation from these predictions and comparisons with state-of-the-art simulations will be used to quantify the impact of non-gravitational physics such as gas cooling and feedback from supernovae and \gls{AGN}, as well as to probe hydrodynamical phenomena induced by hierarchical structure formation, such as the process of mass and energy accretion and distribution though shocks, turbulence, and bulk motions \citep{Kravtsov2012}, plus the relative importance of cosmic rays and magnetic fields \citep{Brunetti2014}. These astrophysical phenomena are the reason that the galaxy-halo connection is so complex and our poor understanding of these phenomena limit our ability to constrain cosmology with both primary \Euclid probes (see \cref{sec:weaklensing,sec:likelihood}) as well as with cluster counts.

As mentioned in \cref{sec:NearbyGal}, \Euclid will have an unrivalled ability to measure low-surface brightness features in the infrared, with \YE, \JE, and \HE\ depths of $28.4\,\mathrm{mag}\,\mathrm{arcsec}^{-2}$ \citep{Scaramella-EP1}, which are sufficient to observe intracluster light out to 100\,kpc in $z\,{\sim}\,2$ protoclusters \citep{werner2023} and to the splashback radius in clusters to $z\ga0.6$ \citep{Gonzalez2021}. Intracluster stars, freed from their host galaxies, are expected to follow the global distribution of dark matter in clusters and may act as a luminous tracer of the dark matter distribution \citep{Montes2019}, which is especially important for the $z>1$ clusters where mass measurements from weak gravitational lensing cannot be made.

\subsection{Protoclusters}
\label{sec:protocluster}

Galaxy clusters in the local Universe are dominated by elliptical galaxies that were formed in the early Universe at $z>2$, often in short-lived intense starbursts \citep[e.g.][]{bower1992}. Therefore, understanding how clusters assembled their mass in the early Universe is of critical importance, because their precursors are expected to contribute significantly to the star-formation rate density at high redshifts
 \citep{chiang2017}. 
 
Although different definitions are used in the literature, \cite{overzier2016} suggests to define a protocluster as `a non-virialised structure in the distant Universe that will finally collapse into a typical local galaxy cluster, 
 a virialised system of a mass larger than $10^{14}\,\si{\solarmass}$.
Although a common definition is useful in many respects, we acknowledge that the very nature of protoclusters (i.e. being defined based on their future fate) prevents a robust and univocal operational definition.

The study of galaxy protoclusters in the distant Universe has been an emerging research field in the past decade \citep[for a review, see][]{overzier2016} and is now reaching maturity, with an increasing number of systematic searches, complemented by serendipitous discoveries. 
Both the \gls{EWS} and \gls{EDS}, together with the exquisite optical/\gls{NIR} imaging data sets, offer a unique opportunity to search for protoclusters in a systematic way and make a big leap forward in this research field (B{\"o}hringer et al., in prep.). To this end, we have developed several tools tailored for the detection of galaxy protoclusters, including fine-tuned variations of the two official cluster-selection algorithms \texttt{AMICO} \citep{bellagamba2018} and \texttt{PZWav} \citep[][also see \cref{sec:clusterfinding}]{gonzalez2014}. All these codes have been applied to real data and \Euclid-like simulations to study the properties of protoclusters as they will be observed in the \Euclid surveys and to test the synergies with other available data-sets.  With such a study we aim to provide guidelines for the detection of these objects, the interpretation of the results, and pave the way for follow-up observations.  More generally, we will study both the physical properties of the protoclusters as whole objects, and the properties of their members. This will allow us to make a complete census of protoclusters at several stages of evolution (which will require a careful synergy with theoretical simulations) and to understand and to understand the physics behind the correlation of galaxy properties with environment (see \cref{sec:env}). 

For the detection of protoclusters we adopt two strategies: (1) a blind search exploring the entire wealth of data; and (2) a search around possible signposts, such as sub-millimetre galaxies \citep[e.g.][]{planck2015SPIRE,planck2016PHz, calvi2023} and/or high-$z$ radio galaxies \citep[e.g.][]{kurk2000,pentericci2000}. In this way it will be possible to study, with rich statistics, any possible differences and biases between the two methods. 
For the protocluster detection, we will rely mostly on photometric redshifts, based mainly on optical and \gls{NIR} imaging
(see \cref{sec:photoz}). Protoclusters could also be revealed through overdensities in the spectroscopic redshift data, out to $z=1.8$ using \ha\ and perhaps also out to $z=2.7$ through the \oiii{4959,5007} doublet. For specific areas on the sky, far-IR and (sub)mm observations are also available, and we will use these observations to reveal the progenitors of elliptical galaxies dominating local galaxy clusters, the dusty starbursts \citep[for a review, see][]{alberts2022}.

\subsection{Transient objects}

Exploration of the time domain is a rapidly growing area of modern astronomy. In particular, time-domain astronomy in the \gls{NIR} is an unexplored frontier. Although the majority of the \gls{EWS} will be visited only once, some fields such as the \glspl{EDF} and self-calibration field are planned to be observed repeatedly, and hence they can be used to search for transient objects. 

Transient surveys in the \gls{NIR} can identify transients that are obscured by dust absorption. It has been suggested that a significant fraction of transients, such as \gls{SNe} and \glspl{TDE}, have been missed by optical transient surveys because of strong absorption in their host galaxies \citep[e.g.][]{2018MNRAS.473.5641K,2023ApJ...948L...5P}. Measuring the rates of dust-obscured transients will allow us to estimate the true event rates of transients that are essential information to uncover their nature.
In addition, some transients are known to be intrinsically bright in \gls{NIR} and such transients can be explored by \Euclid. For example, the extragalactic infrared transient survey conducted by \textit{Spitzer} revealed that there is a population of unusual infrared transients without optical counterparts \citep{2017ApJ...839...88K}. Thanks to the combination of simultaneous VIS and NISP data, \Euclid will be able to carry out a census of transients that are intrinsically luminous in \gls{NIR} in nearby galaxies to understand their origins.

\gls{NIR} transient surveys also allow us to discover high-redshift transients. For example, \Euclid can discover hundreds of Type~Ia \gls{SNe} at $1\la z\la 1.5$, which could lead to  a significant improvement in cosmological parameter estimation if \Euclid can perform a dedicated transient survey for 6~months \citep{2014A&A...572A..80A}. Although such a survey is not currently planned, such a dedicated transient survey could be conducted with \Euclid if there is time available later in the mission. However, even with the current survey plan, long-lasting luminous transients, such as superluminous \gls{SNe} and pair-instability \gls{SNe}, can be discovered out to $z\sim 3.5$ in the \glspl{EDF} \citep{2018A&A...609A..83I,2022A&A...666A.157M}. High-redshift superluminous \gls{SNe} and pair-instability \gls{SNe} allow us to constrain properties of massive stars (expected to be the progenitors of such SNe) at high redshifts \citep{2022A&A...666A.157M,2023MNRAS.519L..32T}. High-redshift superluminous \gls{SNe} may also provide additional cosmological parameter constraints \citep{2018A&A...609A..83I}.

As an indication of the expected numbers of SNe detectable with \Euclid, the weekly observations of the self-calibration field (with an area of approximately $3\,\mathrm{deg}^2$) obtained during the \gls{PV} phase can yield around 40 Type~Ia SNe discoveries, of which about 10 are expected to be at $z>1$. A similar number of core-collapse SNe is also expected to be discovered. During regular operations, the self-calibration field will be observed monthly. Such long-term monthly observations over the full duration of the survey should enable the  discovery of around 500 Type~Ia SNe at $z>1$ and around 200 core-collapse SNe at $z>1$. In addition, up to dozens of superluminous SNe and pair-instability SNe out to $z\sim 4$ may be discovered during the long-term self-calibration field observations, depending on their unknown event rates.

\begin{figure}[hbtp!]
    \centering
    \includegraphics[width=0.495\textwidth]{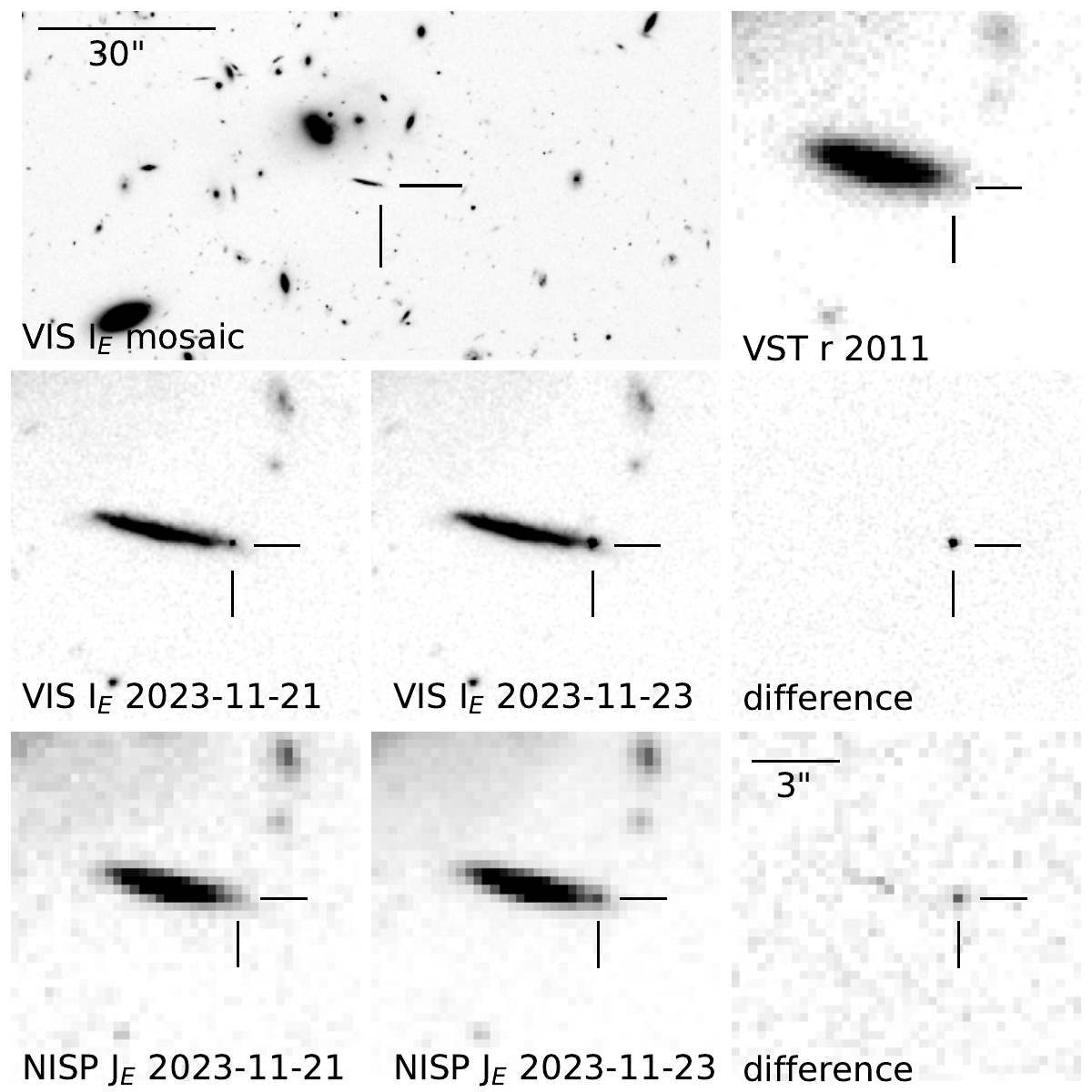}
\caption{
{\it Top row}: The left panel shows a section of a VIS image acquired on 21 November 2023 centred on ${\rm RA} = 09^{\rm h}59^{\rm m}39.872^{\rm s}$, ${\rm Dec} = +02^{\circ}35\arcmin54\farcs129$ (J2000.0), the location of the SN candidate AT~2023adqt (internally called Euclid\_SNT\_2023B). It is close to the galaxy SDSS~J095940.08+023554.6 \citep[$z=0.246$;][]{2012ApJ...753..121K}, which is the likely host.
The right panel shows this galaxy in a deep stacked image in the $r$-band obtained by the \gls{SUDARE} program in 2011 using the VLT Survey Telescope (VST, \citealt{2015A&A...584A..62C}). No source is visible on the SN position.
{\it Middle row}: the SN candidate is clearly visible on two VIS $I_{\scriptscriptstyle\rm E}$ band images  acquired on 21 November 2023 and 23 November 2023 as well as in the corresponding difference image. 
{\it Bottom row}: the SN candidate is not visible in the NISP $J_{\scriptscriptstyle\rm E}$ band image on 21 November 2023, but it appears on 23 November 2023. The difference image clearly shows the SN candidate.
}
\label{fig:stswg_sn_fin}
\end{figure}

An example of an early \Euclid discovery of a transient object is shown in \cref{fig:stswg_sn_fin}. The object with coordinates ${\rm RA} = 09^{\rm h}59^{\rm m}39.872^{\rm s}$, ${\rm Dec} = +02^{\circ}35\arcmin54\farcs129$ (J2000.0) was discovered in \Euclid observations of the COSMOS field taken on 21-23 November 2023. 
The object, a likely SN, brightened significantly during this period, enabling a clear detection in $I_{\scriptscriptstyle\rm E}$ difference images. Subsequent analysis shows that the object is also clearly detected in the three \Euclid NIR bands.
Given that there is no previously known transient reported at this position, this is a new discovery by \Euclid and it is officially named as AT~2023adqt. The host galaxy (SDSS J095940.08+023554.6) has a spectroscopic redshift of $z=0.246$ \citep{2012ApJ...753..121K}. 
Photometry of the transient object yields a brightness of  $I_{\scriptscriptstyle\rm E}=23.64\pm0.06$, $Y_{\scriptscriptstyle\rm E}=25.18\pm0.10$, $J_{\scriptscriptstyle\rm E}=24.60\pm0.09$ and $H_{\scriptscriptstyle\rm E}=24.94\pm0.15$ on 23 November 2023. Its optical to NIR colours are consistent with the rise of a SN at the same redshift, although at this stage it is not possible to determine the SN type. The clear detection and deep NIR photometry, enabling detection even when significantly far from peak brightness, demonstrate \Euclid’s power for the detection and study of transient objects.

\Euclid can also provide important information on transients discovered by the \gls{LSST} \citep{guy2022}. As the \Euclid survey fields overlap with some \gls{LSST} fields, some transients are likely to be observed by both \Euclid and the \gls{LSST}. In such a case, \Euclid can provide complementary \gls{NIR} photometry information to \gls{LSST}. Such \gls{NIR} data can be used to constrain dust production in and around transients, for example. The joint detection of thousands of Type~Ia SNe by \Euclid and LSST would provide a valuable contribution to our understanding of the impact of dust on Type~Ia SN cosmology \citep{2023MNRAS.524.5432B}. \Euclid will also provide some spectroscopic measurements of SNe after scene modelling \citep[e.g.][]{2022A&A...668A..43L}.
For transients discovered by \gls{LSST} and other facilities, the \gls{EWS} and \gls{EDS} can provide essential information such as morphology, infrared photometry, and spectroscopy of their host galaxies. 
Furthermore, red supergiant progenitors of Type~II \gls{SNe} discovered by \gls{LSST} or other transient surveys may be identified in the \Euclid images taken before their explosion.

\subsection{Demographics of cool exoplanets}

From 2027 the NASA \textit{Roman} mission is expected to undertake the Roman Galactic Exoplanet Survey \citep[RGES,][]{2019ApJS..241....3P}, a statistical census of the cool exoplanet regime using the microlensing effect. The cool exoplanet census will be pivotal for testing planet formation models \citep{2021A&A...656A..72B}, and the RGES census will complement that of hot exoplanets obtained by \textit{Kepler} \citep{2010Sci...327..977B}. RGES is the key science driver for the \gls{GBTDS}, one of the three \textit{Roman} core community surveys that will each occupy around $25\%$ of the first 5 years of \textit{Roman} mission. 

The RGES is designed to find around 
1\,400 cool exoplanets down to the mass of Mars \citep{2019ApJS..241....3P}, and to be able to make direct mass measurements for at least half of its sample. Direct planet mass measurements can be achieved through a number of approaches, including: direct measurement of the lens host flux; measurements of finite-source magnification effects in the lightcurve; and measurements of \gls{PSF} distortion arising from relative proper motion between the foreground lensed host and background magnified source star. Together with the event duration, such measurements provide the 
means to break what would otherwise be a three-way degeneracy between the planet mass, distance, and transverse speed.

However, many \textit{Roman} events will have insufficient data to fully break the mass-distance-speed degeneracy, and others may have relatively poor precision due to the challenging measurement of lens-source relative proper motion. Due to the very close alignment of lens and source on the sky, measurements of proper motion rely on the detection of \gls{PSF} skewness, the measurement precision of which improves with the cube of the observation time baseline \citep{2007ApJ...660..781B}.

At the time of writing, \Euclid precursor imaging of the \textit{Roman} fields is under active consideration. The \gls{GBTDS} area can be covered by nine \Euclid  pointings, taking up to 42 hours of time if executed using four standard \gls{ROS} cycles for each pointing (see \cref{sec:wide}).
Due to pointing restrictions imposed by the solar aspect angle, the \gls{GBTDS} can only be accessed by \Euclid near the spring and autumn equinox for periods of up to 23 days. If these observations by \Euclid can be scheduled during the first year of operations, this would extend the proper motion sensitivity baseline for \textit{Roman} from $5\,$ years to $8\,$ years and so improve the planet mass measurement precision by a factor $(8/5)^3 \simeq 4$ for events that rely on proper motion measurements \citep{2022A&A...664A.136B,2023arXiv230610210K}.

\subsection{Solar system objects}

The number of \glspl{SSO} for which the composition can be characterised, has recently increased significantly, thanks to low-resolution reflectance spectroscopy obtained by \Gaia for more than 60\,000 objects
\citep{spoto2018, tanga2023, galluccio2023}. The next \Gaia release will increase this sample further, and the upcoming \gls{LSST} is expected to discover and characterise 5--6\, million \glspl{SSO} with broadband photometry in the optical \citep{lsst2009}. This avalanche of optical data should be contrasted with the shortage of observations in the \gls{NIR}.

To date, the largest corpus of \gls{NIR} broad-band colours for
34\,998 \glspl{SSO} was extracted from the ESO \gls{VISTA} VHS
survey \citep{popescu2016}. Decades of targeted spectral observations in the
\gls{NIR} raised the sample to only about 3000 low-resolution spectra
covering the 0.9--2.4\,\micron~range \citep[see][]{mahlke2022}. 
Low-resolution spectroscopy or broad-band photometry in the 
near-infrared is, however, crucial to disentangle among several compositions
that are otherwise degenerate with only visible data \citep[e.g.][]{demeo2009}.
The \YE, \JE, and \HE~filters of the \Euclid NISP photometer 
(\cref{fig:euclid_passband_comparison}) offer an important complement to
visible colours 
to characterise the surface composition of \glspl{SSO} \citep{carry2018}. However, the geometry of the observations implies high phase angles. Hence, it will be necessary to account for possible phase effects which may
affect colours \citep{mahlke2021, AlvarezCandal2024}

Characterising the distribution of compositions of \glspl{SSO} is key
for understanding the formation of our Solar System \citep{demeocarry2014}.
Among the thousands of exoplanet systems discovered to date,\footnote{\url{https://exoplanetarchive.ipac.caltech.edu}}
the Solar System with its external giant planets is more an
exception than the rule \citep[most systems are composed of multiple super-Earth planets, e.g.][]{Raymond2018}.
Several models have been put forward to explain the peculiar orbital
architecture of the Solar System 
\citep[e.g.][]{walsh, Raymond2017}. These models succeed in reproducing the
dynamical architecture of terrestrial and giant planets, but often diverge
in the dynamical, and more importantly compositional, distribution of \glspl{SSO}.

The current distribution of orbits and compositions is, however,
an evolved version of the primordial distribution, resulting from planetary
formation. Collisions break up bodies, injecting fragments into orbit that
 enhance the fraction of bodies sharing the same composition on similar orbits.
These clumps of fragments are called `dynamical families'
\citep{Hirayama1918}. It is crucial to distinguish between collisional
fragments and planetesimals \citep{delbo2017} to debias the current distribution
of compositions and access the primordial distribution.
An additional complication results from the secular, non-gravitational, 
dynamical evolution of \glspl{SSO} which spread structures through the Yarkovsky effect
\citep[due to the delayed thermal emission of the incoming solar illumination,][]{bottke2001}.
The Yarkovsky effect results from a complex interplay of the physical properties of 
\glspl{SSO} (diameter, spin orientation, surface reflectivity, etc). As such, deciphering
the early history of the Solar System requires the characterisation of the
composition of numerous \glspl{SSO} and of their physical properties.

There are millions of \glspl{SSO}
with apparent magnitudes within the depth of the \gls{EWS} and observable almost at any time on the celestial sphere \citep{grav2011}.
Because the main \Euclid cosmological survey avoids low ecliptic latitudes (\cref{fig:skysurvey}), the fraction of \glspl{SSO} in the survey is, however, small. Nevertheless, considering an early design of the \Euclid survey, it is predicted that about 150\,000 objects will be serendipitously observed over the course of the mission \citep{carry2018}.
Conversely, the \glspl{SSO} detected by \Euclid will belong to populations with large inclination to the ecliptic plane \citep{carruba, novakovic, terai, chen, 2023A&A...673A..93S}, often missed by discovery surveys more typically focusing on the ecliptic \citep[see reports of biases against high inclination by][]{mahlke2018, carry2021}.

The sequence of observations, with four repetitions of 
VIS-\YE, \JE, and \HE\ exposures (\cref{sec:wide}),
is fortunately well matched to the detection of \glspl{SSO} from their motion between frames. While objects from the outer Solar System (Kuiper-belt objects or Centaurs) will appear as point sources, objects closer to the Earth (Jupiter Trojans, asteroids, or near-Earth objects) will produce long trails on \Euclid frames \citep[up to tens of pixels,][]{carry2018}. Therefore, two different methods are used to detect \glspl{SSO}: those
optimised for slow-moving point sources (Nucita et al., in prep.) and those for fast-moving trailed sources \citep{Pontinen20,Pontinen23}.

Going beyond the basic detections, \Euclid's observing sequence allows us to determine the intrinsic colours of \glspl{SSO}, without any bias introduced by the light curve due to their irregular spinning shapes \citep{popescu2016, carry2018}. 
The \glspl{SSO} detected by \Euclid, once combined with colours in the visible (from, e.g. \gls{LSST}), will
provide a large reference sample to study the distribution of compositions in the Solar System.
In addition to merging colours at the catalogue level, the quasi-simultaneous observations of \glspl{SSO} by
\Euclid and \gls{LSST} would provide a direct and accurate measurement of their distances, and hence
dramatically improve the quality of their orbits \citep{Granvik2007,rhodes2017}.
While such observations may happen by pure coincidence, scheduling them would rely on 
coordination with \gls{LSST} \citep{snodgrass2018, guy2022}.

\Euclid's sequence of observations will also provide hour-long light curves, critical for the 
determination of physical properties (rotation period and obliquity mainly). While photometry that is
sparse in time (in which the time interval between measure is larger than the period of the signal) can be successfully used to determine the physical properties of \glspl{SSO} \citep{kaasalainen2004},
the rate of objects for which solutions are found remains limited \citep[under 50\%, see][]{durech2023}.  
The combination of a partial light curve to the sparse photometry allows us to reject degenerate solutions, efficiently improving the solutions \citep{durech2015}.
\Euclid light curves, along with the NISP spectroscopic measurements, will thus be particularly powerful when used together with \gls{LSST} photometry. 

\section{Conclusions and outlook}
\label{sec:conclusions}

\Euclid was successfully launched  on 1 July 2023 into an orbit around L2, which provides the thermally stable environment that is needed to achieve its main objective, namely to measure the growth of structure over a significant fraction of the age of the Universe with unprecedented precision. 
Comparison of the measurements with models of structure formation can shed light on the origin of the accelerated expansion of the Universe, test for deviations from \gls{GR}, examine scenarios for inflation, and robustly explore many other aspects of the $\Lambda$CDM model.

\Euclid can achieve its challenging objectives because it is optimised to measure galaxy clustering and weak gravitational lensing, while ensuring that observational sources of bias remain subdominant. 
As discussed in \cref{sec:survey}, the \gls{EWS} will cover 14\,000\,deg$^2$ of extragalactic sky, combining near-infrared spectroscopy and photometry with diffraction-limited visual imaging, while the \gls{EDS} will
yield deep observations that cover 53\,deg$^2$. Additional deep observations, primarily designed for calibration purposes, provide further opportunities to study the distant Universe. To obtain all these data within its nominal mission span of six years, the spacecraft contains two instruments with a common \gls{FOV} of about 0.54\,deg$^2$.
The VIS instrument \citep[\cref{sec:vis};][]{EuclidSkyVIS} provides the high-resolution optical imaging needed for accurate shape measurements of about 1.5 billion galaxies. It is complemented by \gls{NISP} \citep[\cref{sec:nisp};][]{EuclidSkyNISP}, which provides near-IR imaging and spectroscopy over the same area.

The performance of the spacecraft and the instruments is excellent, but as discussed in \cref{sec:preliminaryCommissioning}, several anomalies were discovered during commissioning.
Fortunately, all of these can be mitigated.
The discovery of high levels of stray light for a wide range of \gls{AA} values led to a complete overhaul of the observing plan. The main consequence of a more restricted range in spacecraft orientation is a modest reduction in survey speed. 

After an initial performance and verification phase, a period of 24 hours was devoted to carrying out observations of targets that could highlight the broad astrophysical potential of \Euclid. The first results from these \gls{ERO} data are presented in an accompanying series of papers \citep{EROData, EROOrion, EROGalGCs, ERONearbyGals, EROFornaxGCs, EROPerseusICL, EROPerseusOverview, EROLensData, EROLensVISDropouts}. These provide concrete validations of the wide range of science objectives listed in this paper, for example highlighting the potential for the study of \gls{LSB} features, or the study of UCD and GC populations.

On  14 February 2024, the scientific survey started and \Euclid commenced its journey to explore the dark Universe, at a rate of about 10\,deg$^2$ per day. First public releases are planned in early 2025 (a single visit over the \glspl{EDF}) and  mid-2026 (about 2500\,deg$^2$ of the \gls{EWS}). The resulting 
high-quality data products that will be released, described in \cref{sec:data}, will allow us to determine cosmological parameters with unprecedented precision using the primary probes
(\cref{sec:cosmology}), and this can be improved further with additional probes (\cref{sec:additional}). Importantly, the impact of \Euclid is not limited to cosmology, and some other applications of the data were highlighted in \cref{sec:legacy}.

The data obtained to date show that 
\Euclid is on track to fulfil the many science goals described in this paper. What we have outlined here, has largely focused on studies where the \Euclid data play a dominant role; however, we expect a far greater impact when combining \Euclid with other data. 
For instance, cross-correlations with data sets at completely different wavelengths, as well as joint analyses with other spectroscopic and imaging surveys, will continue to add value. The deep, high-resolution space-based data covering a large fraction of the extragalactic sky will also enable studies in a huge array of astrophysics topics. While many of these have been described in this paper, there is also the exciting possibility for \Euclid enabling discoveries about our Universe that were completely unanticipated. The \gls{EWS} and \gls{EDS} images and catalogues will be an exceptional database of astronomical sources for decades to come and will be a gold mine for detecting new rare or unknown astronomical phenomena.

\begin{acknowledgements} 
\AckEC
\end{acknowledgements}

\bibliography{Euclid,overview,EROplus}

\begin{appendix}
\section{List of acronyms\label{app:glossary}}
\setglossarystyle{list}
\printglossaries

\end{appendix}

\label{LastPage}
\end{document}